\documentclass{pasj00}
\usepackage{natbib,aas_macros}
\usepackage{eclclass}
\begin{document} 

\SetRunningHead{Hanami et al.}{Dusty IRBGs/LIRGs at $z=0.4-2$}
\Received{2010/09/01}
\Accepted{2011/12/28}

\title{Star Formation and AGN activity in Galaxies classified 
using 
the 1.6 $\mu$m Bump and PAH features at $z = 0.4-2$.}

%
\author{Hitoshi \textsc{Hanami},\altaffilmark{1} 
Tsuyoshi \textsc{Ishigaki},\altaffilmark{1,2}
Naofumi \textsc{Fujishiro},\altaffilmark{3}
Kouichiro \textsc{Nakanishi},\altaffilmark{4}\\ 
Takamitsu \textsc{Miyaji},\altaffilmark{5,6} 
Mirko \textsc{Krumpe},\altaffilmark{6,7}
Keiichi \textsc{Umetsu},\altaffilmark{8} 
Youichi \textsc{Ohyama},\altaffilmark{8} \\ 
Hyun Jin \textsc{Shim},\altaffilmark{9} 
Myungshin \textsc{Im},\altaffilmark{9} 
Hyoung Mok \textsc{Lee},\altaffilmark{9} 
Myung Gyoon \textsc{Lee},\altaffilmark{9} \\
Stephen \textsc{Serjeant},\altaffilmark{10}   
Glenn J. \textsc{White},\altaffilmark{10,11}   
Christopher N. \textsc{Willmer},\altaffilmark{12} 
Tomotsugu \textsc{Goto},\altaffilmark{13} \\  
Shinki \textsc{Oyabu},\altaffilmark{14} 
Toshinobu \textsc{Takagi},\altaffilmark{15} 
Takehiko \textsc{Wada},\altaffilmark{15} \\
and \\ 
Hideo \textsc{Matsuhara}\altaffilmark{15}
} 
\altaffiltext{1}{Physics Section, Iwate University, Morioka, 020-8550, Japan}
\email{hanami@iwate-u.ac.jp}
\altaffiltext{2}{Asahikawa National College of Technology, 2-2
  Syunkodai, Asahikawa, Hokkaido 071-8142}
\altaffiltext{3}{Kyoto Sangyo University, Kyoto, Japan}
\altaffiltext{4}{ALMA/NAOJ office, Santiago, Chili}
\altaffiltext{5}{Instituto de Astronom\'ia, 
Universidad Nacional Aut\'onoma de M\'exico, 
Ensenada, Baja Californioa, M\'exico \\ 
(mailing address: PO Box 439027, San Ysidro, CA92143-9027, USA)}
\altaffiltext{6}{University of California, San Diego, Center for Astrophysics 
\& Space Sciences,\\ 
9500 Gilman Drive, La Jolla, CA 92093-0424, USA}
\altaffiltext{7}{ESO Headquarters, Karl-Schwarzschild-Stra\ss e 2, 
Germany}
\altaffiltext{8}{ASIAA, Taipei, 10617, Taiwan,R.O.C.}
\altaffiltext{9}{Astronomy Program, Department of Physics and
  Astronomy, \\
Seoul National University, Seoul, 151-747, Korea}
\altaffiltext{10}{Department of Physics and Astronomy, The Open University, UK}
\altaffiltext{11}{Space Science $\&$ Technology Division, 
STFC Rutherford Appleton Laboratory, Chilton, Didcot, UK}
\altaffiltext{12}{Steward Observatory, University of Arizona, 
933 North Cherry Avenue, Tucson, AZ 85721, USA}
\altaffiltext{13}{Institute for Astronomy, University of Hawaii, USA}
\altaffiltext{14}{Graduate School of Science, Nagoya University, Furo-cho, 
Chikusa-ku, Nagoya 464-8602, Japan}
\altaffiltext{15}{Institute of Space and Astronautical Science, Japan
  Aerospace Exploration Agency, \\
  3-1-1 Yoshinodai, Chuo-ku, Sagamihara, Kanagawa 252-5210, Japan} 


\maketitle

\begin{abstract}
We have studied the star-formation and AGN activity of massive galaxies in 
the redshift range $z=0.4-2$, which are detected in a deep survey field 
using the AKARI InfraRed (IR) astronomical satellite and {\em Subaru} 
telescope toward the North Ecliptic Pole (NEP). 
The AKARI/IRC Mid-InfraRed (MIR) multiband photometry is used to trace their 
star-forming activities with the Polycyclic-Aromatic Hydrocarbon 
(PAH) emissions, which is also used to 
distinguish star-forming populations from AGN dominated ones and 
to estimate the Star Formation Rate (SFR) derived from 
their total emitting IR (TIR) luminosities. 
In combination with analyses of their stellar components, 
we have studied the MIR SED features of star-forming and AGN-harboring 
galaxies, which we summarize below:   
1) The rest-frame 7.7-$\mu$m and 5-$\mu$m luminosities are good tracers  
of star-forming and AGN activities from their PAH and dusty tori emissions, 
respectively.  
2) For dusty star-forming galaxies without AGN, 
their SFR shows a correlation that is nearly proportional to their stellar 
mass and their specific SFR (sSFR) per unit stellar 
mass increases with redshift.  Extinctions estimated 
from their TIR luminosities are larger than those from their optical SED 
fittings, which may be caused by geometric variations of dust in them.  
3) Even for dusty star-forming galaxies with AGN, SFRs can be derived  
from their TIR luminosities with subtraction of the obscured AGN contribution, 
which indicates that their SFRs were possibly quenched 
around $z \simeq 0.8$ compared with those without AGN.  
4) The AGN activity from their rest-frame 5-$\mu$m luminosity 
suggests that their Super Massive Black Holes (SMBH) 
could already have grown to $\simeq 3 \times 10^8 M_{\odot}$ in most massive 
galaxies with $10^{11} M_{\odot}$ at $z>1.2$, and 
the mass relation between SMBHs and their host galaxies 
has already become established by $z\simeq 1-2$. 
\end{abstract}

\section{Introduction}
\label{sec:intro}
A major issue in observational cosmology is to reconstruct the cosmic 
evolution of star formation and the detailed growth history of the 
present-day galaxies.  
Even though it is not easy to reconstruct the whole path of the 
stellar mass for 
galaxies as a function of the present-day galaxy stellar mass 
$M_{\ast;0}$ and the cosmic time $t$ as $M_{\ast}(M_{\ast;0}, t)$, 
it is essential to obtain knowledge of their Star Formation Rate (SFR) 
as a function $SFR(M_{\ast},t)$.  

Recent near-infrared (NIR) observations suggest that most of the stellar mass 
is contained in massive galaxies, and that this situation had already become 
established at $z \simeq 1$. This appears to have been followed by a rapid 
evolution of the stellar mass density between $z = 1-3$ 
\citep{papovich_stellar_2001, dickinson_evolution_2003, fontana_assembly_2003, 
franx_significant_2003, drory_comparing_2004, fontana_k20_2004, 
glazebrook_high_2004, van_dokkum_stellar_2004, labb_irac_2005, 
fontana_galaxy_2006, papovich_spitzer_2006, yan_stellar_2006, webb_star_2006, 
kriek_direct_2006, pozzetti_vimos_2007, arnouts_swire-vvds-cfhtls_2007, 
vergani_vimos_2008}. These recent studies imply that most of the 
massive galaxies 
must already have formed at $z\simeq 1$, supporting the so-called 
{\it downsizing} scenario \citep{cowie_evolution_1997}. In this scenario, star 
formation shifts from higher to lower mass galaxies as a function of 
cosmic time $t$, although the stellar mass in massive galaxies may not 
directly follow the growth history of their Dark Matter Halos (DMHs) which 
build up hierarchically from less massive systems 
to more massive ones at a later time.  However, several studies of massive 
galaxies at $z < 1$ \citep{brown_evolving_2007, faber_galaxy_2007, 
brown_red_2008} report the evolution from predominantly blue to red 
galaxies, suggesting that the number and stellar mass of galaxies in the 
red-sequence may continue to grow, even after $z = 1$ 
\citep{bell_nearly_2004, zucca_vimos_2006}, 
in agreement with the hierarchical merging scenario.  
Recently, a mass-dependent quenching of star formation 
has been proposed to explain these rapid increase of the number density 
of the massive galaxies relative to lower-mass galaxies at $z=1-2$ 
(Peng et al. 2010; Kajisawa 2011; Brammer et al. 2011).  

The evolutionary path of the $M_{\ast}(M_{\ast;0}, t)$ up to the
present mass $M_{\ast;0}$ can be also reconstructed from knowledge of
the $SFR(M_{\ast},t)$.  Recent studies have confirmed the features for
rapid SFR evolution up to $z \simeq 2$ \citep{brinchmann_mass_2000,
fontana_assembly_2003, prez-gonzlez_spitzer_2005,
feulner_specific_2005, feulner_connection_2005, papovich_spitzer_2006,
iglesias-pramo_uv_2007, zheng_infrared_2007, cowie_integrated_2008, 
damen_star_2009, dunne_star_2009, santini_star_2009,
pannella_star_2009}. However, we have not found a general consensus
concerning the stellar mass dependence for specific SFRs: $sSFR
=SFR/M_{\ast}$ (the SFR per unit stellar mass).  Most studies based on
UV derived SFR's have found evidence to support an anti-correlation
with stellar mass, not only local galaxies, but also the distant
Universe, supporting the {\it downsizing} scenario. It is not easy,
however, to accept that UV estimates are the only way to estimate the SFR 
that will capture all the emission  
originally from new born stars in a galaxy, which 
can be independently estimated from rest-frame luminosities in the UV, IR,
Submillimeter (Submm), and radio wavelength regions by introducing
appropriate calibration factors.  Although the UV luminosity is used
as a proxy of the SFR, its reliability often depends on the use of
large, and rather uncertain corrections for dust extinction.  Most of
the short wavelength photons from star-forming galaxies are
reprocessed through absorption and scattering with dust  
in their star-forming regions, and then re-emitted in the IR-Submm 
wavelength regions. Thus,
the mid/far-IR and their submillimeter emissions become an important tracer
of the total star forming activity in high-$z$ galaxies.

The total IR (TIR) luminosity $L_{IR}$ integrated across the long
wavelength spectral region is a good proxy for the SFR, which is
relatively independent of the dust obscuration
\citep{kennicutt_star_1998}.  Luminous InfraRed Galaxies (LIRGs) and 
Ultra Luminous InfraRed Galaxies (ULIRGs) up to
$z=1-2$, detected at the Mid-InfraRed (MIR) wavelength with the IR
observing satellites as the ISO \citep{mann_observations_2002, 
elbaz_bulk_2002, pozzi_mid-infrared_2004, dennefeld_firback._2005, 
taylor_properties_2005, marcillac_mid_2006, marcillac_star_2006}, 
the AKARI \citep{takagi_multi-wavelength_2007, takagi_polycyclic_2010}, 
and the {\it Spitzer} \citep{egami_spitzer_2004, shapley_ultraviolet_2005, 
caputi_linking_2006, choi_star_2006, marcillac_star_2006, 
marcillac_mid_2006, papovich_spitzer_2006, papovich_spitzer_2006-1, 
reddy_star_2006, elbaz_reversal_2007, papovich_spitzer_2007, 
rocca-volmerange_12_2007, shim_massive_2007, teplitz_measuring_2007, 
brand_spitzer_2008, magnelli_irac_2008, dasyra_0.9_2009, 
hernn-caballero_mid-infrared_2009, murphy_balancingenergy_2009, 
santini_star_2009, strazzullo_deep_2010}, have shown that the SFR
rapidly increases with the redshift, as they
contribute $\simeq 70 \%$ of the cosmic infrared luminosity density
\citep{le_floch_infrared_2005}. The MIR emission from LIRGs/ULIRGs 
is evidence of dusty starburst or AGN surrounded by a dusty
torus. Recently, the LIRGs/ULIRGs are expected to suppress star 
formation at $z=1-2$ with feedback activity from starbursts or AGNs.  
Understanding of the LIRGs/ULIRGs 
at $z=1-2$ can be a key to solve the cosmic 
history of star formation and AGN activities.  
The SFR for galaxies detected with the {\it Spitzer} is 
frequently estimated with the TIR luminosity estimated from the 24
$\mu$m flux \citep{daddi_new_2004, santini_star_2009,
  pannella_star_2009}, which is nearly proportional to their stellar
mass up to $z \simeq 2$. These recent results with the {\it Spitzer}
are also basically consistent with recent results from stacking
analysis for their radio emissions \citep{dunne_star_2009,
  pannella_star_2009} as another SFR estimation immune to dust 
extinctions. These two radio stacking analyses, however, show
a discrepancy in their details as \cite{dunne_star_2009}
report sSFR with a negative correlation to stellar mass while
\cite{pannella_star_2009} report sSFR with stellar mass independence.
Thus, there may be various unknown factors in the SFR
estimations. The TIR luminosity $L_{IR}$ of LIRGs/ULIRGs 
is generally derived 
from the fitting of the observed Spectral Energy Distribution (SED) with
synthetic templates. In the SED fitting, we could sometimes use only a
few bands in the IR wavelength without longer $100-1000~\mu$m
wavelengths, which may cause large uncertainties in the $L_{IR}$
estimation. Obscured AGNs are another notable problem in the $L_{IR}$
estimation. It means that a sample selection without significant AGN
contamination is also important to obtain an accurate $SFR_{IR}$. Thus, 
systematically designed surveys are required to construct samples for
distant star-forming galaxies.

Optical surveys have been successful in selecting numerous Lyman Break
Galaxies (LBGs) as star forming galaxies beyond $z \sim 2$
\citep{steidel_spectroscopy_1996, steidel_spectroscopic_1996, 
steidel_lyman-break_1999, lehnert_luminous_2003, iwata_lyman_2003, 
ouchi_subaru_2003, ouchi_subaru_2004, dickinson_color-selected_2004, 
 sawicki_keck_2006-1, sawicki_keck_2006} and even up to z $\simeq 6$
\citep{yan_high-redshift_2004, stanway_lyman_2003, bunker_star_2004, 
bouwens_galaxy_2004, shimasaku_number_2005}. They may, however, 
fail to detect 
substantial numbers of massive galaxies at $z < 3$, in which the Lyman
break photometric technique does not work well with ground-based
telescopes because of telluric absorption.  Furthermore, even if
galaxies are undergoing intensive star formation, the internal dust
extinction of their UV radiation may reduce the prominence of their
Lyman break features, to appear as LIRGs/ULIRGs in the local
Universe. However, alternative photometric techniques can be used to
select massive high-$z$ galaxies through their 
Balmer/4000~\AA\ break or the 1.6 $\mu$m bump features in their
Spectral Energy Distributions (SEDs), as evidence of the presence of
an evolved stellar population.  We will designate the former and the latter
as 'Balmer' Break Galaxies (BBGs) and as IR 'Bump' Galaxies (IRBGs), 
respectively.  It is known that both star-forming and
passive BBGs in the redshift desert around $z = 2$ can be selected
with a simple and sophisticated two-color selection based on $B,z$, 
and $K$ band
photometry \citep{daddi_new_2004}. This three filter optical-NIR 
technique was designed to detect the Balmer/4000~\AA\ break, nearly
free from the dust extinction which caused some difficulties
for previous `red color' techniques relying on a color diagnostic of
$(R-K)_{Vega} >5-6$, corresponding to $(R-K)_{AB}> 3.35-4.35$
\citep{elston_deep_1988, thompson_surface_1999, daddi_eros_2000, 
firth_las_2002, roche_clustering_2002, roche_nature_2003} or for
$(J-K)_{Vega}>2.3$ corresponding to $(J-K)_{AB}>1.3$
\footnote{$J_{AB}=J_{Vega}+0.89, Ks_{AB} =
Ks_{Vega}+1.84$} \citep{pozzetti_k20_2003,
franx_significant_2003}. With the near-IR (NIR) observations from the 
{\it Spitzer} and from the AKARI, it has become possible to select IRBGs at
$z>1$ with NIR colors at $> 3$ $\mu$m of the 1.6 $\mu$m bump
\citep{simpson_detection_1999, sawicki_1.6_2002,
berta_contribution_2007}. These optical-NIR observations, detecting 
the Balmer/4000~\AA\ break and the 1.6 $\mu$m bump, have been also useful to
trace how stellar mass is correlated with the absolute NIR luminosity.

Although optical-NIR multi-band photometry has the potential to select
a sample for star-forming massive galaxies, it may not be completely
free from the AGN contamination. MIR multi-band photometry tracing PAH
emissions at 6.2, 7.7 and 11.3 $\mu$m and Silicate (Si) absorption at 10~$\mu$m
(rest-frame) from dusty star-forming regions, can be used to
discriminate between the star-forming activity in LIRGs/ULIRGs and that of
AGN dominated galaxies. Thus, systematic multi-wavelength observations 
that include observations at $> 2~\mu$m are important for studies the
star-forming activity of massive high-$z$ galaxies, and in looking
back at the evolution of their SFRs. In order to connect the history
of stellar mass growth in massive galaxies traced in optical-NIR
wavelength with ground based telescopes and that of star formation
traced in MIR wavelength observed outside of the atmosphere, we
undertook multi-wavelength surveys of deep and wide areas with 
$\simeq 0.38$ and $\simeq 5.8$ deg$^2$, respectively,  
in a field close to the North
Ecliptic Pole (NEP), which were extensively observed at the IR wavelength
with the first Japanese infrared astronomical satellite AKARI.  

In this paper, we attempt to reconstruct the cosmic history of dusty massive
galaxies up to $z \simeq 2$, selected estimated from optical-IR images in the
NEP survey field. We report on: 1) photometric redshift $z_{phot}$, 
stellar mass $M_{\ast}$, and $SFR_{UV;cor}$, which are estimated with 
their optical-NIR SED fittings, 
2) the distinction between dusty starburst and AGN 
dominated galaxies using the MIR SED feature, 3) 
TIR luminosities of starburst dominated galaxies 
with the subtraction of the obscured AGN contribution, 
from which we can derive $SFR_{IR+UV}$ complementary to the $SFR_{UV:cor}$, 
and 4) IR luminosities of the dusty tori of AGN dominated galaxies, which 
show evolutionary trends in the connection between AGN and star forming 
activities. 

In section~\ref{sec:data}, we summarize the data we used; 
in section~\ref{sec:photo}, we present
the photometry for $z'$-band detected galaxies 
in our multi-wavelength dataset; 
in section~\ref{sec:irbg}, we also report AKARI/IRC detected IRBGs 
classified using the redshifted 1.6-$\mu$m IR bump in the NIR bands, 
which are divided into three redshifted subgroups;  
in section~\ref{sec:pah}, we present AKARI/IRC MIR detected 
galaxies which correspond to LIRGs/ULIRGs and 
classify them into dusty starbursts and AGN from    
their MIR SEDs with/without PAH features in the MIR bands; 
in section~\ref{sec:ir_lum}, 
we derive the TIR luminosities 
for the MIR detected galaxies; 
in section~\ref{sec:evol_sfr}, we estimate their 
star formation rates and extinctions;   
in section~\ref{sec:evol_pop}, 
we reconstruct their intrinsic stellar populations 
with extinction correction;  
in section~\ref{sec:evol_agn}
we also study the AGN activities in the MIR detected galaxies; 
in section~\ref{sec:discuss}, 
we discuss some expectations for on-going and future  
observations; 
finally in section~\ref{sec:concl}, 
we summarize the conclusions of our study.  

In this paper we adopt $H_0=70$ km s$^{-1}$ Mpc$^{-1}$,
$\Omega_m=0.3$, $\Omega_{\Lambda}=0.7$ throughout.  All magnitudes are
represented in the AB system (e.g. Oke 1974).

\section{The data set}
\label{sec:data}  
We have used mainly the data sets obtained in 
the AKARI Deep imaging survey  
with all of the available bands of the Infra Red Camera (IRC) on the AKARI
\citep{onaka_infrared_2007} near the North Ecliptic Pole (NEP). We
define the $N2,~N3,~N4,~S7,~S9W,~S11,~L15,~L18W$, and~$L24$ bands as
having central wavelengths at 2, 3, 4, 7, 9, 11, 15, 18, and 24 $\mu$m
respectively. 
These are reported in this paper as the IR photometric
AB magnitude of an object observed in the various IRC bands. Details of
the data reduction of the NEP Deep and Wide surveys have been
presented by \cite{wada_akari/irc_2008} and \cite{lee_nature_2007}
(see also \cite{hwang_optical_2007} and Ko et al. 2012), respectively.  
We have used the same IRC scientific data 
for the AKARI NEP Deep field as described in detail by 
\cite{wada_akari/irc_2008}, 
which include mosaicing them together 
and making pointing and distortion corrections,
as well as correction to mitigate detector artifacts.  

The AKARI NEP Deep survey area has been covered by observations in the
near-UV $u^{\ast}$ band with the CFHT/MegaCam(M-Cam)
\citep{boulade_megacam:new_2003}; the optical $B, V, R, i'$, and $z'$
bands with the Subaru/Suprime-Cam (S-Cam)
\citep{miyazaki_subaru_2002}; and in the NIR $J$ and $K_s$ bands with
the KPNO(2.1m)/Flamingos(FLMG) \citep{elston_flamingos_2006}. 
Details of the Subaru/S-Cam, CFHT/M-Cam, 
and KPNO/FLMG data reduction can be found in
\citet{imai_j-_2007}, Takagi et al. (2012), and Ishigaki et al. (2012). 
In this study, we have selected $\simeq 56000$ galaxies 
with $z' < 26.4$ and S/N$ > 3$.  
For these sources, we estimate their multi-wavelength
photometric characteristics, adding ground-based near-UV, optical, and
NIR imaging data to the AKARI/IRC data as described in the following
Section. For simplification, we use $u, i, z$, and $K$ as $u^{\ast},
i', z'$ and $K_s$, respectively for representing colors derived from the
respective filters.  
 
\section{Photometry}
\label{sec:photo}  

\begin{table}
\caption{Summary of photometric parameters.}
\begin{tabular}{lrrr}
\hline 
\hline 
Band & FWHM    & FWHM     &  Ap.D. \\ 
     & (pixel) & (arcsec) &  (arcsec) \\ 
\hline
\hline 
$u^{\ast}$     & 4.9 & 0.90 & 3.0 \\
$B,V,R,i',z'$ & 5.2 & 1.00 & 2.0 \\
$J$(SE)       & 2.7\footnotemark[$a$] & 1.60 & 3.0 \\
$J$(NE)       & 3.1\footnotemark[$a$] & 1.90 & 3.0 \\
$J$(SW)       & 3.0\footnotemark[$a$] & 1.80 & 3.0 \\
$J$(NW)       & 3.5\footnotemark[$a$] & 2.10 & 4.0 \\
$K_s$(SE)     & 2.7\footnotemark[$a$] & 1.60 & 3.0 \\
$K_s$(NE)     & 3.0\footnotemark[$a$] & 1.80 & 3.0 \\
$K_s$(SW)     & 3.7\footnotemark[$a$] & 2.20 & 4.0 \\
$K_s$(NW)     & 3.2\footnotemark[$a$] & 1.90 & 4.0 \\
N2,N3,N4      & 3.1 & 4.50 & 8.0 \\
S7,S9W,S11    & 2.4 & 5.60 & 10.0 \\
L15,L18W,L24  & 2.8 & 6.70 & 10.0 \\
\hline
\multicolumn{4}{@{}l@{}}{\hbox to 0pt{\parbox{85mm}{\footnotesize
\footnotemark[$a$] The S-Cam FoV of $34' \times 27'$ is covered with 
4 FLMN FoV, in which each covers the $25' \times 30'$ area. \\
}\hss}} 
\end{tabular}
\label{tab:fwhm}
\end{table} 

\begin{table}
\caption{Summary of limiting magnitudes and photometric corrections.}
\begin{tabular}{llrrr}
\hline 
\hline 
Band & $m_{lim.}$\footnotemark[$a$]  & $\Delta m_{a.c.}$\footnotemark[$b$]  
          & $\Delta m_{g.e.}$\footnotemark[$c$]  & $\Delta m_{s.o.}$\footnotemark[$a$]  \\ 
     & [AB]              & [AB] & [AB] &  [AB] \\ 
\hline
\hline 
$u^{\ast}$  & 26.31 & -0.06 & -0.20 & +0.20 \\
$B$         & 28.25 & -0.24 & -0.18 & +0.00 \\
$V$         & 27.57 & -0.24 & -0.13 & -0.02 \\
$R$         & 27.41 & -0.24 & -0.11 & -0.11 \\
$i'$        & 27.14 & -0.24 & -0.09 & -0.07 \\
$z'$        & 26.40 & -0.24 & -0,06 & -0.13 \\
$J$(SE)     & 22.28 & -0.22 & -0.04 & +0.02 \\
$J$(NE)     & 22.22 & -0.27 & -0.04 & +0.02 \\
$J$(SW)     & 22.23 & -0.28 & -0.04 & +0.02 \\
$J$(NW)     & 21.83 & -0.20 & -0.04 & +0.02 \\
$K_s$(SE)   & 21.66 & -0.22 & -0.02 & -0.09 \\
$K_s$(NE)   & 21.58 & -0.22 & -0.02 & -0.09 \\
$K_s$(SW)   & 21.34 & -0.22 & -0.02 & -0.09 \\
$K_s$(NW)   & 21.20 & -0.19 & -0.02 & -0.09 \\
N2          & 21.89 & -0.24 & -0.01 & -0.08 \\
N3          & 22.24 & -0.24 & -0.01 & +0.12 \\
N4          & 22.49 & -0.24 & -0.01 & +0.13 \\
S7          & 20.29 & -0.20 &  & \\
S9W         & 20.00 & -0.22 &  & \\
S11         & 19.82 & -0.24 &  & \\
L15         & 19.60 & -0.27 &  & \\
L18W        & 19.60 & -0.26 &  & \\
L24         & 18.60 & -0.27 &  & \\
\hline
\multicolumn{5}{@{}l@{}}{\hbox to 0pt{\parbox{85mm}{\footnotesize
\footnotemark[$a$] $m_{lim.}$ is the 
limiting magnitude.  \\
\footnotemark[$b$] $\Delta m_{a.c.}$\ is 
the aperture correction in magnitude.  \\
\footnotemark[$c$] $\Delta m_{g.e.}$\ is 
the magnitude correction for galactic extinction in the field. \\
\footnotemark[$d$] $\Delta m_{s.o.}$\ is 
the magnitude correction for systematic offset with 
the SED calibration for the stellar objects. \\
}\hss}} 
\end{tabular}
\label{tab:photmt}
\end{table}

\subsection{Optical and NIR photometry for ground-based data}
\label{subsec:ground}

Object detection and photometry were made using {\tt SExtractor} 2.3.2
(Bertin \& Arnouts 1996). The $z'$ S-Cam image was used to detect
optical objects, defined as a source with a 5 pixel connection above
the 2 $\sigma$ noise level.  We derived the Full Width Half Maximum (FWHM) 
of the Point Spread Function (PSF) from the photometric images 
of stellar objects as point sources (see their selection 
scheme in subsection~\ref{subsec:stargal}), which are summarized in 
table~\ref{tab:fwhm}.  In the process to make a catalogue, 
the reason why the $z'$-band was 
chosen as the detection band is the following: 
1) the FWHM of the PSF $1"$ in the $z'$ image with the S-Cam is good enough to 
prevent the confusion,  
2) the $z'$ photometry is one of the essential bands in the star-galaxy 
selection with the $BzK$ color as shown in subsection~\ref{subsec:stargal}, 
3) most of the galaxies of interest as detected by the AKARI; 
N3 Red Galaxies (N3Rs) and MIR bright N3Rs (MbN3Rs), and optical-NIR selected 
galaxies as BBGs are clearly identified in the $z'$ image.  
See the details about the N3Rs, the MbN3Rs, and BBGs in subsections 
~\ref{subsec:irbg} and ~\ref{subsec:mir_sed_sb_agn}, and 
in appendix~\ref{app_sec:bbg}), respectively.   
The catalogue of 
detected objects to $z'< 26.4$ with a 2$''$ diameter aperture was generated 
over the whole 23$'$ $\times$ 32$'$ area of the S-Cam $z'$ image.  

After selecting $\sim$ 56000 $z'$ sources in the area observed with
the FLMG, from the $\sim$ 66000 sources detected in the S-Cam $z'$
images, we performed aperture photometry in the near-UV $u^{\ast}$,
optical $B, V, R, i', z'$, and NIR $J, K_s$ images, fixed at the
positions of the $z'$-detected sources. The photometry on the $B, V,
R, i'$, and $z'$ images was performed with a small 2$''$ diameter
aperture with the {\tt SExtractor}.  The photometry on the $u^{\ast}$
and the $J$ and $K_s$ images was performed with larger 3$''$ and 4$''$
diameter apertures with the {\sf photo} of the IRAF task 
since these image qualities in off-center regions are not good due to 
the aberrations compared with those of the S-Cam.  
The larger aperture photometry reduces the uncertainties in the photometry 
and the alignment errors in the World Coordinate System (WCS) 
from these aberrations.  The errors of the ground-based 
optical and NIR photometry come mainly from the sky background and its 
fluctuation. We have estimated them from a Monte Carlo simulation with
the same method as that of Ishigaki et al. (2012), in which we
measured the counts in randomly distributed apertures and fit the
negative part of the count histogram with a Gaussian profile.  Thus,
the 3$\sigma$ limiting magnitudes $m_{\lim}$ were derived from the
background estimation, as summarized in table~\ref{tab:photmt}.
 
The final integrated magnitudes for all filters were obtained after
applying an aperture correction $\Delta m_{a.c.}$ based on empirical
PSFs, which are constructed from the photometry for stars in each
image band.  All of the optical and NIR magnitudes were further
corrected for a galactic extinction of $E(B-V)=0.041$, based on
observations of the survey area taken from \citet{schlegel_maps_1998}
and using the empirical selective extinction function of 
\citet{cardelli_relationship_1989}. 
 
\subsection{Infrared photometry for {\sl AKARI/IRC} images}  
\label{subsec:akari}

As shown above, we have produced a multiband merged catalogue from the
ground-based telescope images for the $z'$-detected objects in the
field. As shown in table~\ref{tab:photmt}, the depth of $N2$ images
obtained with the AKARI/IRC is comparable to that of $K_s$ images with
the KPNO/FLMG, although the images of all other bands at longer
wavelengths are shallower than those of $z'$.  
In fact, from 5819 sources extracted with the {\tt SExtractor} in the N3 band  
with a 5 pixel connection above the 2 $\sigma$ noise level, 
$92\%$ of them can be identified as 5332 counterparts of $z'$-detected sources 
without confusion.   
Thus, we have performed the photometry for all of the AKARI/IRC 
bands on the $z'$-detected sources.  

The IRC photometry on the $z'$-detected sources was 
performed with 8$''$ and 10$''$ diameter apertures for NIR ($N2, N3$, and 
$N4$) and MIR ($S7, S9W, S11, L15, L18W$, and $L24$) bands, 
respectively, with the IRAF task {\sf photo}.

The final integrated magnitudes in the IRC photometry were obtained
after applying an aperture correction based on the growth curves with
empirical PSFs, constructed from photometry of stars in each band
image, which is basically similar to 
the scheme descirbed in Takagi et al. (2012).  
The aperture corrections for the IRC are summarized in
table~\ref{tab:photmt}, along with those of the ground-based near-UV,
optical, and NIR magnitudes.  

In order to reduce the source confusion due to the fact that the IRC
PSFs are notably worse than those in the ground-based observed images, we
have performed IRC photometry with several different apertures in each IRC
band, and compared these magnitudes, excluding objects with a large
magnitude difference from the IRC counterparts in the $z'$-detected
catalogue. In the photometry in $N2, N3$, and $N4$ bands, we have
optimized the constraint for the IRC/NIR counterparts as $ N3(6")-N3(8")
< 0.51 $ between $6''$ and $8''$ diameter photometry in the $N3$
band, adding another constraint of $Ks-N2<1.0$ at 2~$\mu$m. 

For MIR sources detected with SN3 at least in one MIR band,  
there were many more sources with confusion compared with the NIR sources.  
We could select $\sim$ 800 
MIR bright N3 Red galaxies (MbN3Rs) and $\sim$ 900 MIR 
marginally-detected N3 Red galaxies (MmN3Rs) as 
shown in section~\ref{sec:pah}.  
However, 
we have taken confirmed samples of $\sim 600$ MbN3Rs and $\sim 600$ MmN3Rs 
after excluding NIR counterparts of the MIR sources having pairs 
within 6$"$, in order to correctly select the 
candidates of LIRGs at $z>0.4$ as discussed in section~\ref{sec:pah}.  
Roughly two thirds of these MbN3Rs and MmN3Rs were also 
identified as sources listed in another MIR source catalogue 
reported by Takagi et al. (2012), which are almost the same 
after taking the difference in their detection limits into consideration,   
as the former was selected with $SN>3$ at least in one MIR band 
while the latter was with $SN>5$.   
Concerning these NIR and MIR detected galaxies, we could confirm 
with our eyes that it can work to mostly exclude confusion 
from neighborhood sources within
a circleregion corresponding to the IRC PSF radius around each $z'$ source.  

\subsection{Photometric check and systematic offsets}
\label{subsec:offset}

In order to further check the photometric zero points, we have
performed SED fitting with $ u^{\ast}BVRi'z'JK_s $ and AKARI $N2, N3,$ and 
$N4$ photometric data for the stellar objects.  It is well known that
galactic stars can be clearly selected using a two-color criterion
\citep{daddi_new_2004, kong_wide_2006, lane_colour_2007,
quadri_multiwavelength_2007, blanc_multiwavelength_2008}:
\begin{equation}  
BzK_{\ast} \equiv (z-K)_{\rm AB} - 0.3(B-z)_{\rm AB} < -0.3 \; ,  
\label{eq:bzk_star}
\end{equation}  
where we have taken $-0.3$ as a parameter in the right side, which
defines a line separating stars and galaxies that is slightly larger
than Daddi's original value of $-0.5$, to reduce the contamination of
stars in our galaxy sample.  We have used the $BzK$ color criterion
to select the photometric calibrators as stellar objects with
additional criteria for the magnitudes $z' > 19$ and stellarity
parameter CLASS\_STAR $ > 0.95$ in {\tt SExtractor} (see also
subsection \ref{subsec:stargal}).  
We have also excluded IRC/NIR confused image sources by visual examination.  

We have adopted SED templates from the stellar library of
Bruzual-Persson-Gunn-Stryker (BPGS)
Atlas\footnote{http://www.stsci.edu/hst/observatory/cdbs/bpgs.html},
based on \citet{gunn_stellar_1983}, which is similar to the
\citet{pickles_stellar_1998} stellar 
spectra\footnote{http://www.ifa.hawaii.edu/users/pickles/AJP/hilib.html},
and \citet{lejeune_standard_1997}. In order to include the $N4$
magnitudes in our fitting of SEDs to the stellar objects with the BPGS
Atlas, we have extended the library SEDs beyond wavelengths of   
$\lambda >2.5~\mu$m with a 
Rayleigh-Jeans law. The best fit is sought by means of $\chi^2$
minimization, among all templates in the stellar library after varying
their normalization. We have estimated the offsets in each of the bands, 
which are seen as systematic discrepancies between the original observed
magnitudes and the best-fitted library magnitudes.  These systematic
offsets are also summarized in table~\ref{tab:photmt}.  
Since the derived offsets are less than $0.2$, basically we can 
confirm the accuracy of the photometric zero points in all the bands. 
In the comparison between the spectroscopic and photometric redshifts,  
the result for the systematic offset is slightly 
better than that without this.  Thus, we use the offset.  

\subsection{Star-galaxy separation} 
\label{subsec:stargal}

As discussed in subsection~\ref{subsec:offset}, 
the colors estimated with the BPGS
star atlas overlap with the observed stellar sequence appearing
in the $BzK$ color region (see figure~\ref{fig:bzk_spec_phot} 
in the appendix).  Thus, we can also use the $BzK$ criterion 
of equation~(\ref{eq:bzk_star})  
as a robust method for star-galaxy separation. For sources
detected with $SN>3$, up to $K_s=20$, we have excluded objects
satisfying the following conditions; 1) equation~(\ref{eq:bzk_star}) and 2)
stellarity parameter CLASS\_STAR$>0.95$ in {\tt SExtractor}.   In order
to separate stars and galaxies for all of the $z'$-detected objects up to
an optical limiting magnitude $z'=26$ that are fainter than the above
$K_s$ detected sources, we performed SED fittings for them  
with not only the BC03 models \citep{bruzual_stellar_2003} for galaxies 
 but also the BPGS star atlas for stellar objects, 
including those corrected with Allen's Milky Way extinction curves.  
We extracted objects as stellar objects, which were fitted better by 
the latter than the former, from catalogue of $z'$-detected galaxies  
(see also subsection~\ref{subsec:photz}).  
These stellar objects appear in the sequences of the BPGS star atlas 
in BBG color-color diagrams (see not only figure~\ref{fig:bzk_spec_phot},
but also figures~\ref{fig:urj_spec_phot}, and \ref{fig:uvi_spec_phot} 
in appendix~\ref{app_sec:bbg}).  

\section{Photometric redshifts}
\label{sec:photz}

\begin{figure}[ht]
\begin{minipage}{0.99\linewidth}
\includegraphics[width=\linewidth]
{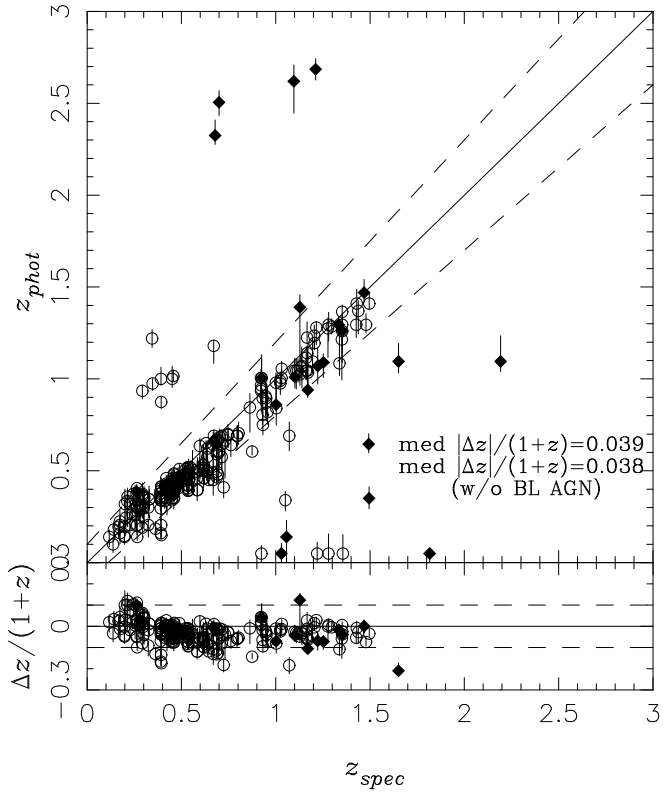}
\end{minipage}
\caption{Comparison between spectroscopic and photometric redshifts 
for $z'$-detected galaxies with spectroscopic observations.  
Filled diamonds and open circles represent 
objects with and without broad line emissions in their 
spectra, respectively.   
Dashed lines represent a deviation with $\Delta z/(1+z_{spec})=0.1$. 
}
\label{fig:specz_photz}
\end{figure}  

\begin{figure}[ht]
\begin{minipage}{0.99\linewidth}
\includegraphics[angle=0,width=\linewidth]
{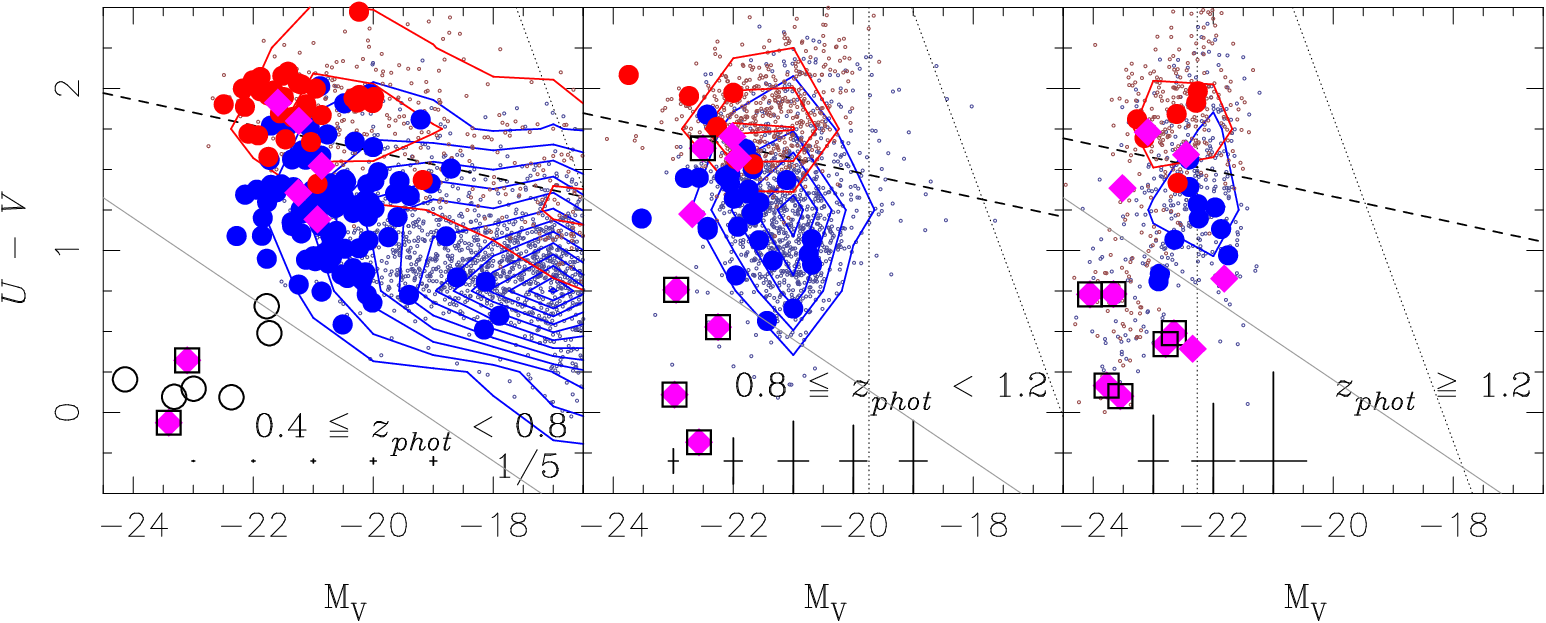}
\end{minipage}
\begin{minipage}{0.99\linewidth}
\includegraphics[angle=0,width=\linewidth]
{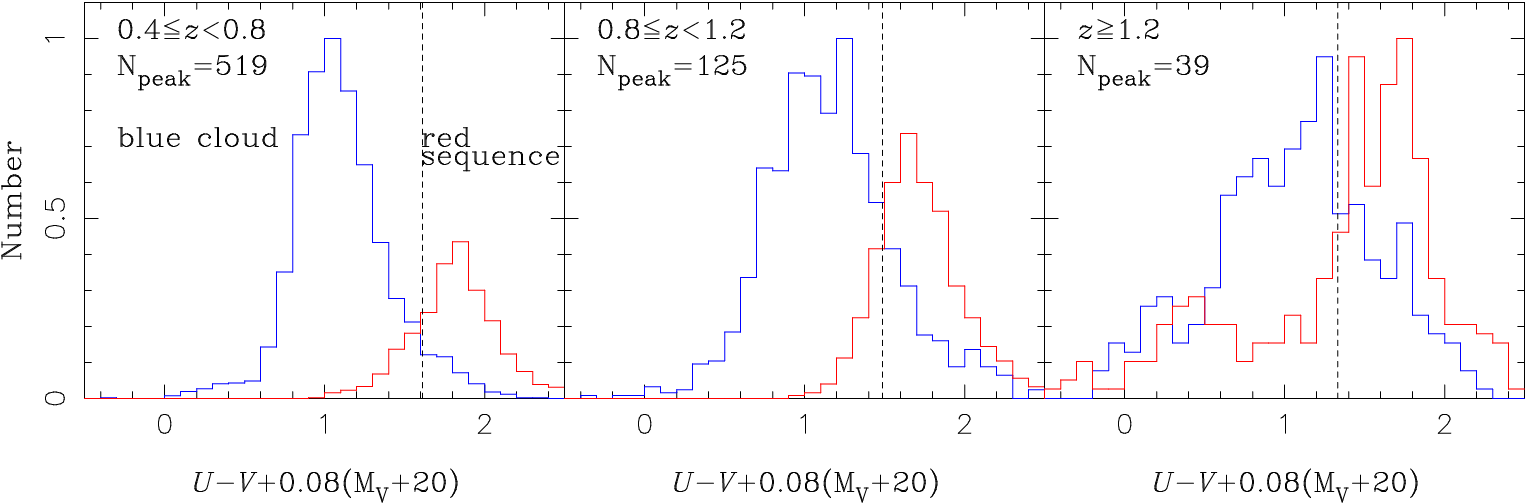}
\end{minipage}
\caption{Rest-frame color $U-V$ for the spectroscopic sample 
and $z'$-detected galaxies. 
Top: Rest-frame Color Magnitude Diagram (CMD) 
of absolute $V$-band magnitude vs. $U-V$ color.   
Objects in redshift ranges of 
$0.4 \le z < 0.8$, $0.8 \le z < 1.2$, and $1.2 \le z$  
are represented from the left to the right. Large symbols represent the 
spectroscopic sample.  Red and blue symbols represent lower and higher 
SFR populations with sSFR $< 0.1$~Gyr$^{-1}$~ and 
sSFR $ > 0.1$~Gyr$^{-1}$~ derived from the SED fitting, respectively.  
Magenta diamonds represent BL AGNs, in which diamonds in squares  
represent outlier BL AGNs with 
$\Delta z/(1+z_{spec})>0.2$ in the redshift comparison.  
Small gray dots represent the remainder in the $z'$-detected 
galaxies.  Red and blue contours represent 
the distribution of lower and higher SFR population in the $z'$-detected 
galaxies with sSFR $< 0.1$~Gyr$^{-1}$~ and sSFR  
$> 0.1$~Gyr$^{-1}$, respectively. 
Dashed lines represent Bell's color boundary between star-forming and passive 
populations.  Plain black crosses around $U-V \simeq -0.2$ 
represent typical errors of estimated colors 
and mass.  The vertical solid and steep slope dense dotted lines represent 
the limiting absolute $V$ and $U$-band magnitudes at the mean redshift, 
respectively.  Open circles represent a population categorized as type 1 QSOs 
at $z=0.2-0.7$ in the catalogue 
of the SDSS/SWIRE sample by Hatziminaoglou et al. (2005).  
Bottom: Color distributions of the $z'$-detected galaxies.  
The color is scaled as $(U-V)+0.08(M_V+20)$.  
Red and blue lines represent lower and higher SFR populations with sSFR 
$< 0.1$~Gyr$^{-1}$~ and sSFR$ > 0.1$~Gyr$^{-1}$~ derived from 
the SED fitting, respectively.  
}
\label{fig:mv_uvcol_zdet_spec}
\end{figure}  

\subsection{Photometric and spectroscopic redshifts}
\label{subsec:photz}

For all of the $z'$-detected galaxies remaining after the star-galaxy
separation procedure as described in subsection~\ref{subsec:stargal}, 
we simultaneously obtained the photometric redshift $z_{phot}$ and the 
absolute magnitude $M_{H;opt}$ corresponding to the stellar mass $M_{\ast}$ 
(see also appendix~\ref{app_sec:smass}), by applying a 
photometric redshift code {\it hyperz}
\citep{bolzonella_photometric_2000}. This is based on comparison of
the observed ground-based $u^{\ast}BV$$Ri'z'$$JK_s$ photometry with a
grid of stellar population synthesis models produced from 
\citet{bruzual_stellar_2003}.  From this BC03 model we prepared SED
templates of a total of nine star-forming history models with 
exponential decaying star formation (SF) as $SFR(t) \propto \exp (-t/\tau)$, 
with time scales for the star formation activity of
$\tau=0.1,0.3,1,2,3,5,15, \mbox{and}~30$~Gyr, and we varied the age $t$ from
0 to 20 Gyr, and included the Constant Star Formation (CSF) model.  
All of the models are for solar metallicity.  With the variations of
$E(B-V)=0, 0.1, 0.2, 0.3, 0.4, 0.5,$ and $0.6$, we additionally
applied Allen's Milky Way, Seaton's Milky Way, LMC, SMC, and
Calzetti's extinction curves, to obtain the extinction $A_V$.  
In selecting the best fitting one from five extinction curves for each object, 
firstly we determine a median photometric redshift $z_m$ 
in five of them related to these extinction models, 
and secondly we have used one model with reduced $\chi^2$ minimization 
as the photometric redshift should be within 0.2$(1+z_m)$ around the median 
$z_m$.  We adopted a Salpeter IMF from 0.1 to 100 M$_\odot$
throughout this work.  

In the AKARI NEP Deep field, we also obtained spectra 
with the KeckII/DEIMOS, 
the Subaru/FOCAS, and the MMT/Hectospec for 420, 57, and 62 objects, 
which were selected from objects with $R<24, <24$, and $<22$ mag, 
respectively.   
The KeckII/DEIMOS and the Subaru/FOCAS observations partially covered 
the NEP Deep field while the MMT/Hectospec observations covered 
the NEP Deep field by two, 1 deg$^{2}$ multi-object configurations.  
More details on the MMT/Hectospec observations can be 
found in Ko et al. (2012).  They are mostly detected as IR sources with the 
AKARI \citep{takagi_polycyclic_2010}.  
We have obtained 292 spectra for the $z'$-detected galaxies, 
in which we can determine secure redshifts for 230 objects are 
detected with more than two lines or clearly with [OII] line.  
They include 27 galaxies harboring AGN with Broad Line 
(BL) emissions, which are spectroscopically classified as BL AGNs.     
Figure~\ref{fig:specz_photz} shows the comparison between 
the spectroscopic and the photometric redshifts of the 230 objects and their 
redshift discrepancy $\Delta z= z_{phot} -z_{spec}$.  The median accuracies of 
$\langle \Delta z \rangle/(1+z_{spec})$ are 0.041 and 0.038 for the 
spectroscopic samples without and with excluding the BL AGNs, 
respectively.  
In the spectroscopic sample, there are the 22 and 52 outliers 
with $\Delta z/(1+z_{spec})>0.2$ and $>0.1$, 
in which 11 and 12 out of these outliers are BL AGNs, respectively.    
Thus, the photometric redshifts 
are consistent with their spectroscopic ones not only for the samples  
except BL AGNs but also for even half of the BL AGNs.  
The outlier BL AGNs except one are confirmed as an Extremely Blue 
Object (EBO) in the rest-frame Color-Magnitude Diagrams (CMDs), 
which are selected with a criterion:   
\begin{equation}
(U-V) < -0.25(M_V+22.0) +0.7 \; ,  
\label{eq:ebo}
\end{equation}
as shown in figure \ref{fig:mv_uvcol_zdet_spec}.  
If EBOs are type 1 AGNs, this trend is reasonable 
since SED models of the AGNs were not included  
in SED fittings for the photometric redshift estimations.  

The AKARI MIR colors can alternatively 
extract candidates associated with dusty AGN activities, in which we 
spectroscopically observed 14 agn-MbN3Rs and 2 s/a-MbN3Rs 
(see the details about them in subsection~\ref{subsec:mir_sed_sb_agn}).  
We could confirm that 8 out of 
14 agn-MbN3Rs and 2 out of 2 s/a-MbN3Rs show the discrepancy 
with $\Delta z/(1+z_{spec})>0.1$, respectively.  All of 
these outliers are also identified as BL AGNs, 
in which 7 out of 8 the agn-MbN3Rs and 2 out of 2 the s/a-MBN3Rs 
are classified as the EBOs, respectively.  
As long as their redshifts are correctly obtained with spectroscopy, 
the overlap of the EBOs with the agn-MbN3Rs suggests that 
the dust emission detected as the agn-MbN3Rs should coexist 
with optically blue emission classified as the EBOs in all of them.  
This supports not only the above expectations that AGNs are harbored 
in the EBOs and the agn-MbN3Rs but also a picture of a dusty torus 
surrounding AGN which can explain the variation of their SEDs 
as frequently proposed in the AGN unified theory.  

We also checked the redshift discrepancy $\Delta z$ in various populations 
subclassified not only with the spectroscopy, but also with NIR and MIR 
photometry as N3Rs and MbN3Rs, which are summarized 
in appendix~\ref{app_sec:photz_var}.  
Even though most of the spectroscopic samples are mainly selected from dusty 
populations with heavy extinction as MbN3Rs detected in the AKARI MIR 
photometry, their median accuracies of $\langle \Delta z \rangle/(1+z_{spec})$ 
are less than 0.05 for any subclassified populations of N3Rs and MbN3Rs 
except agn-MbN3Rs.  Their photometric redshifts can be 
estimated well from the SED fitting with their major stellar emission 
of the host galaxies, in which the extinctions might not be a serious 
matter.  Thus, the estimated photometric redshifts even  
for most of the galaxies can be used to reconstruct approximately their 
actual redshifts within $\Delta z/(1+z_{spec}) < 0.05$.  

\subsection{Reconstructions of rest-frame magnitudes and colors}  
\label{subsec:mv_uvcol}

From the derived photometric redshifts $z_{phot}$ with the optical-NIR SED 
fittings, we can also estimate rest-frame magnitudes and colors of the 
$z'$-detected galaxies.  For an object at a redshift $z$, a magnitude 
$m_{\lambda_0}(z)$ at a rest-frame wavelength $\lambda_0$ is 
approximately interpolated from the observed magnitudes as
\begin{eqnarray} 
 m_{\lambda_0}(z) & = & \frac{  \left[ \lambda_r -\lambda(z) \right] m_b  
                               + \left[\lambda(z) -\lambda_b \right] m_r }
                       {\lambda_r -\lambda_b} \; , \\ 
 \lambda(z) & = & \lambda_{0}(1+z) \; ,               
\label{eq:mag_intr}
\end{eqnarray} 
where $m_r$ and $\lambda_r$ ($m_b$ and $\lambda_b$) are the observed
magnitudes and central wavelengths of the band neighboring the red-side
(blue-side) of the observing wavelength $\lambda(z)$.  
By using the rest-frame magnitudes derived with $z_{phot}$, 
CMDs for the $z'$-detected galaxies can be 
obtained.  The CMD fundamentally is almost the same as 
the color-mass diagram, which is a powerful tool for studying galaxy evolution 
since it shows the bimodal galaxy distribution as early types concentrate on a
tight red sequence while late types distribute in a blue dispersed
cloud. This has been studied not only in the local Universe with the SDSS 
\citep{blanton_estimating_2003} but also in the distant Universe with 
surveys for $z<1$ \citep{bell_nearly_2004, faber_galaxy_2007, 
borch_stellar_2006} and for $z>1$ \citep{pannella_star_2009, 
brammer_dead_2009}.  

Figure~\ref{fig:mv_uvcol_zdet_spec} shows a  
CMD of the absolute $V$-band magnitude $M_V$ and $U-V$ color.  
We display rest-frame $V$-detected galaxies on the CMD, and 
excluded objects with less photometric accuracy of $SN<2$  
in the rest-frame $V$ band.  
The rest-frame $U-V$ color straddles the Balmer/4000~\AA\ break and  
is also essential to introduce classifications for 
their stellar populations in IRBGs, MbN3Rs, MmN3Rs, and BBGs as 
discussed in the following sections and the appendix.      
This interpolation scheme, applied for the $z'$-detected galaxies 
with $z = z_{phot}$,  can be used 
to derive their $M_V$ and rest-frame $U-V$ colors.  
In figure~\ref{fig:mv_uvcol_zdet_spec}, we can see that the distribution 
of galaxies is bimodal up to $z\simeq 1$ as previously shown 
\citep{bell_nearly_2004, faber_galaxy_2007, borch_stellar_2006}.     
In figure~\ref{fig:mv_uvcol_zdet_spec}, we can see that 
the red sequence and the blue cloud are separated with a criterion:  
\begin{equation} 
(U-V) = 1.15 - 0.31z - 0.08 (M_V -5 \log h +20) \; , 
\label{eq:mv_uvcol_red_blue}
\end{equation}
which is introduced for the definition of red sequence galaxies 
by \citet{bell_nearly_2004} and represented as dashed lines on the CMDs 
in figure~\ref{fig:mv_uvcol_zdet_spec}.  
Thus, our catalogue of $z'$-detected galaxies 
with photometric redshifts is basically the same as 
those of other blank sky surveys.  

\section{Classification with IR bump} 
\label{sec:irbg}   

From the sample of $z'$-detected galaxies including the Balmer Break Galaxies
(BBGs; see the details in appendix~\ref{app_sec:bbg}), we
extracted IR classified subpopulations with the AKARI/IRC
photometry.  The NIR IRC photometry can select IR bump Galaxies
(IRBGs) which have a bump around 1.6 $\mu$m with the AKARI 
\citep{takagi_multi-wavelength_2007} and the {\em Spitzer} 
\citep{berta_eso-spitzer_2008}.  Both the IRBGs and the BBGs 
are characterized mainly by the emission from the
stellar components.   In this section, we will present three redshifted 
populations selected by criteria of the IR bump at 1.6 $\mu$m 
combining the Balmer break.  In fact, the selection with the 1.6 $\mu$m 
can reduce the contamination from the low-z interlopers 
in the IRBGs at $z>0.4$, which is also useful to classify MIR-bright 
populations.  We also try to select AGN candidates as outliers with NIR 
colors redder than normal IRBGs.     

\subsection{AKARI NIR color-color diagram and 1.6 $\mu$m bump} 
\label{subsec:irbg}  

\begin{table}[t]
\begin{center}
  \caption{The Parameters for IRBG Selections}
\label{tab:irbg_select}
  \begin{tabular}{lccccc}
  \hline
  \hline
  IRBGs   & $ f_{\scriptsize{N23}} $ 
          & $ \min_{\scriptsize{N34}} $ 
          & $ \max_{\scriptsize{N34}} $ 
          & $z_{\mbox{\scriptsize{min}}}$ 
          & $z_{\mbox{\scriptsize{max}}}$ 
  \\
  \hline
  N3R & $-0.3$ & --     & --     & $0.4$ & -- \\
  \hline
  N23  & $-0.3$ & --     & $-0.5$ & $0.4$ & $0.8$ \\
  N34  & $-0.3$ & $-0.5$ & $-0.1$ & $0.8$ & $1.3$ \\
  N4R  & $-0.3$ & $-0.1$ & --     & $1.3$ & $-$ \\
  \hline
  \hline
  \end{tabular} 
\end{center}
\end{table}

\begin{figure}[ht]
\begin{minipage}{0.99\linewidth}
\includegraphics[width=\linewidth]
{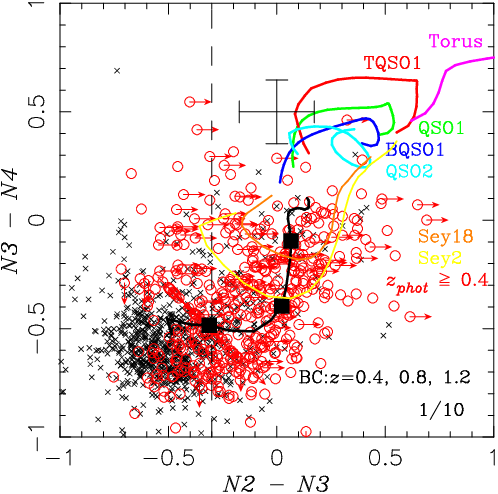}
\end{minipage}
\begin{minipage}{0.49\linewidth}
\includegraphics[width=\linewidth]
{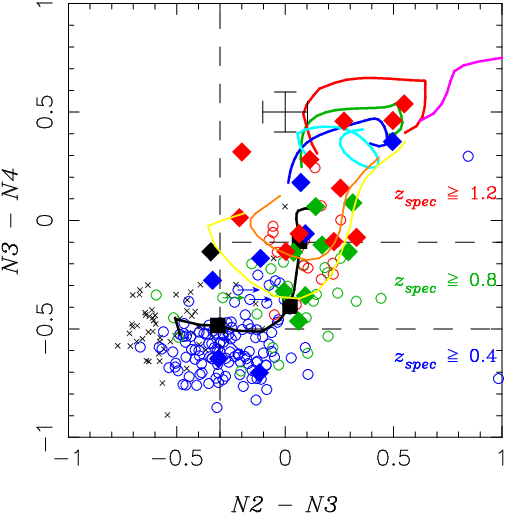}
\end{minipage}
\begin{minipage}{0.49\linewidth}
\includegraphics[width=\linewidth]
{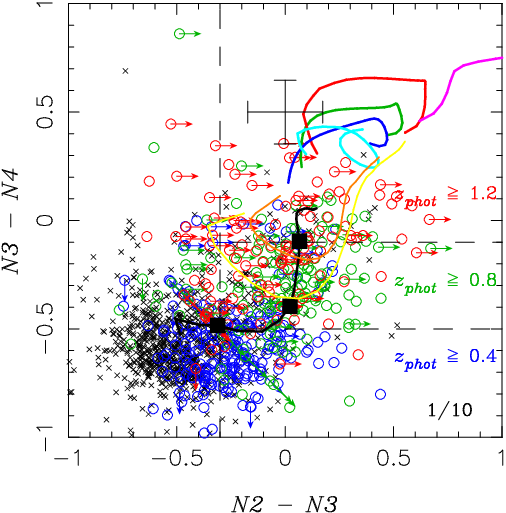}
\end{minipage} 
\caption{Two-color diagram of $N2-N3$ vs. $N3-N4$ in the field.  
Symbols of cross bar at $(N2-N3,N3-N4)=(0.1,0.5)$ represent typical 
errors of the color.  
Black line is superimposed as a redshifted track of a BC03 model galaxy 
with $\tau=1$Gyr and age$=1$Gyr, on which the squares are related 
to the redshifts of $z=0.4, 0.8,$ and $1.2$ from lower-left to upper-right.  
Other superimposed green, red, blue, cyan, and magenta lines are tracks 
of QSO1, TQSO1, BQSO1, QSO2, and dusty torus IR templates, respectively.  
Redshift from $z=0$ to $z=2$ corresponds to 
the upper-right to the lower-left along each line.   
Top: For objects detected in the $N2, N3,$ and $N4$ bands.  
Arrows represent objects which are fainter than the threshold limit  
in the $N2$ or $N4$ band.     
Red open circles  and black crosses represent the objects 
at $z_{phot}\geqq 0.4$ and $z_{phot}<0.4)$, respectively.  
Bottom-Left: For the spectroscopic samples. Blue, green, and red symbols  
represent the objects at $ 0.4 \le z_{spec}< 0.8$, 
$0.8 \le z_{spec}< 1.2$, and $ z_{spec} \geqq 1.2$, respectively. 
Diamonds represent BL AGNs.        
Bottom-Right: The same as on the left for the photometric samples.
}
\label{fig:n234_spec_phot}
\end{figure}  

A bump centered at 1.6~$\mu$m in the rest-frame is a major SED feature
of galaxies except for the youngest ones for ages $<10$ Myr, in which
stellar components are dominated by type M stars.  The 1.6~$\mu$m IR
bump is produced by the combination of the spectral peak in
black-body radiation from low-mass cool stars as M stars and a minimum
in the opacity of H$^-$ ion present in their stellar atmospheres with
molecular absorptions \citep{john_continuous_1988}.  Universally appearing 
in most galaxies, the IR bump is one of the most
useful SED features in their photometric measurements 
\citep{simpson_detection_1999, sawicki_1.6_2002}.  

The AKARI NIR photometry in $N2N3N4$ bands can trace the IR bump at 
$z=0.4-2$.  We have shown the redshift tracks of the BC model galaxies with 
$\tau=1$Gyr and age$=1$Gyr on a two-color diagram of $N2-N3$ vs. $N3-N4$ in
figures~\ref{fig:n234_spec_phot}, where the solid squares on the tracks 
represent the redshifts at $z=0.4, 0.8$, and 1.2 from the bottom-left to the
top-right.  The tracks are mostly parallel to the $N2-N3$ axis up to $z
\simeq 0.8$, bending upward around $z \simeq 0.8$, and are vertical along
the $N3-N4$ axis at $z \simeq 0.8-1.2$.  
This trend can be understood as the excess of redshifted IR
bump is on the $N2$ side in the $N3$ band at $z \simeq 0.4-0.8$, the $N4$
side in the $N3$ band at $z \simeq 0.8-1.2$, and in the $N4$ band at $z >1.2$.  
The redshift is the dominating factor in determining the AKARI NIR color 
tracks of galaxies without AGNs while the galaxy type is less effective.  

First, we extract $N3$-detected ones as $N3$ Red galaxies (N3Rs) possibly 
at $z>0.4$ from the $z'$-detected galaxies with a criterion: 
\begin{equation} 
N2-N3> f_{\scriptsize{N23}} \; . 
\label{eq:n234} 
\end{equation}  
Second, we can also subclassify the N3Rs into N23, N34, and N4 bumpers, 
which are defined as IRBGs of their IR bump detected 
with the $N2$ and $N3$ bands, the $N3$ and $N4$ bands, and the $N4$ band 
in three redshift ranges of $z \simeq
0.4-0.8$, $z\simeq 0.8-1.2$, and $z > 1.2$, respectively.  
The color criteria are represented as 
\begin{equation} 
N2-N3> f_{\scriptsize{N23}} \; \cap \; \; 
\mbox{min}_{\scriptsize{N34}} \le N3-N4 < \mbox{max}_{\scriptsize{N34}} \;, 
\label{eq:n23_n34_n4}
\end{equation}
where the parameters of $f_{\scriptsize{N23}}, \min_{\scriptsize{N34}}$, and 
$\max_{\scriptsize{N34}}$ are given in table~\ref{tab:irbg_select}.  

The bottom-left of figure~\ref{fig:n234_spec_phot} represents a two-color
diagram for the spectroscopic sample in the field, in which most of the 
galaxies at $z_{spec} \simeq 0.4-0.8, 0.8-1.2$, and $>1.2$ appear as 
N23, N34, and N4 bumpers, respectively.  The bottom-right of 
figure~\ref{fig:n234_spec_phot} also represents another two-color 
diagram for the photometric sample in the field, in which the NIR 
color of galaxies at $z_{phot} \simeq 0.4-0.8, 0.8-1.2$, and $>1.2$ 
distribute around the regions of N23, N34, and N4 bumpers, 
respectively.  
   
When a galaxy harbors an AGN, however, IR emission from the dusty torus 
around the AGN affect the NIR colors.  
We have also superimposed tracks of redshifted QSO SED templates in the 
SWIRE library~\footnote{http://www.iasf-milano.inaf.it/~polletta}; 
three types of optically-selected type 1 QSOs (QSO1, TQSO1, and BQSO1) 
and two models of type 2 QSOs (QSO2 and Torus).  All of them show 
dominant emissions from the dusty torus at 
IR wavelength $\lambda >1~\mu$m.  
Thus, the NIR color criterion from equation~(\ref{eq:n234}) can be still 
effective for selecting not only galaxies at $z>0.4$ but also AGN candidates.  
Furthermore, all the QSO tracks are overlapping with the region 
of the N4 bumpers.  In fact, most of the BL AGNs, selected with 
the spectroscopy, appear around the region of N4 bumpers.  
Thus, the N3Rs with $N3-N4>-0.1$ may include not only 
the N4 bumpers of IRBGs but also these AGN candidates, 
which will be also confirmed with the MIR classified AGNs 
in section~\ref{sec:pah}.  We will designate the N3Rs with $N3-N4>-0.1$ 
as N4 Red galaxies (N4Rs).     


\subsection{N3Rs subclassified with Balmer break} 
\label{subsec:irbg_bbg}  

\begin{figure}[ht] 
\begin{minipage}{0.49\linewidth}
\includegraphics[width=\linewidth]
{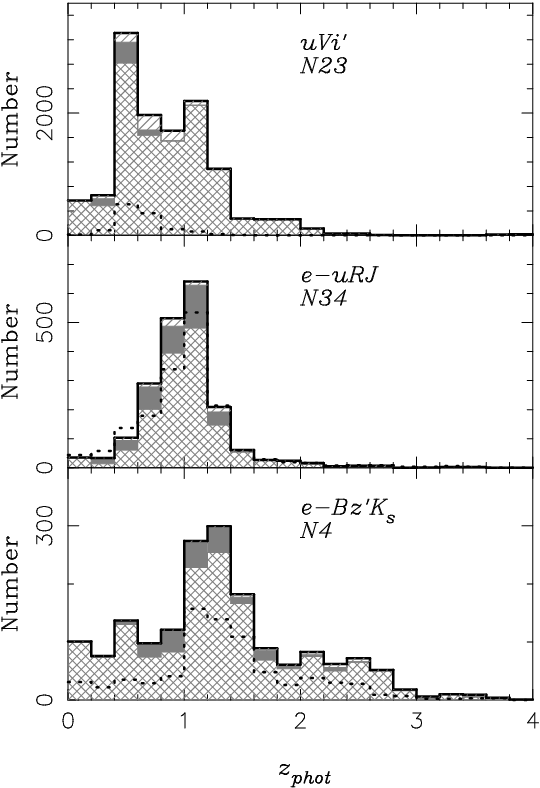}
\end{minipage}
\begin{minipage}{0.49\linewidth}
\includegraphics[width=\linewidth]
{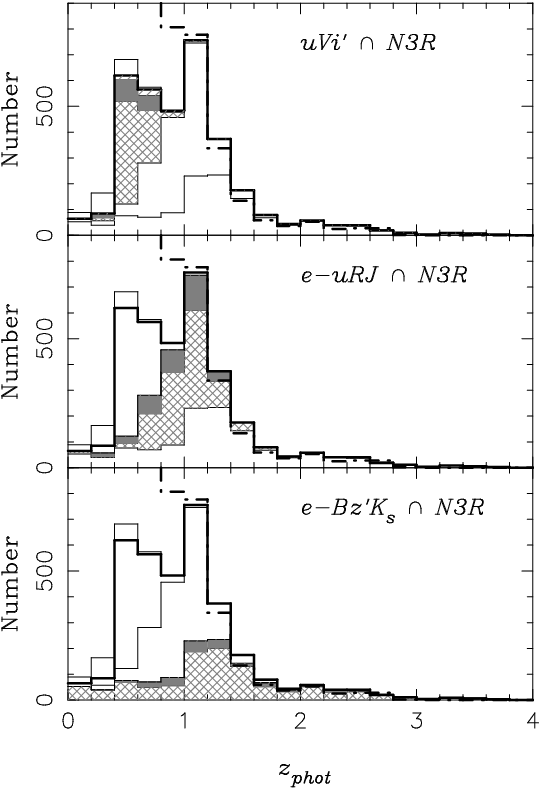}
\end{minipage}
\caption{The photometric redshift distributions for 
exclusively classified BBGs; 
e-BzKs, exclusively classified e-uRJs 
from the remainings after exclusion 
of the e-BzKs, and exclusively classified uVis 
from the remainings after exclusion of both the 
e-BzKs and the exclusively classified e-uRJs, 
from bottom to top.  Shaded, single-hatched, and cross-hatched area 
represent p-, qp-, and s-BBGs.   
Left: For exclusively classified BBGs and IRBGs.  
Dotted lines represent N23 bumpers, N34 bumpers, and N4Rs, from 
top to bottom.  
Right: For N3 Red BBGs. 
Dashed, thin solid and thick solid lines represent 
the rest-frame $V$-detected galaxies, the N3Rs, all the N3 Red BBGs, 
respectively.  
}
\label{fig:photzhist_bbgn3r}
\end{figure}  

\begin{figure}
\begin{minipage}{0.99\linewidth}
\includegraphics[angle=0,width=\linewidth]
{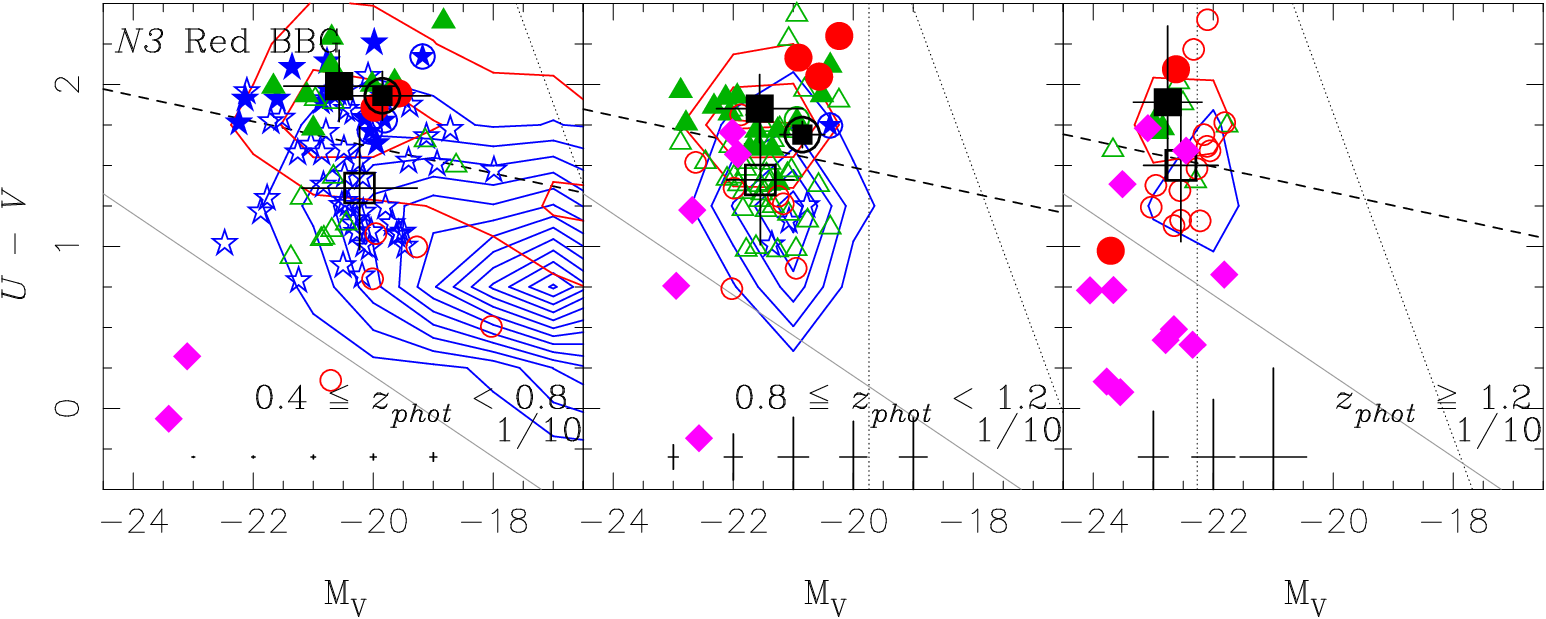}
\end{minipage}
\caption{The same as figure~\ref{fig:mv_uvcol_zdet_spec} for N3 Red BBGs.  
The plots for $0.4<z<0.8$, $0.8<z<1.2$, and $1.2<z$ from left to right.    
In general, colored open, solid, and circled symbols 
represent objects classified as s-, p-, and qp-BBGs, respectively. 
Basically, red circles, green triangles, and blue stars  
represent objects selected with the two-color criteria 
as e-BzKs, e-uRJs, and uVis, respectively. 
Black-large open, solid, and circled symbols represent mean values 
for the s-, p-, and qp-BBGs, respectively.  
Crosses on their black symbols represent their standard 
deviations.  
Only a part of samples are plotted to reduce crowding in the diagram, whose 
fractions are presented as numbers in the right-bottom region of all figures.
}  
\label{fig:mv_uvcol_n234}
\end{figure}

As shown in figure~\ref{fig:n234_spec_phot}, 
both spectroscopic and photometric redshifts of the N3Rs
are almost consistent with the NIR color criterion 
$N2-N3 >-0.3$ of the IR bump at $z>0.4$.  
As mentioned in subsection~\ref{subsec:photz},  
the photometric redshifts were derived from only the ground-based data 
without the AKARI NIR data, which means that the selection with the 
IR bump is independent of the photometric redshifts.  
Thus, the color criterion with $N2-N3> -0.3$ 
is useful in order to exclude low redshift interlopers at $z <0.4$ 
in the IRBGs as long as they are detected in the $N2$ (or $K$) and $N3$ bands 
as discussed in subsection~\ref{subsec:irbg}.  
Furthermore, adding the AKARI/NIR photometry 
to the ground-based optical-NIR photometry can salvage the BBGs 
in the IRBGs as shown in appendix~\ref{app_subsec:exbbg}.  
These are the merits in using the AKARI/NIR photometry with the ground-based 
photometry.   

On the other hand, even though the IR bump criteria 
of N23 and N34 bumpers, and N4Rs correspond to 
three redshift intervals as shown in figure~\ref{fig:n234_spec_phot}, 
they cannot be robust for the redshift subclassification 
since the N4Rs possibly include AGNs as discussed in 
subsection~\ref{subsec:nir_mbn3r}.  
In redshift subclassifications, optical color criteria with the Balmer break  
are more reliable than those of the $N3-N4$ colors with the IR bump.    
As described in appendix~\ref{app_sec:bbg}, 
we have generalized the color criteria of 
BzKs at $1.4<z<2.5$ with $BzK$ filter set \citep{daddi_new_2004} 
to select BBGs at $0.8<z<1.8$ and $0.4<z<0.8$ by using $uRJ$ 
and $uVi$ filter sets, called as uRJs and uVis, respectively. 
Furthermore, we have selected extended BBGs as e-BzKs and e-uRJs 
with the $N3$ band photometry which is the deepest one in the IRC bands 
(see appendix~\ref{app_subsec:exbbg}).  
However, these e-BzKs, e-uRJs, and uVis overlap each other.  
Thus, we introduced an exclusive 
subclassification of BBGs as first picked with 
the $BzK$ criteria, second applied the $uRJ$ 
criteria for those remaining after exclusion of the BzKs, and finally applied 
the $uVi$ criteria for those remaining after exclusion of the uRJs 
and the BzKs. 
As shown on the left in figure~\ref{fig:photzhist_bbgn3r}, 
the redshift distributions of the exclusively classified BBGs except the uVis 
are similar to those of IRBGs as the uRJs and the BzKs correspond to 
the N34 bumpers and the N4Rs, respectively.   
Most counterparts of the MIR bright populations at $z>0.4$, 
detected with the AKARI MIR photometry in the field, 
are identified as the N3Rs.  
Thus, hereafter, we will take a subclassification of the N3Rs into 
the star-forming (s-), passively evolving (p-), and quasi passively evolving 
(qp-) BBGs with the exclusively subclassification.   
We will name these BBGs exclusively subclassified 
from the N3Rs as N3 Red BBGs (N3RBBGs); N3RBzKs, N3RuRJs, and N3RuVis.  
The redshift distributions of N3 Red BBGs are also shown 
on the right in figure~\ref{fig:photzhist_bbgn3r}.   
Figure~\ref{fig:mv_uvcol_n234} shows the CMD 
of $M_V$ and $U-V$ for these N3 Red BBGs.   
We can see that the subclassification of the s-, p-, and qp-N3RBBGs 
is consistent with that using  
Bell's color boundary in figure~\ref{fig:mv_uvcol_n234}. 

Since the $N3$ band photometry is the deepest one in the IRC bands, 
as discussed in appendix~\ref{app_subsec:exbbg},   
the N3 Red BBGs salvaged with the $N3$ photometry 
are almost the same selected population 
as the rest-frame $V$-detected galaxies at $z>0.4$.  
It can be confirmed as both of them show a similar redshift 
distribution as shown in figure~\ref{fig:photzhist_bbgn3r}.    
The rest-frame $V$-detected galaxies are a kind of referenced sample 
as frequently plotted with the referenced contours on the CMDs 
in figures~\ref{fig:mv_uvcol_zdet_spec} and ~\ref{fig:mv_uvcol_n234} (see also 
figures ~\ref{fig:mv_uvcol_bbg} and ~\ref{fig:mv_uvcol_exbbg} 
in the appendix).  Thus, the N3 Red BBGs are suitable in the 
study for their stellar populations by using the CMD even in MIR bright 
populations as seen in section~\ref{sec:pah}.   

\section{Classifications with MIR SEDs} 
\label{sec:pah}

The MIR multi-band photometry with the IRC 
has identified MIR-bright populations in the N3Rs as detected with 
more than an SN3 in 
MIR bands and classfied into starbursts or AGNs 
with two MIR color diagrams, which are called as MbN3Rs.  The
IRC MIR photometry can detect Poly-Aromatic Hydrocarbon (PAH) emissions
from dusty star-forming regions for a subgroup in the IRC detected
MbN3Rs.  The subgroup with the PAH features can be classified with the MIR 
colors as starbursts, which are overlapped with the PAH luminous galaxies 
in some companion papers \citep{takagi_multi-wavelength_2007, 
takagi_polycyclic_2010}.  Another subgroup of the MbN3Rs without any 
PAH features or with weak ones is possibly AGN dominant. 
Some of them may overlap with Power-Law
Galaxies (PLGs) and IR Excess Galaxies (IREGs), which are 
already classified with the {\em Spitzer} as AGN-dominant
populations.  The former has MIR SEDs well fitted by a power law
\citep{lacy_obscured_2004, stern_mid-infrared_2005,
alonso-herrero_infrared_2006, donley_spitzer_2007}.  The latter shows
the large infrared-to-UV/optical flux ratio
\citep{daddi_multiwavelength_2007, dey_significant_2008,
fiore_unveiling_2008, polletta_obscuration_2008}, in which some 
may be identified as sources with heavily obscured optical emissions.  
The MbN3Rs and their relatives at $z>0.4$ detected in the field have typically 
TIR luminosities with $>10^{11}$L$_{\odot}$, which correspond to those of 
LIRGs/ULIRGs as shown in appendix~\ref{app_sec:zdist}.    
In this section, thus, we will mainly study the MbN3Rs and their relatives  
by analyzing the MIR and NIR SEDs to reveal star forming and AGN 
activities of LIRGs/ULIRGs at $z>0.4$ in the following sections.  

\subsection{Starburst and AGN in MIR bright N3Rs} 
\label{subsec:mir_sed_sb_agn}  

\begin{figure}[ht]
\centering 
\begin{minipage}{0.49\linewidth}
\includegraphics[angle=0,width=\linewidth]
{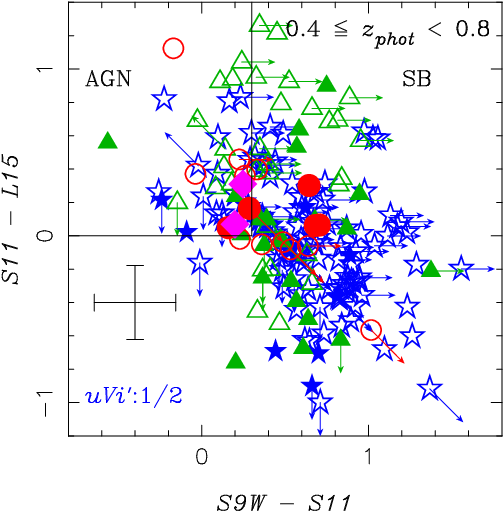}
\end{minipage}
\begin{minipage}{0.49\linewidth}
\includegraphics[angle=0,width=\linewidth]
{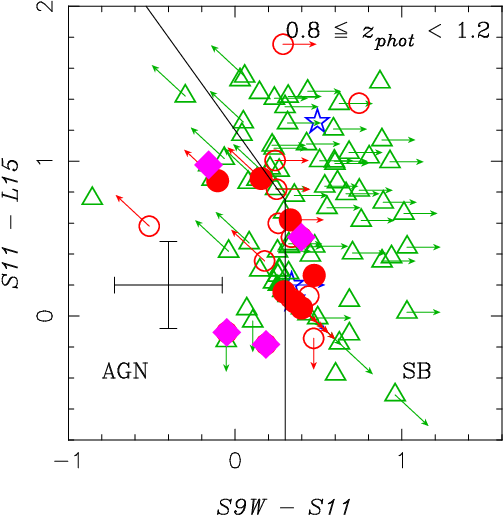}
\end{minipage}
\begin{minipage}{0.49\linewidth}
\includegraphics[angle=0,width=\linewidth]
{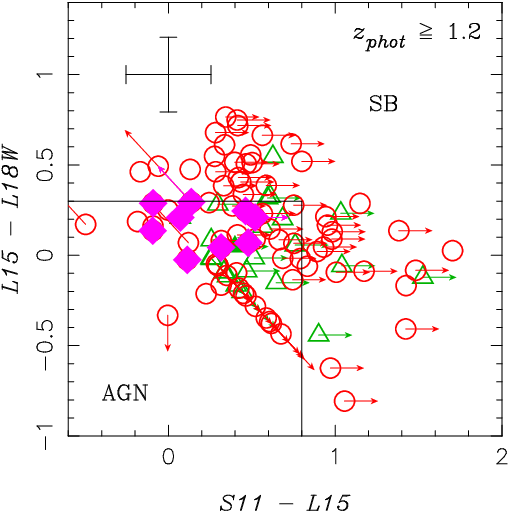}
\end{minipage}
\caption{MIR color-color diagrams for the MbN3Rs, which are also 
subclassified into s/p/qp-BBG/N3Rs and BL AGNs.  
Blue stars, green triangles, and red circles represent 
s(p)-uVis, s(p)-(uRJs+uRN3s+BRN3s), and
s(p)-(BzKs+uzN3s+BzN3s), respectively.  
Diamonds represent the spectroscopic BL AGNs.  Top-Left: The $S9W-S11$
vs. $S11-L15$ diagram for the observed MbN3Rs at $0.4 \le z <0.8$.   
Top-Right: The same as the Top-Left, for the MbN3Rs at $0.8 \le z <1.2$ 
with the tracks in $0.8 \le z<1.4$.  Bottom: The $S11-L15$ 
vs. $L15-L18W$ diagram for the MbN3Rs at $ 0.8 \le z < 1.2$ 
with the tracks in $1.0 <z<2.8$.}
\label{fig:col_col_mir_mbn3r}
\end{figure}  

\begin{figure}[ht]
\begin{minipage}{0.49\linewidth}
\includegraphics[angle=0,width=\linewidth]
{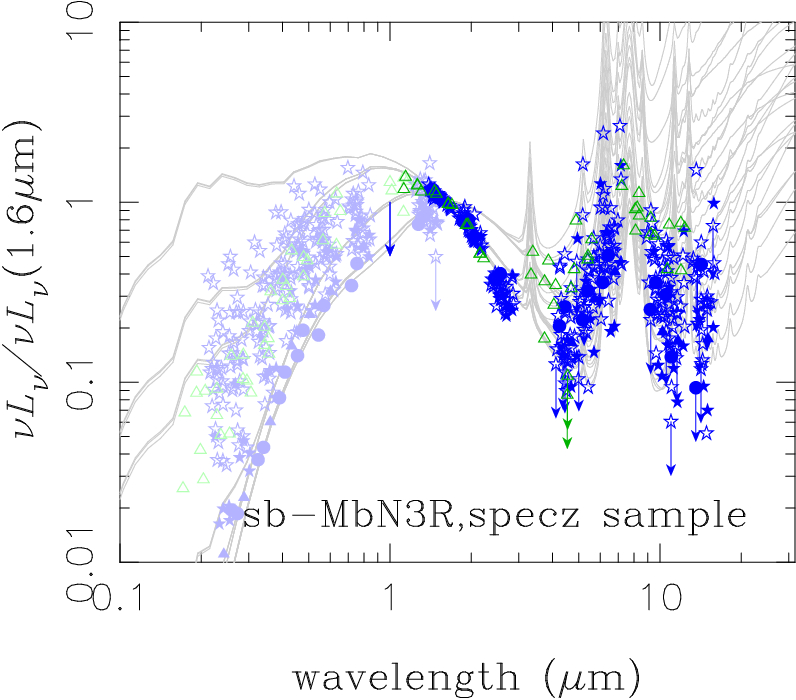}
\end{minipage}
\begin{minipage}{0.49\linewidth}
\includegraphics[angle=0,width=\linewidth]
{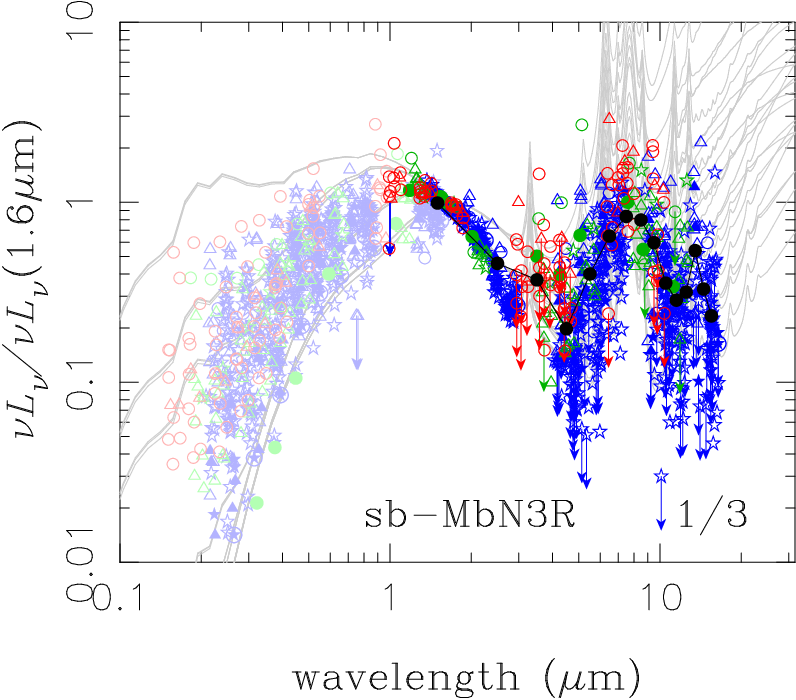}
\end{minipage}
\begin{minipage}{0.49\linewidth}
\includegraphics[angle=0,width=\linewidth]
{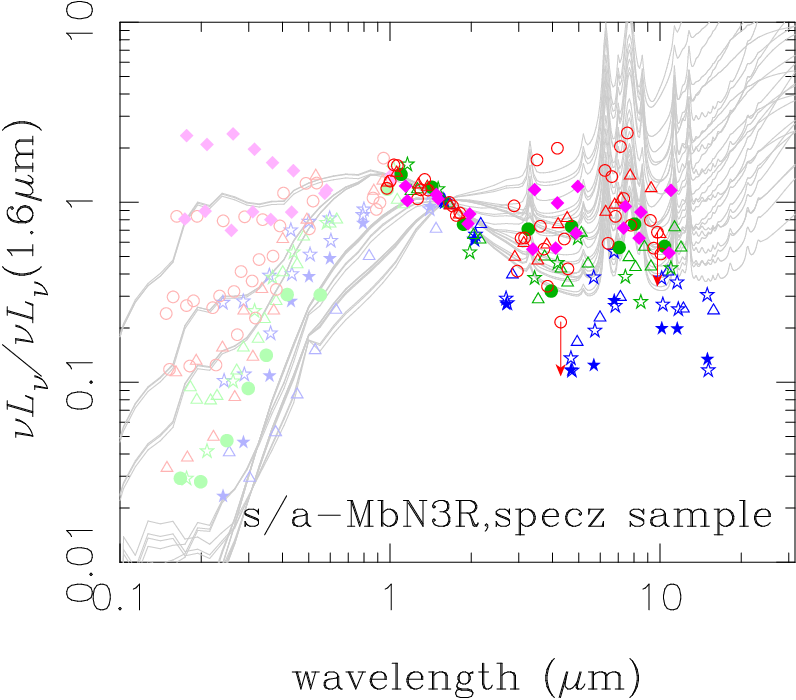}
\end{minipage}
\begin{minipage}{0.49\linewidth}
\includegraphics[angle=0,width=\linewidth]
{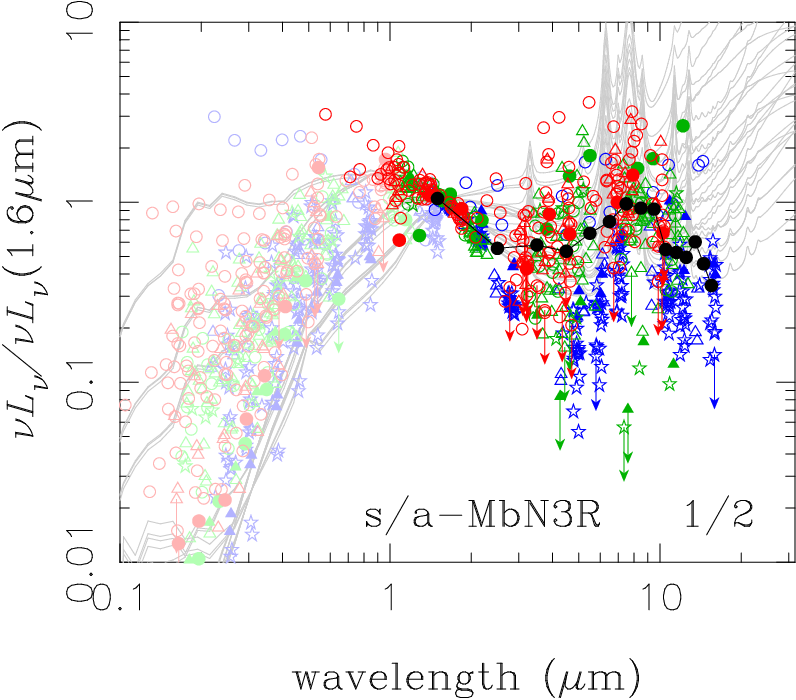}
\end{minipage}
\begin{minipage}{0.49\linewidth}
\includegraphics[angle=0,width=\linewidth]
{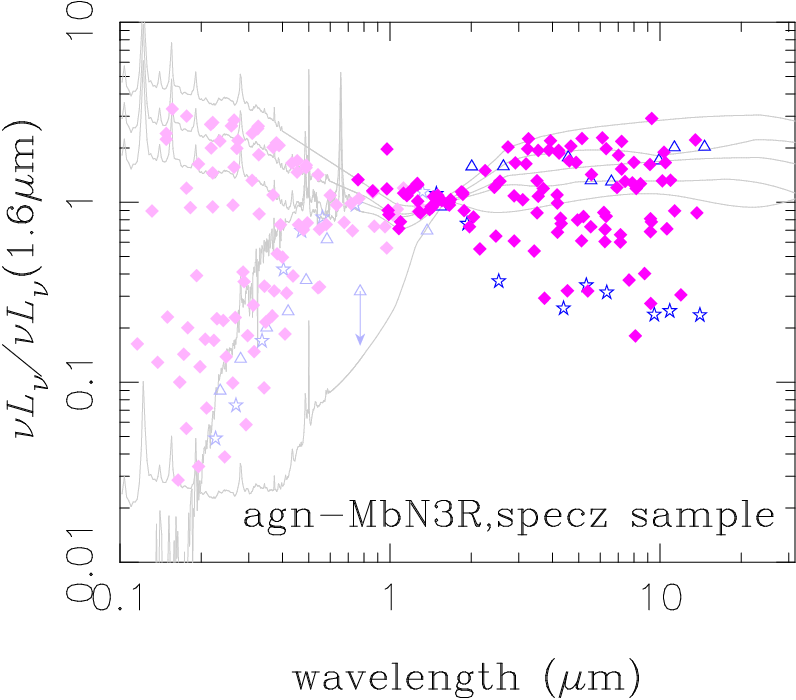}
\end{minipage}
\begin{minipage}{0.49\linewidth}
\includegraphics[angle=0,width=\linewidth]
{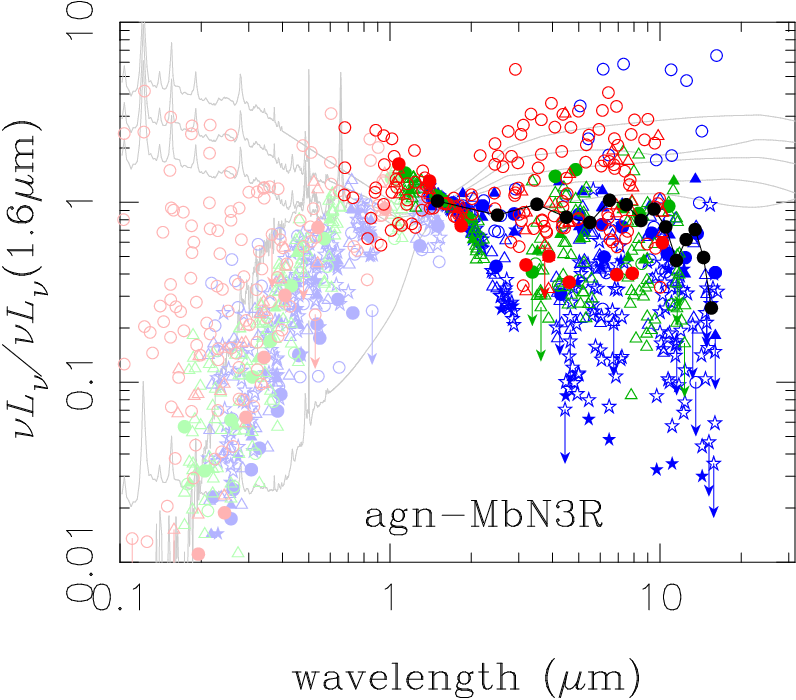}
\end{minipage}
\caption{Rest-frame SED normalized at 1.6~$\mu$m for the MbN3Rs.
Symbols are the same as figure~\ref{fig:col_col_mir_mbn3r} except for 
their colors.  Blue, green, and red symbols 
represent the redshifts of $0.4 \le
z <0.8$, $0.8 \le z < 1.2$, and $z \ge 1.2$, respectively.  
Dark and faint symbols represent AKARI IRC photometric data and ground-based 
photometric data, respectively.  
The averaged observed MIR SEDs are represented 
with solid lines in the figures on the right.  
Top-Left: 
For the sb-MbN3Rs in the spectroscopic sample.  Superimposed lines 
represent dusty starburst SED templates of the S\&K models used 
in the MIR SED fitting.  
Top-Right: The same as on the left for the sb-MbN3Rs 
in the photometric sample.  
Middle-Left: The same as the Top-Left 
for the s/a-MbN3Rs in the spectroscopic sample.  Superimposed lines 
represent SEDs composed of the S\&K models and the dusty torus model in the 
SWIRE SED library with the ratio of 0.3 to 0.7 at 5$~\mu$m.  
Middle-Right: The same as on the left 
for the s/a MbN3Rs in the photometric sample.  
Bottom-Left: The same as the Top-Left for the agn-MbN3Rs 
in the spectroscopic sample.  Superimposed lines 
represent the SWIRE SED templates of dusty torus, QSO2, BQSO1, TQSO1, and 
QSO1 from top to below at the long wavelength.
Bottom-Right: The same as on the left 
for the agn-MbN3Rs in the photometric sample.}
\label{fig:sed16_mbn3r}
\end{figure}  

\begin{figure}[ht]
\begin{minipage}{0.49\linewidth}
\includegraphics[angle=0,width=\linewidth] 
{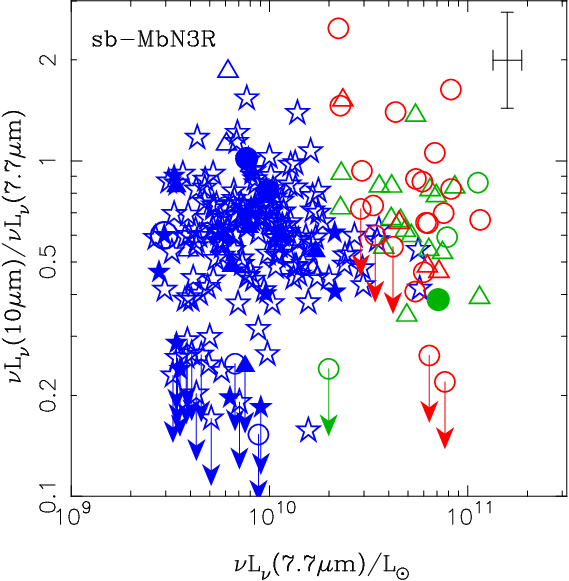} 
\end{minipage}
\begin{minipage}{0.49\linewidth}
\includegraphics[angle=0,width=\linewidth]
{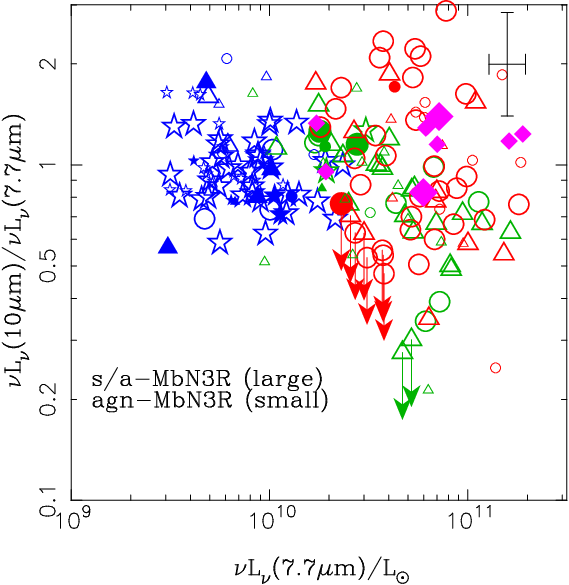} 
\end{minipage}
\begin{minipage}{0.49\linewidth}
\includegraphics[angle=0,width=\linewidth]
{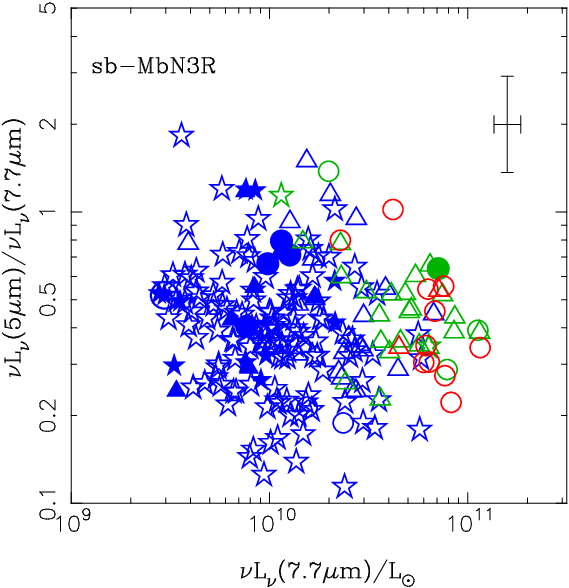} 
\end{minipage}
\begin{minipage}{0.49\linewidth}
\includegraphics[angle=0,width=\linewidth]
{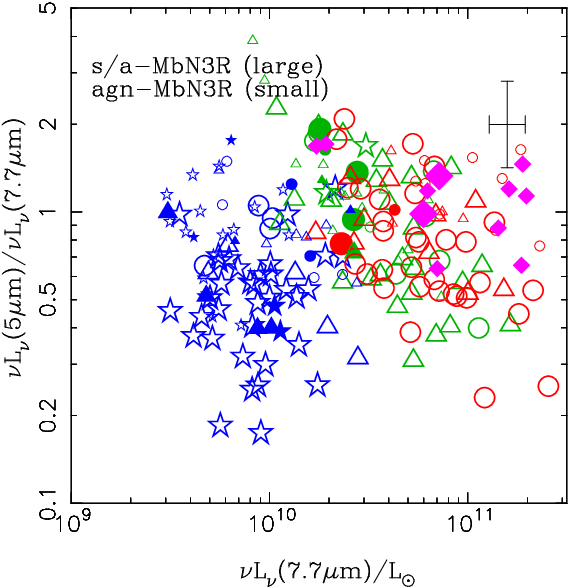}
\end{minipage}
\caption{Correlation between 
the monochromatic luminosity $\nu L_{\nu\; 7.7}$ at the rest-frame 
7.7-$\mu\mbox{m}$ and the ratio of monochromatic luminosities.  
Symbols are the same as figure~\ref{fig:col_col_mir_mbn3r} except for their 
colors and sizes. The blue, green, and red symbols 
represent the redshifts of $0.4 \le
z <0.8$, $0.8 \le z < 1.2$, and $z \ge 1.2$, respectively.    
Small symbols represent the agn-MbN3Rs classified with the IR SED fitting.   
Top: Ratio
of the monochromatic luminosity $\nu L_{\nu\;10}$ at the rest-frame
10~$\mu$m to the $\nu L_{\nu\; 7.7}$. 
Bottom: The ratio of
the monochromatic luminosity $\nu L_{\nu\; 5}$ at the rest-frame
5~$\mu$m to the $\nu L_{\nu\; 7.7}$ for the MbN3Rs.  
Left: For the sb-MbN3Rs.  
Right: For the s/a- and the agn-MbN3Rs.  
Large and small symbols represent the s/a-
and agn-MbN3Rs, respectively.  
}
\label{fig:lum77_pahratio_mbn3r}
\end{figure}  

\begin{figure}[ht]
\begin{minipage}{0.49\linewidth}
\includegraphics[width=\linewidth]
{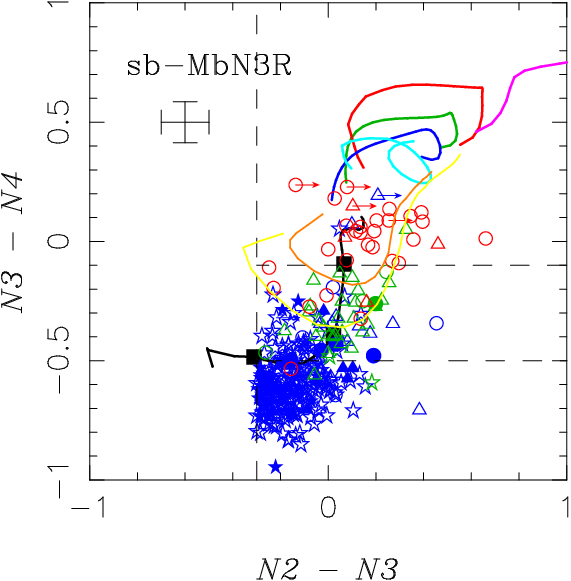}
\end{minipage}
\begin{minipage}{0.49\linewidth}
\includegraphics[width=\linewidth]
{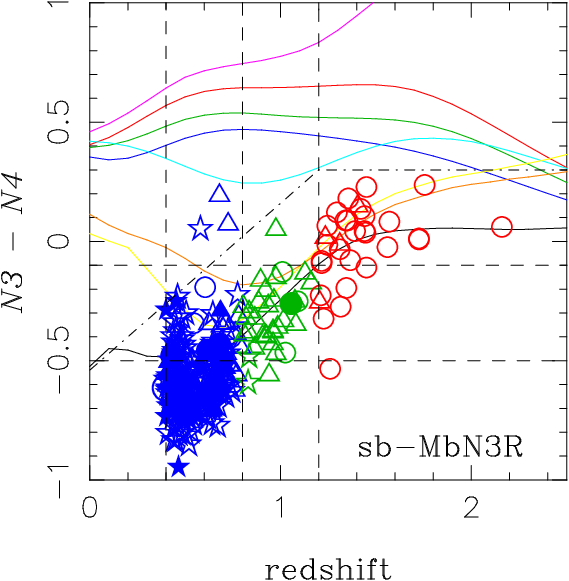}
\end{minipage} 
\begin{minipage}{0.49\linewidth}
\includegraphics[width=\linewidth]
{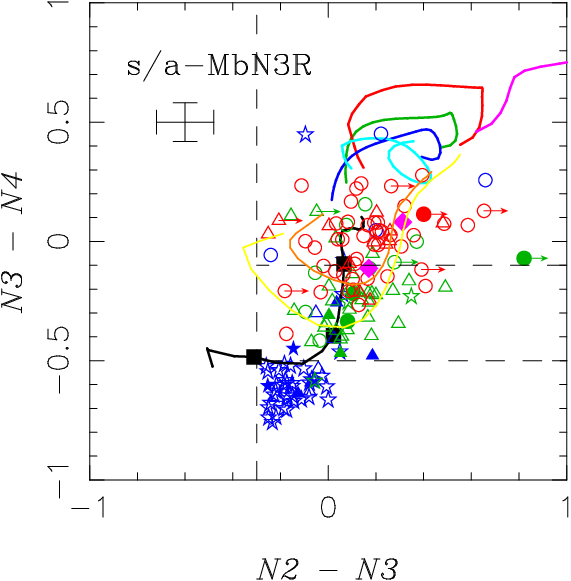}
\end{minipage}
\begin{minipage}{0.49\linewidth}
\includegraphics[width=\linewidth]
{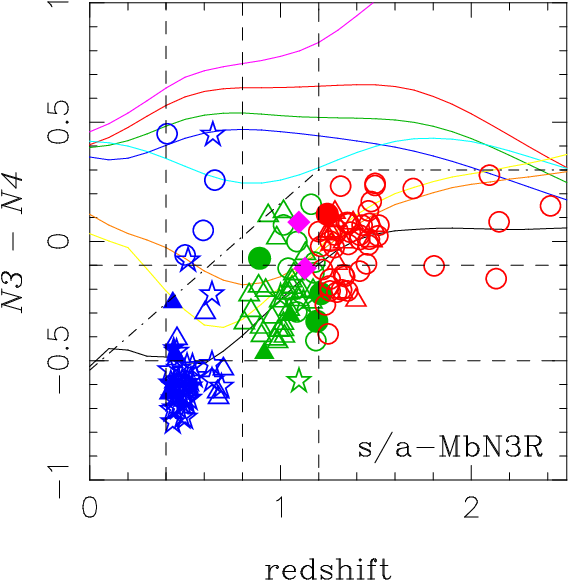}
\end{minipage}
\begin{minipage}{0.49\linewidth}
\includegraphics[width=\linewidth]
{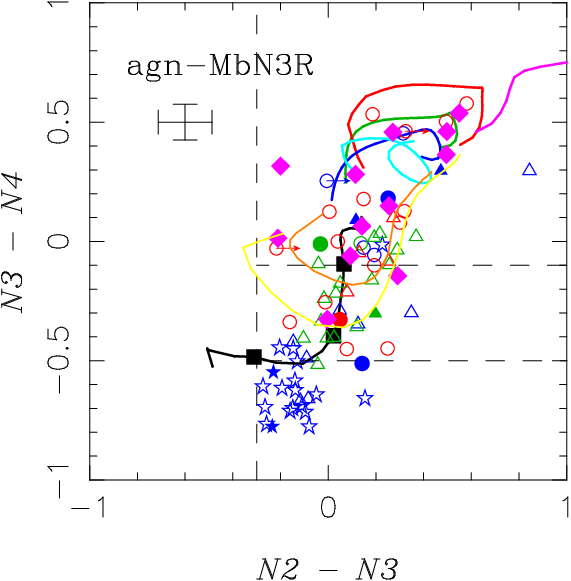}
\end{minipage}
\begin{minipage}{0.49\linewidth}
\includegraphics[width=\linewidth]
{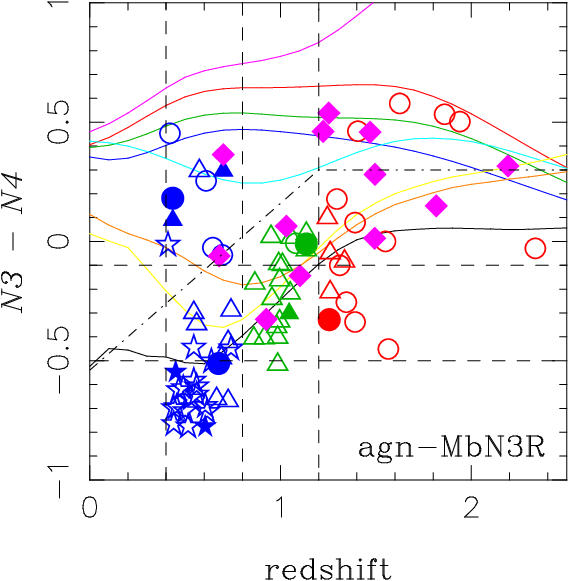}
\end{minipage}
\caption{Left: The same as figure~\ref{fig:n234_spec_phot} for the MbN3Rs.  
Symbols are the same as figure~\ref{fig:col_col_mir_mbn3r} except for their 
colors. Blue, green, and red symbols represent objects at 
the redshifts of $0.4 \le z <0.8$, $0.8 \le z < 1.2$, and $z \ge 1.2$, 
respectively.    
Right: Redshift vs. $N3-N4$ color.  
Top: For the sb-MbN3Rs.  Middle: For the s/a-MbN3Rs.  
Bottom: For the agn-MbN3Rs.
}
\label{fig:n234_mbn3r}
\end{figure}  

\begin{figure}[ht]
\begin{minipage}{0.32\linewidth}
\includegraphics[width=\linewidth]
{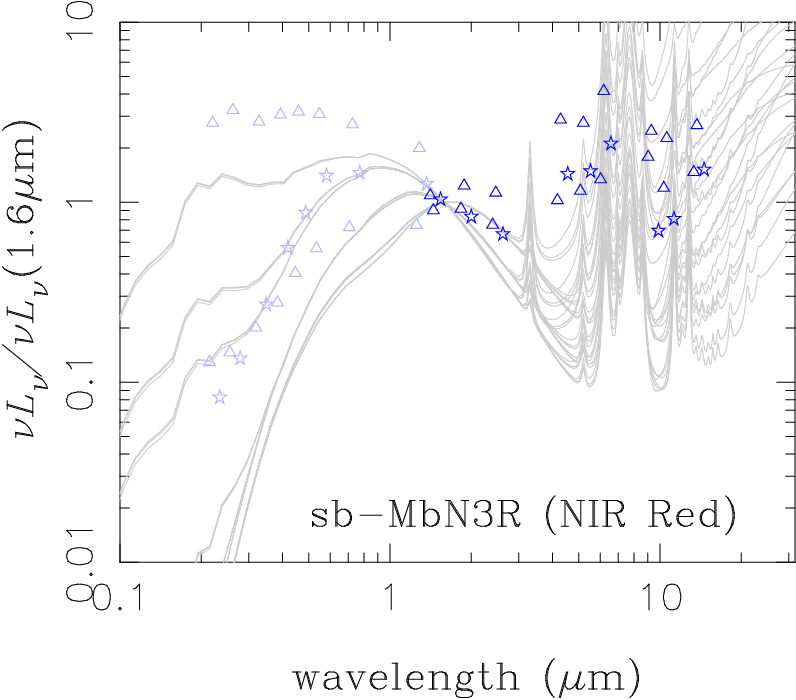}
\end{minipage}
\begin{minipage}{0.32\linewidth}
\includegraphics[width=\linewidth]
{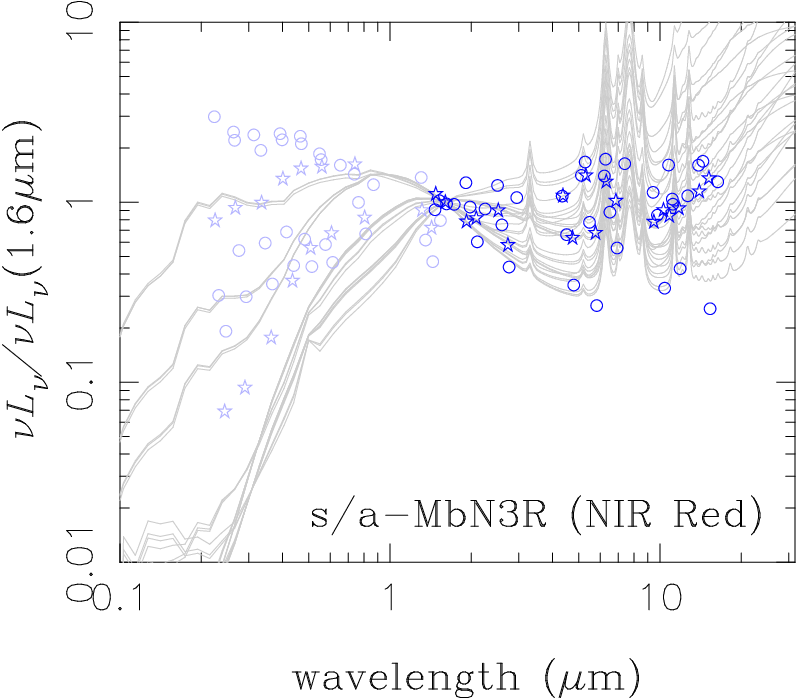}
\end{minipage}
\begin{minipage}{0.32\linewidth}
\includegraphics[width=\linewidth]
{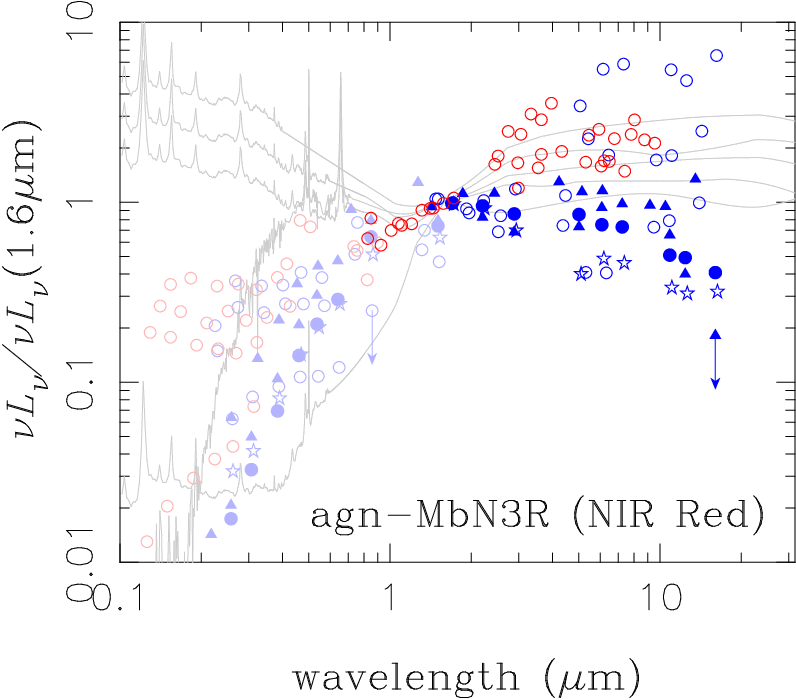}
\end{minipage}
\caption{The same as figure~\ref{fig:sed16_mbn3r} for the MbN3N4Rs.  
Left: For the sb-MbN3N4Rs.  Middle: For the s/a-MbN3N4Rs.  
Right:For the agn-MbN3N4Rs.  
}
\label{fig:sed16_mbn3n4r}
\end{figure}  

The AKARI/IRC MIR photometry, in the $S7$, $S9W$, $S11$, $L15$, $L18W$,
and $L24$ bands, is capable of tracing the features of PAH emissions at
3.3, 6.2, 7.7 and 11.3 $\mu$m and Si absorption at 10~$\mu$m 
(rest-frame) from dusty starbursts in the MbN3Rs.  
The MIR two-color diagrams of 
$S9W-S11$ vs. $S11-L15$ ($S9W/S11/L15$) and $S11-L15$ vs. $L15-L18W$
($S11/L15/L18W$) are useful to trace the redshifted PAH features as
the strong PAH 7.7 $\mu$m emission (the dip between PAH 6.2 and 7.7~$\mu$m) 
enters into the $L15 (S11)$ band for higher-redshifted dusty
starbursts in a redshift range of $z=0.4-2.6$.  In order to
characterize MIR SED features for the MbN3Rs, we present their observed
colors on two-color diagrams of $S9W/S11/L15$ and $S11/L15/L18W$ as
shown in figure~\ref{fig:col_col_mir_mbn3r}.  
In order to classify them in the MIR two-color diagram.  
we have selected $\sim 600$ MbN3Rs  
from the N3Rs with the detection of $SN>3$ in the MIR 
bands.  

Dusty starburst MbN3Rs (s-MbN3Rs) are distinguished from AGN
dominated ones (agn-MbN3Rs) with solid lines in 
figure~\ref{fig:col_col_mir_mbn3r} as the boundary between the two
populations, which are derived from the redshifted tracks of their MIR SEDs 
(see figure~\ref{fig:col_col_mir_model} in the appendix).  
On the other hand, the agn-MbN3Rs are selected 
with MIR color criteria in each of the redshift interval classes,
which are 
\begin{eqnarray}
(S9W-S11) < 0.3 \; \cap \; (S11-L15) > 0.0 
\label{eq:lirg1}
\end{eqnarray} 
at $z=0.4-1.0$, 
\begin{eqnarray}
(S9W-S11) < 0.3 \; \cap  \nonumber \\ 
\; (S11-L15) < -1.5(S11-L15) +1.2 
\label{eq:lirg2}
\end{eqnarray} 
at $z=0.8-1.4$,  
\begin{eqnarray}
(S11-L15) < 0.8 \; \cap \; (L15-L18W) < 0.3 
\label{eq:lirg3}
\end{eqnarray}
at $1.0<z<2.8$.  Assuming that the MbN3Rs are at photometric redshift
$z_{phot}$ derived in subsection~\ref{subsec:photz}, the
classification criteria at $z=0.4-1.0$, $z=0.8-1.4$, and $1.0<z<2.8$
are applied for the MbN3Rs with the photometric redshifts of $0.4 \le 
z_{phot} < 0.8$, $0.8 \le z_{phot} < 1.2$, and $ 1.2 \le z_{phot} $, 
respectively, as shown in figure~\ref{fig:col_col_mir_mbn3r}.  
The three redshift intervals 
roughly correspond to N3 Red uVis (N23 bumpers), 
uRJs (N34 bumpers), and BzKs (N4Rs) in N3 Red BBGs (IRBGs), 
respectively.  Even with substitution of the photometric selections as the 
BBGs (IRBGs) for the redshift selections, 
the MIR two-color diagrams can be also 
applied to the distinction between dusty starbursts and AGNs.   

We have also fitted this AKARI/IRC photometry for the s-MbN3Rs 
observed at the $7-24~\mu$m wavelength with the SED templates of starbursts
\citep{siebenmorgen_dust_2007} (hereafter S\&K model, see also  
appendix~\ref{app_subsec:ir_sed_models}).  
Even though we can roughly
distinguish s-MbN3Rs and agn-MbN3Rs only with the AKARI MIR colors as their
MIR emissions are dominant from dusty starburst and AGN, some AGNs may
still coexist with dusty starbursts as hidden in the s-MbN3Rs.  
In order to classify the s-MbN3Rs into two
subgroups of starburst dominants and starburst$/$AGN coexisting,
we performed the MIR SED fitting with composite SED models, in which
the dusty torus model in the SWIRE library 
is added to the dusty starburst S\&K model with the mixing rates of
$10,20,30,40,50,60,70,80,90$, and  $100\%$ in the rest-frame 5-$\mu$m 
luminosity.  We have selected sb-MbN3Rs and s/a-MbN3Rs  
with the mixing rates of the dusty torus $<50\%$ and $\ge50\%$ 
as starburst-dominants and starburst/AGN coexisting, respectively, 
derived from the MIR SED fitting.  
The MIR SED fitting results for the MbN3Rs are consistent 
with their simple MIR color
classification as MbN3Rs fitted well with the high AGN mixing rate which  
overlaped with the agn-MbN3Rs classified by the MIR color.
Figure~\ref{fig:sed16_mbn3r} shows the SEDs normalized at 1.6 $\mu$m
for sb-, s/a-, and agn-MbN3Rs where we had estimated the rest-frame
monochromatic magnitudes (luminosities) with their spectroscopic and 
photometric redshifts.  For the photometric sb-, s/a-, and agn-MbN3Rs, 
their averaged MIR SEDs are also derived as shown 
with black lines on the right side. 

Rest-frame SEDs of the sb-MbN3Rs, as shown in figure~\ref{fig:sed16_mbn3r}, 
show deep dips around 5 and 10 $\mu$m and a strong excess around 8 $\mu$m with 
weak ones around 3 and 11 $\mu$m.   
The features agree well with SEDs of dusty starburst 
S\&K models overplotted as thin lines.  It indicates not only the consistency 
for this starburst classification with MIR colors and MIR SED fittings 
but also accuracy for their photometric redshifts.   
On the other hand, the features are not so clear for the agn-MbN3Rs.  
The 5 $\mu$m dip corresponds to the transition between the stellar 
emission and the dust emission from their star-forming regions.    
The strong excess around 8 $\mu$m with weak ones around 3 and 11 $\mu$m
and the dip around 10 $\mu$m can be also naturally 
explained with the PAH 7.7, 3.3, and 11.3~$\mu$m emissions 
and Si 10~$\mu$m absorption in typical MIR emission 
from their dusty star-forming regions.  

Figure~\ref{fig:lum77_pahratio_mbn3r} shows similar trends to those which 
appeared in figure~\ref{fig:sed16_mbn3r}
as the dips at rest-frame 5 and 10~$\mu$m 
are deeper in the sb-MbN3Rs than in the s/a- and agn-MbN3Rs.  The
trends can be also naturally explained as MIR SEDs of s/a-MbN3R consist
of a smooth power-law or convex component adding to an original
component with deep concave dips at 5 and 10 $\mu$m,  which agree 
well with the superimposed SEDs composed of dusty starburst S\&K models 
and the dusty torus model in the 
SWIRE SED library with the ratio of 0.3 to 0.7 at 5~$\mu$m.  
The composition of starburst and dusty torus 
is typically characterized by the monochromatic luminosity ratio of
s/a-MbN3Rs shown in figure~\ref{fig:lum77_pahratio_mbn3r}.  This is
consistent with a picture of an obscured AGN coexisiting with 
dusty star-forming region in an s/a-MbN3R which contributes to 
the additional smooth component and
the emission with the 5 and 10 $\mu$m dips in the MIR SED, respectively.  

\subsection{NIR colors of MbN3Rs}
\label{subsec:nir_mbn3r}

The NIR color diagram of figure~\ref{fig:n234_mbn3r} for the MbN3Rs 
can also support the existence of the dusty torus hidden in the agn- 
and s/a-MbN3Rs with the excess in the $N4$ band.  
The agn- and s/a-MbN3Rs even at $z<1$ appear in the region of $N3-N4>-0.1$, 
which is the same trend as for 
the BL AGNs and expected from the tracks of the dusty torus and QSO models 
as shown in figure~\ref{fig:n234_mbn3r}.  
For stellar compornent with the IR bump, the mean $N3-N4$ color track, 
represented as solid lines in the right plots of figure~\ref{fig:n234_mbn3r},   
can be roughly approxinated as a function $N3N4(z)$ of $z$:  
\begin{equation} 
 N3N4(z)  = \left\{ \begin{array}{ll}  
                    0.74 z -0.84  &  (0.4 \le z<1.2)  \\
                    0.048         &  (z\ge 1.2)  \; 
                    \end{array} 
            \right. \; .
\label{eq:n3n4}
\end{equation} 
Thus, we have selected AGN candidates with an NIR color criterion:    
\begin{equation} 
 N3-N4  > N3N4(z) + \Delta N34 \;, 
\label{eq:n3n4r}
\end{equation} 
where we have taken a typical $N3-N4$ color error $\Delta(N3-N4)$ 
into account and introduced $\Delta N34 = 3 \Delta(N3-N4)$.   
The galaxies, selected with this additional NIR color criterion 
from the N3Rs, will be called N3N4 Red galaxies (N3N4Rs).  
The boundary of the NIR color criterion is shown as 
dotted dash lines in the right plots of figure~\ref{fig:n234_mbn3r}, where 
$ \Delta(N3-N4) = 0.1 $ as a typical $N3-N4$ color error 
for the MbN3Rs, which selects MIR bright N3N4Rs (MbN3N4Rs).  
As shown in figure~\ref{fig:sed16_mbn3n4r}, the SEDs of 
agn- and s/a-MbN3N4Rs are similar to spectroscopically confirmed BL AGNs 
in agn-MbN3Rs.  
The redder color $N3-N4$ in the agn- and s/a-MbN3N4Rs is consistent 
with the existence of a dusty torus, which dominates stellar 
components of the host.  
On the other hand, the SEDs of sb-MbN3N4Rs are 
still similar to those of dusty starbursts as most of the sb-MbN3Rs.  
The sb-MbN3N4Rs, even with the redder color $N3-N4$, 
may be still dusty starburst dominant.  Thus, the MIR SED classification 
for the sb-MbN3Rs is robust to distinguish dusty starbursts from AGNs 
while the NIR color classification can be effective only to select 
bright obscured AGNs. 

\subsection{MIR marginally-detected N3Rs}
\label{subsec:mcn3r}  

\begin{figure}[ht]
\centering 
\begin{minipage}{0.32\linewidth}
\includegraphics[angle=0,width=\linewidth]
{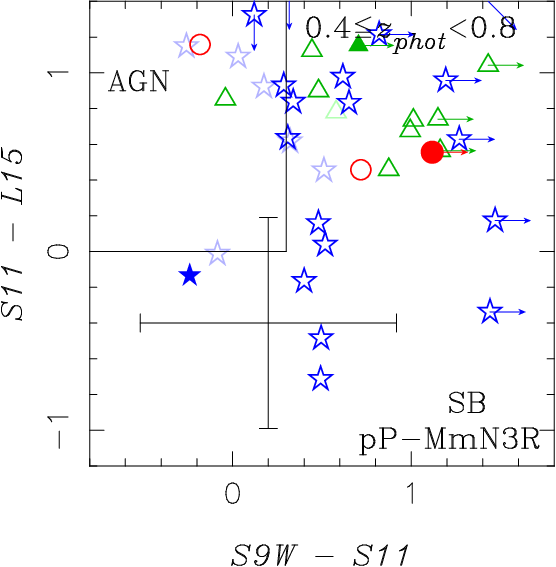}
\end{minipage}
\begin{minipage}{0.32\linewidth}
\includegraphics[angle=0,width=\linewidth]
{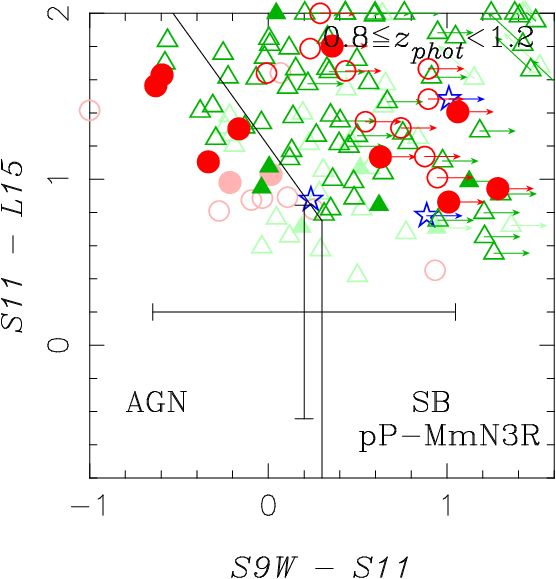}
\end{minipage}
\begin{minipage}{0.32\linewidth}
\includegraphics[angle=0,width=\linewidth]
{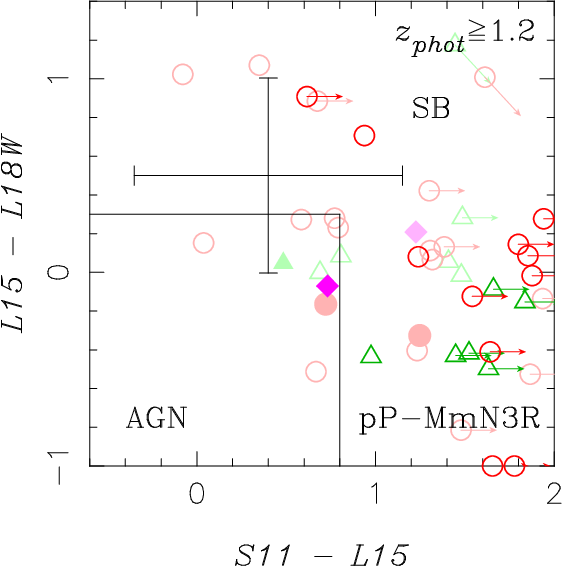}
\end{minipage}
\caption{The same as figure~\ref{fig:col_col_mir_mbn3r} for the pP-MmN3Rs. 
Dark and faint symbols represent the sb- and non sb- pP-MmN3Rs, respectively.}
\label{fig:col_col_mir_ppmcn3r}
\end{figure}  

\begin{figure}[ht]
\begin{minipage}{0.32\linewidth}
\includegraphics[angle=0,width=\linewidth]
{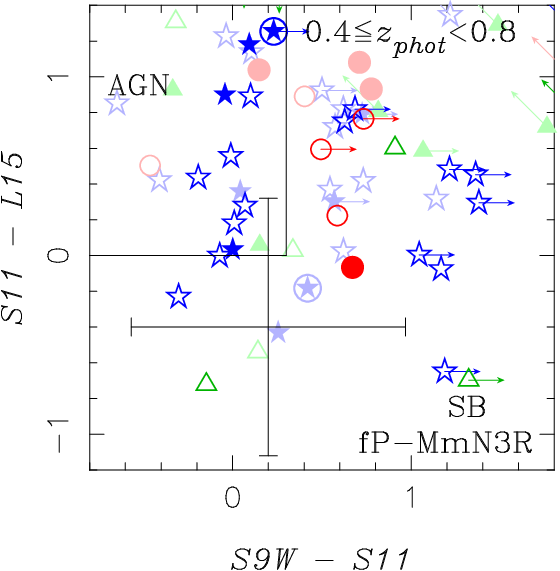}
\end{minipage}
\begin{minipage}{0.32\linewidth}
\includegraphics[angle=0,width=\linewidth]
{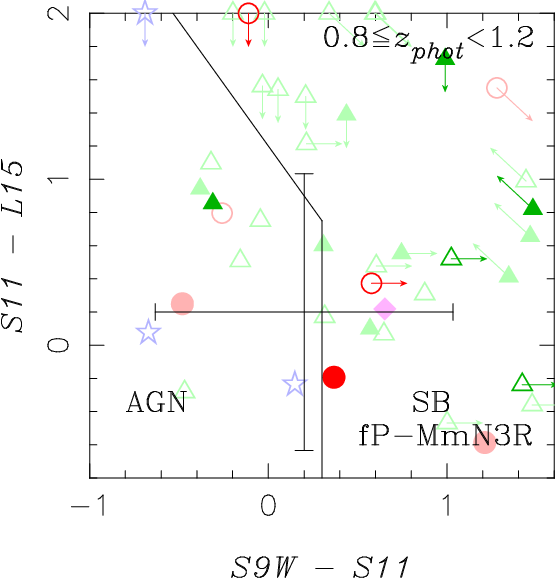}
\end{minipage}
\begin{minipage}{0.32\linewidth}
\includegraphics[angle=0,width=\linewidth]
{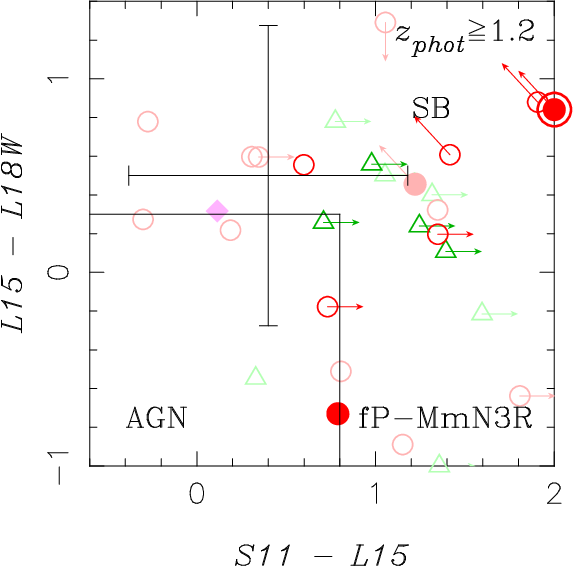}
\end{minipage}
\caption{The same as figure~\ref{fig:col_col_mir_ppmcn3r} for the fP-MmN3Rs.  
Dark and faint symbols represent sb- and non sb-fP-MmN3Rs, respectively.}
\label{fig:col_col_mir_fpmcn3r}
\end{figure}  

\begin{figure}[ht]
\begin{minipage}{0.49\linewidth}
\includegraphics[angle=0,width=\linewidth]
{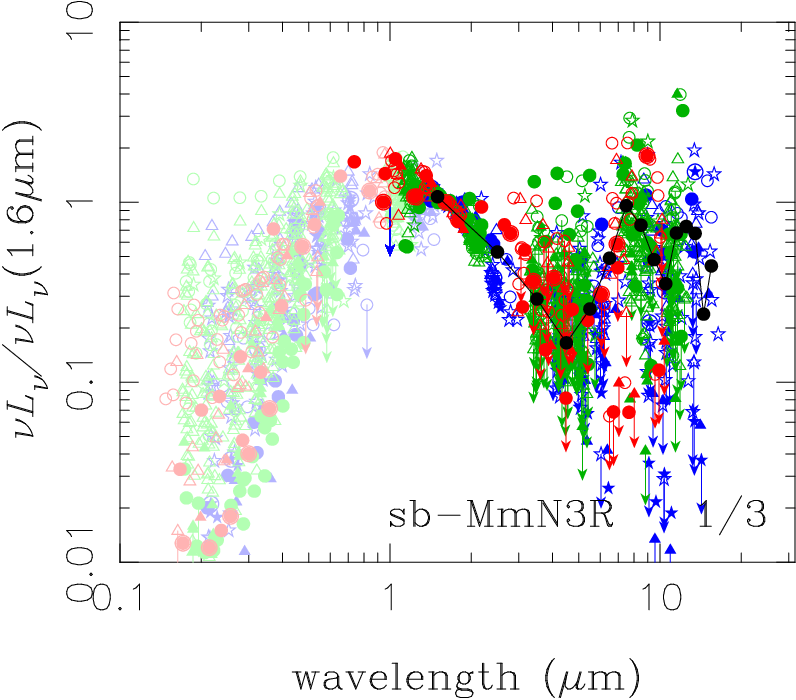}
\end{minipage}
\begin{minipage}{0.49\linewidth}
\includegraphics[angle=0,width=\linewidth]
{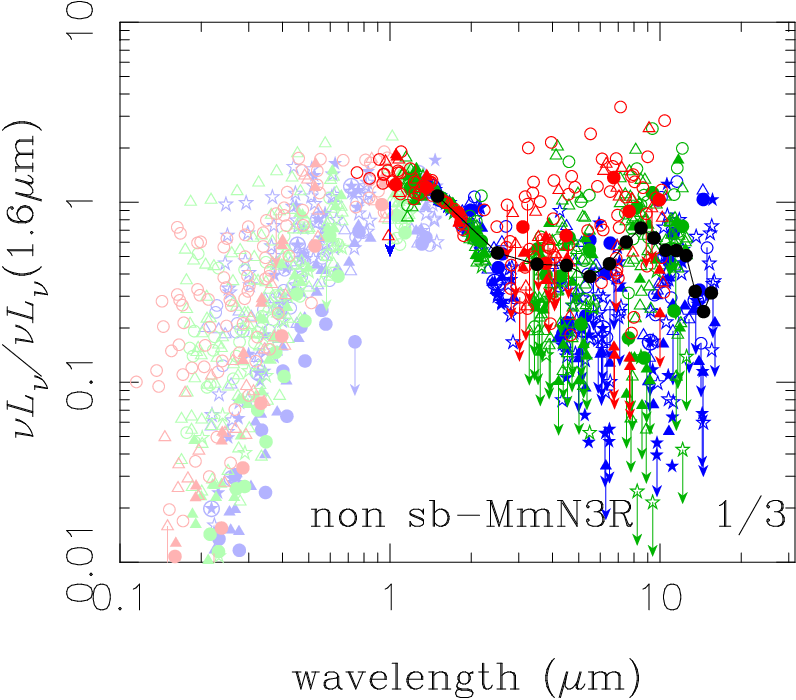}
\end{minipage}
\caption{The same as figure~\ref{fig:sed16_mbn3r} for the MmN3Rs.  
Left: For the sb-MmN3Rs.  
Right: For the non sb-MmN3Rs.}
\label{fig:sed16_mcn3r}
\end{figure}  

\begin{figure}[ht]
\begin{minipage}{0.49\linewidth}
\includegraphics[angle=0,width=\linewidth] 
{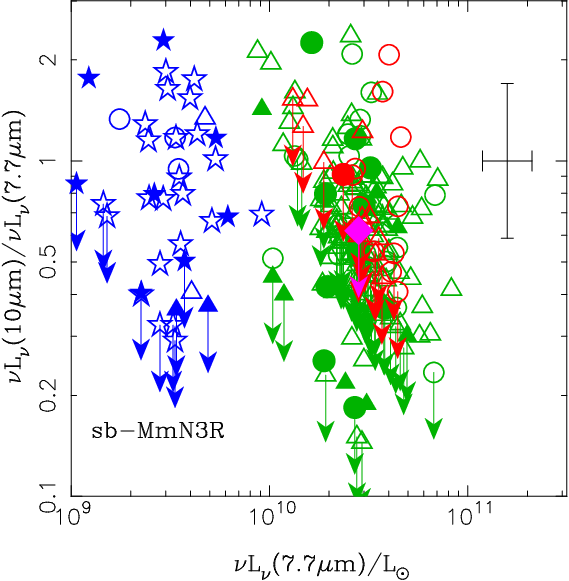} 
\end{minipage}
\begin{minipage}{0.49\linewidth}
\includegraphics[angle=0,width=\linewidth]
{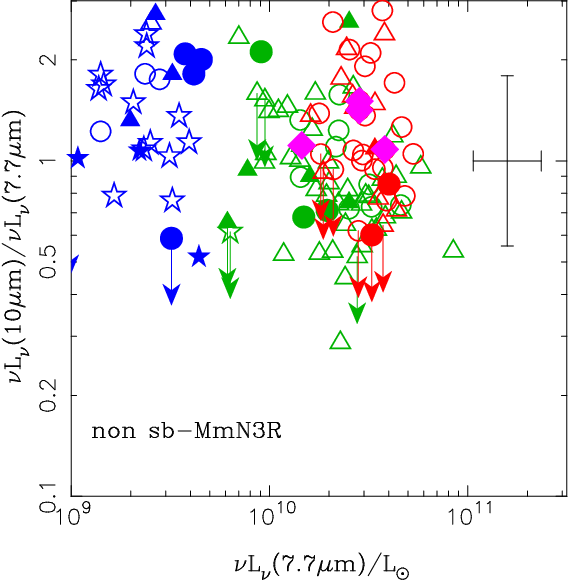} 
\end{minipage}
\begin{minipage}{0.49\linewidth}
\includegraphics[angle=0,width=\linewidth]
{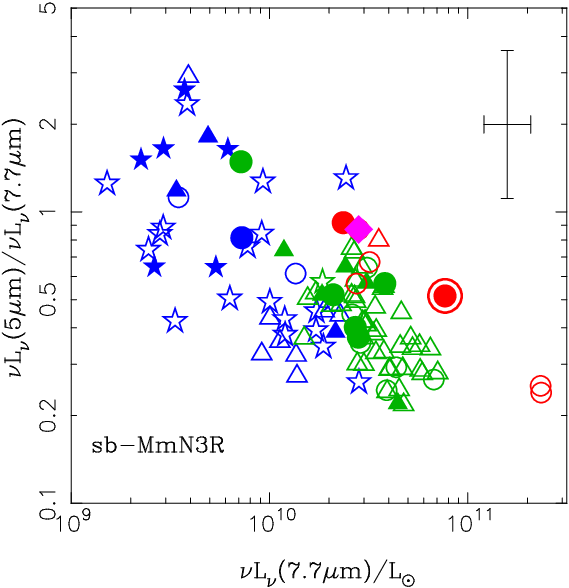} 
\end{minipage}
\begin{minipage}{0.49\linewidth}
\includegraphics[angle=0,width=\linewidth]
{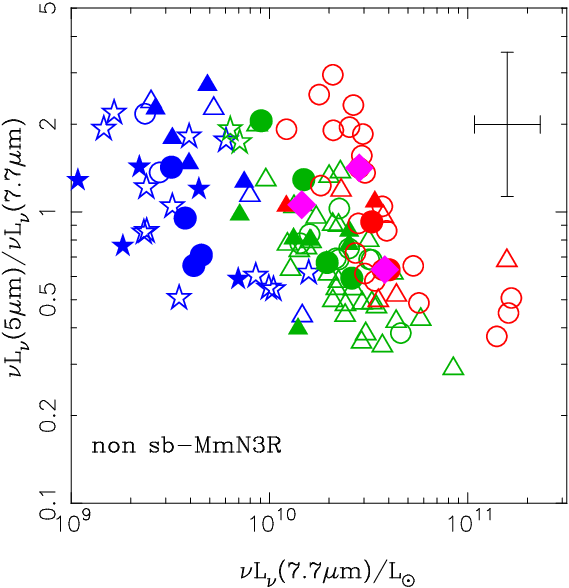}
\end{minipage}
\caption{The same as figure~\ref{fig:lum77_pahratio_mbn3r} for 
the sb- and non sb-MmN3Rs.
}
\label{fig:lum77_pahratio_mcn3r}
\end{figure} 

\begin{figure}[ht]
\begin{minipage}{0.49\linewidth}
\includegraphics[width=\linewidth]
{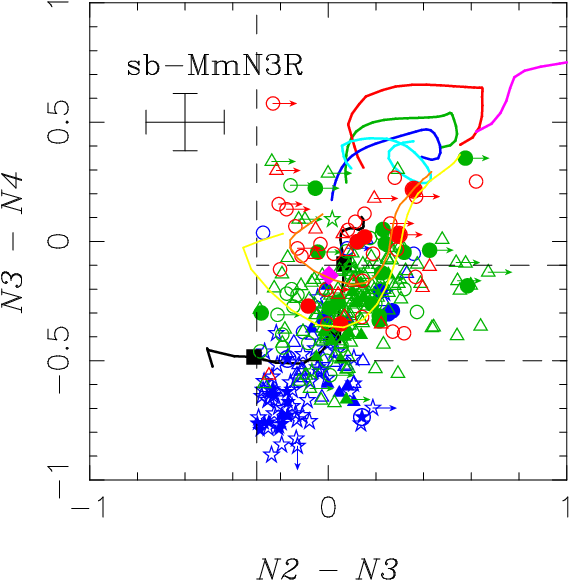}
\end{minipage}
\begin{minipage}{0.49\linewidth}
\includegraphics[width=\linewidth]
{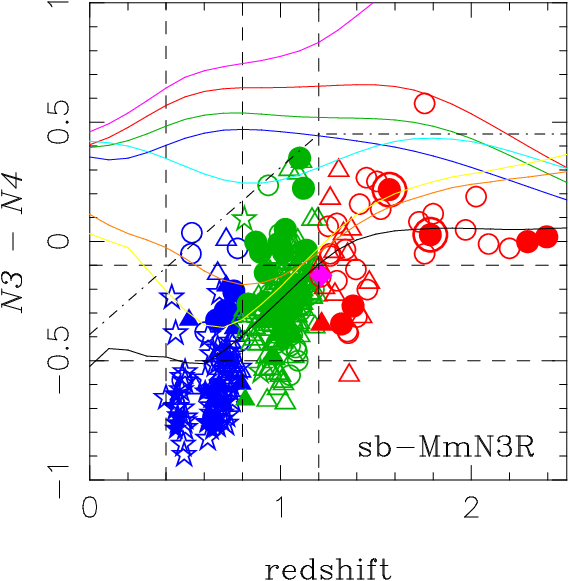}
\end{minipage}
\begin{minipage}{0.49\linewidth}
\includegraphics[width=\linewidth]
{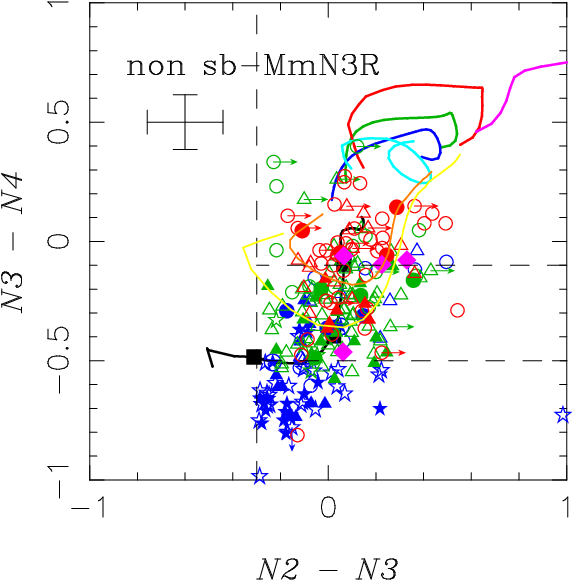}
\end{minipage}
\begin{minipage}{0.49\linewidth}
\includegraphics[width=\linewidth]
{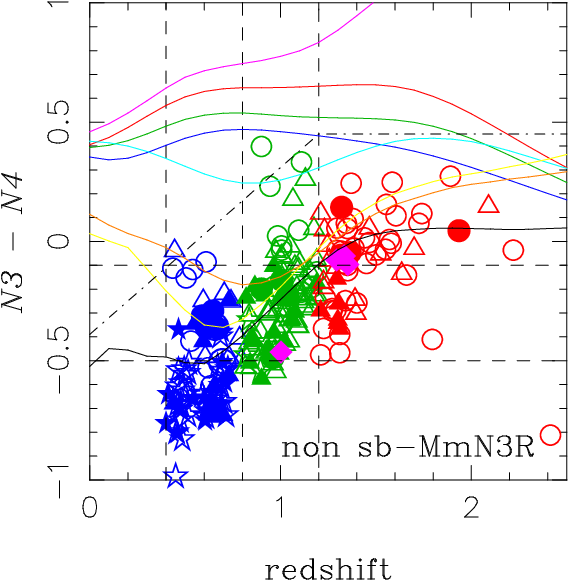}
\end{minipage}
\caption{The same as figure~\ref{fig:n234_mbn3r} for the MmN3Rs.  
}
\label{fig:n234_mcn3r}
\end{figure}  

\begin{figure}[ht]
\begin{minipage}{0.49\linewidth}
\includegraphics[width=\linewidth]
{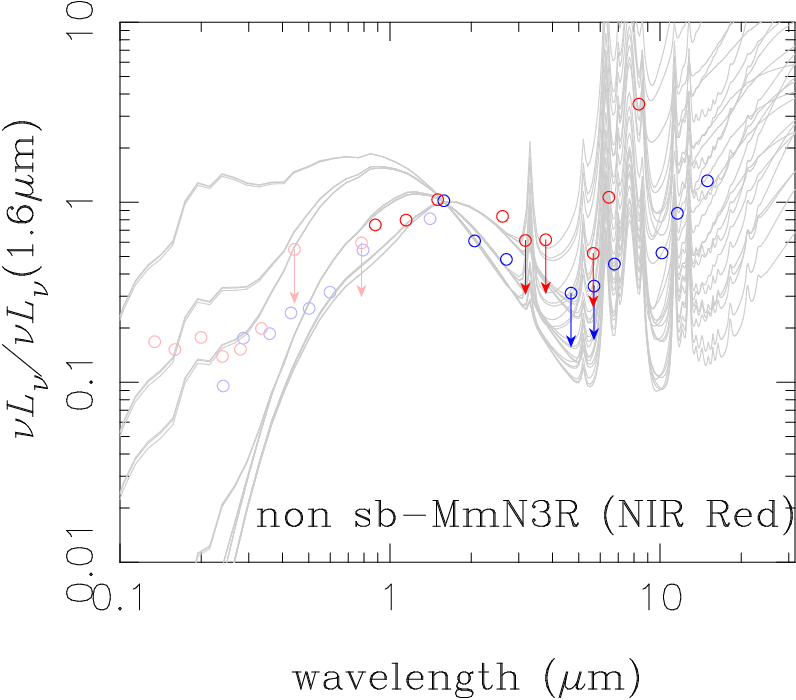}
\end{minipage}
\begin{minipage}{0.49\linewidth}
\includegraphics[width=\linewidth]
{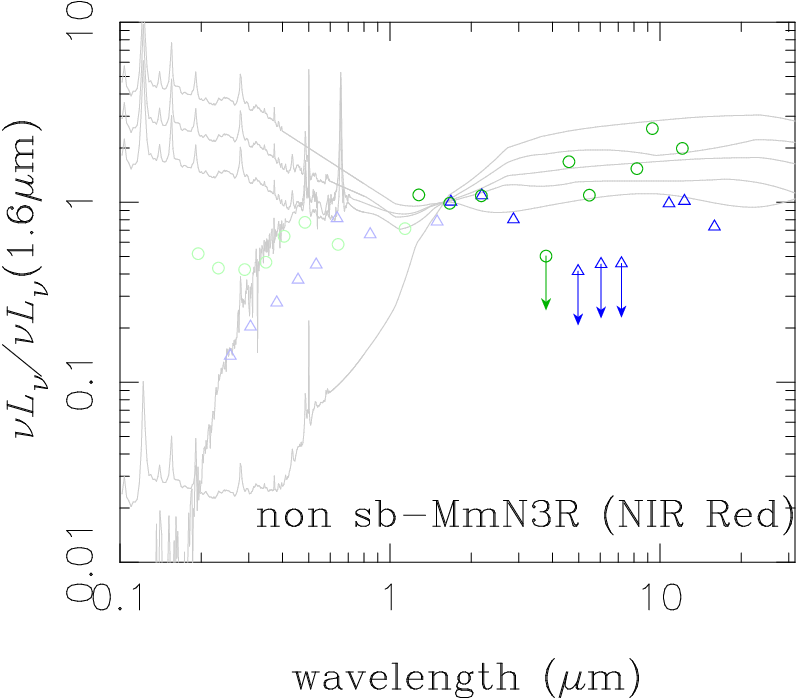}
\end{minipage}
\caption{The same as figure~\ref{fig:sed16_mbn3r} for the MmN3N4Rs.  
}
\label{fig:sed16_mcn3n4r}
\end{figure} 

We should remark that the MIR multi-band photometry with the IRC
can select unique samples for studying star formation and AGN
activity as not only the MbN3Rs but also sources detected only in some of
the S7, S9W, S11, L15, L18W, and L24 MIR bands.  
As shown above, the S11, L15, L18W, and L24 of the IRC are effective in 
detecting the 7.7 $\mu$m PAH emission from dusty starbursts at $z \ge 0.4$.  
Even though we have selected $\sim600$ MbN3Rs 
with the detection of $SN>3$ in more than two MIR bands 
in subsection~\ref{subsec:mir_sed_sb_agn}, 
we have also excluded MIR sources in the N3Rs as 
detected only in one or two MIR bands with detection of $SN>3$ 
and in other MIR bands with detection of $SN<3$, 
which are called MIR 
marginally-detected N3Rs (MmN3Rs), 
whose MIR magnitudes and colors can be still estimated 
as the MmN3Rs are mimics of the MbN3Rs.  
We selected $\sim$600 MmN3Rs, 
which are subclassified into $\sim$320 possible PAH emitting (pP-)MmN3Rs 
and $\sim$270 faint PAH emitting (fP-)MmN3Rs.  
The former can be PAH 7.7 $\mu$m   
emitters as detected in the S11, L15, and L18 bands at $z=0.4-0.65, 
0.65-1.5$, and $>1.5$, respectively, 
while the latter even have MIR emission 
but possibly not related to the PAH emissions.  
Thus, the pP-MmN3Rs and fP-MmN3Rs could be candidates for dusty starbursts 
and AGN mixtures or less-active star-forming with and without 
the PAH emissions in their selection criteria, respectively.  

Following the MIR color classifications 
in subsection~\ref{subsec:mir_sed_sb_agn}, we checked    
the difference between the pP- and fP-MmN3Rs on the MIR color-color 
diagrams as shown in figures~\ref{fig:col_col_mir_ppmcn3r} 
and \ref{fig:col_col_mir_fpmcn3r}.  
Even though the errors in the MIR colors of the MmN3Rs are large, 
most of the pP-MmN3Rs appear in the area of dusty starburst in 
the MIR color-color diagrams, which are similar to the s-MbN3Rs.  
On the other hand, 
the fP-MmN3Rs are more dispersed on the diagrams, which suggests that 
the fP-MmN3Rs mainly consist of star-forming and AGN mixtures.  
In order to classify the MmN3Rs into two
subgroups of starburst and non starburst, 
we also performed the MIR SED fitting with the same composite SED models 
as described in subsection~\ref{subsec:mir_sed_sb_agn}.  
We selected sb- and non sb-MmN3Rs with the mixing rates of the
dusty torus $< 50\%$ and $\ge 50\%$ in MIR SED fittings 
as starburst-dominants and possibly non starburst, 
which are represented as dark and thin symbols, respectively,  
in figures~\ref{fig:col_col_mir_ppmcn3r} and \ref{fig:col_col_mir_fpmcn3r}.  
Then, we categorized these starburst dominant 
MmN3Rs (sb-MmN3Rs) as mimics of sb-MbN3Rs.    

In fact, $\sim 240$ out of $\sim 320$ pP-MmN3Rs are sb-pP-MmN3Rs, 
while only $\sim 110$ out of $\sim 270$ fP-MmN3Rs are 
sb-fP-MmN3Rs.  Thus, the classifications from the MIR colors for  
pP-MmN3Rs and fP-MmN3Rs are roughly consistent with those from 
the MIR SED fittings for sb-MmN3Rs and non sb-MmN3Rs.  
Figure~\ref{fig:sed16_mcn3r} also shows averaged MIR SEDs of 
sb-MmN3Rs and non sb-MmN3Rs.  We can see that 
the averaged MIR SED of the sb-MmN3Rs 
have dusty starburst features 
with deep dips around 5 and 10 $\mu$m and excesses around 8 and 11 $\mu$m 
in the SEDs, 
which are similar to the sb-MbN3Rs as shown in figure~~\ref{fig:sed16_mbn3r}. 
We can also see these features in the luminosity ratios of 
$\nu L_{\nu,10}/\nu L_{\nu,7.7}$ and $\nu L_{\nu,5}/\nu L_{\nu,7.7}$ as shown in 
figure~\ref{fig:lum77_pahratio_mcn3r}.  
On the other hand, the averaged MIR SED and the luminosity ratio 
of the nonsb-MmN3Rs are similar to those of the s/a-MbN3Rs. 
We also tried to find AGN candidates hidden in the MmN3Rs with 
the NIR color criterion of equation~(\ref{eq:n3n4r}) as applied in the 
MbN3Rs in subsection~\ref{subsec:mir_sed_sb_agn}.   
With $ \Delta(N3-N4) = 0.15 $ as a typical $N3-N4$ color error 
for the MmN3Rs, the boundary of the criterion is represented again as 
dotted-dashed lines in figure~\ref{fig:n234_mbn3r}, which select MmN3N4Rs.  
As shown in figure~\ref{fig:sed16_mcn3n4r}, 
the sb- and nonsb-MmN3N4Rs are similar to the sb- and agn-MbN3N4Rs, 
respectively.  

Thus, the distinction between starbursters and AGNs with features in MIR/NIR 
SED is basically valid even in the MmN3Rs the same as the MbN3Rs.  Thus, 
we can treat sb-MmN3Rs as mimics of s-MbN3Rs in the following sections.     

\subsection{MIR faint N3Rs and BBGs}
\label{subsec:mfn3r}  

From the opposite point of view to subsections~\ref{subsec:mir_sed_sb_agn} 
and~\ref{subsec:mcn3r}, weak emission or non-detection 
in these MIR IRC bands are still important constraints for selecting galaxies 
with low extinction or AGNs with weak dusty torus emission.  
These MIR faint sources, having less than the limiting magnitudes  
of those in table~\ref{tab:photmt} in all the MIR bands (S9, S11, L15, L18W, 
and L24), can be classified into a subclass population 
of N3Rs, which we call as MIR faint N3Rs (MfN3Rs) 
as alternatives to not only the MbN3Rs but also the MmN3Rs.  
The MfN3Rs may in general also consist of star-forming and passive populations 
in the BBGs with weak dust emission.  
For example, passive uVis, uRJs, and BzKs without the 7.7 $\mu$m PAH emission 
at $0.4 \le z <0.8$, $0.8 \le z<1.2$, and $1.2 \le z$ 
are expected to be faint in the S11, S11/L15, and L15/L18W bands,   
respectively.  Thus, the MfN3Rs can be subclassified into 
s-and p-BBGs with the two-color diagram.  

The MfN3Rs at $z>1.2$ form a limited sample in our shallow 
survey depth in the $J,K$ and $N3$ bands, and they are mostly identical 
to MIR faint BBGs (MfBBGs),  MfN3Rs at $z<1.2$ are a part of MIR faint 
uVis (MfuVis).  
In order to study the evolution and mass dependence 
in star formation, these MfN3Rs and MfBBGs   
are also important as reference samples compared 
with the MbN3Rs and MmN3Rs for analyzing their evolutionary 
features of SFR, extinction, and $M_{\ast}$ in section~\ref{sec:evol_sfr}.   
We will sometimes refer to MfBBGs as MfBBGs not classified as MfN3Rs  
in the following.  

\subsection{Comparison with X-ray and Radio Observations}
\label{subsec:x-ray_radio} 

Before closing this section, we will quickly compare the 
MIR SED classification with recent results at X-ray and radio wavelengths.  
AGNs being harbored in the agn-MbN3Rs and the s/a-MbN3Rs 
are also indicated from 
a preliminary look at our recent Chandra X-ray Observatory (CXO) data 
on this field (Krumpe et al. in prep.; Miyaji et al. in prep), which
show that $\sim$50\% of the agn-MbN3Rs and $\sim$15\% of the s/a-MbN3Rs 
have counterparts of X-ray sources.  Thus, the trend for the 
X-ray sources also support that the MIR SED classification is effective 
for selecting AGNs.  Furthermore, as shown in figure~\ref{fig:sed16_mcn3r}, 
the nonsb-MmN3Rs are possibly harboring AGNs with dusty starbursts.    
If it is true, their X-ray emission can be detected with 
stacking analysis of their CXO data.  Thus, the CXO data can be effective 
to study AGN activities not only for the MbN3Rs but also for the MmN3Rs. 
 
We have also made a radio observation with the WSRT at 1.4~GHz of 
$\simeq 1.7$~deg$^2$ covering the field, which 
found $\simeq 500$ sources with $>0.1$~mJy in the field
~\citep{white_deep_2010}.  The radio source counts are consistent with 
a two population model with radio loud sources and 
radio faint sources with $<$ 1 mJy, in which it was 
assumed that the former and the latter are powered by AGNs and 
star-forming galaxies as in previous studies of other deep fields.  
The radio loud and the sub-mJy populations possibly overlap with 
the agn- and s-MbN3Rs, respectively.  However, as the s-MbN3Rs were   
subclassified to the starburst dominates (sb-MbN3Rs) and 
the starburst/AGN co-existence(s/a-MbN3Rs) 
in subsection~\ref{subsec:mir_sed_sb_agn}, 
the sub-mJy radio populations can be also subclassified to 
starburst dominant and starburst/AGN co-existence populations.  
Thus, it will be interesting to compare the MIR populations 
with the X-ray and radio sources, which will be studied in a future paper.  

\section{Calorimetric studies with MIR SEDs}
\label{sec:ir_lum} 

For the s-MbN3Rs and the sb-MmN3Rs classified as dusty starbursts, 
their MIR SEDs with PAH emissions and Si absorption 
should contain rich information not only as distinguished from AGNs 
but also about the physical properties in their 
star-forming regions such as their bolometric luminosities, SFRs, and 
extinction, which were derived by calorimetric schemes in 
MIR SED analysis.   

\subsection{Bolometric Luminosity and SFR}  
\label{subsec:sfr_ssfr}  

As long as AGN activity is negligible in a galaxy, 
young stars are dominant sources for the emission.  
Thus, the contribution of SFRs are nearly equal to the bolometric luminosity 
for star-forming 
galaxies.  We can estimate the total emitting power with an SED analysis 
for multi-wavelength photometric data and convert it to the SFR.    
SFRs are derived not only from intrinsic UV continuum
luminosity by assuming an extinction law for all the $z'$-detected
galaxies, but also from IR luminosity for the s-MbN3Rs and the sb-MmN3Rs.

When massive stars are not obscured by dust, SFR on a
star-forming region can be directly traced with their UV continuum
luminosity in the rest-frame wavelength 1500$\AA$.  Thus, the UV luminosity
can be translated into the SFR by the relation:
\begin{equation}
SFR_{UV} 
\left[ \mbox{M}_\odot{\rm yr}^{-1} \right] = \frac{L_{\nu}(1500\mbox{\AA})} 
{8.85\times10^{27} [{\rm erg\ s^{-1} Hz^{-1}}] } \;. 
\label{eq:madau}
\end{equation}  

When massive stars are born in dust obscured star-forming regions, their UV
radiation is reprocessed into IR emission.  In this case, 
the SFR can be traced mainly with their TIR luminosity $L_{IR}$.  
Even without using their TIR luminosity, as frequently applied, 
the intrinsic UV continuum luminosity $L_{UV;cor}$ 
can be reconstructed with a correction for the classical dust extinction 
$A_{UV;opt}$ from the observed UV luminosity $L_{UV;obs}$
(see also the details in subsection~\ref{subsec:extinctions}), and derive
$SFR_{UV;cor}$ from $L_{UV;cor}$ with equation~(\ref{eq:madau}) and/or 
directly optical-NIR SED fittings with synthesis models.  
The scheme for $SFR_{UV;cor}$ 
can be applied even for all the $z'$-detected galaxies including 
MIR faint galaxies.  

For the s-MbN3Rs and the sb-MmN3Rs, on the other hand, 
with integrating a fitted model SED from 8 to 1000 $\mu$m in the 
rest-frame with correcting for the photometric redshift $z_{phot}$,  
we can obtain the total IR (TIR) luminosity $L_{IR}$ of the emission 
from the dusty star forming regions.  
According to \citet{kennicutt_star_1998}, thus, the SFR can be estimated 
from $L_{IR}$ as
\begin{eqnarray}
SFR_{IR} \left[ \mbox{M}_{\odot} \mbox{yr}^{-1} \right] =\frac{L_{IR}}
{ 2.2 \times 10^{44} \left[ \mbox{erg s}^{-1} \right] } \; , 
\end{eqnarray}  
where we assumed the Salpeter IMF in the optical-NIR SED fitting 
as shown in subsection~\ref{subsec:photz}.  
In general, the total SFR in a galaxy is derived as a sum:  
\begin{eqnarray}
SFR_{IR+UV} = SFR_{IR} + SFR_{UV;obs} \; ,    
\end{eqnarray}
since dust can partially obscure galaxies.  It is
noted that ${\rm SFR}_{UV;obs}$ is derived from the observed UV luminosity
$L_{UV;obs}$ without extinction correction.  

\begin{figure}[ht]
\begin{minipage}{0.49\linewidth}
\includegraphics[angle=0,width=\linewidth]
{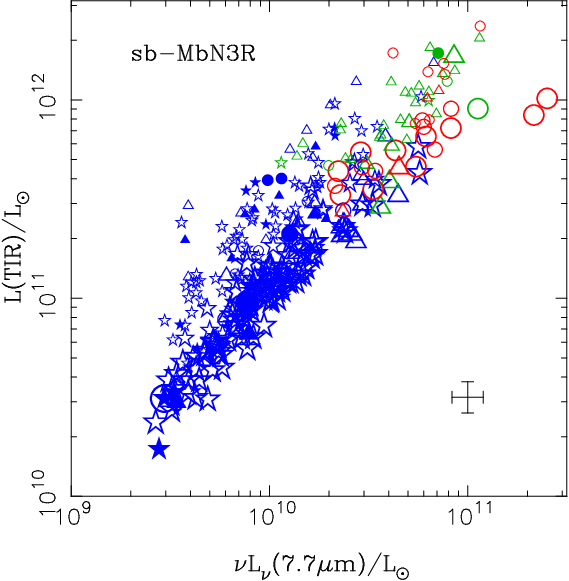} 
\end{minipage}
\begin{minipage}{0.49\linewidth}
\includegraphics[angle=0,width=\linewidth]
{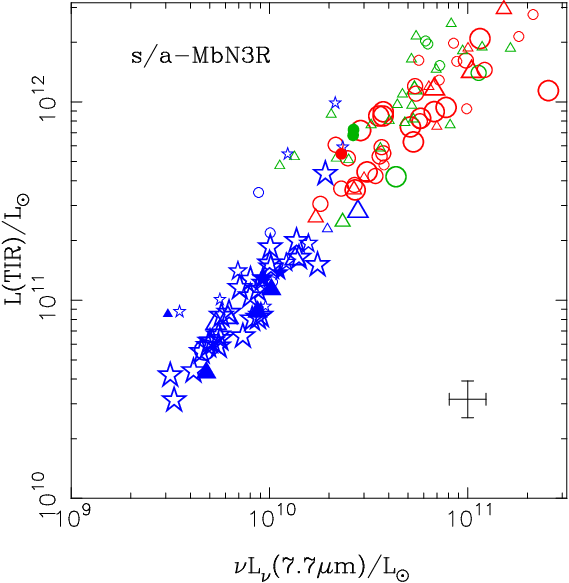} 
\end{minipage}
\begin{minipage}{0.49\linewidth}
\includegraphics[angle=0,width=\linewidth]
{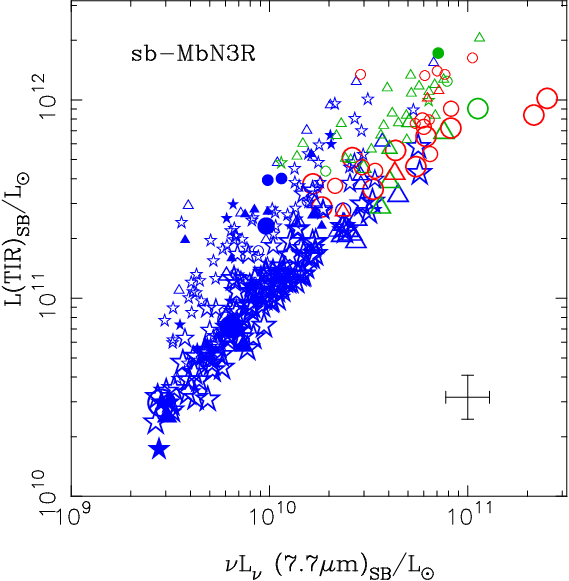} 
\end{minipage}
\begin{minipage}{0.49\linewidth}
\includegraphics[angle=0,width=\linewidth]
{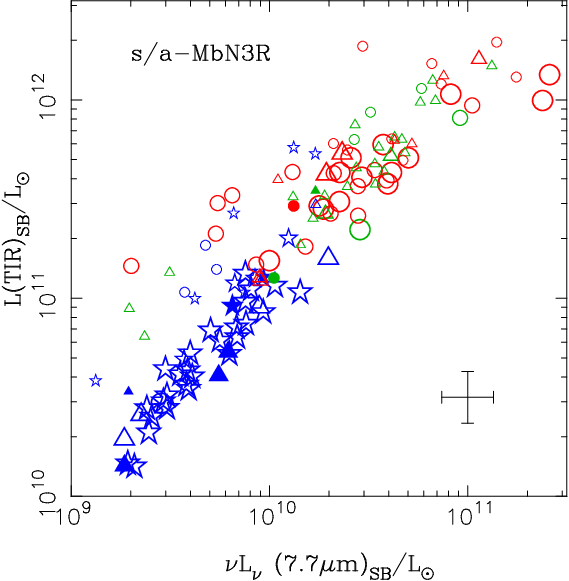} 
\end{minipage}
\caption{Rest-frame 7.7-$\mu$m monochromatic luminosity
$\nu L_{\nu\; 7.7~\mu\mbox{m}}$ vs. the TIR luminosity $L_{IR}$ derived from 
the IR SED fittings for the s-MbN3Rs.  
Symbols are the same as in figure~\ref{fig:col_col_mir_mbn3r} except for their 
colors and sizes. Blue, green, and red symbols represent 
objects at the redshifts of $0.4 \le
z <0.8$, $0.8 \le z < 1.2$, and $z \ge 1.2$, respectively.  
Large, medium, and small symbols indicate the results which 
fit well with the S\&K model SEDs of $A_{V;SK}=2.2, 6.7$, and
$17.9$, respectively.     
Top: The $L_{IR}$ derived without
excluding the contribution from the hidden AGN.  Bottom: The $L_{IR}$
derived with excluding the contribution from the hidden AGN.  Left:
For the sb-MbN3Rs.  Right: For the s/a-MbN3Rs.  }
\label{fig:lum77_tir}
\end{figure}  

\begin{figure}[ht]
\begin{minipage}{0.49\linewidth}
\includegraphics[angle=0,width=\linewidth]
{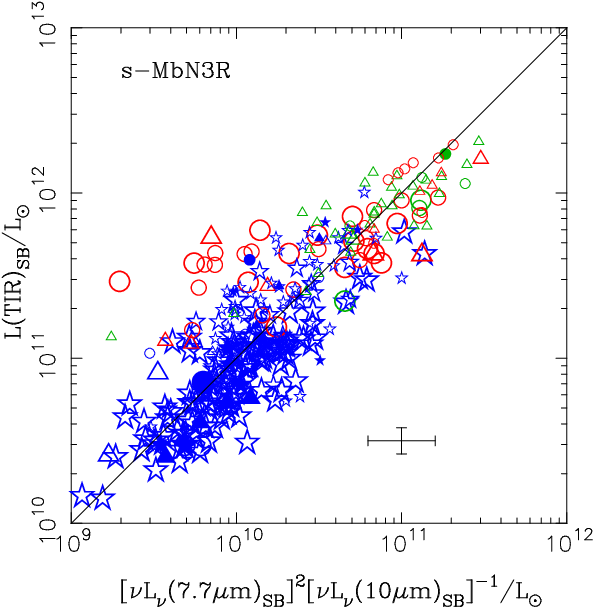} 
\end{minipage}
\begin{minipage}{0.49\linewidth}
\includegraphics[angle=0,width=\linewidth]
{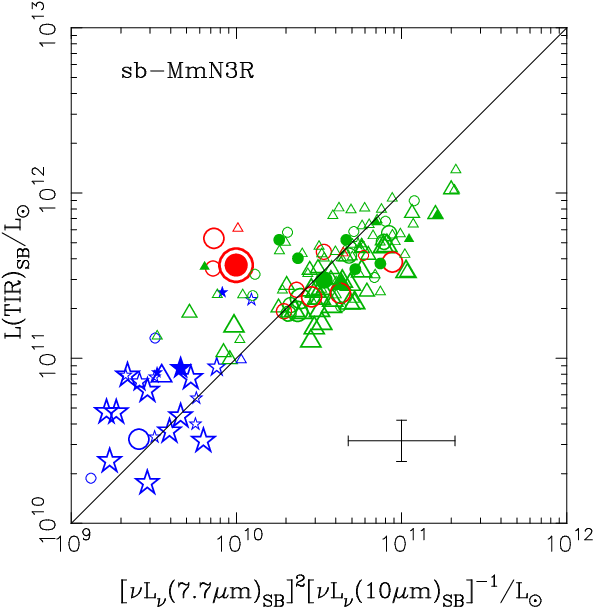} 
\end{minipage}
\caption{Relation between TIR luminosity $L_{IR}$
estimated from the S\&K model and the rest-frame luminosities of 
$L_{\nu;7.7~\mu m}$ and $L_{\nu;10~\mu m}$.  
Symbols are the same as in figure~\ref{fig:col_col_mir_mbn3r} 
except for their colors and sizes.  
Blue, green, and red symbols represent objects at 
the redshifts of $0.4 \le
z <0.8$, $0.8 \le z < 1.2$, and $z \ge 1.2$, respectively. 
Large, medium, and small symbols indicate the results which fit well 
with the S\&K model SEDs of $A_{V;SK}=2.2, 6.7$, and
$17.9$, respectively. 
The line represents 
the relation approximated with equation~(\ref{eq:lum7_lum10_lumir}).  
}
\label{fig:lum7_lum10_lumir}
\end{figure}  

\subsection{Total IR Luminosities from MIR SED}
\label{subsec:tir}

By means of the S\&K models as a library of SED templates required 
in MIR SED fittings,  we could derive TIR luminosity from the fitted 
IR SEDs for the s-MbN3Rs.  
Thus, we could obtain a relation between 
the rest-frame monochromatic 7.7~$\mu$m luminosity $\nu L_{\nu 7.7}$ 
and the TIR luminosity $L_{IR}$ for all the s-MbN3Rs including both sb-MbN3Rs 
and s/a-MbN3Rs as shown in figure~\ref{fig:lum77_tir}.  The top and bottom 
diagrams in figure~\ref{fig:lum77_tir} indicates the results in the $L_{IR}$
estimation without and with the subtraction of the obscured AGN
contribution, respectively.  
Before the subtraction of the AGN contribution, 
TIR luminosity range of the s/a-MbN3Rs is larger than those of the sb-MbN3Rs 
as shown in the top diagrams in figure~\ref{fig:lum77_tir}.  
After the subtraction, the discrepancy in TIR luminosities between 
the sb- and s/a-MbN3Rs was reduced.  This suggests that the subtraction 
is effective to derive TIR luminosities even for the s/a-MbN3Rs, which 
is also essential to estimate their SFRs 
as discussed in subsection~\ref{subsec:sfr}.   

Even though the TIR luminosity correlates well with the rest-frame
7.7-$\mu$m monochromatic luminosity $\nu L_{\nu 7.7}$ as shown in
figure~\ref{fig:lum77_tir}, this estimation of the TIR luminosity with
the MIR SED fitting seems to be still a kind of extrapolation to
obtain the Far InfraRed (FIR) emission around $\simeq 1000~\mu$m from
the NIR/MIR side only up to $\simeq 24~\mu$m with the best-fitted
model SED.  However, we would like to remark that it can be more than
an extrapolation scheme since the AKARI MIR multi-band photometry 
can certainly trace not only the PAH emission but also the
Si absorption feature as already shown in
figures~\ref{fig:sed16_mbn3r} and~\ref{fig:lum77_pahratio_mbn3r}.    
The absorption feature depends on the optical depth in galaxies 
and corresponds to the TIR luminosity as discussed in the following.  

We can see multi-sequences in figure~\ref{fig:lum77_tir}.  
The splitting into the multi-sequences is 
mainly caused by the difference in the ``mean'' optical depth, which can
be characterized as an SED model parameter of ``visual extinction'' 
$A_{V;SK}$ at the edge of dusty starburst nucleus in the S\&K model, which 
is indicated by the symbol size.  In MIR SED fittings with 
the S\&K model library, we have taken the model SEDs with
$A_{V;SK}=2.2, 6.7$, and $17.9$ corresponding to large, medium, and small 
symbols in figures, respectively.  
Even though $A_{V;SK}$ is not equivalent to the observed visual
extinction $A_V$ assuming spherical geometry for star forming regions
in the S\&K model, this parameter $A_{V;SK}$ is quantitatively related to 
the optical depth of the observed dusty starbursts as the model MIR SEDs 
with larger $A_{V;SK}$ show a deeper Si $10~\mu$m absorption feature
since it is induced with the dust self-absorption process 
in dense dusty regions.  
In fact, the s-MbN3Rs, with not only PAH emissions but also 
Si $10~\mu$m absorption, have various 
ratios of $\nu L_{\nu;10}/ \nu  L_{\nu;7.7} $ between 10 and 7.7 $\mu$m 
as already shown in figure~\ref{fig:lum77_pahratio_mbn3r}.  
Then, the observed ratio $\nu L_{\nu;10}/ \nu  L_{\nu;7.7} $ contains the 
information about the ``mean'' optical depth related to 
the denseness and the size of a dusty star-forming region, 
which is parametrized with 
the $A_{V;SK}$ in the S\&K models (see the details in 
appendix~\ref{app_subsec:ir_sed_models}).  

Thus, the MIR SED fitting with the S\&K model can analyze a variational 
degree of ``mean'' optical depth possibly related to the Si absorption depth.
The ``mean'' optical depth is another
influential parameter for estimating the TIR luminosity from the MIR SED 
fitting.  
Thus, it is expected that their TIR luminosity is correlated
not only with the rest-frame monochromatic 7.7~$\mu$m PAH luminosity 
$\nu  L_{\nu;7.7}$ 
but also with the Si absorption depth at least for the selected
starburst samples of the s-MbN3Rs and the sb-MmN3Rs as 
\begin{eqnarray}
L_{IR;SK} \simeq 10 \times \left( \frac{\nu L_{\nu 10}}{\nu L_{\nu 7.7}}
\right)^{\alpha} \times \nu L_{\nu 7.7}
\;,    
\label{eq:lum7_lum10_lumir}
\end{eqnarray}  
where the units of $L_{IR}, \nu L_{\nu 7.7}$ and $\nu L_{\nu 10}$ is
$\left[ \mbox{erg s}^{-1} \right]$ and $\alpha \simeq -1$.  
This approximate formulation can work well as shown 
in figure~\ref{fig:lum7_lum10_lumir}.   

Even though the S\&K models can reproduce MIR SEDs of the s-MbN3Rs and 
the sb-MmN3Rs well, they are produced with some simplifying assumptions 
such as 
a spherical geometry with constant dust distribution for the modeled 
starburst region.  Even though we could confirm the accuracy of the  
estimation for the TIR luminosity with complementary observations 
at longer wavelengths than MIR as FIR and submillimeter as remarked in 
section~\ref{sec:discuss},  we have not yet obtained them in the field.  
At this moment, we can not rule out a possibility that the TIR luminosity 
$L_{IR;SK}$, estimated with the fitted S\&K models, 
includes a systematic offset.   Thus, we will introduce a 
factor $f_{SK}$ in the conversion 
from the model estimated $L_{IR;SK}$ to the real $L_{IR}$:  
\begin{equation}   
L_{IR} = f_{SK} L_{IR;SK} \; .   
\label{eq:lumir_lumsk}
\end{equation}   
As long as the proportional factor $f_{SK}=0.3-1$, the estimation of $L_{IR}$ 
is consistent with previous work.  
For example, taking $f_{SK} \simeq 1$, we can roughly reproduce a relation
$L_{IR} \simeq 17.2 L_{5-8.5~\mu\mbox{m}}$ by \cite{elbaz_bulk_2002}.  
We found an offset between the S\&K model and the C\&E model, in which 
the TIR luminosity with the C\&E model can be reproduced 
with $f_{SK} \simeq 0.3$ (see figure~\ref{fig:tir_sk_ce} in the appendix).  
In the following, we take $f_{SK} = 1 $ as the default value.  

\subsection{Classical and calorimetric extinctions}  
\label{subsec:extinctions}  

\begin{figure}[ht]
\begin{minipage}{0.49\linewidth}
\includegraphics[angle=0,width=\linewidth]
{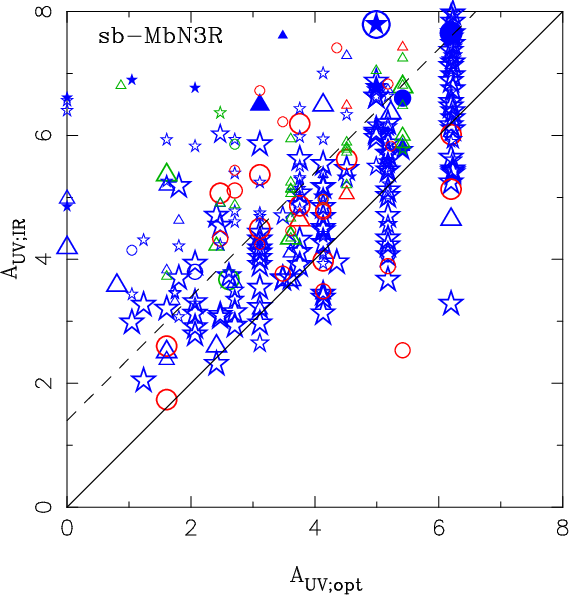}
\end{minipage}
\begin{minipage}{0.49\linewidth}
\includegraphics[angle=0,width=\linewidth]
{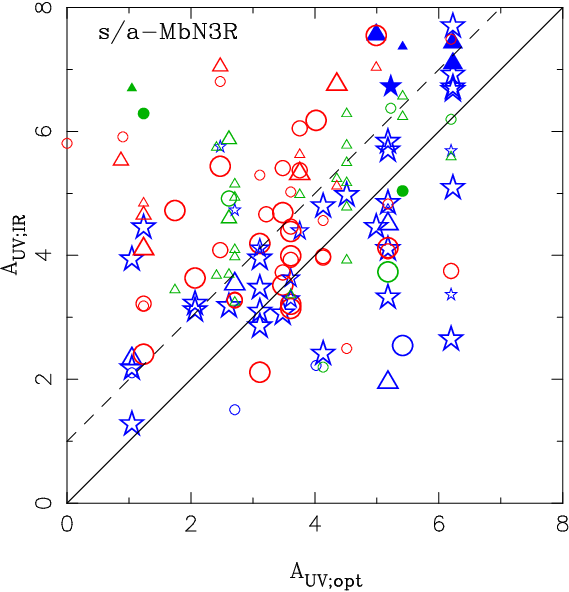}
\end{minipage}
\caption{Correlation between two extinctions at 1500\AA\ derived from 
the optical-NIR SED fitting and the luminosity ratio of the total to observed 
UV as  $A_{UV;opt}$ vs. the $SFR_{IR+UV}/SFR_{UV,obs}$.  
Symbols are the same as in figure~\ref{fig:lum7_lum10_lumir}. 
Dashed line represents a proportional law with 
equation~(\ref{eq:auv_opt_ir}), obtained with fitting the sample.  
Left: For the sb-MbN3Rs.  Right: For the s/a-MbN3Rs.}
\label{fig:auv_ir2uv_mbn3r}
\end{figure}

\begin{figure}[ht]
\begin{minipage}{0.49\linewidth}
\includegraphics[angle=0,width=\linewidth]
{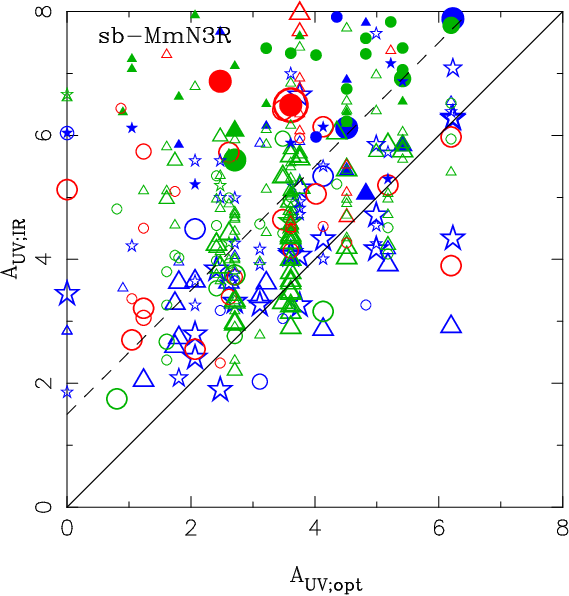}
\end{minipage}
\begin{minipage}{0.49\linewidth}
\includegraphics[angle=0,width=\linewidth]
{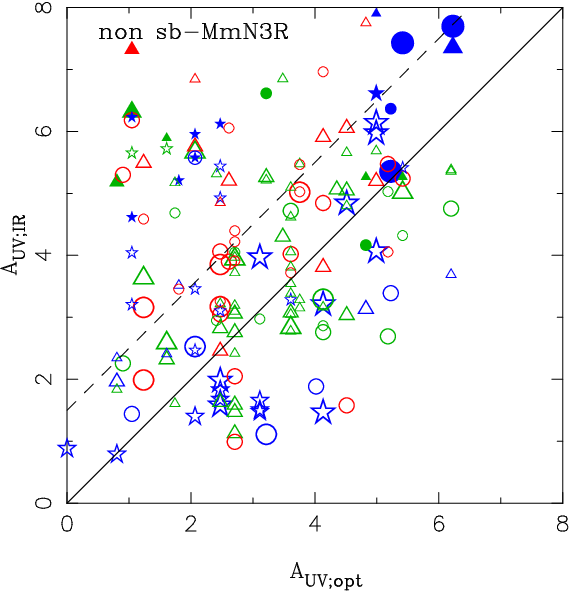}
\end{minipage}
\caption{The same as figure~\ref{fig:auv_ir2uv_mbn3r} for the MmN3Rs.  
Dashed line represents a proportional law with 
equation~(\ref{eq:auv_opt_ir}), obtained with fitting the sample.  
Left: For the sb-MmN3Rs.  Right: For the nonsb-MmN3Rs.}
\label{fig:auv_ir2uv_mcn3r}
\end{figure}

\begin{figure}[ht]
\begin{minipage}{0.49\linewidth}
\includegraphics[angle=0,width=\linewidth]
{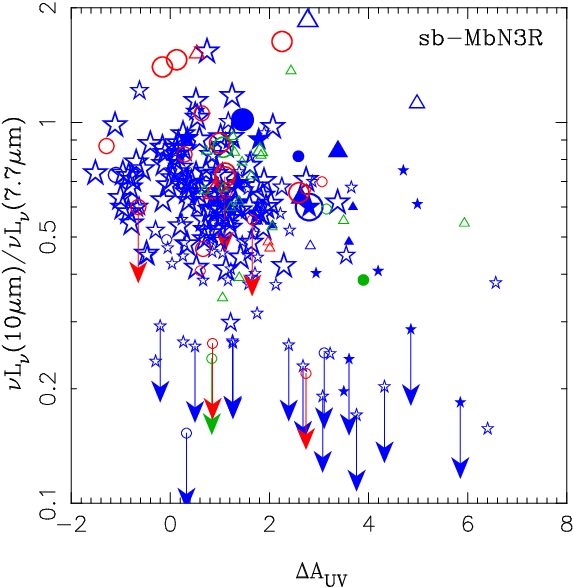}
\end{minipage}
\begin{minipage}{0.49\linewidth}
\includegraphics[angle=0,width=\linewidth]
{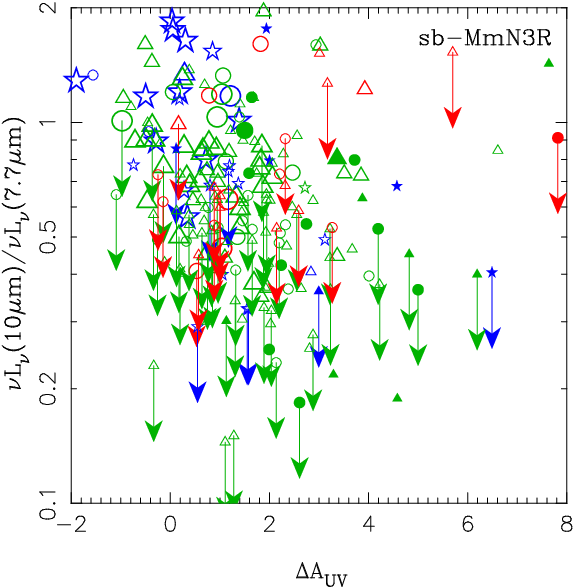}
\end{minipage}
\caption{Extinction difference $\Delta A_{UV}$ vs. the 
monochromatic luminosity ratio of $\nu L_{\nu 10}/\nu L_{\nu 7.7}$, where 
$\Delta A_{UV} = A_{UV;IR}- A_{UV;opt}$ is derived from the calorimetric   
estimated extinction 
$A_{UV;IR}=2.5\log (SFR_{IR+UV}/SFR_{UV;obs})$ and the classical 
extinction $A_{UV;opt}$ .  
Symbols are the same as in figure~\ref{fig:lum7_lum10_lumir}. 
}
\label{fig:dauv_lum10_77}
\end{figure}

\begin{table}[t]
\begin{center}
  \caption{Fitting results with equation~(\ref{eq:auv_opt_ir}). 
}
  \begin{tabular}{lccc}
  \hline
  \hline
     & $a_{A}$ 
     & $\Delta A_{UV}$ 
     & residuals \\
     & 
     & 
     & (RMS) \\
  \hline
sb-MbN3R     & $0.52\pm0.05$    & $3.3\pm0.2$  & $1.51$ \\
             & fixed 1          & $1.4\pm0.1$  & $1.69$ \\
s/a-MbN3R    & $0.41\pm0.08$    & $3.2\pm0.3$  & $1.47$ \\
             & fixed 1          & $1.0\pm0.1$  & $1.74$ \\
sb-MmN3R     & $0.22\pm0.09$    & $4.6\pm0.3$  & $2.34$ \\
             & fixed 1          & $2.0\pm0.1$  & $2.57$ \\
non sb-MmN3R & $0.29\pm0.12$    & $3.7\pm0.4$  & $2.46$ \\
             & fixed 1          & $1.5\pm0.1$  & $2.68$ \\
  \hline
  \hline
\multicolumn{4}{@{}l@{}}{\hbox to 0pt{\parbox{85mm}{\footnotesize
Values are means and $\pm$ standard deviations.
}}} 
  \end{tabular} 
\label{tab:auv_opt_ir}
\end{center}
\end{table}

In this subsection, we will derive extinctions at UV wavelength
$\simeq 1500$\AA\ with two schemes and compare them.  
One scheme is with optical-NIR SED analysis, deriving classical UV extinction 
$A_{UV,opt}$ related to common visual extinction $A_V$.  
Another is with the IR SED analysis,
deriving calorimetric extinction $A_{UV,IR}$ 
as a ratio of total emission to observed UV emission.

For all BBGs/IRBGs including galaxies even without MIR detections,
visual extinction $A_V$ can be derived from the multicolor
optical-NIR SED fitting with stellar population synthesis models as
the BC03 model.    
In a dust-screen model, the dust extinction magnitude $A_{\lambda}$
is proportional to the optical depth at a wavelength $\lambda$ to
characterize the difference between intrinsic flux $F_{int}(\lambda)$
and observed flux $F_{obs}(\lambda)$ as
\begin{eqnarray}
F_{obs}(\lambda) = F_{int}(\lambda) 10^{-0.4 A_{\lambda}} 
\; .  
\end{eqnarray}  
With selecting a reddening curve of $k(\lambda)$ from extinction models: 
the Milky way, the LMC, the SMC and Calzetti law, extinction $A_{\lambda}$
at a wavelength $\lambda$ can be represented with the visual
extinction $A_V$ or the color excess $E(B-V)$ as
\begin{eqnarray}
A_{\lambda} = \frac{k(\lambda)}{R_{V}} \cdot A_V = k(\lambda) E(B-V)   
\;.    
\label{eq:a_lambda}
\end{eqnarray}  
where $R_V=3.1, 2.72$, and 4.05 is for both the Milky way and the LMC, for
the SMC, and for the Calzetti law, respectively, and the reddening
curves are normalized as $k(B)-k(V)=1$.  
In the following, we have taken the best-fitted reddening curve 
in the optical-NIR SED fittings for each galaxy to convert the obtained 
$A_{V;opt}$ to $A_{UV;opt}=A_{1500}$ at $\lambda=$1500\AA\ as a typical 
``optically observed'' UV extinction.  

Even though the above estimation of $A_{V;opt}$ has been frequently 
applied in studying extinction of distant galaxies, it should be 
recalled that it is based on a simple dust-screen model, which 
cannot guarantee the correct description of the radiation fields 
in nuclear starbursts 
expected in the s-MbN3Rs and the sb-MmN3Rs.  
Thus, we also tried to estimate calorimetric extinction 
$A_{UV;IR}$ for the s-MbN3Rs and the sb-MmN3Rs.   
If the dust and stellar components are mixed well in a galaxy 
as nuclear starbursts in local LIRGs/ULIRGs, extinction
estimations with the IR SED analysis may be more accurate than the
common scheme for the classical extinction $A_{UV;opt}$ from visual
extinction $A_V$.  Total bolometric luminosity $L_{IR+UV}$ from the
star-forming regions in s-MbN3Rs and sb-MmN3Rs can be estimated as
$L_{IR+UV}=L_{IR}+L_{UV;obs}$, which is the sum of the TIR luminosity
$L_{IR}$ derived from the MIR photometric SED analysis in 
section~\ref{sec:ir_lum} and the observed UV luminosity
$L_{UV;obs}$ not corrected for dust extinction derived from
optical-NIR photometric SED fitting.  Using $L_{IR+UV}$ and
$L_{UV;obs}$, we can derive an alternative extinction 
with a calorimetric scheme as 
\begin{equation}
A_{UV;IR}=2.5 \log \left( \frac{L_{IR+UV}}{L_{UV;obs}} \right) \; ,  
\label{eq:auv_ir} 
\end{equation}
where we have taken the SFR ratio as the luminosity ratio 
$(L_{IR+UV}/L_{UV;obs}) \simeq (SFR_{IR+UV}/SFR_{UV;obs})$.
Figures~\ref{fig:auv_ir2uv_mbn3r} and ~\ref{fig:auv_ir2uv_mcn3r} 
show the comparison between the calorimetric extinction $A_{UV;IR}$ and 
the classical extinction $A_{UV;opt}=A_{1500}$ for the s-MbN3Rs and the 
sb-MmN3Rs, respectively, 
where we have taken $f_{SK}=1$ in the estimation of $L_{IR}$.  
We can see that the calorimetric extinction $A_{UV;IR}$ with the 
IR SED analysis shows a proportional trend for the classical extinction 
$A_{UV;opt}$, which suggests both extinctions are different, yet almost 
the same.  

However, there are some offsets between $A_{UV;IR}$ and 
$A_{UV;opt}$ as shown in figures~\ref{fig:auv_ir2uv_mbn3r} and
~\ref{fig:auv_ir2uv_mcn3r}.  We have fitted the trend with a form as    
\begin{equation}
A_{UV;IR} = a_A A_{UV;opt} + \Delta A_{UV} \; .    
\label{eq:auv_opt_ir}
\end{equation} 
As the fitting results are summarized in table~\ref{tab:auv_opt_ir},   
mean offsets of sb-MbN3Rs, s/a-MbN3Rs, and sb-MmN3Rs are significant  
compared with the RMS of residuals while those of non sb-MmN3Rs are not. 
The result suggests that the offsets of sb-MbN3Rs, s/a-MbN3Rs, 
and sb-MmN3Rs may be systematic ones.  
With fixing $a_A=1$ as a proportionality between $A_{UV;IR}$ and $A_{UV;opt}$ 
assumed for sb-MbN3Rs, s/a-MbN3Rs, and sb-MmN3Rs, the 
systematic offsets are $ \langle \Delta A_{UV} \rangle = 1.4, 1.0$, and 2.0, 
respectively, 
which are represented with a dashed line as shown in 
figures~\ref{fig:auv_ir2uv_mbn3r} and ~\ref{fig:auv_ir2uv_mcn3r}.  
There are two ways to understand the systematic offset; 1) 
the $A_{UV;IR}$ with the S\&K model tends to overestimate,  
or 2) the $A_{UV;opt}$ with the classical extinction laws tends 
to underestimate.  At least for s-MbN3Rs, the former recommends to take 
$f_{SK} \simeq 0.3$ while the latter suggests 
$f_{SK} \simeq 1$.  On-going or near-future observations 
in submillimeter and FIR wavelengths can determine the parameter $f_{SK}$ 
as discussed in section~\ref{sec:discuss}.      

The MIR SED fitting with the S\&K model 
derives not only $A_{UV;IR}$ from $L_{IR;SK}$ but also 
visual extinction $A_{V;SK}$ in the model parameters. 
In figures~\ref{fig:auv_ir2uv_mbn3r} and ~\ref{fig:auv_ir2uv_mcn3r}, 
the smaller and larger symbols represent galaxies fitted 
with larger and smaller $A_{V;SK}$, respectively.  
We can see another trend as sb-MbN3Rs with 
a larger $A_{V;SK}$ has a larger offset in extinction $\Delta A_{UV}$ due
to the variation of $A_{UV;IR}$ even in the same $A_{UV;opt}$ sample.       
This variation of $A_{UV;IR}$ can be caused by that of interstellar 
density, which also determines a parameter $A_{V;SK}$ in the S\&K models.  
As discussed in subsection~\ref{subsec:tir},  
the $A_{V;SK}$ characterizes the monochromatic luminosity
ratio between rest-frame 10 and 7.7 $\mu$m;$\nu L_{\nu 10}/\nu L_{\nu 7.7}$ 
as the effect on ``mean'' optical depth in 
$L_{IR}/\nu L_{\nu 7.7}$ represented in equation~(\ref{eq:lum7_lum10_lumir})
 (see also figures~\ref{fig:lum77_tir} and ~\ref{fig:lum7_lum10_lumir}).  
Then, it is expected that we can see some correlations between 
$\Delta A_{UV}$ and $\nu L_{\nu 10}/\nu L_{\nu 7.7}$.     
In figure~\ref{fig:dauv_lum10_77}, indeed, we can see that
sb-MbN3Rs with a larger $A_{V;SK}$ have not only a larger $\Delta A_{UV;IR}$,
but also a deeper 10-$\mu$m decrement, which may be caused by the
Si self-absorptions as remarked in subsection~\ref{subsec:mir_sed_sb_agn}.  

These trends can be naturally explained by different characteristics of 
the extinctions of $A_{UV;opt}$ and $A_{UV;IR}$ as follows.  
The calorimetric extinction $ A_{UV;IR}$ may trace the dust extinction in 
dense dust-obscured areas in star-forming regions 
as the dust column density becomes 
large enough to cause the Si self-absorption. In contrast, the classical
extinction $A_{UV;opt}$ tends to trace relatively lower optical depth
areas such as optical emissions which are mainly detected from the 
lower optical depth areas.  Thus, the extinction difference $\Delta A_{UV}$ may
be related to the geometrical variations in their star-forming regions
and dust distributions in the s-MbN3Rs and sb-MmN3Rs.   

\section{Evolution of SFRs and extinctions} 
\label{sec:evol_sfr}  

As examined in section~\ref{sec:ir_lum}, we could derive the TIR luminosity and 
the extinction for s-MbN3Rs and sb-MmN3Rs at $z=0.4-2$, which can be applied 
to estimate the SFR.  In this section, we will reconstruct 
their evolutionary features and mass dependence 
of extinction and SFR in galaxies up to $z=2$.  

\begin{figure}[ht]
\begin{minipage}{0.49\linewidth}
\includegraphics[angle=0,width=\linewidth]
{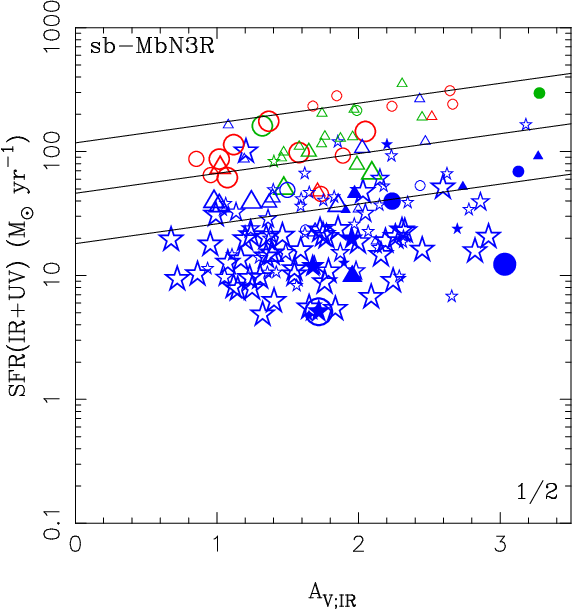}
\end{minipage}
\begin{minipage}{0.49\linewidth}
\includegraphics[angle=0,width=\linewidth]
{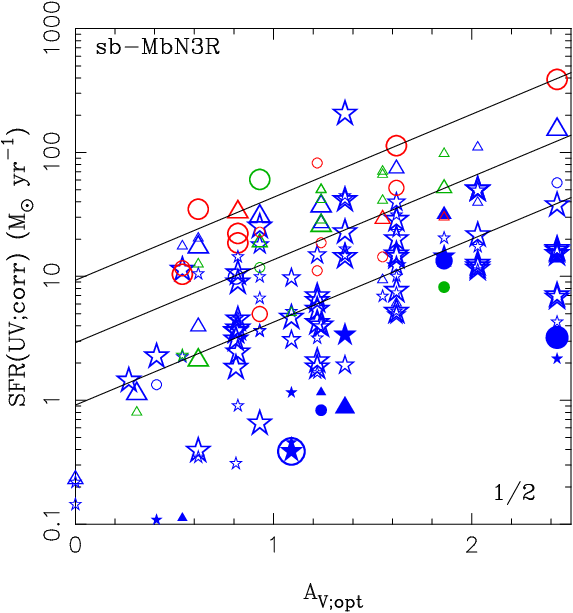}
\end{minipage}
\begin{minipage}{0.49\linewidth}
\includegraphics[angle=0,width=\linewidth]
{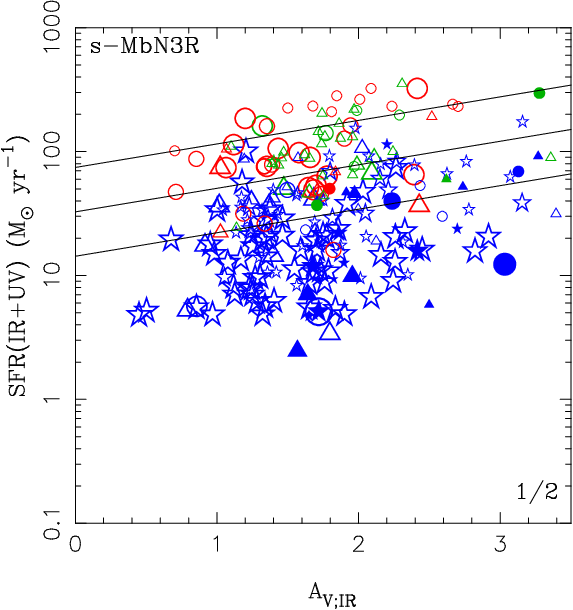}
\end{minipage}
\begin{minipage}{0.49\linewidth}
\includegraphics[angle=0,width=\linewidth]
{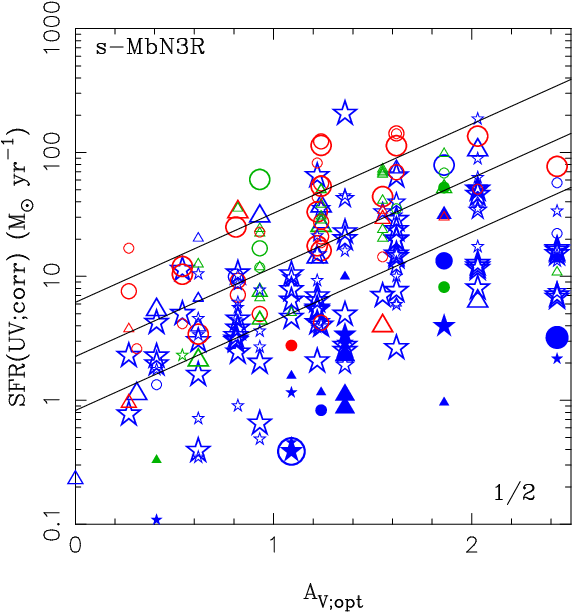}
\end{minipage}
\caption{Relation between visual extinction $A_V$ and $SFR$ for the s-MbN3Rs 
and the sb-MbN3Rs.  Symbols are the same as 
in figure~\ref{fig:lum7_lum10_lumir}. 
Solid lines represent the fitted results with equation~(\ref{eq:sfr_av_z_sm}) 
for $M_{\ast}=10^{11}M_{\odot}$ at $z=1.5, 1.0$, and $0.6$ from top to bottom.   
Top-Left: For $A_{V;IR}$ and $SFR_{IR+UV}$ of sb-MbN3Rs. 
Top-Right: For $A_{V;opt}$ and $SFR_{UV:corr}$ of sb-MbN3Rs. 
Bottom-Left: The same as the Top-Left for s-MbN3Rs.  
Bottom-Right: The same as the Top-Right for s-MbN3Rs. 
}
\label{fig:av_sfr_smbn3r}
\end{figure}

\begin{figure}[ht]
\begin{minipage}{0.49\linewidth}
\includegraphics[angle=0,width=\linewidth]
{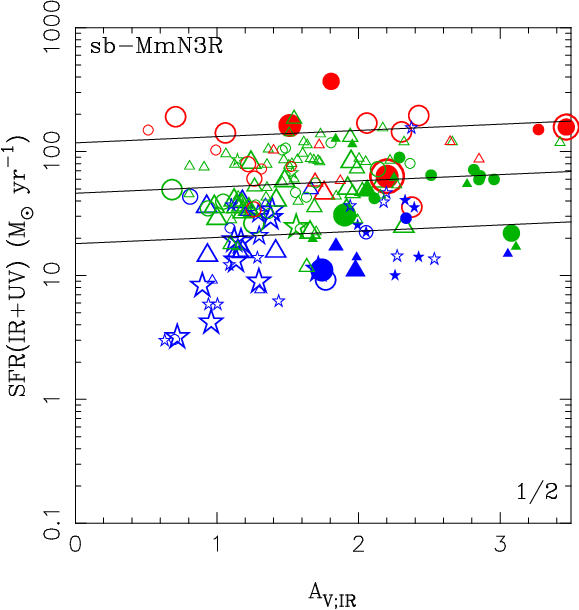}
\end{minipage}
\begin{minipage}{0.49\linewidth}
\includegraphics[angle=0,width=\linewidth]
{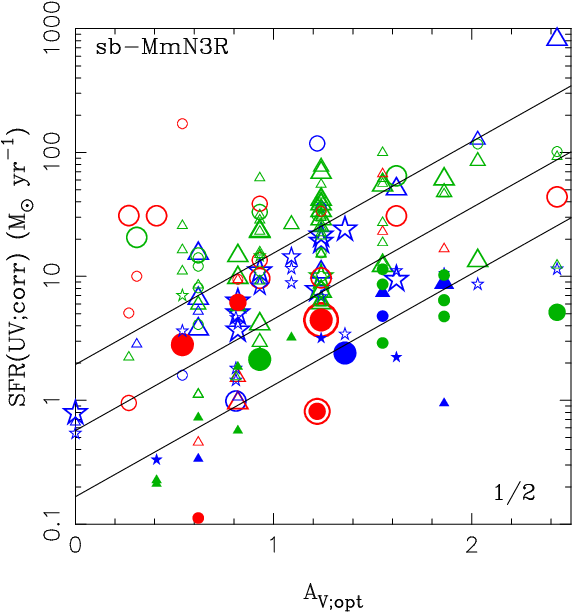}
\end{minipage}
\caption{The same as figure~\ref{fig:av_sfr_smbn3r} for sb-MmN3Rs. 
Left: For $A_{V;IR}$ and $SFR_{IR+UV}$. 
Right: For $A_{V;opt}$ and $SFR_{UV:corr}$.  
}
\label{fig:av_sfr_smcn3r}
\end{figure}

\begin{figure}[ht]
\begin{minipage}{0.49\linewidth}
\includegraphics[angle=0,width=\linewidth]
{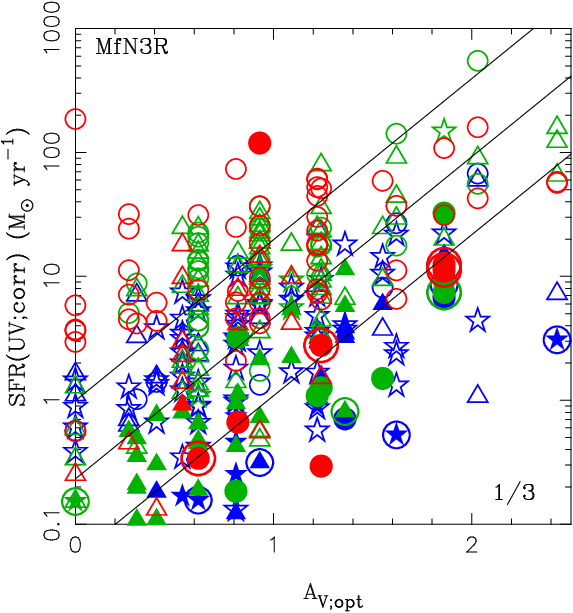}
\end{minipage}
\begin{minipage}{0.49\linewidth}
\includegraphics[angle=0,width=\linewidth]
{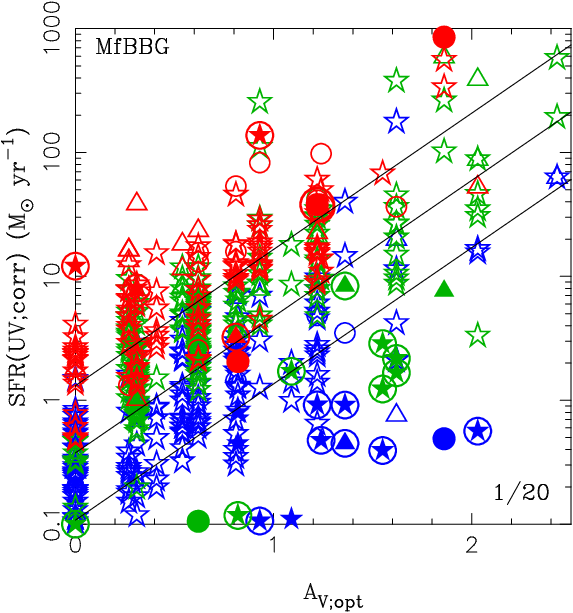}
\end{minipage}
\caption{Relation between visual extinction $A_V$ and $SFR_{UV:corr}$ 
for the MfN3Rs and the MfBBGs.  
Solid lines represent the fitted results with equation~(\ref{eq:sfr_av_z_sm}) 
for $M_{\ast}=10^{10}M_{\odot}$ at $z=1.5, 1.0$, and $0.6$ from top to bottom.   
Left: For the MfN3Rs. 
Right: For the MfBBGs. 
}
\label{fig:av_sfr_mirf}
\end{figure}

\begin{table}[t]
\begin{center}
  \caption{Fitting results with equation~(\ref{eq:sfr_av_z_sm}). 
}
  \begin{tabular}{lcccc}
  \hline
  \hline
     & $\log(\frac{SFR_0}{M_{\odot}\mbox{yr}^{-1}})$ 
     & $\alpha$ 
     & $\beta$ 
     & $\gamma$ 
  \\
  \hline
$A_{V;IR}$ & & & & \\
sb-MbN3R  & $0.4$    & $0.16$    & $4.2$    & $0.17$ \\
          & $\pm0.1$ & $\pm0.02$ & $\pm0.2$ & $\pm0.04$ \\
s-MbN3R   & $0.4$    & $0.19$    & $3.7$    & $0.13$ \\
          & $\pm0.1$ & $\pm0.02$ & $\pm0.2$ & $\pm0.04$ \\
sb-MmN3R  & $0.4$   & $0.05$    & $4.2$    & $0.16$ \\
          & $\pm0.1$ & $\pm0.02$ & $\pm0.2$ & $\pm0.03$ \\
$A_{V;opt}$ & & & & \\
sb-MbN3R & $-1.1$   & $0.67$    & $5.2$    & $-0.19$ \\
        & $\pm0.2$ & $\pm0.05$ & $\pm0.5$ & $\pm0.08$ \\
s-MbN3R  & $-1.0$   & $0.72$    & $4.5$    & $-0.21$ \\
        & $\pm0.1$ & $\pm0.04$ & $\pm0.4$ & $\pm0.06$ \\
sb-MmN3R & $-1.9$   & $0.9$    & $5.5$    & $-0.42$ \\
       & $\pm0.3$ & $\pm0.1$ & $\pm0.8$ & $\pm0.11$ \\
MfN3Rs & $-3.3$   & $1.3$    & $6.5$    & $-0.71$ \\
       & $\pm0.2$ & $\pm0.1$ & $\pm0.4$ & $\pm0.06$ \\
MfBBGs & $-2.3$   & $1.1$    & $5.6$    & $-0.19$ \\
       & $\pm0.2$ & $\pm0.1$ & $\pm0.1$ & $\pm0.06$ \\
  \hline
  \hline
\multicolumn{5}{@{}l@{}}{\hbox to 0pt{\parbox{85mm}{\footnotesize
Values are means $\pm$ standard deviations.
}}}
  \end{tabular} 
\label{tab:extinctions}
\end{center}
\end{table}

\subsection{Correlation among SFR, extinction, redshift, and mass}  
\label{subsec:sfr_ext_z_smass}  

Figures~\ref{fig:av_sfr_smbn3r}, ~\ref{fig:av_sfr_smcn3r}, 
and~\ref{fig:av_sfr_mirf} show the correlation between $A_V$ and SFR for 
s-MbN3Rs, sb-MmN3Rs, MfN3Rs, and MfBBGs.  
For all the subsamples,  we have fitted  
correlations among $A_V$, SFR, $z$ and $M_{\ast}$ with a form:  
\begin{equation}
\frac{SFR}{SFR_0} = 10^{\alpha A_V} (1+z)^{\beta} 
\left(\frac{M_{\ast}}{10^{11} M_{\odot}}\right)^{\gamma} \; ,  
\label{eq:sfr_av_z_sm}
\end{equation} 
where ${SFR_0}$ is a normalization parameter for the SFR 
in the unit of $M_{\odot} \mbox{yr}^{-1}$.  
Their best-fitting planes are determined 
in the space of 
$A_V$, $\log (1+z)$, $\log M_{\ast}$, and $\log SFR$ as summarized 
in table~\ref{tab:extinctions} and shown 
in figures~\ref{fig:av_sfr_smbn3r}, ~\ref{fig:av_sfr_smcn3r}, 
and~\ref{fig:av_sfr_mirf}.     
For the s-MbN3Rs, we can estimate both classical visual extinction 
$A_{V}$ with the SED fitting and calorimetric visual extinction 
$A_{V;IR}$ converted from the calorimetric UV extinction $A_{UV;IR}$ with 
equation~(\ref{eq:a_lambda}) for the best-fitted reddening curve 
in the optical-NIR SED fitting.  

We should remember that 
$SFR_{IR+UV}$ and $A_{V;IR}$ are estimated from their MIR SED fittings  
while $SFR_{UV;corr}$ and $A_{V;opt}$ are from their optical SED fittings.  
This denotes that the trends of $SFR_{IR+UV}$ and $A_{V;IR}$ 
are independent of those of $SFR_{UV;corr}$ and $A_{V;opt}$.  
As shown in figure~\ref{fig:av_sfr_smbn3r}, however, 
the correlation between $SFR_{IR+UV}$ and $A_{V;IR}$ 
is similar to that between $SFR_{UV;corr}$ and $A_{V;opt}$.  
Furthermore, the trends, among $SFR_{UV;corr}$, 
$A_{V;opt}$ and $z_{photz}$, except $M_{\ast}$, are similar to each other 
not only in the MbN3Rs and the MmN3Rs but also in all the subsamples including 
the MfN3Rs and MfBBGs as shown in figure~\ref{fig:av_sfr_smbn3r} 
 (see also table~\ref{tab:extinctions}).  
All of them show the same trends; 1) larger SFR, larger extinction $A_V$, and 
2) lower redshift, larger extinction $A_V$.  The second trend, increasing 
$A_V$ with the cosmic time, suggests 
that increasing dust has been 
induced with chemical enrichment in evolving galaxies.    

\subsection{Mass dependence and evolution of SFR}
\label{subsec:sfr}

\begin{figure}[t]
\begin{minipage}{0.49\linewidth}
\includegraphics[angle=0,width=\linewidth]
{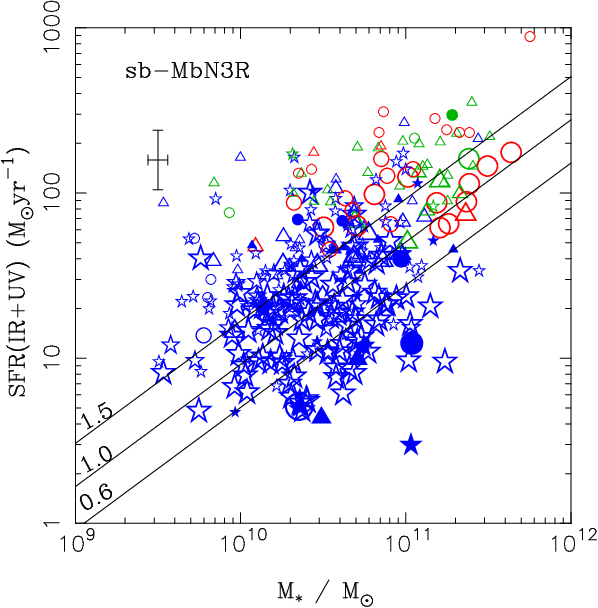}
\end{minipage}
\begin{minipage}{0.49\linewidth}
\includegraphics[angle=0,width=\linewidth]
{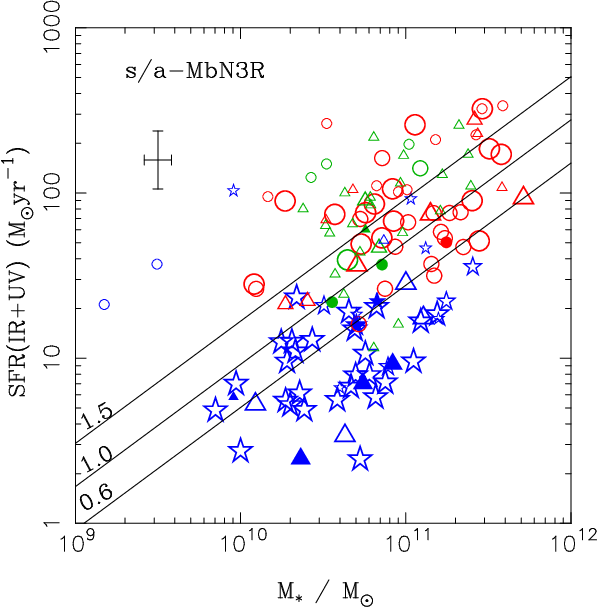}
\end{minipage}
\caption{Stellar mass M$_{\ast}$ vs. $SFR_{IR+UV}$ derived
with the TIR and observed UV luminosities.  
The lines are the same as in figure~\ref{fig:smass_sfr_all} 
(see the text in detail).  
Symbols are the same as in figure~\ref{fig:lum7_lum10_lumir}.  
Left: For the sb-MbN3Rs.  Right: The same as on the left for
the star-forming components in the s/a-MbN3Rs. }
\label{fig:smass_sfr_ir_smbn3r}
\end{figure}  

\begin{figure}[t]
\begin{minipage}{0.49\linewidth}
\includegraphics[angle=0,width=\linewidth]
{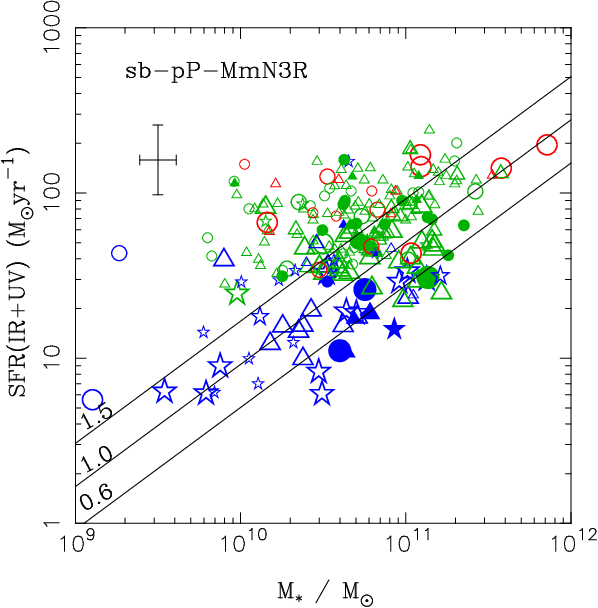}
\end{minipage}
\begin{minipage}{0.49\linewidth}
\includegraphics[angle=0,width=\linewidth]
{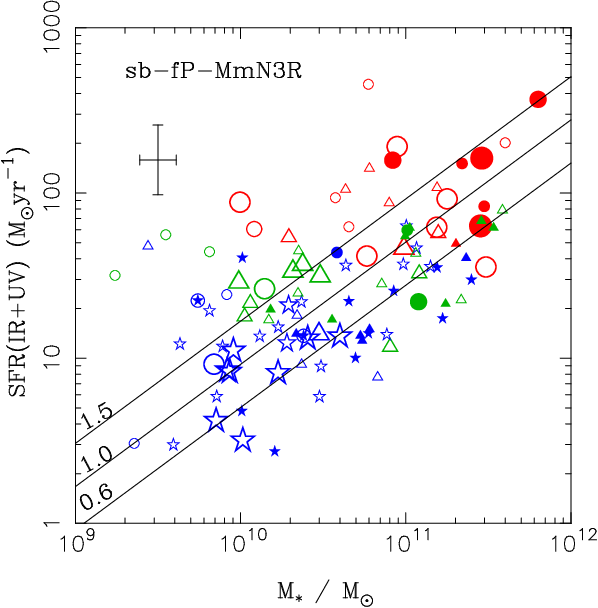}
\end{minipage}
\caption{The same as figure~\ref{fig:smass_sfr_ir_smbn3r} for the sb-MmN3Rs.  
Left: For the sb-pP-MmN3Rs.  Right: For the sb-fP-MmN3Rs.}
\label{fig:smass_sfr_ir_smcn3r}
\end{figure}  

\begin{figure}[t]
\begin{minipage}{0.99\linewidth}
\includegraphics[angle=0,width=\linewidth]
{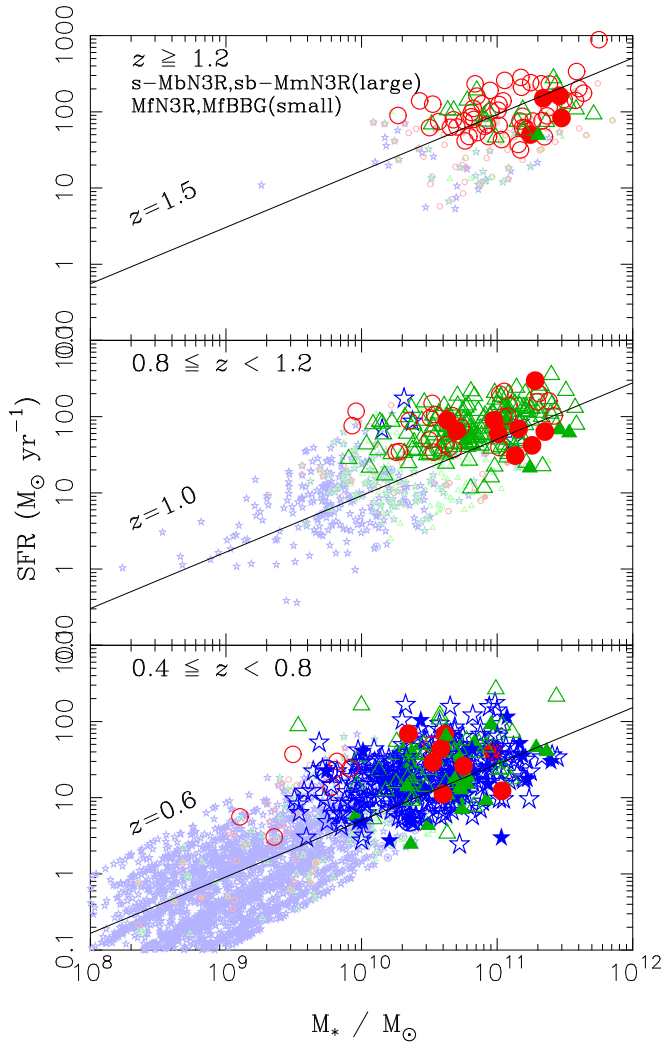}
\end{minipage}
\caption{Stellar mass M$_{\ast}$ vs. SFR for all the star-forming 
galaxies; the sb-MmN3Rs, MfN3Rs, and MfBBGs.   
The lines represent the fitting result (see the text in detail).  
Symbols are the same as in figure~\ref{fig:col_col_mir_mbn3r},  
except for the large dark and small thin ones represent the 
MIR-detected star-forming N3Rs (s-MbN3Rs and sb-MmN3Rs) and 
the MIR faint star-forming galaxies (star-forming MfN3Rs and MfBBGs), 
respectively.  
Top: For $z \geqq 1.2$.  Middle: For $ 0.8 \le z < 1.2$
Bottom: For $ 0.4 \le z < 0.8$. }
\label{fig:smass_sfr_all}
\end{figure}  

Figures~\ref{fig:smass_sfr_ir_smbn3r} and~\ref{fig:smass_sfr_ir_smcn3r} 
show $SFR_{IR+UV}$ to $M_{\ast}$ for the sb- and s/a-MbN3Rs and sb-MmN3Rs.  
We can see correlations between $SFR_{IR+UV}$ and $M_{\ast}$, 
which are more obvious for the sb-MbN3Rs than for the s/a-MbN3Rs.  
The correlation between $SFR_{IR+UV}$ and $M_{\ast}$ in the sb-MbN3Rs
has also an evolutionary trend.  
The $SFR_{IR+UV}$-$M_{\ast}$ correlation in the s/a-MbN3Rs is 
more scattered to the lower SFR side from that in the sb-MbN3Rs, 
in which the decrement becomes more obvious in lower redshifts $z < 0.8$ 
even though the highest mass group around $10^{11}$M$_{\odot}$ 
at higher redshifts $z \ge 0.8$ tends to mostly follow that in the sb-MbN3Rs.  
They indicate that star formation activities in the s/a-MbN3Rs 
were quenched at $z < 0.8$ compared 
with those in the sb-MbN3Rs.  The indicated connection between 
quenching star formation and AGN activity around 
$z \simeq 0.8$ will be discussed with the evolution of 
the stellar populations in section~\ref{sec:evol_pop}.    

The evolutionary trend 
and mass dependence on the SFR can be seen more clearly 
in figure~\ref{fig:smass_sfr_all},  for all the star-forming 
galaxies including the s-MbN3Rs, the sb-MmN3Rs, the star-forming 
MfN3Rs and the MfBBGs,  which are plotted from the top to the 
bottom in redshift intervals of $0.4 \le z<0.8, 0.8
\le z<1.2$, and $z\ge 1.2$.  
The star-forming MfN3Rs and MfBBGs are selected as 
the population in the region of ``blue cloud'' in the $(U-V)-\Delta_{UV}$ 
color as remarked in subsection~\ref{subsec:mfn3r}, which excludes    
extremely young objects fitted with the BC03 models of age $t<0.1$ Gyr.  

The mean $SFR$ for all the star-forming 
galaxies including both the s-MbN3Rs, sb-MmN3Rs, MfN3Rs, and MfBBGs   
is presented as an approximate function of $M_{\ast}$ and $z$:
\begin{eqnarray}
SFR \left[ M_{\odot} yr^{-1} \right] \simeq  
SFR_{11}
\left( \frac{M_{\ast}}{10^{11}\mbox{M}_{\odot}} \right)^{\alpha} 
(1+z)^{\beta}\; , 
\label{eq:sfr_sm_z}
\end{eqnarray}  
where the best-fitting plane in the space of 
$\log M_{\ast}$, $\log (1+z)$, and $\log SFR$ is determined as 
$SFR_{11}/M_{\odot} yr^{-1} = 7.8$, $\alpha=0.74 \pm 0.01$, 
and $\beta=2.7 \pm 0.1$.   
In figures~\ref{fig:smass_sfr_ir_smbn3r}, ~\ref{fig:smass_sfr_ir_smcn3r},  
and ~\ref{fig:smass_sfr_all}, 
the solid lines at $z=0.65, 1.0$, and 1.5 represent the best-fitting 
result.   
Roughly taking $\alpha \simeq 1$ and $\beta \simeq 3$ for
simplification, the growth time scale of the star-forming galaxies 
is typically derived as 
\begin{eqnarray}
\frac{M_{\ast}}{SFR}  \left[ \mbox{yr} \right] \simeq  10^{10} (1+z)^{-3}\; ,  
\label{eq:t_sf}
\end{eqnarray}  
which means the inverse of the sSFR.  
These SFR evolutionary properties and mass dependence are
consistent with previous work for star-forming galaxies selected 
with general schemes \citep{noeske_star_2007, santini_star_2009,
dunne_star_2009, pannella_star_2009}.  

\section{Evolution of stellar populations}
\label{sec:evol_pop} 

\begin{figure}[ht]
\begin{minipage}{0.99\linewidth}
\includegraphics[angle=0,width=\linewidth]
{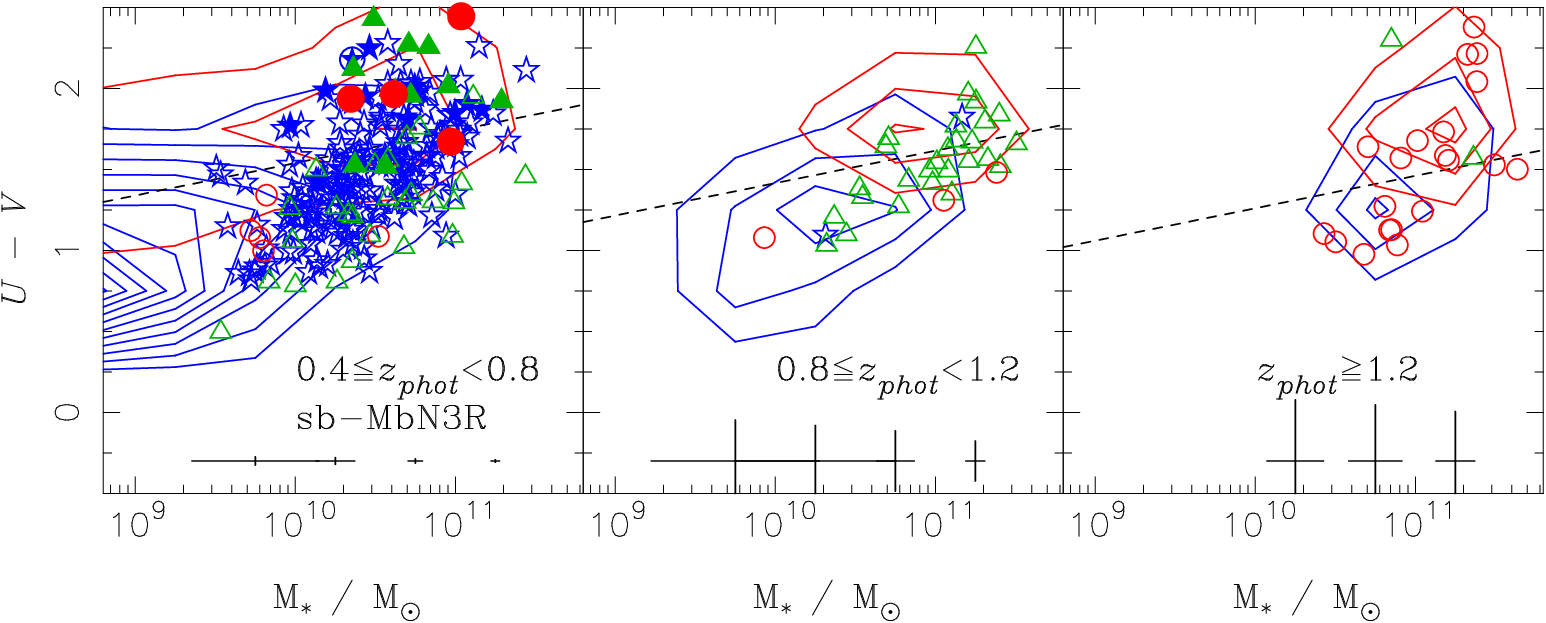}
\end{minipage}
\begin{minipage}{0.99\linewidth}
\includegraphics[angle=0,width=\linewidth]
{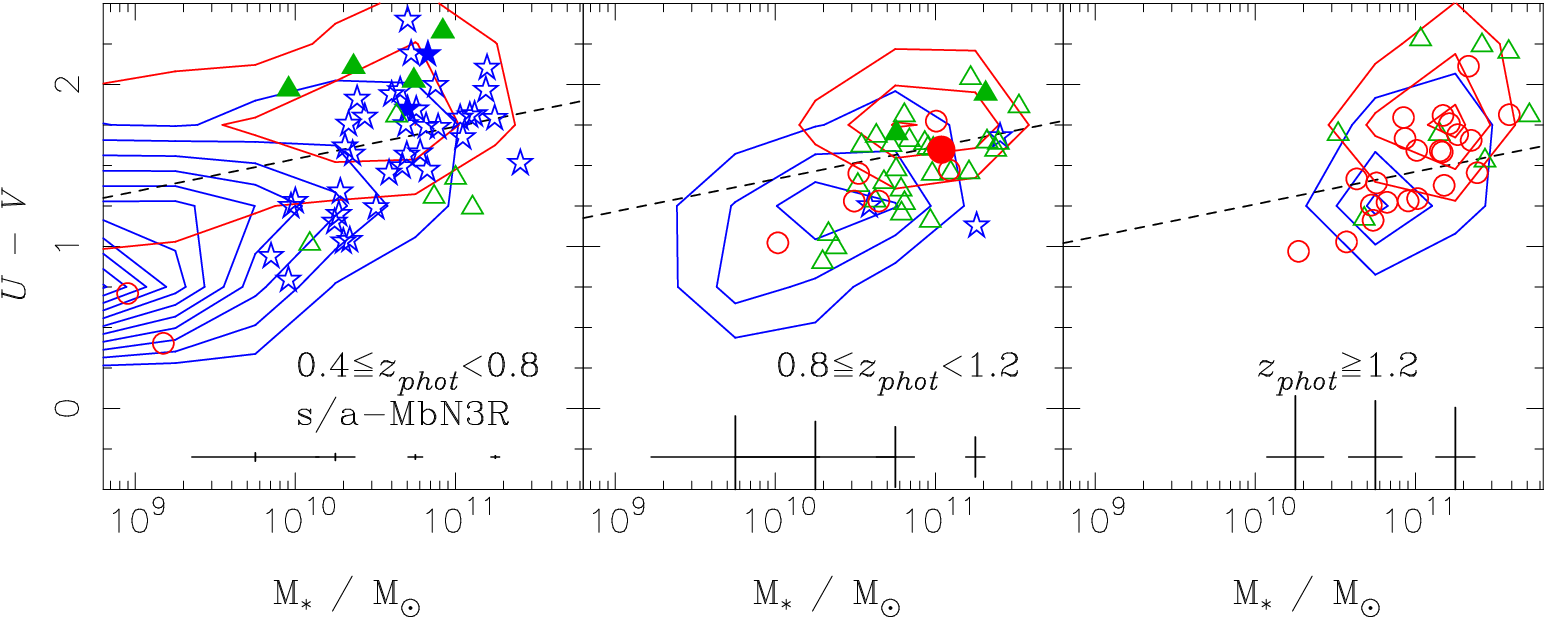}
\end{minipage}
\begin{minipage}{0.99\linewidth}
\includegraphics[angle=0,width=\linewidth]
{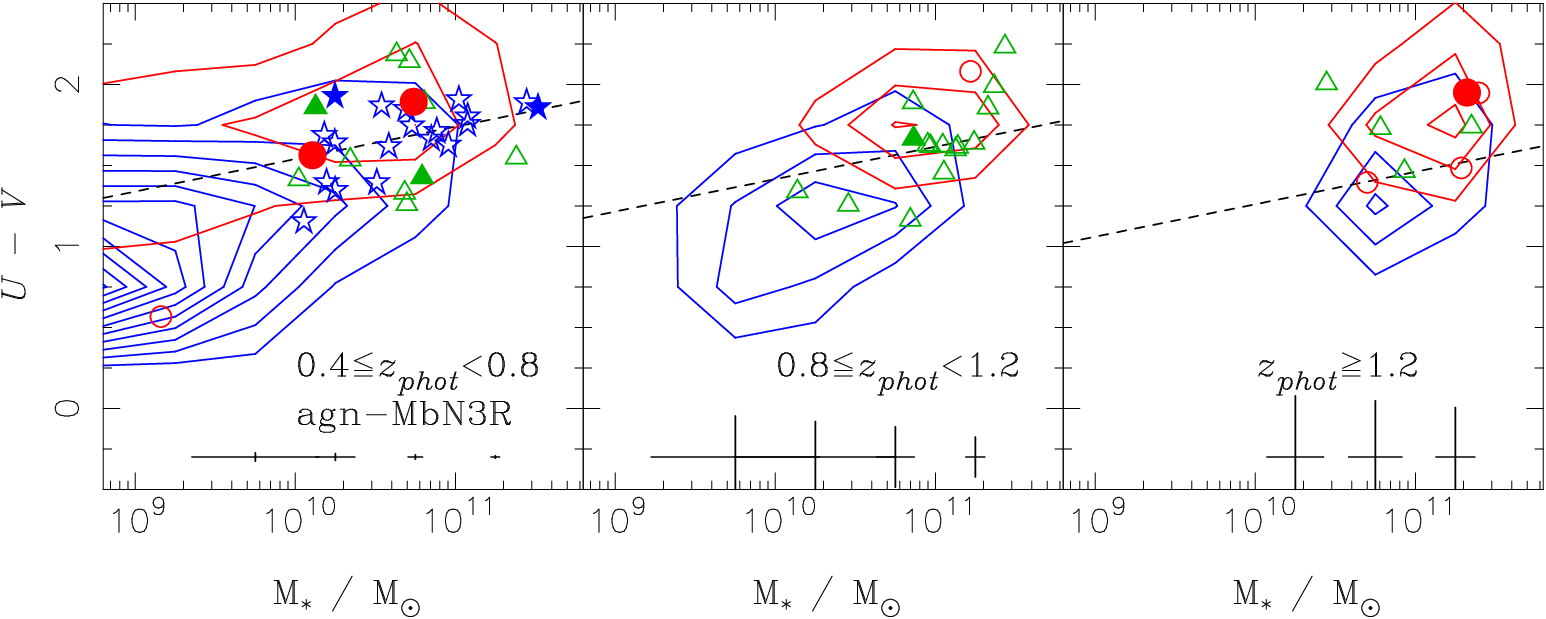}
\end{minipage}
\caption{Stellar mass and rest-frame color diagram of $M_{\ast}$ vs. $U-V$ 
for the MbN3Rs excluded BL AGNs and EBOs.  
The color is not corrected for dust extinction.  Typical 
errors of the mass and color are represented with crosses at the bottom.  
Large open and filled symbols and filled symbols in circles 
represent s-,p-, and qp-BBGs, respectively, 
which are the same as in figure~\ref{fig:col_col_mir_mbn3r}. 
The plots are for $0.4 \le z<0.8$, $0.8 \le z<1.2$, and $z \geqq 1.2$ 
from left to right.  Red and blue contours represent the rest-frame $V$ 
detected galaxies with $sSFR<0.1 \mbox{Gyr}^{-1}$ and 
$sSFR>0.1 \mbox{Gyr}^{-1}$, respectively.
The dotted line corresponds to Bell's critical line dividing the 
photometric samples into 
the blue cloud and the red sequence.   
Top: Large colored and small gray symbols represent the sb-MbN3Rs 
and the $z'$-detected galaxies, respectively.   
Middle: The same as the Top for the s/a-MbN3Rs.  
Bottom: The same as the Top for the agn-MbN3Rs. 
}
\label{fig:smass_uvcol_mbn3r}
\end{figure}  

\begin{figure}[ht]
\begin{minipage}{0.99\linewidth}
\includegraphics[angle=0,width=\linewidth]
{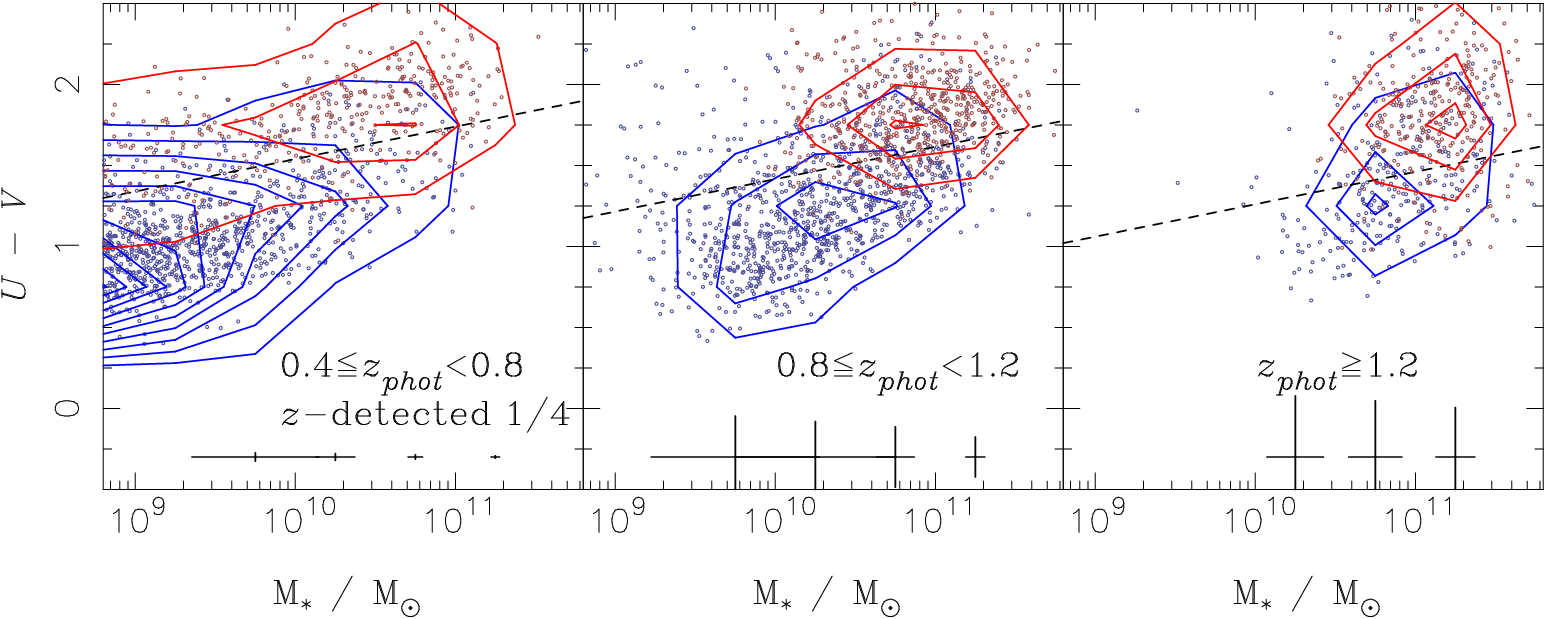}
\end{minipage}
\begin{minipage}{0.99\linewidth}
\includegraphics[angle=0,width=\linewidth]
{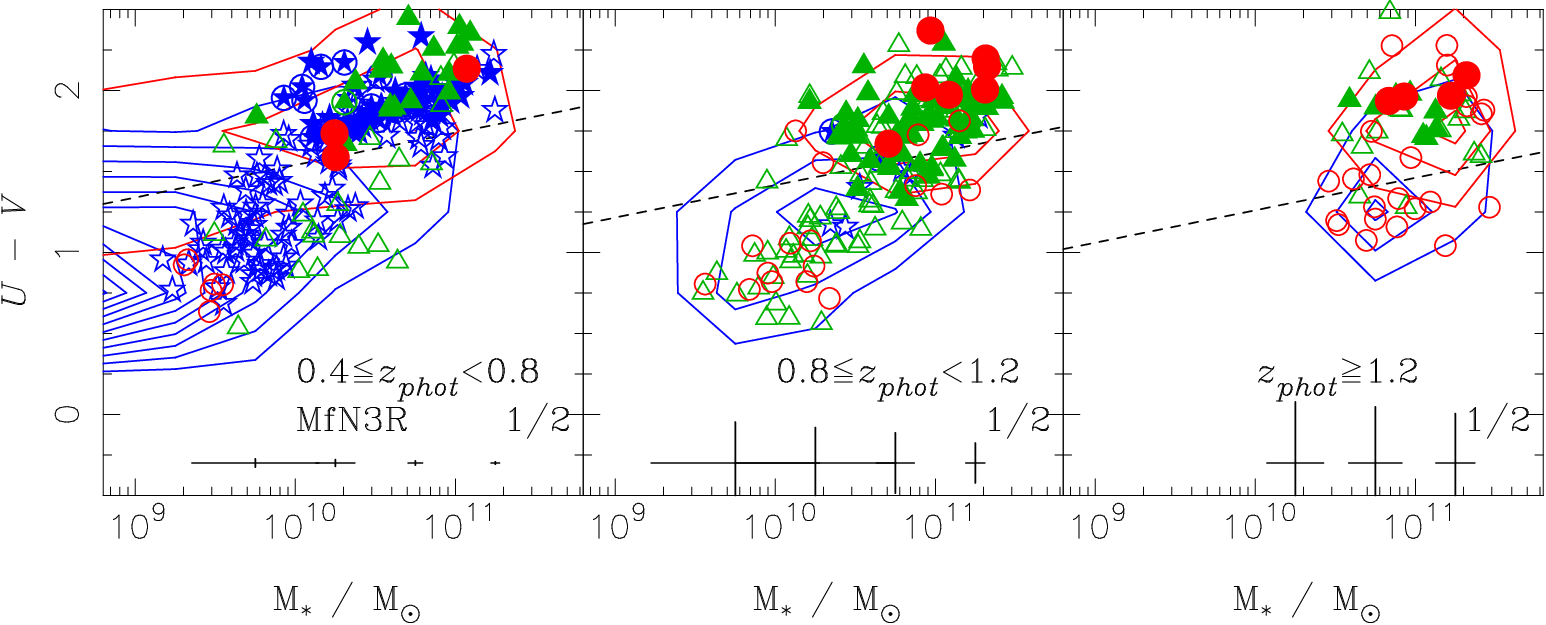}
\end{minipage}
\caption{
The same as figure~\ref{fig:smass_uvcol_mbn3r} for the 
the rest-frame $V$ detected galaxies and the MfN3Rs.  
The red and blue small dots with contour 
represent the rest-frame $V$ detected galaxies 
with $sSFR < 0.1 \mbox{Gyr}^{-1}$ and 
$sSFR > 0.1 \mbox{Gyr}^{-1}$, respectively. 
Top: For the rest-frame $V$ detected galaxies.  
Bottom: For the MfN3Rs.   
 }
\label{fig:smass_uvcol_zdet_mfn3r}
\end{figure}  

\begin{figure}[ht]
\begin{minipage}{0.99\linewidth}
\includegraphics[angle=0,width=\linewidth]
{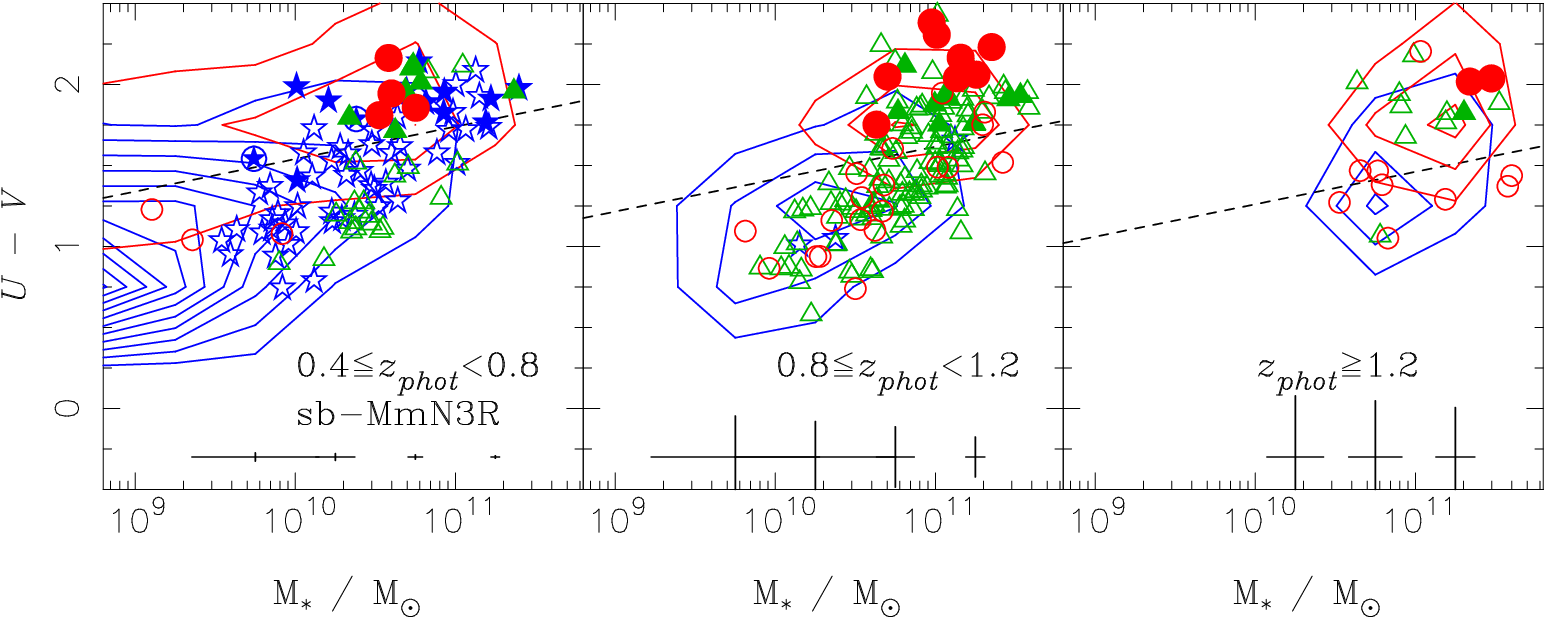}
\end{minipage}
\begin{minipage}{0.99\linewidth}
\includegraphics[angle=0,width=\linewidth]
{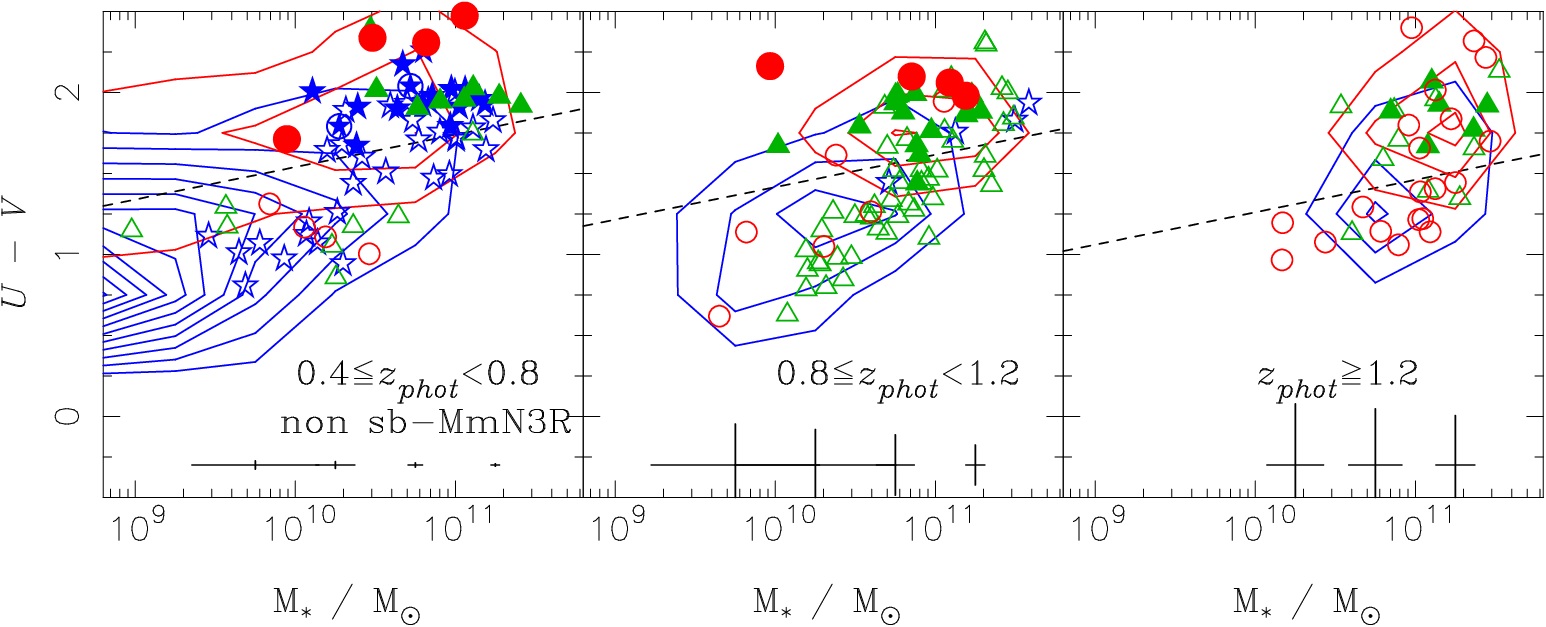}
\end{minipage}
\caption{The same as figure~\ref{fig:smass_uvcol_mbn3r} 
for the MmN3Rs. 
Top: Large colored symbols represent the sb-MmN3Rs.   
Bottom: The same as the Top for the nsb-MmN3Rs.  
}
\label{fig:smass_uvcol_mcn3r}
\end{figure}  

\begin{figure}[ht]
\begin{minipage}{0.99\linewidth}
\includegraphics[angle=0,width=\linewidth]
{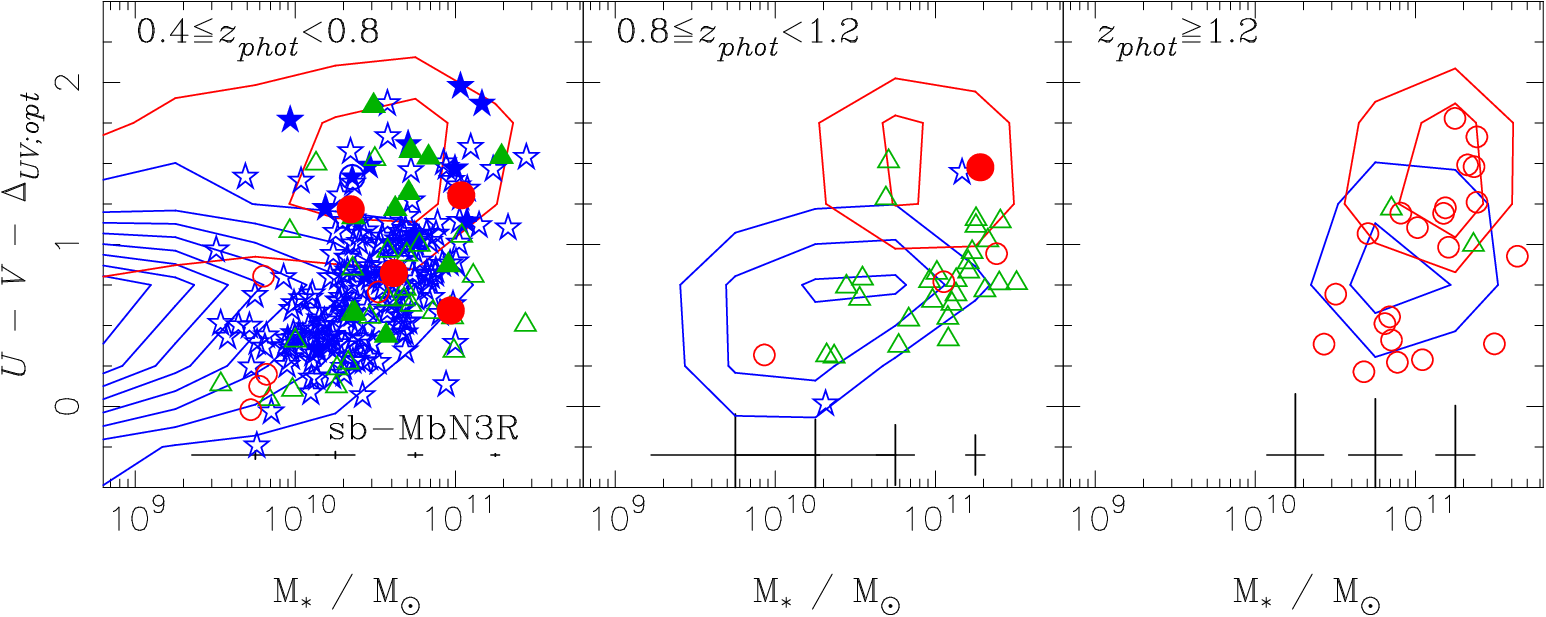}
\end{minipage}
\begin{minipage}{0.99\linewidth}
\includegraphics[angle=0,width=\linewidth]
{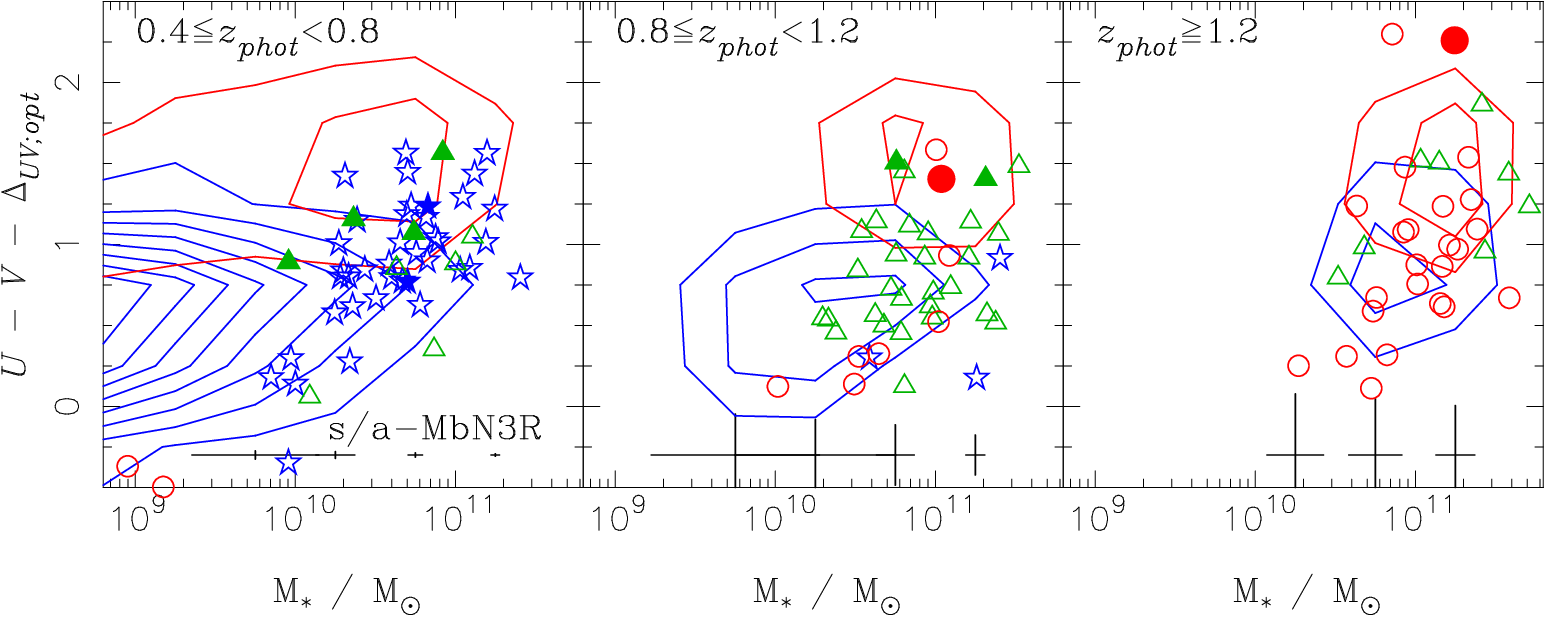}
\end{minipage}
\begin{minipage}{0.99\linewidth}
\includegraphics[angle=0,width=\linewidth]
{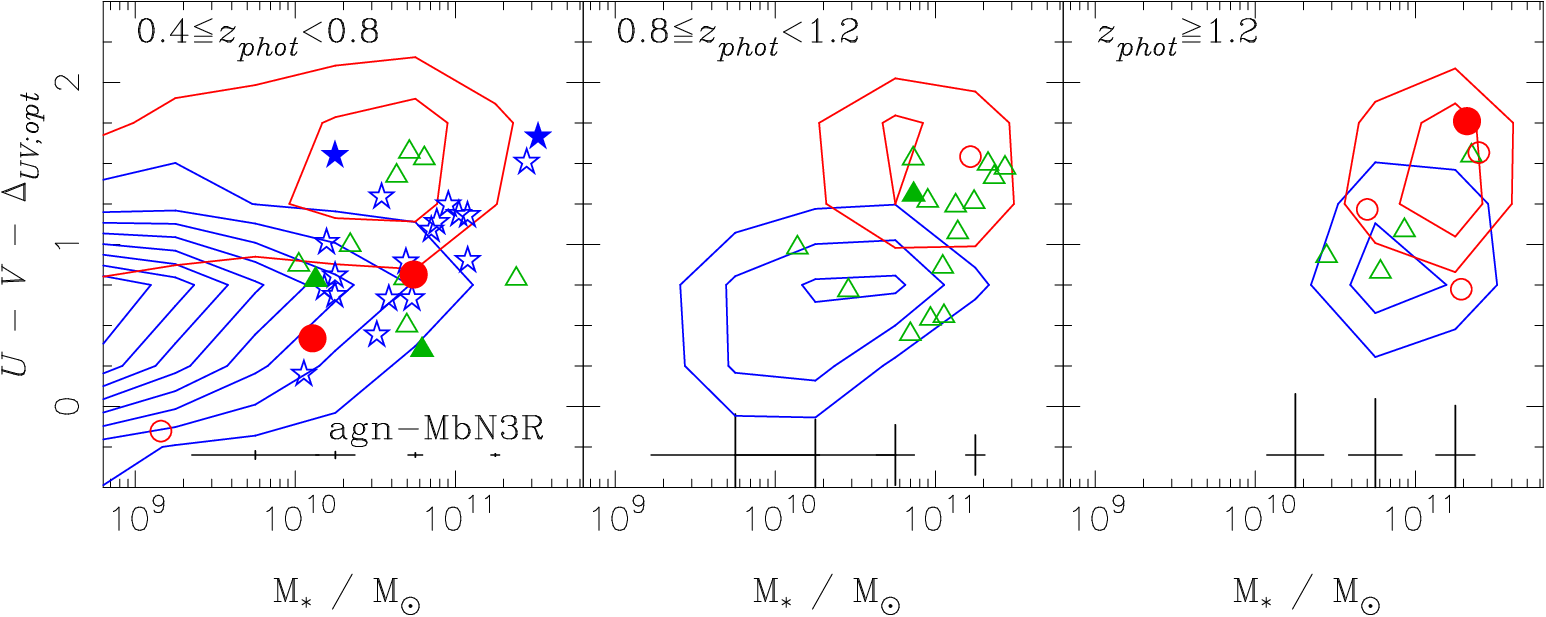}
\end{minipage}
\caption{The same as figure~\ref{fig:smass_uvcol_mbn3r} 
for color $U-V-\Delta_{UV}$ corrected using the extinction $A_{UV;opt}$, 
which is derived from the optical-NIR SED fitting. 
The sb-,s/a-, and agn-MbN3Rs are plotted from the top to the bottom.     
}
\label{fig:smass_uvcol_redc2_mbn3r}
\end{figure}  

\begin{figure}[ht]
\begin{minipage}{0.99\linewidth}
\includegraphics[angle=0,width=\linewidth]
{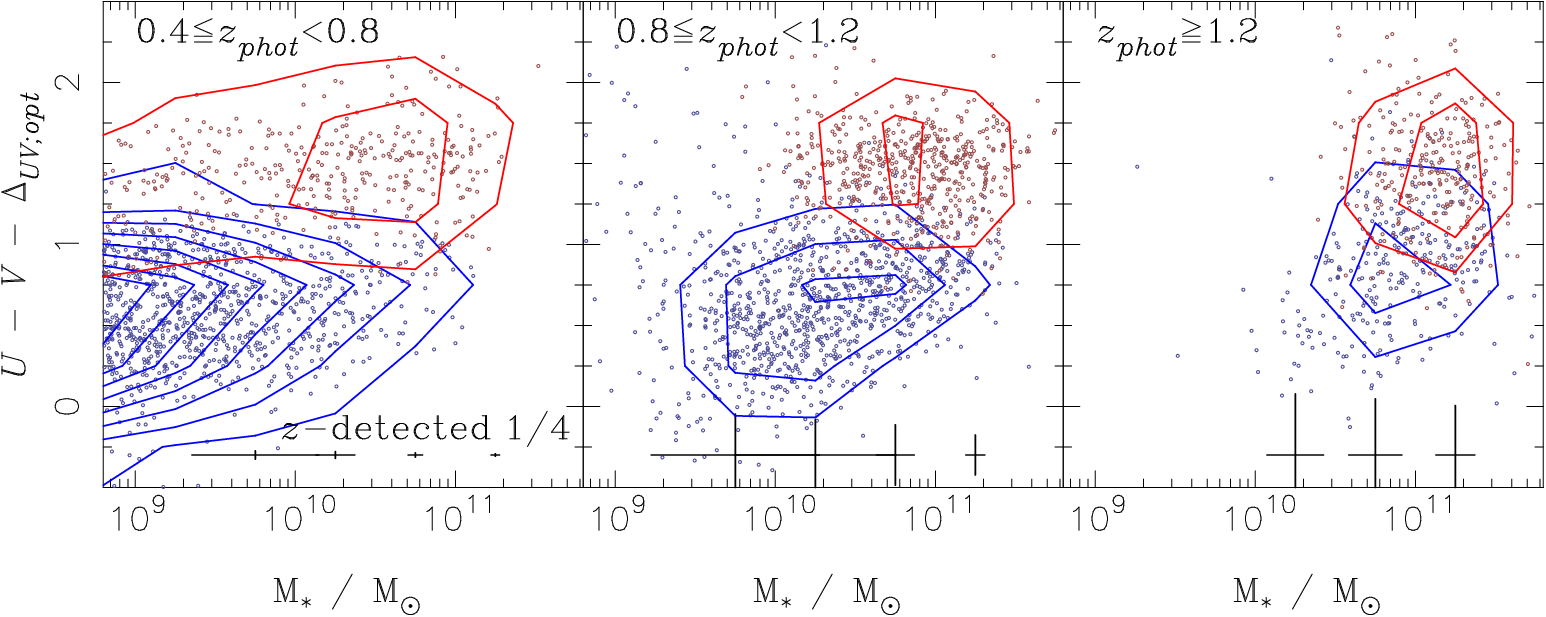}
\end{minipage}
\begin{minipage}{0.99\linewidth}
\includegraphics[angle=0,width=\linewidth]
{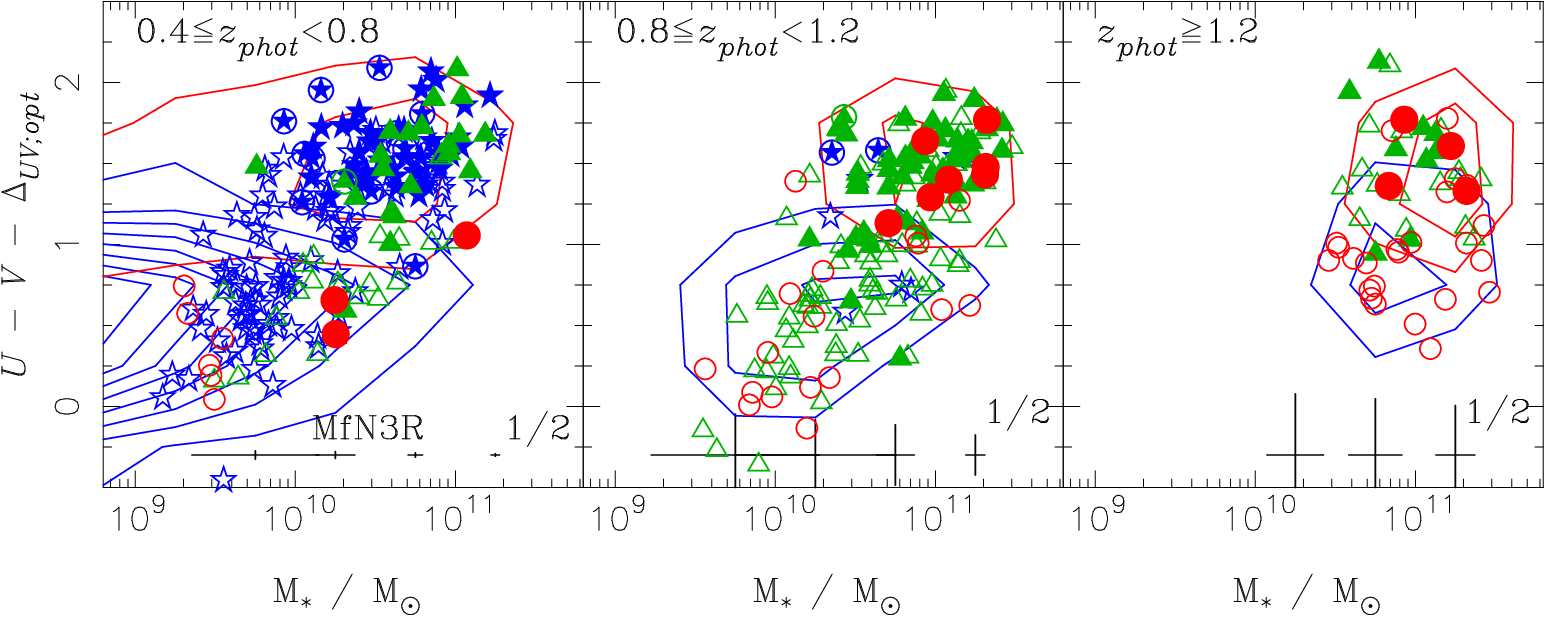}
\end{minipage}
\caption{The same as figure~\ref{fig:smass_uvcol_zdet_mfn3r} 
for color $U-V-\Delta_{UV}$ 
corrected using the extinction $A_{UV;opt}$.  }
\label{fig:smass_uvcol_redc2_zdet_mfn3r}
\end{figure}  

\begin{figure}[ht]
\begin{minipage}{0.99\linewidth}
\includegraphics[angle=0,width=\linewidth]
{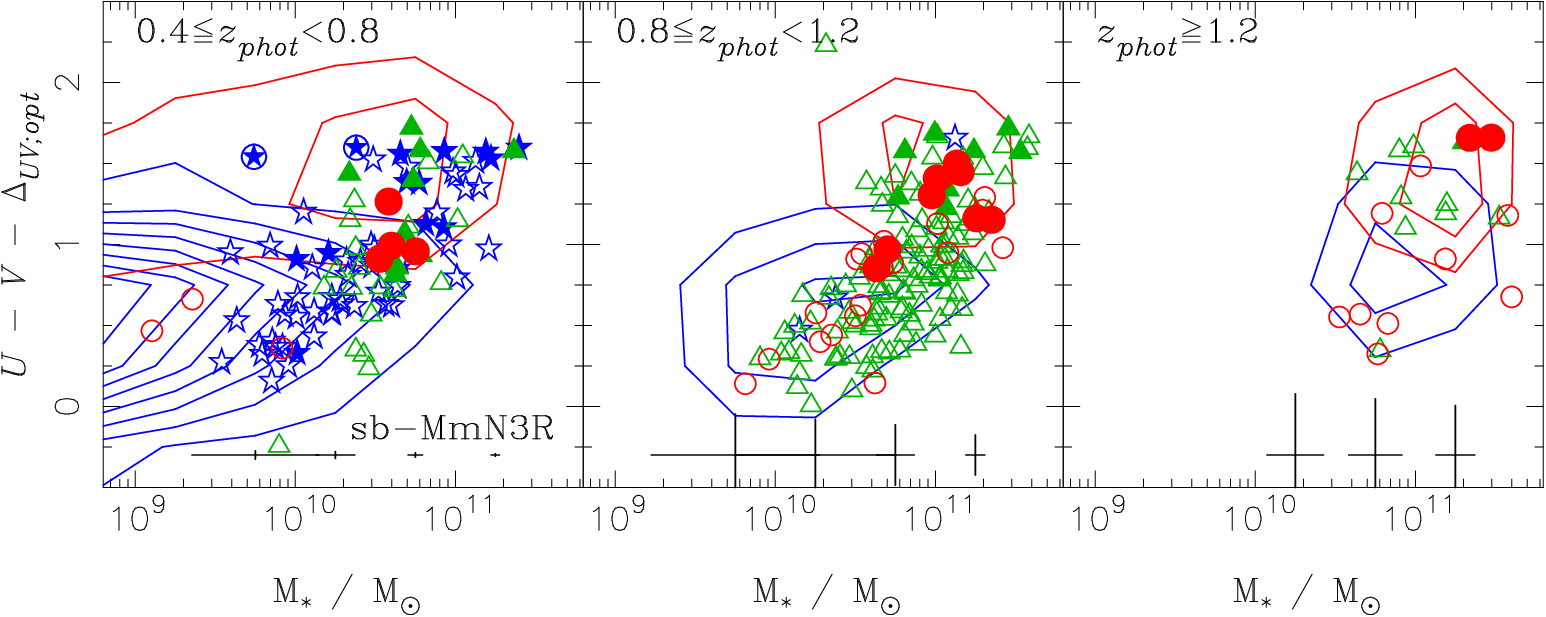}
\end{minipage}
\begin{minipage}{0.99\linewidth}
\includegraphics[angle=0,width=\linewidth]
{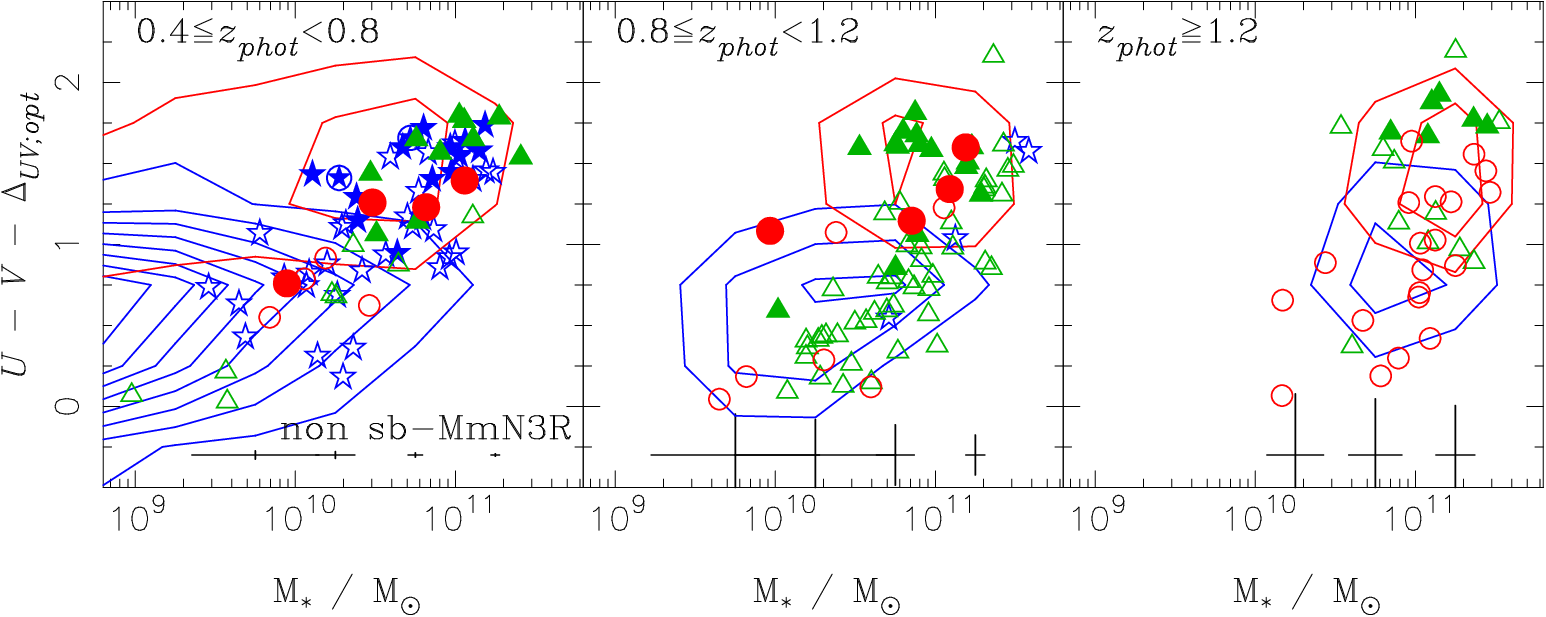}
\end{minipage}
\caption{
The same as figure~\ref{fig:smass_uvcol_mcn3r} 
for color $U-V-\Delta_{UV}$ 
corrected using the extinction $A_{UV;opt}$.  
}
\label{fig:smass_uvcol_redc2_mcn3r}
\end{figure}  

\begin{figure}[ht]
\begin{minipage}{0.99\linewidth}
\includegraphics[angle=0,width=\linewidth]
{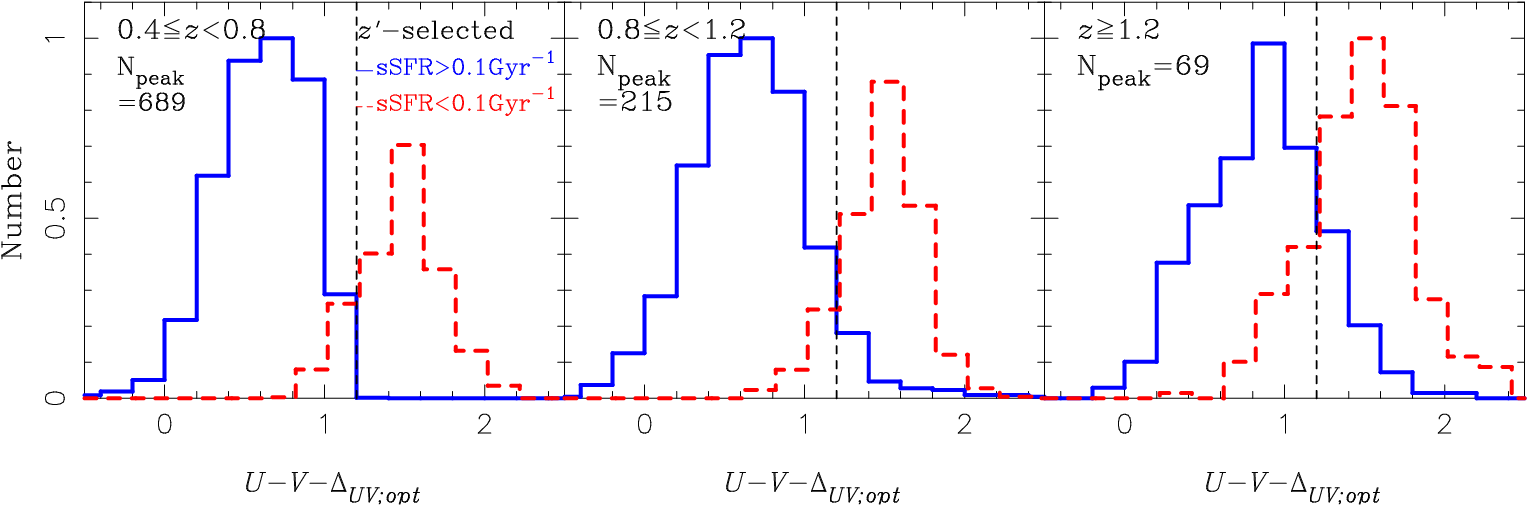}
\end{minipage} 
\begin{minipage}{0.99\linewidth}
\includegraphics[angle=0,width=\linewidth]
{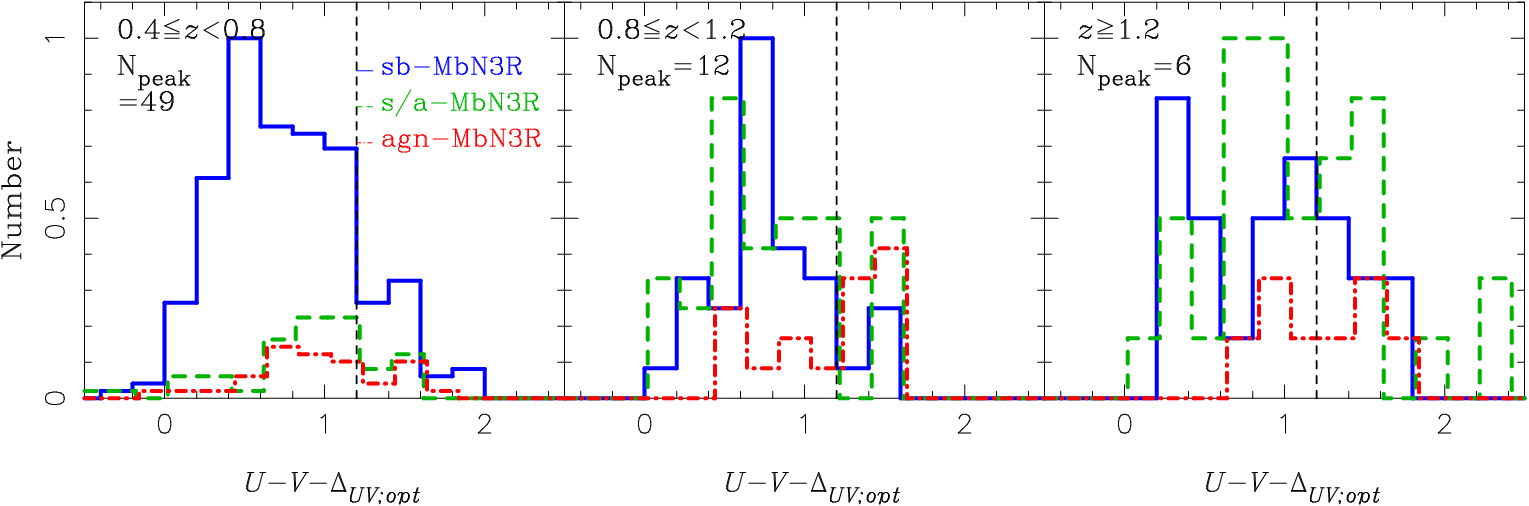}
\end{minipage} 
\caption{
Histograms of the corrected color distribution with $A_V$ derived from 
optical SED fittings.  Height of histograms is normalized with the maximum 
number, which is displayed on the upper-left side.  
Top: For the rest-frame $V$ detected galaxies, which are classified into 
two groups with sSFR larger and less than $0.1\mbox{Gyr}^{-1}$, which 
corresponds to a color $U-V-\Delta = 1.2$.   
Bottom: For the sb-,s/a-, and agn-MbN3Rs.  
}
\label{fig:uvcolhist}
\end{figure}  

\begin{figure}[ht]
\begin{minipage}{0.99\linewidth}
\includegraphics[angle=0,width=\linewidth]
{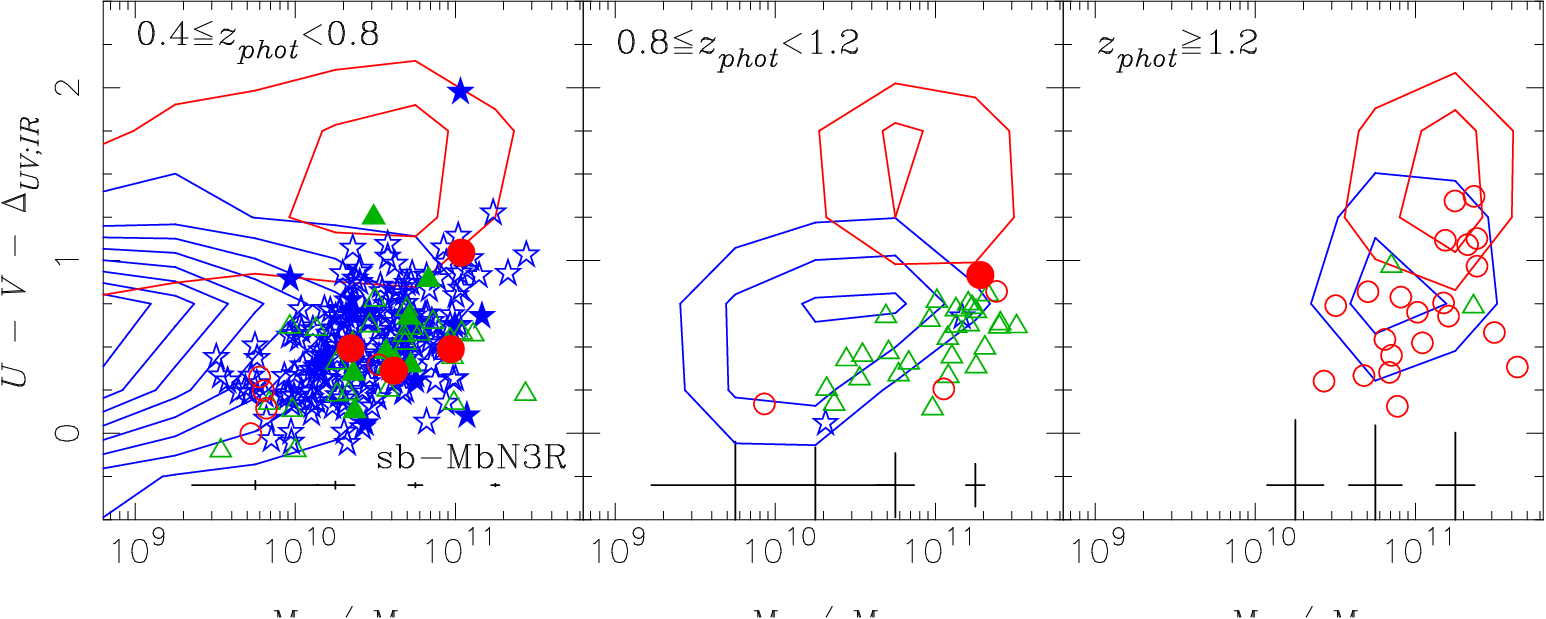}
\end{minipage}
\begin{minipage}{0.99\linewidth}
\includegraphics[angle=0,width=\linewidth]
{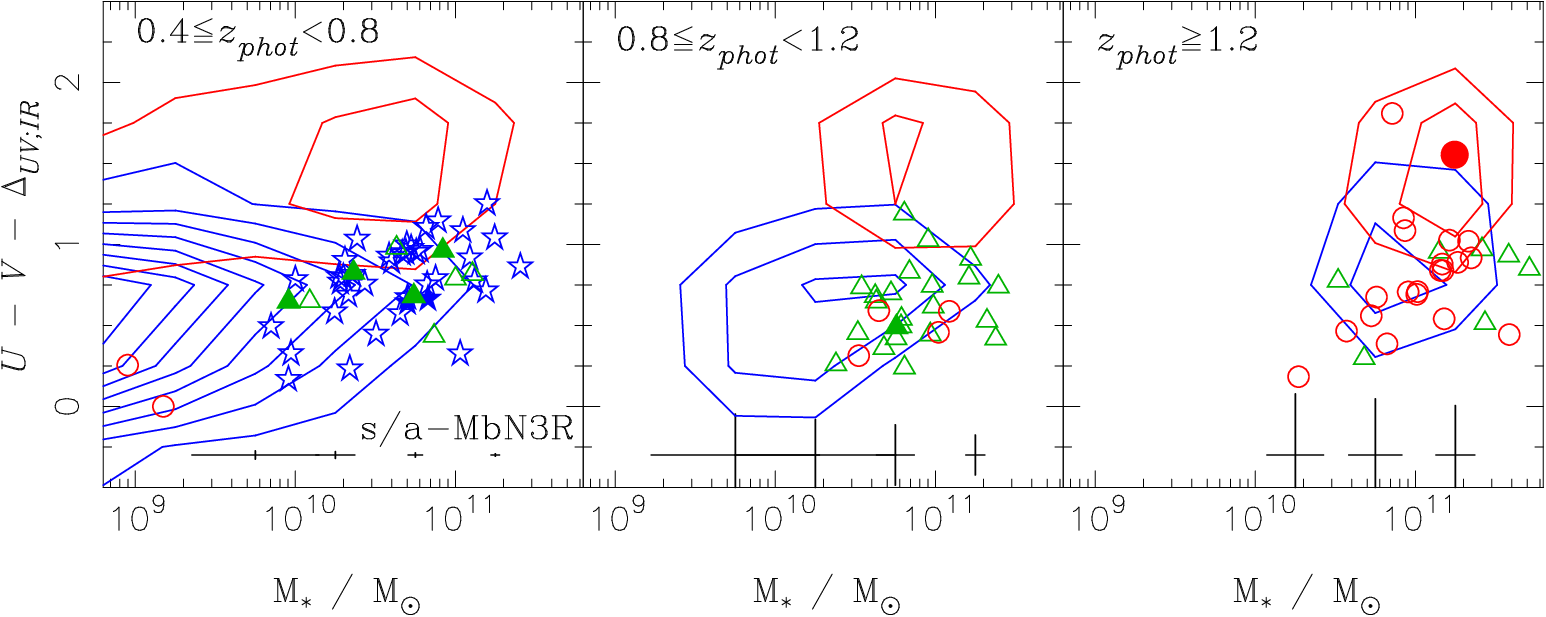}
\end{minipage}
\caption{
The same as figure~\ref{fig:smass_uvcol_redc2_mbn3r} 
for color $U-V-\Delta_{UV}$ 
corrected using the extinction $A_{UV;IR}$, which is derived 
with equation~(\ref{eq:auv_ir}).    
Top: For the sb-MbN3Rs.  Bottom: For the s/a-MbN3Rs, 
its remaining dusty starburst component can be extracted and then used 
to derive $SFR_{IR+UV}$ by fitting its MIR SED with the S\&K models.  
}
\label{fig:smass_uvcol_redc_mbn3r}
\end{figure}  

\begin{figure}[ht]
\begin{minipage}{0.99\linewidth}
\includegraphics[angle=0,width=\linewidth]
{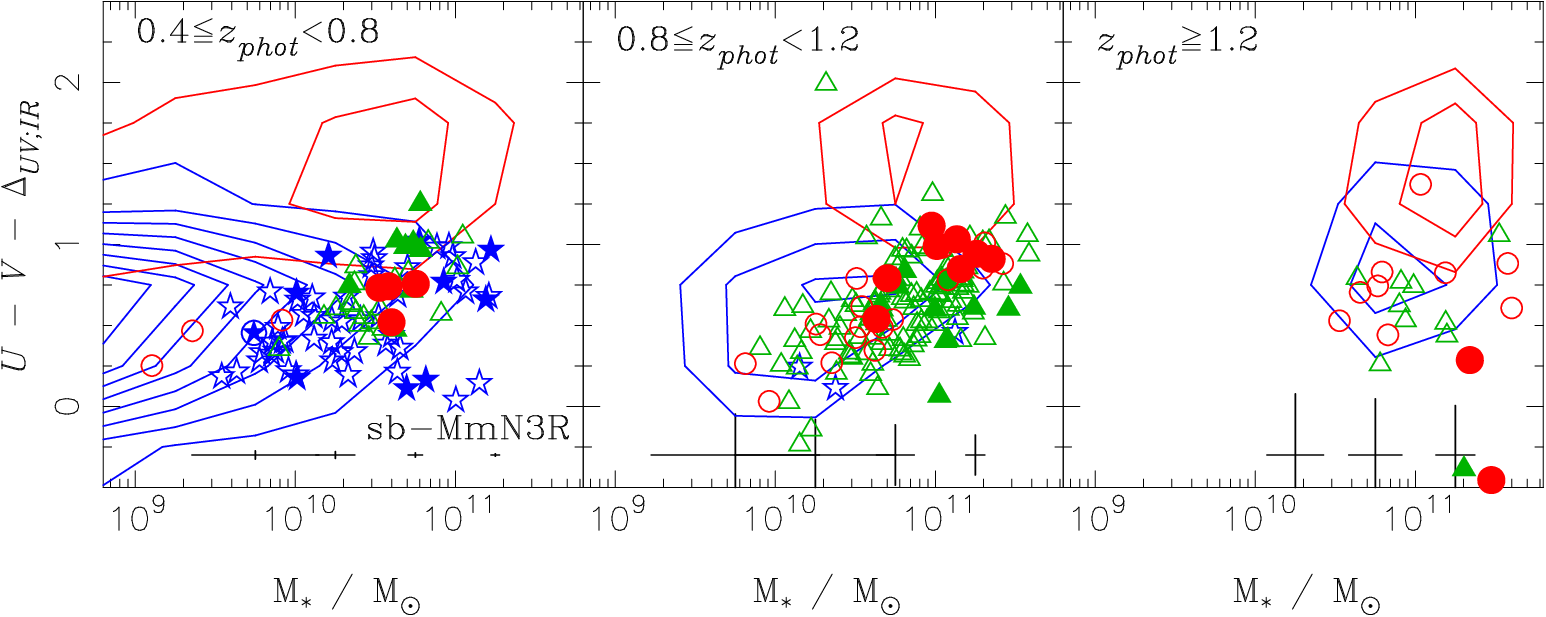}
\end{minipage}
\begin{minipage}{0.99\linewidth}
\includegraphics[angle=0,width=\linewidth]
{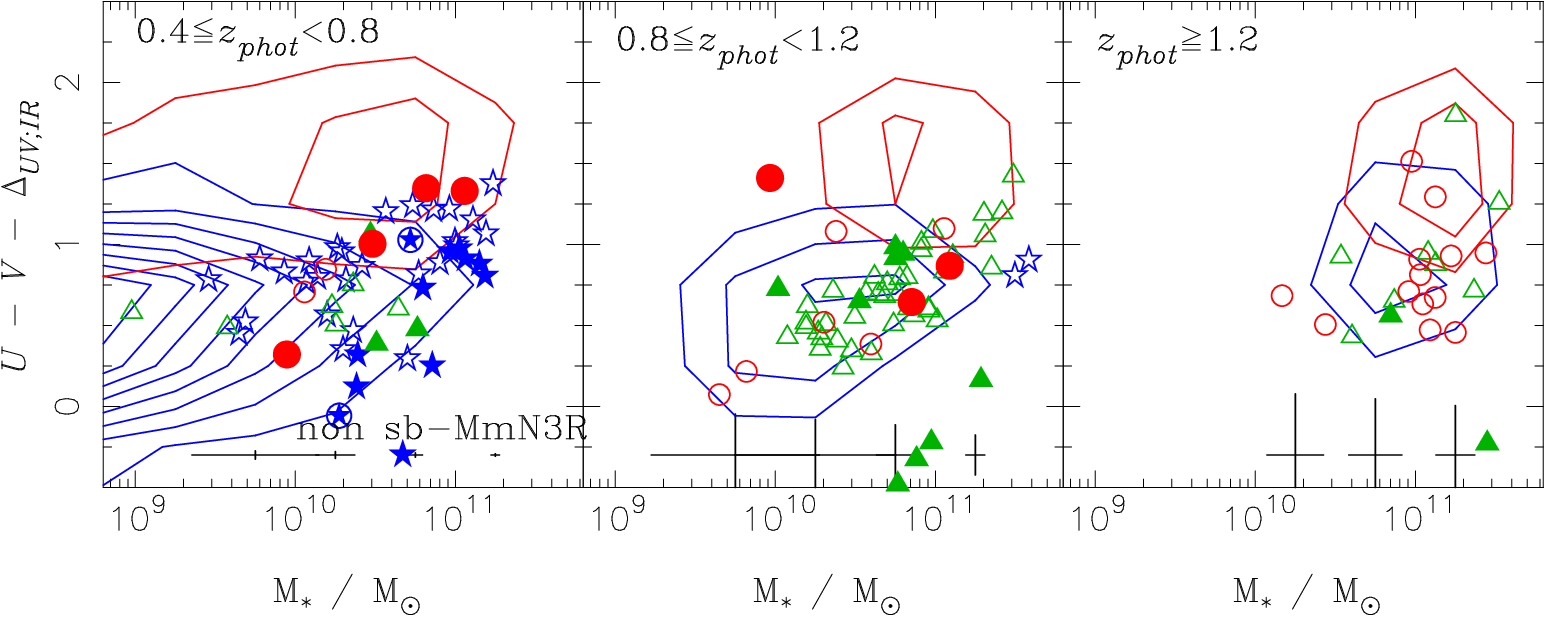}
\end{minipage}
\caption{
The same as figure~\ref{fig:smass_uvcol_redc_mbn3r} for the MmN3Rs.  
Top: For the sb-MmN3Rs.  
Bottom: For the non sb-MmN3Rs. 
}
\label{fig:smass_uvcol_redc_mcn3r}
\end{figure}  

\subsection{Rest-frame optical color distributions}  
\label{subsec:smass_uvcol}
	
The CMD is a powerful tool for studying galaxy evolution since it
shows the bimodal galaxy distribution as early types that concentrate in a
tight red sequence while late types distribute in a blue dispersed
cloud, which have been studied not only in the local Universe with the SDSS 
\citep{blanton_estimating_2003} but also in the distant Universe with 
the surveys for $z<1$ \citep{bell_nearly_2004, faber_galaxy_2007, 
borch_stellar_2006} and for $z>1$ \citep{pannella_star_2009, 
brammer_dead_2009}.  
Spitzer results have also shown that rest-frame colors of MIPS-detected 
objects, which may overlap with the MbN3Rs and the MmN3Rs, 
fall between the red sequence and the blue cloud up to $z=2$ 
\citep{cowie_integrated_2008, salim_mid-ir_2009, brammer_dead_2009}.  

Thus, we also study rest-frame $U-V$ colors for MbN3Rs and MmN3Rs 
in the field from their magnitudes interpolated 
with equation~(\ref{eq:mag_intr}).  Figure~\ref{fig:smass_uvcol_mbn3r} 
shows the $U-V$ color without extinction correction for the stellar mass
$M_{\ast}$ for the sb-, s/a-, and agn-MbN3Rs from top to bottom, 
where the left, middle, and right figures correspond to the
redshift intervals of $0.4 \le z<0.8$, $0.8 \le z<1.2$, and $z \ge 1.2$, 
respectively, and overplotted contours represent the rest-frame 
$V$ detected galaxies, which are also shown  
with the MfN3Rs in figure~\ref{fig:smass_uvcol_zdet_mfn3r}. 

The well-known ``red and dead'' population in a red sequence has a typical
color of $U-V \simeq 2$, which can be the reddest ones even in the
local Universe with low $A_V$ values.  However, redder populations of
$U-V>2$ can clearly be seen in all the diagrams not only for 
the MbN3Rs but also for the MmN3Rs and even for the MfN3Rs 
as shown in figures~\ref{fig:smass_uvcol_mbn3r}, 
~\ref{fig:smass_uvcol_mcn3r} and 
~\ref{fig:smass_uvcol_zdet_mfn3r}.  
It suggests that not only the MbN3Rs but also the MmN3Rs and most 
of the MfN3Rs suffer from dust extinction, which is already 
expected from the results in section~\ref{sec:evol_sfr}.  
Thus, we have taken into account color corrections for these diagrams.  
Figures~\ref{fig:smass_uvcol_redc2_mbn3r},  
~\ref{fig:smass_uvcol_redc2_zdet_mfn3r}, and 
~\ref{fig:smass_uvcol_redc2_mcn3r}   
represent intrinsic stellar color $(U-V)-\Delta_{UV}$ corrected for 
extinction using $A_{UV;opt}$ derived from the optical-NIR SED fitting.  
In the top diagram of figure~\ref{fig:smass_uvcol_redc2_mbn3r}, 
we can see that the
bimodal color distribution of the MbN3Rs persists even up to $z \simeq 
1.2$, which is also consistent with the BBG classification as p-BBGs 
mainly appear in the region of a red sequence.  
As discussed in appendix~\ref{app_subsec:exbbg}, 
unfortunately, 
our $K_s$ photometry is not deep enough to obtain a complete sample up to 
$z \ge 1.2$, which may be one of the reasons why the bimodal feature at 
$z \ge 1.2$ is not so clear compared with the results of 
\citet{brammer_dead_2009}.  

Figure~\ref{fig:smass_uvcol_redc_mbn3r} also shows the intrinsic colors of
stellar components in the sb- and s/a-MbN3Rs, 
in which the colors are corrected with
$A_{UV;IR}$.  Few of them remain in the red sequence region, which comes 
from the effect of the extinctions as discussed in 
subsection~\ref{subsec:extinctions}: 
$\langle A_{UV;IR} \rangle \simeq \langle A_{UV;opt} \rangle + \langle \Delta 
A_{UV} \rangle $ with $\langle \Delta A_{UV} \rangle \sim 0.5$ as long as 
$ f_{SK} =1$. Furthermore, we cannot clearly see the bimodality 
in figure~\ref{fig:smass_uvcol_redc_mbn3r} compared with 
figure~\ref{fig:smass_uvcol_redc2_mbn3r}.   
The same trends can be also confirmed for the MmN3Rs as shown 
in figures~\ref{fig:smass_uvcol_redc2_mbn3r} and
~\ref{fig:smass_uvcol_redc_mcn3r}.  
Even if we have $\langle \Delta A_{UV} \rangle \sim 0$ by  
taking the proportional factor as $f_{sk} \simeq 0.3$, we just obtain a   
bluer mean color by adding $\simeq 0.5$ without recovering the bimodality.  
It suggests that it is appropriate to correct the color of a whole galaxy 
in the classical way with $A_{UV;opt}$ even for dusty systems as 
the MbN3Rs and the MmN3Rs.  Thus, we will discuss the evolution of stellar 
populations along with the results corrected with $A_{UV;opt}$ as  
figure~\ref{fig:smass_uvcol_redc2_mbn3r} in the following. 

We can see the evolutionary trends for sb-, s/a-, and agn-MbN3Rs  
in figures~\ref{fig:smass_uvcol_redc2_mbn3r} 
and~\ref{fig:uvcolhist} and will summarize as 
follows.  In all redshifts, the sb-MbN3Rs mainly stay in the area 
$U-V-\Delta < 1$.  It can be interpreted as they are on-going in 
star formation.       
On the other hand, the agn-MbN3Rs start to appear even around  
$U-V-\Delta < 1$ at $z \ge 1.2$ with large scatters in a mass range 
$> 3 \times 10^{10}M_{\odot}$, and 
relatively concentrate around $U-V-\Delta \sim 1$ at $0.8 \le
z <1.2$ as shown in the bottom diagrams of 
figure~\ref{fig:smass_uvcol_redc2_mbn3r}.  
The trend, as starbursts are bluer than AGNs at $0.8 \le z <1.2$, 
can be seen not only for the MbN3Rs but also for the MmN3Rs in figure
~\ref{fig:smass_uvcol_redc2_mcn3r}.  
Figures~\ref{fig:smass_uvcol_redc2_mbn3r} 
and~\ref{fig:smass_uvcol_redc2_mcn3r} also show that 
the sb-MmN3Rs are relatively redder than the sb-MbN3Rs 
in the same stellar mass.  It is reasonable 
as the former with less luminous IR emissions may have smaller SFRs 
and quench their SF activities more effectively than the latter.      

As shown in the middle of figure~\ref{fig:smass_uvcol_redc2_mbn3r},  
the s/a-MbN3Rs have a similar stellar population of the sb-MbN3Rs 
at $z \ge 0.8$ 
even though some of them were $U-V-\Delta > 1$.    
Thus, a dusty star formation cannot suffer from the obscured AGN 
activity at $z \ge 0.8$ at least.  
On the other hand, at $z<0.8$, it is similar to the agn-MbN3Rs at 
around $U-V-\Delta \sim 1$.  Passive populations such as the p-BBGs and 
on the red sequence are relatively major 
in s/a-MbN3Rs at $ 0.8 \le z < 1.2 $.  This suggests the co-existence of 
the obscured AGN with passive stellar populations in their hosts, 
which are older than those in the sb-MbN3Rs. 
Thus, the stellar populations become already passive in
massive galaxies such as the s/a-MbN3Rs 
of $\simeq 10^{11}$M$_{\odot}$ with obscured AGNs
even at $z \simeq 1$, and the obscured AGNs are activated in
less-massive galaxies of $< 10^{11}$M$_{\odot}$ mainly with star
formation at $z<0.8$.  These indications from the CMDs are consistent 
with the indication about quenching SFRs of the s/a-MbN3Rs at $z<0.8$   
from figures~\ref{fig:smass_sfr_ir_smbn3r} and~\ref{fig:smass_sfr_ir_smcn3r}.   
It suggests that the evolution of AGN activity
is more rapid than that of star-formation around $z = 1-2$ and that
the co-evolutionary link between AGN activity and star formation in
galaxies can be critical around $z \simeq 1$.  This seems to remind us of  
a scenario where AGN activity may quench the star formation 
in the distant universe to explain the {\it downsizing}.

\section{Evolution of AGN activities} 
\label{sec:evol_agn} 

\subsection{AGN activity traced at rest-frame 5 $\mu$m } 
\label{subsec:agn_5}

\begin{figure}[ht]
\begin{minipage}{0.99\linewidth}
\includegraphics[angle=0,width=\linewidth]
{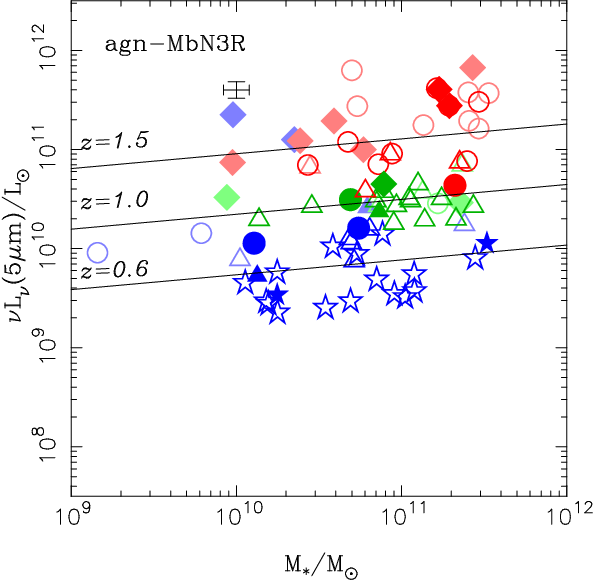}
\end{minipage}
\begin{minipage}{0.49\linewidth}
\includegraphics[angle=0,width=\linewidth]
{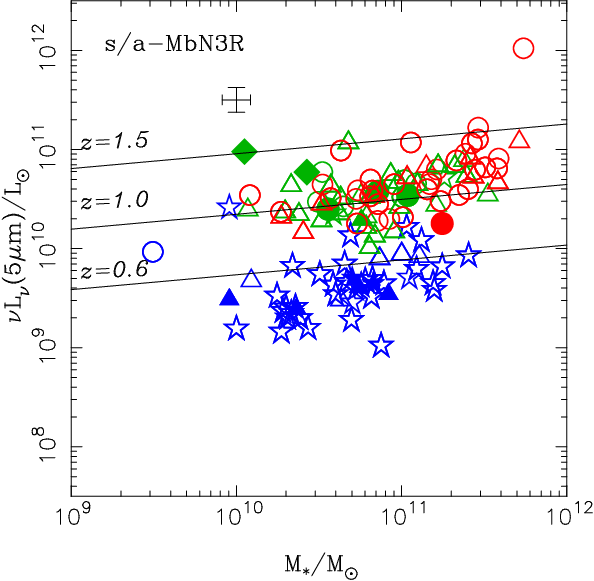}
\end{minipage}
\begin{minipage}{0.49\linewidth}
\includegraphics[angle=0,width=\linewidth]
{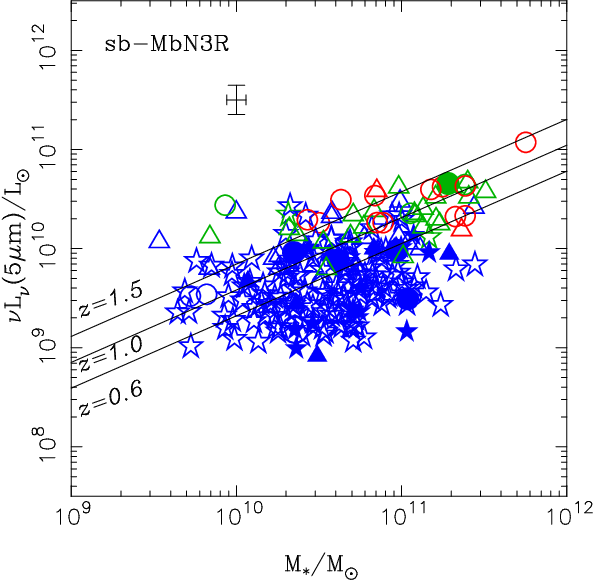}
\end{minipage}
\caption{Correlation between stellar mass $M_{\ast}$ vs. 
$\nu L_{\nu}(5~\mu \mbox{m})$ at
rest-frame 5~$\mu$m for the MbN3Rs.  
Symbols represent the same as in figure~\ref{fig:col_col_mir_mbn3r} 
except for the colors.  The blue,
green, and red symbols represent objects at 
the redshifts of $0.4 \le z <0.8$, $0.8 \le z <
1.2$, and $ z \ge 1.2$, respectively.  Diamonds represent 
the spectroscopic BL AGNs.  Dark and faint symbols 
represent objects with 
reduced $\chi^2<0.5$ and $\chi^2 >0.5$ in the MIR SED fitting, respectively.  
Top: For the agn-MbN3Rs. 
Bottom-Left: For the s/a-MbN3Rs. Bottom-Right: For the sb-MbN3Rs. }
\label{fig:smass_lum5_mbn3r}
\end{figure}  

\begin{figure}[ht]
\begin{minipage}{0.49\linewidth}
\includegraphics[angle=0,width=\linewidth]
{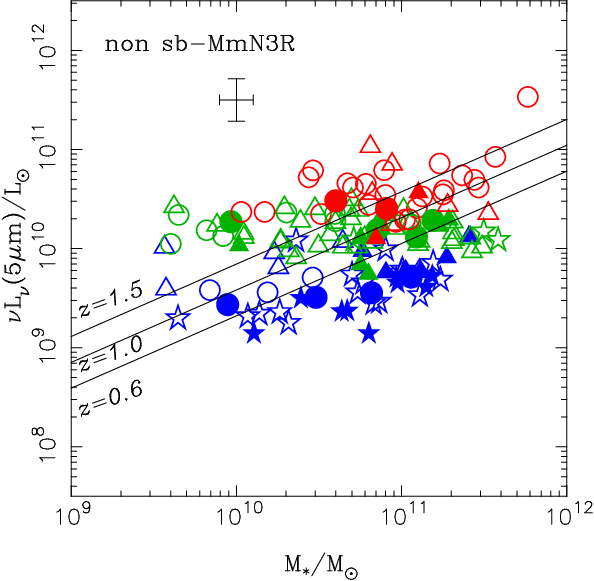}
\end{minipage}
\begin{minipage}{0.49\linewidth}
\includegraphics[angle=0,width=\linewidth]
{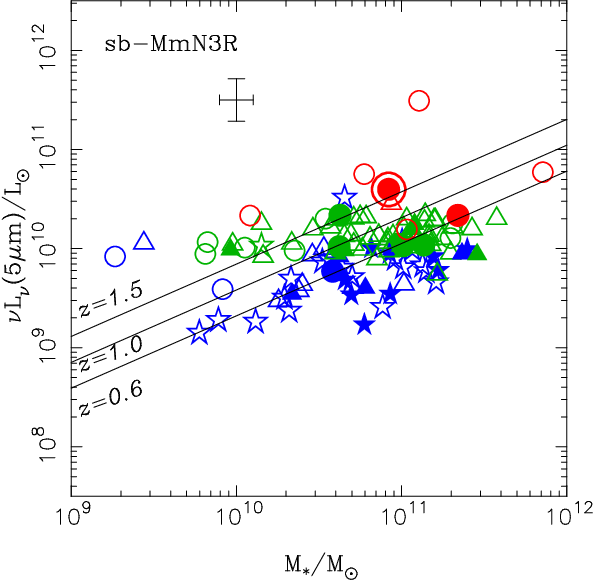}
\end{minipage}
\caption{The same as figure~\ref{fig:smass_lum5_mbn3r} for the MmN3Rs.  
Left: For the non sb-MmN3Rs. Right: For the sb-MmN3Rs. }
\label{fig:smass_lum5_mcn3r}
\end{figure}  

Figure~\ref{fig:smass_lum5_mbn3r} shows the rest-frame absolute
monochromatic luminosities $\nu L_{\nu;5}$ at 5 $\mu$m to their
stellar mass $M_{\ast}$ derived with equation~(\ref{eq:mag_intr}),
for the agn-MbN3Rs, the sb-MbN3Rs and the s/a-MbN3Rs. 
We can see an evolutionary trend as 
the agn-MbN3Rs at higher redshifts are more luminous 
at 5 $\mu$m.   
The mass and evolutionary trend for $\nu L_{\nu;5}$ is
roughly summarized as:
\begin{eqnarray} 
\nu L_{\nu;5}
\left[ L_{\odot} \right]  \simeq  L_{11} 
\left( \frac{M_{\ast}}{10^{11} M_{\odot}} \right)^{\alpha} (1+z)^{\beta} \; ,  
\label{eq:agn_sm_z}  
\end{eqnarray}

In figure~\ref{fig:smass_lum5_mbn3r},  
the data points of the MbN3Rs are plotted 
in redshift intervals of $0.4 \le z<0.8, 0.8
\le z<1.2, 1.2 \le z$, respectively.
The solid line for agn- and s/a-MbN3Rs 
at $z=0.6, 1.0$, and 1.5 represents the best-fitted plane   
of $\log(L_{11}/L_{\odot}) = 8.9 \pm 0.1 $, $\alpha= 0.15 \pm 0.08$, 
and $\beta=6.3 \pm 0.4 $ in equation~(\ref{eq:agn_sm_z}).    
On the other hand, the solid lines for sb-MbN3Rs 
represent the plane of $\alpha= 0.74$ and $\beta=2.7$, which are 
the same expornents of equation~(\ref{eq:sfr_sm_z}) 
for star-forming activities  
in subsection~\ref{subsec:sfr_ssfr}.   
Thus, the 5-$\mu$m emissions of the sb-MbN3Rs are dominant from their dusty
star-forming regions. 

The trends for the 5-$\mu$m luminosities do not depend on
their selections since the $\nu L_{\nu;5}$ of the MbN3Rs is larger than
the limiting luminosities for their detection as $\simeq 10^9, 6
\times 10^9$, and $2 \times 10^{10} L_{\odot}$ at $z = 0.5, 1.0$, and
1.5, respectively. Thus, the emission process at the rest-frame 5~$\mu$m 
for the agn- and the s/a-MbN3Rs should be different from the
dusty star-forming origin for the sb-MbN3Rs.  This supports the supposition 
that MIR emissions from the agn-MbN3Rs are dominated by obscured AGN.  
Thus, it is expected that the 5-$\mu$m luminosity $\nu L_{\nu;5}$ may dominate
in the bolometric luminosity of the obscured AGN $L_{bol}$ for the
s/a- and agn-MbN3Rs.

The evolutionary trend of $\nu L_{\nu;5}$ in the agn-MbN3Rs 
is more rapid than that of SFRs in the sb-MbN3Rs.  
The $\nu L_{\nu;5}$ in the agn-MbN3Rs at $0.8 \le z<1.2$ and
$0.4 \le z <0.8$ also becomes less scattered than those at $z \ge
1.2$, which can be attributable to quenching the AGN activities around $z
\simeq 1$.  

The analysis with $\nu L_{\nu;5}$ can also apply for the MmN3Rs as shown in 
figure~\ref{fig:smass_lum5_mcn3r} even though they are biased 
since most of them are detected in critical threshold 
at the rest-frame 5 $\mu$m.  
The sb-MmN3Rs and the non sb-MmN3Rs show roughly a similar trend to those 
of the sb-MbN3Rs and the s/a-MbN3Rs compared with the agn-MbN3Rs, which 
suggests that AGN activities are not dominant in the MmN3Rs.  

\subsection{Super Massive Black Holes in AGNs} 
\label{subsec:ms_bh}

Consider accretion onto the Super Massive Black Hole (SMBH) with a
mass $M_{BH}$ for an AGN.  The bolometric luminosity $L_{bol}$ of the
emission from the AGN is characterized by an Eddington ratio
$\lambda$ as $L_{bol} = \lambda L_{Edd}$ and the Eddington luminosity
$L_{Edd}$: 
\begin{eqnarray}
L_{Edd} \left[ erg\; s^{-1} \right] 
= 1.26 \times 10^{38} \left( \frac{M_{BH}}{M_{\odot}} \right) 
 \;.  
\end{eqnarray} 
Introducing a luminosity ratio $f_5 = \nu L_{\nu;5} /L_{bol}$, we can
obtain a constraint for the mass of the SMBH from the 5-$\mu$m
luminosity $\nu L_{\nu;5}$ as 
\begin{eqnarray} 
\frac{M_{BH}}{3.06 \times 10^6 M_{\odot}} & = 
& \lambda^{-1} f_5^{-1} \frac{\nu L_{\nu;5}}
{10^{11} L_{\odot}} \; . 
\label{eq:bhms_lum5}  
\end{eqnarray} 
We can see $\nu L_{\nu;5} \simeq 10^{12}L_{\odot}$, and $3 \times 10^{10}
L_{\odot}$ at maximum, and minimum, respectively, 
for the agn-MbN3Rs of $M_{\ast} \simeq
10^{11} M_{\odot}$ at $z>1.2$ in figure~\ref{fig:smass_lum5_mbn3r}. 
It means that SMBHs had already grown in a mass range of $ \simeq 10^6
\lambda^{-1} f_5^{-1} M_{\odot} - 3 \times 10^7 \lambda^{-1} f_5^{-1}
M_{\odot} > 10^6 M_{\odot} $ in a host of $M_{\ast} \simeq 10^{11}
M_{\odot}$ selected as the agn-MbN3Rs at $z>1.2$.  Furthermore, 
as long as $\lambda f_5 > 10^{-1}$, the BH mass in a maximum 5-$\mu$m
luminous agn-MbN3R of $M_{\ast} \simeq 10^{11} M_{\odot}$ at $z>1.2$
does not exceed $ 3 \times 10^8 M_{\odot}$, which is a similar value
attributable to the relation between the mass of SMBH and the stellar mass in
the local spheroid as derived by \citet{hring_black_2004}:  
\begin{eqnarray} 
\log{M_{BH}} & = & -4.12 + 1.12(\log M_{\ast}) 
\; . 
\label{eq:haring}  
\end{eqnarray} 
If the mass relation between SMBH and host was already established 
at $z>1.2$ as this equation~($\ref{eq:haring}$) in the local Universe, then,  
the AGN luminosity in the agn-MbN3Rs at $z>1.2$ could be 
near the Eddington limit $\lambda \simeq 1$, which is consistent with  
consideration about   
SMBH growth history as follows.   

Introducing an energy conversion efficiency $\epsilon= L_{bol}/(\dot{M}c^2)$ 
for mass accretion with $\dot{M}$ onto an SMBH,
the 5-$\mu$m luminosity $\nu L_{\nu;5}$ also implies a constraint for
a mass growth ratio $\dot{M}_{BH}$ of SMBHs as
\begin{eqnarray} 
\frac{\dot{M}_{BH}}{6.8 \times 10^{-3} M_{\odot} yr^{-1}} = 
(1-\epsilon) \epsilon^{-1} 
 f_5^{-1} \frac{\nu L_{\nu;5}} { 10^{11} L_{\odot} } \; ,    
\label{eq:bhacc}  
\end{eqnarray}
where we have used $\dot{M}_{BH}=(1-\epsilon)\dot{M}$ and 
$\nu L_{\nu;5} = f_5 L_{bol}$.  
The thin-disk accretion model has predicted   
the efficiency $\epsilon$ should be in the range $\simeq
0.04-0.31$, depending on the spin of SMBH. 
The growth time scale of SMBHs is also represented as 
\begin{eqnarray} 
\frac{M_{BH}}{\dot{M}_{BH}} \left[ \mbox{yr} \right] 
= 4.5 \times 10^8 \lambda^{-1}   
\frac{\epsilon}{(1-\epsilon)}\;.   
\label{eq:t_bh}  
\end{eqnarray}  
Then, the SMBH growth time could be less than 
the time of $\simeq 2.5$ Gyr from z=2 to z=1 
for $\lambda /\epsilon > 0.2$, which can be satisfied as long as it is 
near the Eddington limit with $\lambda > 0.06 $ for assuming the thin-disk 
accretion model with $\epsilon=0.04-0.31$.  Thus, it is consistent with the
suggestion that SMBHs had already grown in the maximum 5-$\mu$m
luminous agn-MbN3Rs of $M_{\ast} \simeq 10^{11} M_{\odot}$ at $z>1.2$ 
and their AGN luminosity could be near the Eddington limit $\lambda \simeq 1 $
as discussed above.

\subsection{AGN activity and star formation}

Even though the sb-MbN3Rs and the agn-MbN3Rs are different populations,
we could obtain simple formulae representing the activity of star
formation and obscured AGN in the MbN3Rs up to $z \simeq 2$ with
functions of their stellar mass and redshift as
equations~(\ref{eq:sfr_sm_z}) and~(\ref{eq:agn_sm_z}). The 5-$\mu$m
luminosity of the agn-MbN3Rs is rapidly quenching as
equation~(\ref{eq:agn_sm_z}) compared with that of sb-MbN3Rs as
equation~(\ref{eq:sfr_sm_z}). It means that the evolutionary time scale
of AGN activity is shorter than that of star formation for the MbN3Rs
selected with the AKARI MIR photometry.

The rapid evolutionary trend of AGNs in the MbN3Rs was already
expected from its color of stellar population as discussed in
section~\ref{sec:evol_pop}.  Thus, this evolutionary trend in
the time scales, which distinguishes between the agn-MbN3Rs and the sb-MbN3Rs,
is consistent with the evolutionary feature of the MbN3Rs around
$z\simeq 1$ in their distribution on the color-mass diagram as shown
in figure~\ref{fig:smass_uvcol_redc_mbn3r}.  

\section{Discussions} 
\label{sec:discuss} 

\begin{figure}[ht]
\begin{minipage}{0.54\linewidth}
\includegraphics[angle=0,width=\linewidth]
{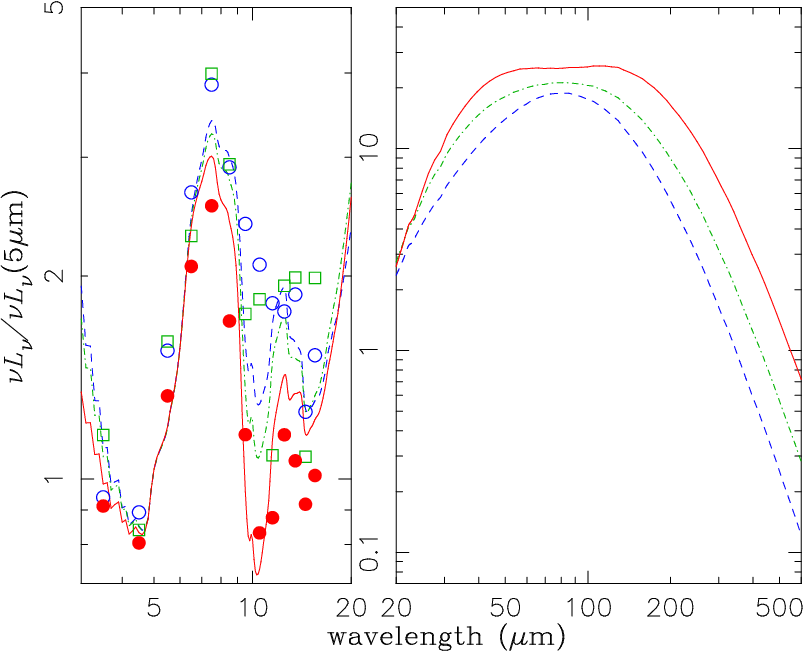}
\end{minipage}
\begin{minipage}{.44\linewidth}
\includegraphics[angle=0,width=\linewidth]
{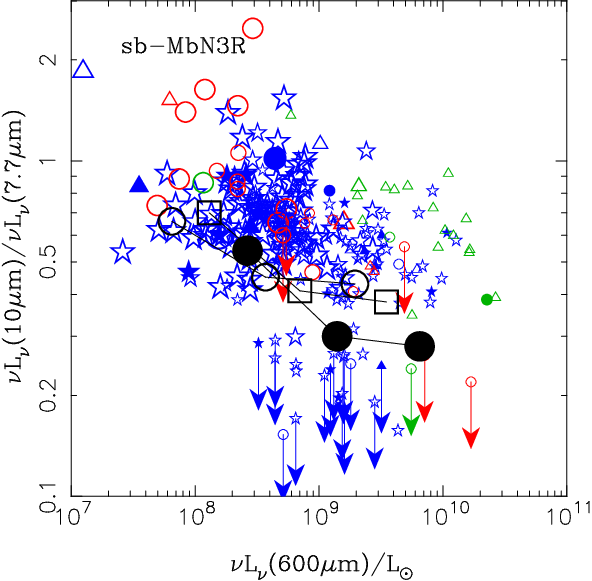}
\end{minipage}
\caption{Left: Averaged rest frame IR-Submm SEDs for the sb-MbN3Rs. 
Red filled circles, green open squares, and blue open circles represent 
monochromatic luminosity $\nu L_{\nu}$ 
averaged for the sb-MbN3Rs which fit well with large, medium, and small  
optical depth S\&K models of 
$A_{V;SK}=17.9, 6.7$, and 2.2, respectively.  
Red solid, green dotted, and blue dashed lines represent  
model SEDs with $A_{V;SK}=17.9, 6.7$, and 2.2, respectively. 
The model SEDs have been convoluted within a 3~$\mu$m wavelength 
corresponding to a mean width of the AKARI/IRC bands.    
Right: Expected monochromatic luminosity at rest-frame 
$600~\mu\mbox{m}$; $\nu L_{\nu}(600~\mu\mbox{m})$ vs. monochromatic 
luminosity ratio at rest-frame $10~\mu\mbox{m}$ to $7.7~\mu\mbox{m}$;    
$\nu L_{\nu}(10~\mu\mbox{m})/\nu L_{\nu}(7.7~\mu\mbox{m})$ for a case with
TIR luminosity $L_{IR}=5 \times 10^{11}$ L$_{\odot}$.  
Black filled circles, open squares, and open circles represent 
monochromatic luminosities 
$\nu L_{\nu}(10~\mu\mbox{m})/\nu L_{\nu}(7.7~\mu\mbox{m})$ from 
the convoluted S\&K models 
with $A_{V;SK}=17.9, 6.7$, and 2.2, respectively.  
We have taken other S\&K model parameters as  
$L^{tot}=10^{10.1}, 10^{11.1}, \mbox{and} 10^{12.1}$ 
L$_{\odot}$ from right to left and fixed $L_{OB}/L^{tot}=0.6$.     
}
\label{fig:lum10_77_600}
\end{figure}  

\begin{figure}[ht]
\begin{minipage}{0.49\linewidth}
\includegraphics[angle=0,width=\linewidth]
{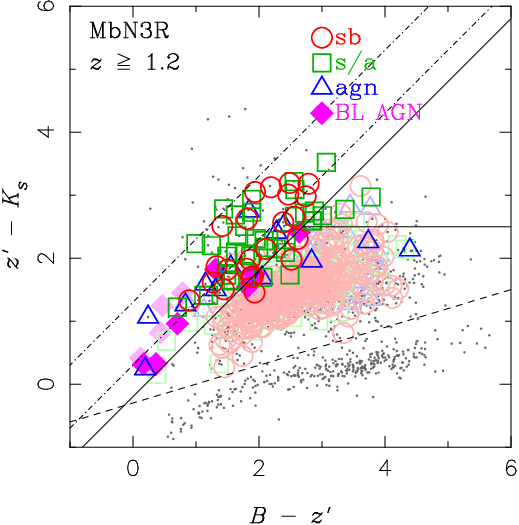}
\end{minipage} 
\begin{minipage}{0.49\linewidth}
\includegraphics[angle=0,width=\linewidth]
{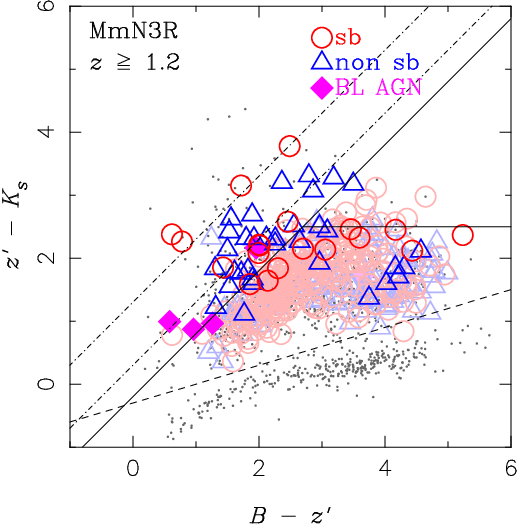}
\end{minipage} 
\caption{$BzK$ color diagram for the MbN3Rs and MmN3Rs. 
Large dark and faint symbols represent galaxies at $z\ge1.2$ and $z<1.2)$, 
resepectively.  
The small dots represent the MIR faint galaxies with $K<21$.  
Left: For the MbN3Rs. 
Red open circles, green open squares, blue open triangles, 
and magenta filled diamonds represent sb-, s/a-, agn-MbN3R, and BL-AGN, 
respectively.   
Right: For the MmN3Rs.  
Red open circles, green open squares and magenta filled diamonds 
represent sb-, non sb-MmN3R, and BL-AGN, respectively.  
}
\label{fig:bzk_mbcn3r}
\end{figure}  

\begin{figure}[ht]
\begin{minipage}{0.49\linewidth}
\includegraphics[angle=0,width=\linewidth]
{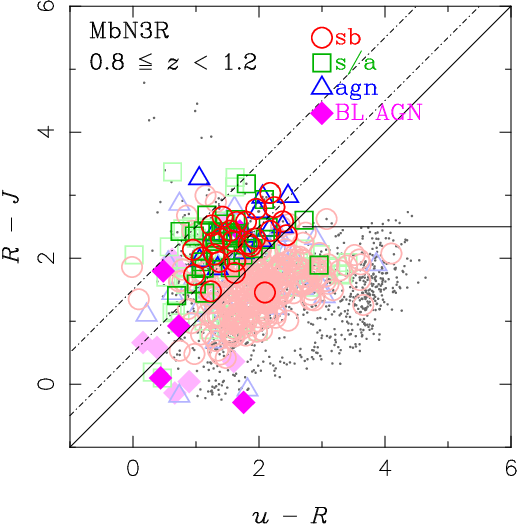}
\end{minipage} 
\begin{minipage}{0.49\linewidth}
\includegraphics[angle=0,width=\linewidth]
{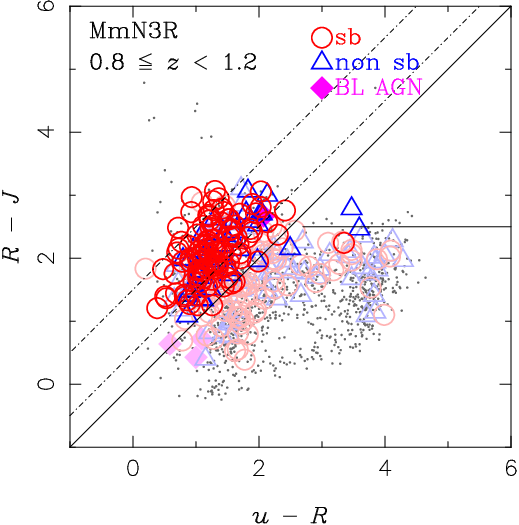}
\end{minipage} 
\caption{The same as figure~\ref{fig:bzk_mbcn3r} 
for the $uRJ$ color diagram. 
Large dark and faint symbols represent galaxies at $0.8 \le z <1.2$ and 
$z<0.8$, respectively.  
Small dots represents the MIR faint galaxies with $J<22$.  
}
\label{fig:urj_mbcn3r}
\end{figure}  

\begin{figure}[ht]
\begin{minipage}{0.49\linewidth}
\includegraphics[angle=0,width=\linewidth]
{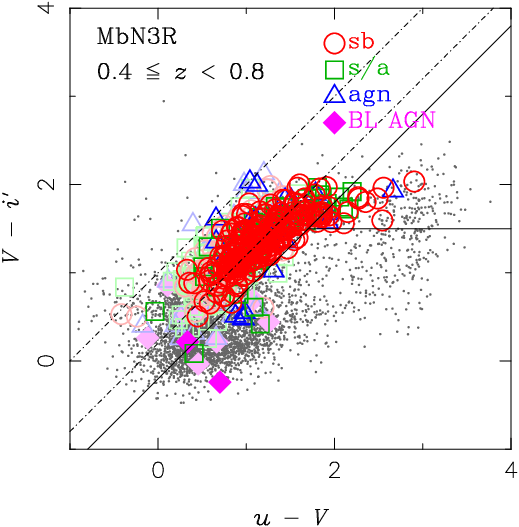}
\end{minipage} 
\begin{minipage}{0.49\linewidth}
\includegraphics[angle=0,width=\linewidth]
{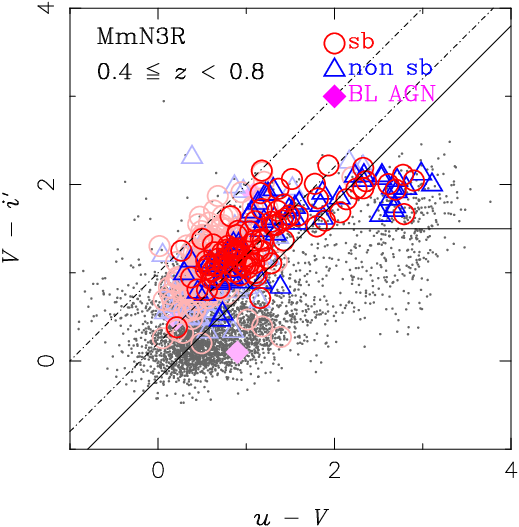}
\end{minipage} 
\caption{The same as figure~\ref{fig:bzk_mbcn3r} 
for the $uVi$ color diagram.}
\label{fig:uvi_mbcn3r}
\end{figure}

\begin{figure}[ht]
\begin{minipage}{0.49\linewidth}
\includegraphics[angle=0,width=\linewidth]
{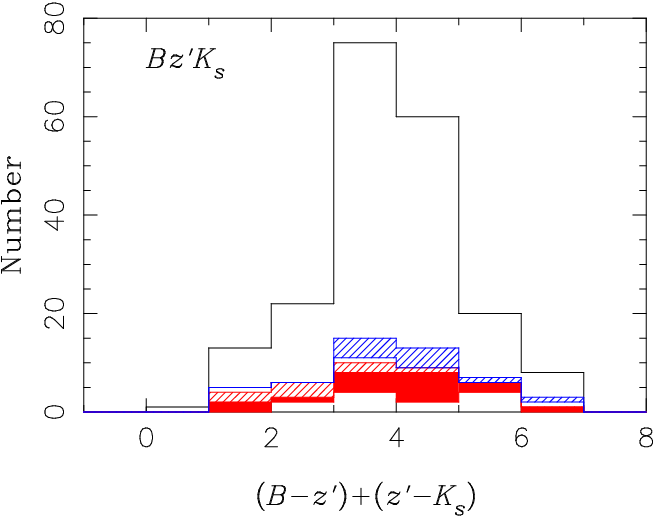}
\end{minipage} 
\begin{minipage}{0.49\linewidth}
\includegraphics[angle=0,width=\linewidth]
{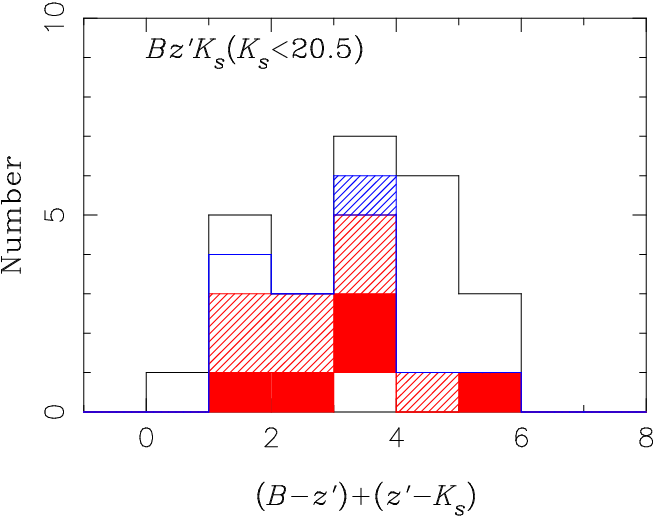}
\end{minipage} 
\begin{minipage}{0.49\linewidth}
\includegraphics[angle=0,width=\linewidth]
{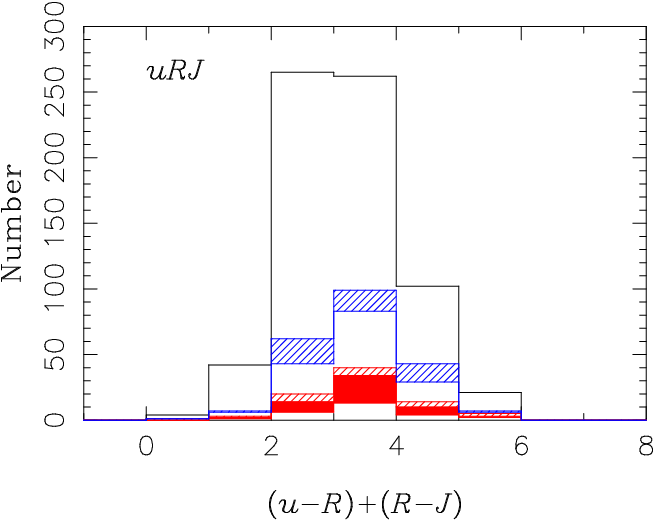}
\end{minipage} 
\begin{minipage}{0.49\linewidth}
\includegraphics[angle=0,width=\linewidth]
{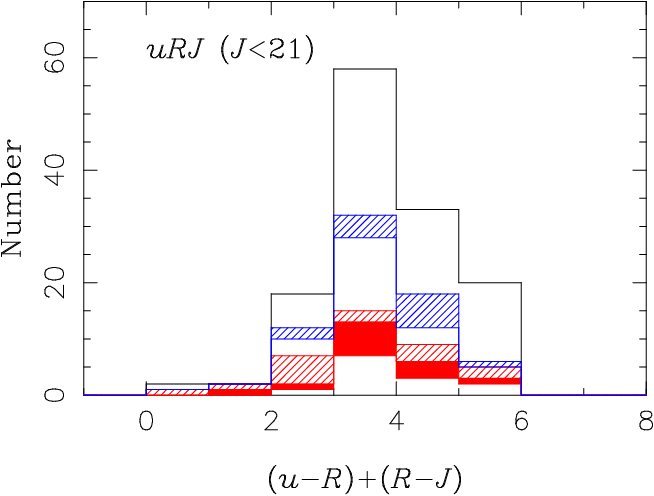}
\end{minipage} 
\caption{
Color distribution of the MbN3Rs along the strip between the dashed 
lines in figures~\ref{fig:bzk_mbcn3r} and ~\ref{fig:urj_mbcn3r}.  
Top solid line histogram indicates 
the number of galaxies detected in the NIR band. 
Red slashed, filled, and outline histograms represent agn-, s/a-, and 
sb-MbN3Rs, respectively.  
Blue slashed and outline histograms represent and non sb-MmN3Rs and sb-MmN3Rs, 
respectively.  
}
\label{fig:bbghist_mbcn3r}
\end{figure}  

\subsection{FIR and Submillimeter Follow-ups}  

As discussed in subsection~\ref{subsec:extinctions}, 
the s-MbN3Rs and the sb-MmN3Rs with a lower 
ratio of monochromatic luminosity at rest-frame $10~\mu\mbox{m}$ 
to that at $7.7~\mu\mbox{m}$; 
$\nu L_{\nu}(10~\mu\mbox{m})/\nu L_{\nu}(7.7~\mu\mbox{m})$  
tend to have larger extinctions 
assuming denser dust distributions (see in figure~\ref{fig:dauv_lum10_77}).  
Larger extinctions, indicated with 
$\nu L_{\nu}(10~\mu\mbox{m})/\nu L_{\nu}(7.7~\mu\mbox{m})$, can make  
reprocessed photon emissions dominant at longer wavelengths. 
This trend can be confirmed in the left plot of figure~\ref{fig:lum10_77_600},  
which shows the average IR/Submm SEDs for the s-MbN3Rs fitted well 
with large, medium, and small optical depth parameters in S\&K model;  
$A_{V;SK}=17.9$, 6.7, and 2.2, respectively.  
The s-MbN3Rs with a larger optical depth tend to show 
not only a deeper dip at $10~\mu$m but also a larger flux in FIR and 
submillimeter wavelengths.  This feature can be confirmed directly with 
observations with FIR devices such as 
the Photodetector Array Camera (PACS) and Photometric Imaging Receiver (SPIRE)
on board {\it Herschel} and submillimeter telescopes as MAMBO/IRAM and ALMA.   
Figure~\ref{fig:lum10_77_600} shows the expected monochromatic luminosity 
at rest-frame $600~\mu\mbox{m}$; $\nu L_{\nu}(600~\mu\mbox{m})$ vs.   
$\nu L_{\nu}(10~\mu\mbox{m})/\nu L_{\nu}(7.7~\mu\mbox{m})$ 
as the trend between extinctions and FIR/Submm wavelength excess.

Observations with ground-based submillimeter telescopes as SCUBA and MAMBO 
have reported that so-called high-z SMGs tend to have SEDs reproduced 
with lower dust temperatures when compared with local 
ULIRGs (e.g. \cite{chapman_population_2004}), which suggests that 
detection of the SMGs with the ground-based submillimeter telescopes 
could be biased and there are missing 
populations of star-forming galaxies with hotter dust 
as Optically Faint Radio Galaxies (OFRGs). 
Recently the PACS and SPIRE/{\it Herschel} observations have revealed that 
ULIRGs at $z\simeq2$ detected with the MIPS/{\it Spitzer} consist of  
both cold and warm populations corresponding to the SMGs and 
the OFRGs (\cite{magdis_herschel_2010}; Elbaz et al. 2011), respectively.  
Even though our samples of MbN3Rs and MmN3Rs may not overlap with  
the {\it Herschel} detected ULIRG samples since our samples 
were detected in the MIR bands and correspond to LIRGs at $z\simeq 1$, 
the variety of $\nu L_{\nu}(10~\mu\mbox{m})/\nu L_{\nu}(7.7~\mu\mbox{m})$  
suggests that LIRGs at $z\simeq 1$ may consist of various populations 
with cold and warm dusts similar to the ULIRGs at $z\simeq 2$ such as the 
SMGs.  The forthcoming and future observations with {\it Herschel} and 
 submillimeter telescopes will confirm 
these features as predicted in figure~\ref{fig:lum10_77_600}. 
     
Adding these photometric data at these longer wavelengths for the MbN3Rs can 
also serve as a strong constraint in the SED analysis with 
direct estimation of their TIR luminosities,  which 
can overcome an ambiguity left in deriving absolute TIR luminosity by 
introducing the proportional factor $f_{SK}$ in section~\ref{sec:ir_lum}. 
Furthermore, this makes a distinction between starburst dominant and 
AGN dominant with the FIR/Submm luminosity with the MIR one,   
which is an alternative way to determine the MIR color criteria for 
selecting s- and agn-MbN3Rs in this work.  The alternative selection 
for AGNs can be also effective to study the census of AGN populations.  
As shown in subsection~\ref{subsec:mir_sed_sb_agn}, 
the CXO found that some agn- and s/a-MbN3Rs are associated with X-ray 
sources.  On the other hand, there are still some populations 
in the agn- and s/a-MbN3Rs escaping from the X-ray detection with the CXO.   
This suggests that obscured AGNs in the agn- and s/a-MbN3Rs 
are classified to two subpopulations associated with and without X-ray sources. 
The former and the latter are possible candidates of obscured AGNs in the 
Compton-thin range of $N_H \le 10^{24} \mbox{cm}^{-2}$ and  
the Compton-thick (CT) range of $N_H > 10^{24} \mbox{cm}^{-2}$, 
respectively.  Thus, the comparison of MIR-FIR/Submm SEDs between 
the CT and non-CT AGN candidates is expected to supply constraints for 
physical states around the SMBHs in the obscured AGNs.   

\subsection{Optical colors of MbN3Rs and MmN3Rs}  

Most of the MbN3Rs and MmN3Rs are also classified as BBGs.  
Then, we have  
also compared the sb-, s/a-, and agn-MbN3Rs with their color 
on the two-color diagrams related to the BBG criteria 
derived in appendix~\ref{app_sec:bbg}. 
Figures~\ref{fig:bzk_mbcn3r}, ~\ref{fig:urj_mbcn3r}, and ~\ref{fig:uvi_mbcn3r} 
show that the s-MbN3Rs and the sb-MmN3Rs mainly concentrate around 
$(B-z,z-K) \simeq (2,2.5)$, $(u-R,R-J) \simeq (2,2.5)$, 
and $(u-V,V-i) \simeq (1.5,1.5)$, while the agn-MbN3Rs are relatively 
more dispersed than the s-MbN3Rs.  
These $\cal{B}\mu\cal{R}$ colors of the s-MbN3Rs and the sb-MmN3Rs are 
consistent with those expected from the reddenings with heavy extinctions 
as shown by \cite{daddi_new_2004} and in appendix~\ref{app_sec:bbg}. 
We detected $\sim 60$ s-BzKs and $\sim 100$ s-uRJs with $K<21.5$ and $J<22$   
in the color region within $\Delta (B-K) \simeq \pm 0.5 $ and  
$\Delta (u-J) \simeq \pm 0.5 $ around 
$(B-z,z-K) \simeq (2,2.5)$ and $(u-R,R-J) \simeq (2,2.5)$, respectively,   
where $\sim 15$ and $\sim 40$ out of these s-BzKs (s-uRJs) are classified 
as s-MbN3Rs as shown on the left in figure~\ref{fig:bbghist_mbcn3r}.  
Thus, roughly, we can detect dusty starbursts at $z \sim 1-2$ 
with a frequency one fifth of s-BzKs and one third of s-uRJ.  
For NIR bright BBGs with $K<20.5$ and $J<21$, the fractions of MbN3Rs and 
MmN3Rs in BBGs increase as shown on the right in 
figure~\ref{fig:bbghist_mbcn3r}.  
In particular, roughly half of the uRJs with $J<21$ 
around $(u-R,R-J) \simeq (2,2.5)$ are LIRGs around $z\sim 1$, 
which are selected in the NEP deep field.  We should note that 
this NEP deep field is a blank field.  
Thus, these kinds of BBGs constrained with NIR brightness and colors  
in any blank field can be mostly detected as LIRGs around $z \sim 1-2$,  
which are interesting targets for a future unbiased survey for galaxies 
with on-going, up-coming, or future FIR/Submm telescopes as {\em Herschel} 
and ALMA.  These future studies will be also essential to confirm the 
trends for the populations marginally detected with the AKARI 
as the MmN3Rs. 

\section{Conclusions} 
\label{sec:concl} 

We have studied BBGs/IRBGs in three 
redshift ranges of $z=0.4-0.8, 0.8-1.2$, and $>1.2$ with  
MIR multiband photometry, which revealed their star-forming/AGN
activities with/without PAH emissions deriving SFR from
their TIR luminosities.  

\begin{itemize}
\item The MIR colors and SED analysis can classify MIR bright  
N3 Red galaxies (MbN3Rs) into three classes; 
starburst dominates (sb-MbN3Rs), starburst/AGN co-existence(s/a-MbN3Rs), 
and AGN dominates (agn-MbN3Rs).  
The ratio of rest-frame luminosities at 5 and 10 $\mu$m to that at 
$7.7~\mu$m is systematically higher in the s/a-(agn-)MbN3Rs than in the 
sb-(s/a-)MbN3Rs, 
which is consistent with the picture of the s/a-MbN3Rs as the 
starburst/AGN co-existent subpopulation in the dusty star-forming MbN3Rs 
(s-MbN3Rs).   
\item MIR 
marginally-detected N3Rs (MmN3Rs) can be also classified  
into two classes; sb-MmN3Rs as mimics of sb-MbN3Rs, and non sb-MmN3Rs.  
\item Their rest-frame 7.7-$\mu$m luminosity is a good tracer of
TIR luminosity as the PAH emission dominates for the sb- and s/a-MbN3Rs and 
sb-MmN3Rs which are classified 
as dusty star-forming galaxies even up to $z=2$.  
\item The $SFR_{IR+UV}$ derived from the total bolometric luminosity
$L_{IR+UV}$ of the sb-MbN3Rs/MmN3Rs shows a correlation that is nearly
proportional to the stellar mass $M_{\ast}$, which is 
consistent with 
that of the extinction corrected UV star formation ratio $SFR_{UV;corr}$.  
\item The $SFR_{IR+UV}$ of the s/a-MbN3Rs tends to be smaller than 
that of the sb-MbN3Rs at $z<0.8$.  
\item The IR-derived specific SFR of the sb-MbN3Rs/MmN3Rs rises 
with the redshift at all stellar masses. 
\item The ratio of the rest-frame luminosity at $10~\mu$m to that at 
$7.7~\mu$m of the sb-MbN3Rs/MmN3Rs may trace the optical depth due to the 
dust extinction.  
\item The AGN activity could be traced with the rest-frame 5-$\mu$m
luminosity for the agn- and the s/a-MbN3Rs. The evolutionary trend
$(1+z)^6$ of the AGN luminosity for the agn-MbN3Rs is more rapid than
the $(1+z)^3$ of SFR for the sb-MbN3Rs/MmN3Rs.
\item SMBHs in the galaxies have grown to $\simeq 3
\times 10^8 M_{\odot}$ in maximum luminous populations with
$10^{12}L_{\odot}$ and $10^{11} M_{\odot}$ of the MIR selected AGNs at
$z>1.2$, which suggests the mass relation between the SMBH and its
host is already established at $z\simeq 1-2$ and 
the same as that at the present day.
\end{itemize}  

\bigskip
This study is based on data mostly collected with the AKARI IR astronomical 
satellite and at the Subaru Telescope, which are operated 
by the ISAS/JAXA and the National Astronomical Observatory of Japan, 
respectively.  We would like to thank their staff 
for their invaluable assistance.  We thank B. Weiner and C. Papovich 
for their help in the spectroscopic observation with the Hectospec/MMT 
under TSIP program of National Optical Astronomy Observatory.  
We are grateful to an anonymous referee 
for his/her comments which have helped to clarify the paper.  
HH thanks P. Langman, Iwate University, for proofreading the English writing.  
This research was mostly supported for HH by the Grant-in-Aid for Scientific
Research (21340042) from the Japan Society for the Promotion of
Science (JSPS), and partly supported for MI by the Korea Science 
and Engineering Foundation (KOSEF) grant no. 2010-0000712 
from the Korea government (MEST).  

\appendix 

\section{Classification with Balmer/4000 \AA\  breaks}
\label{app_sec:bbg}

In order to identify galaxies at various redshift intervals, we have
generalized the two-color technique for detecting the Balmer/4000 \AA\  
break, which is a major feature in the SED of galaxies.  The Balmer
and 4000 \AA\ break are often treated as a single feature, owing to
their similar wavelengths and the overlapping populations of galaxies
having these features. However, the breaks originate due to different
physical processes, and behave differently as a function of age and
population. Both breaks result from absorption in stellar
atmospheres. The `classical' Balmer break at 3648 \AA\ marks the
termination of the hydrogen Balmer series, and is strongest in A-type
stars. This means that it is the most prominent feature for
intermediate ages of $0.3-1.0$ Gyr in a Single Stellar Population
(SSP).  The Balmer break strength, $D_B$, is defined as the ratio of the flux
density $F_{\nu}$ in the 3500-3650 \AA\ and 3800-3950 \AA\ bands around
the break \citep{balogh_differential_1999}.  On the other hand, the
4000 \AA\ break arises because of an accumulation of absorption lines
of mainly, ionized metals.  Its strength is defined using an index,
$D_n(4000)$, that is based on the continuum regions, and the ratio of
the flux density $F_{\nu}$ in the 3850-3950 \AA\ and 4000-4100 \AA\
bands.  As the opacity increases with decreasing stellar temperature,
the 4000 \AA\ break strength, $D_n(4000)$, increases for older ages,
and is largest in old and metal-rich stellar populations. However, the
Balmer break strength $D_B$ does not monotonically increase with age,
instead reaching a maximum at intermediate ages 
\citep{kriek_direct_2006}. The metallicity has a more minor influence
for ages less than 1 Gyr \citep{bruzual_stellar_2003}. Thus, both
break features around 4000 \AA\ include information about the stellar
populations in a galaxy. Although these breaks are individually
unresolved from broad-band photometry, the degenerate Balmer/4000 \AA\ 
break still provides a robust feature in the SEDs with galaxies of an age 
$t>500$ Myr.  In the following subsections, we will demonstrate that the
Balmer/4000 \AA\ break can be used to select galaxies at various
redshift intervals of interest to this work.  

\begin{table}
\begin{center}
  \caption{The Parameters for BBG Selection}
\label{tab:bbg_select}
  \begin{tabular}{lccccccc}
  \hline
  \hline
  Bands & $ f_{\scriptsize{{\cal B}\mu{\cal R}}} $ 
        & $ g_{\scriptsize{{\cal B}\mu{\cal R}}} $ 
        & $ h_{\scriptsize{{\cal B}\mu{\cal R}}} $ 
        & $z_{\mbox{\scriptsize{min}}}$ & $z_{\mbox{\scriptsize{max}}}$ 
  \\
\hline
  $uVi$  &  $-0.2$  & 1.5 & 1.0 & $0.4$ & $1.2$ \\
  $uRJ$  &  0.0     & 2.5 & 1.0 & $0.8$ & $1.8$ \\
  $BzK$  & $-0.2$   & 2.5 & 1.0 & $1.4$ & $2.5$ \\
\hline 
  $BRN3$ & $0.7$   & 3.0 & 3.0 & $0.8$ & $1.8$ \\
  $uRN3$ & $0.0$   & 3.0 & 2.0 & $0.8$ & $1.8$ \\
  $BzN3$ & $-0.2$   & 2.5 & 1.4 & $1.4$ & $2.5$ \\
  $uzN3$ & $-0.5$   & 2.5 & 1.2 & $1.4$ & $2.5$ \\
  \hline
  \end{tabular} 
\end{center}
\end{table}

\begin{figure}[ht]
\begin{minipage}{0.49\linewidth}
\includegraphics[width=\linewidth]
{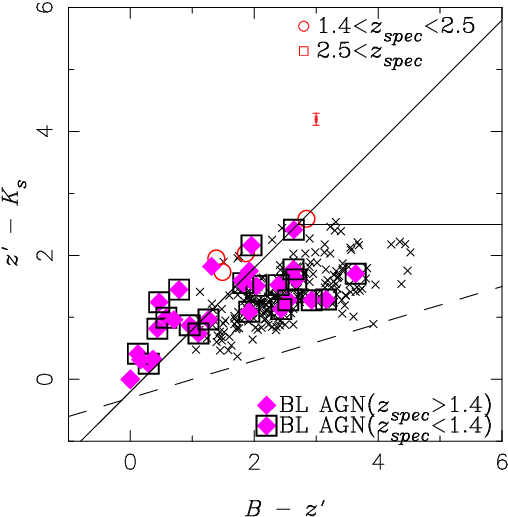}
\end{minipage}
\begin{minipage}{0.49\linewidth}
\includegraphics[angle=0,width=\linewidth]
{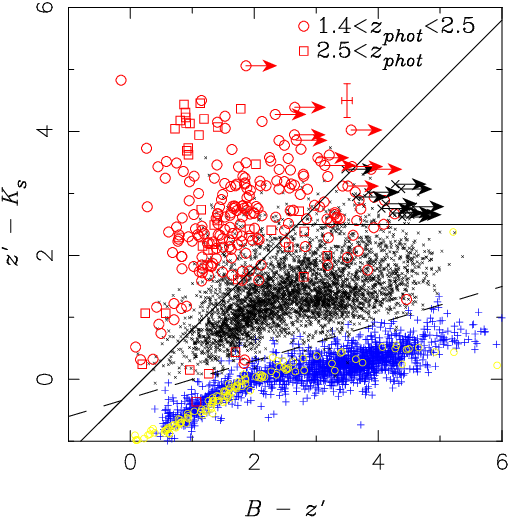}
\end{minipage}
\caption{$BzK$ diagram in the field. 
Symbols of cross bar at $(B-z',z'-K_s)=(3,4.5)$ represent  
typical errors of the color.  
Magenta diamonds represent BL AGNs, in which squared diamonds 
represent lower-redshifted ones at $z<1.4$.  
Left: For the spectroscopic samples, red circles represent sources 
at $z_{spec}=1.4-2.5$.  Right: The same as on the Left 
for the photometric samples, where
blue cross and yellow circles represent the stellar objects
selected with the scheme described in subsection~\ref{subsec:stargal}, 
respectively, 
and the reference stars of the BPGS star atlas.  Red circles and
squares represent sources at $z_{phot}=1.4-2.5$, and at $z_{phot}>2.5$
respectively. }
\label{fig:bzk_spec_phot}
\end{figure}  

\begin{figure}[ht]
\begin{minipage}{0.49\linewidth}
\includegraphics[width=\linewidth]
{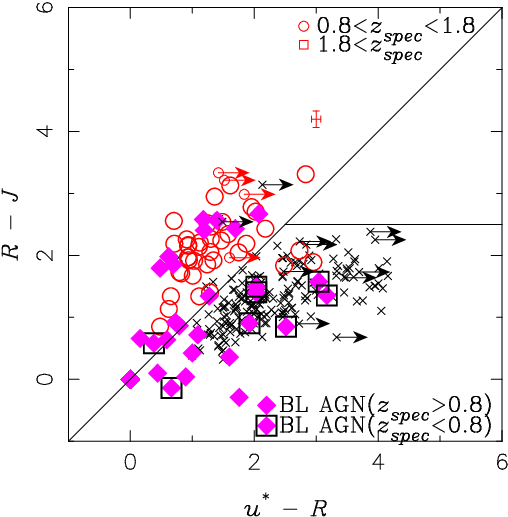}
\end{minipage}
\begin{minipage}{0.49\linewidth}
\includegraphics[width=\linewidth]
{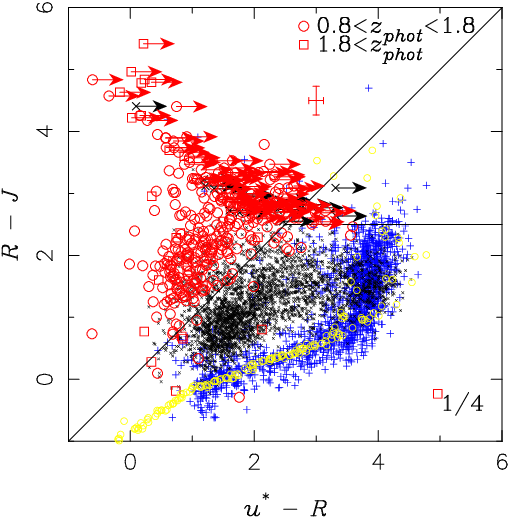}
\end{minipage}
\caption{The same as figure~\ref{fig:bzk_spec_phot} for the uRJs. 
Symbols of cross bar at $(u^{\ast}-R,R-J)=(3,4.5)$ represent  
typical errors of the color.  
Red circles and squares represent sources at $z_{phot}=0.8-1.8$
and at $z_{phot}>1.8$, respectively. 
Left: For the spectroscopic samples. 
Right: For the photometric samples. 
Only one fourth of all photometric sources 
detected in the $uRJ$ bands are plotted to reduce crowding in the diagram. }
\label{fig:urj_spec_phot}
\end{figure}

\begin{figure}
\begin{minipage}{0.49\linewidth}
\includegraphics[width=\linewidth]
{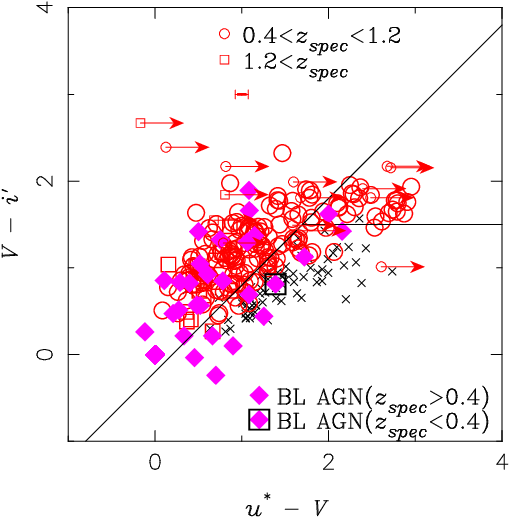}
\end{minipage}
\begin{minipage}{0.49\linewidth}
\includegraphics[width=\linewidth]
{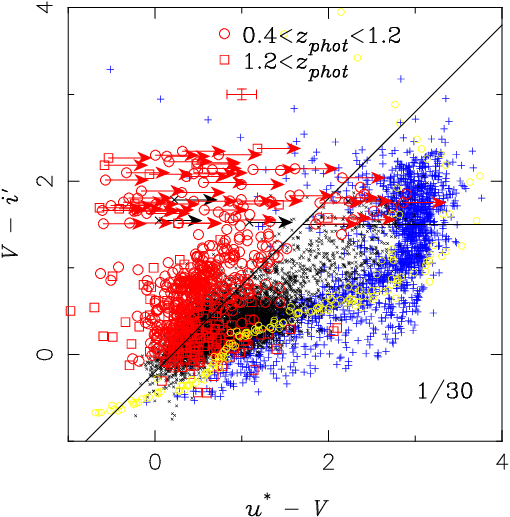}
\end{minipage}
\caption{The same as figure~\ref{fig:bzk_spec_phot} for the uVis. 
Symbols of cross bar at $(u^{\ast}-V,V-i')=(1,3)$ represent typical 
errors of the color.  
Red circle and square represent sources at $z_{spec}=0.4-1.2$ 
and $z_{spec}>1.2$, respectively. 
Left: For the spectroscopic sample.   
Right: For the photometric sample. Only one thirtieth of 
all photometric sources detected 
in the $uVi$ bands are plotted  to reduce crowding in the diagram. }
\label{fig:uvi_spec_phot}
\end{figure}  

\begin{figure}[ht]
\begin{minipage}{0.49\linewidth}
\includegraphics[width=\linewidth]
{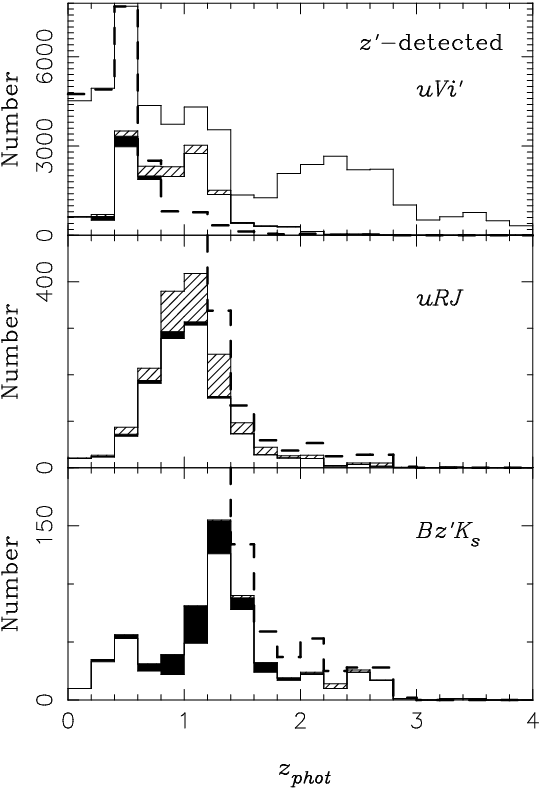}
\end{minipage}
\begin{minipage}{0.49\linewidth}
\includegraphics[width=\linewidth]
{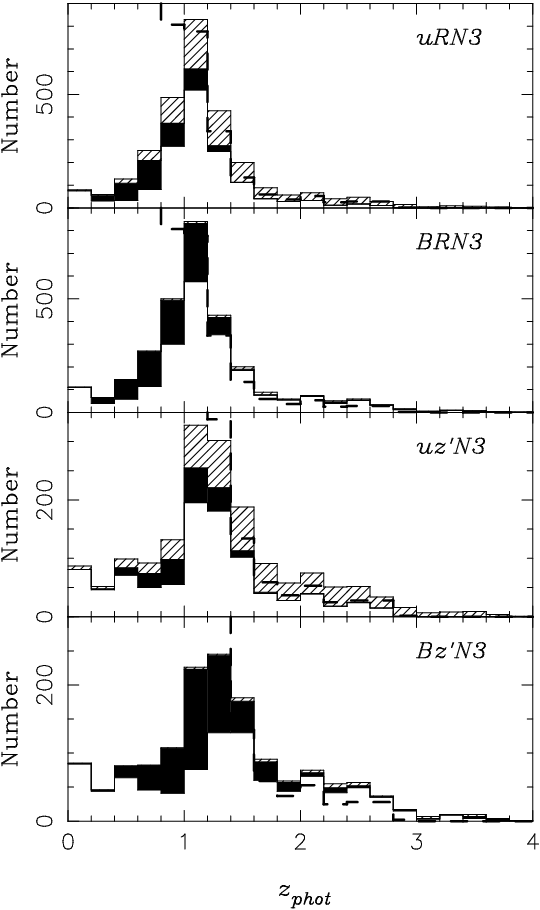}
\end{minipage}
\caption{Photometric redshift distributions for the BBGs in the field.  
The white, shaded, and dark area represent 
the s-, qp-, and p-BBGs, respectively.    
Left: For the classical BBGs. 
Right: The same as on the left for the extended BBGs.  
Left-Top: Lower, upper solid line and dashed line histogram represent the uVis, 
the $z'$-detected galaxies, and the rest-frame $V$-detected galaxies, 
respectively.  
Left-Middle: For the uRJs. 
Left-Bottom: For the BzKs.  
Right: For the uRN3s.  Right-Upper-Middle: For the BRN3s.  
Right-Lower-Middle: For the uzN3s.  Right-Bottom: For the BzN3s.  
}
\label{fig:photzhist_bbg_exbbg}
\end{figure}  

\begin{figure}[ht]
\begin{minipage}{0.99\linewidth}
\includegraphics[angle=0,width=\linewidth]
{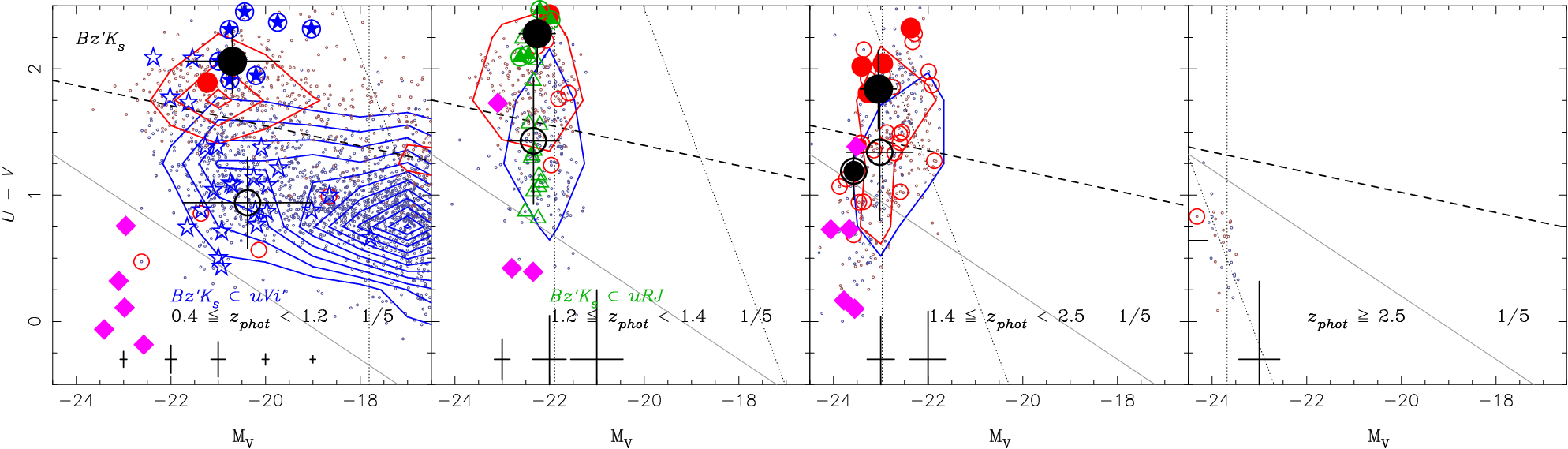}
\end{minipage}
\begin{minipage}{0.99\linewidth}
\includegraphics[angle=0,width=\linewidth]
{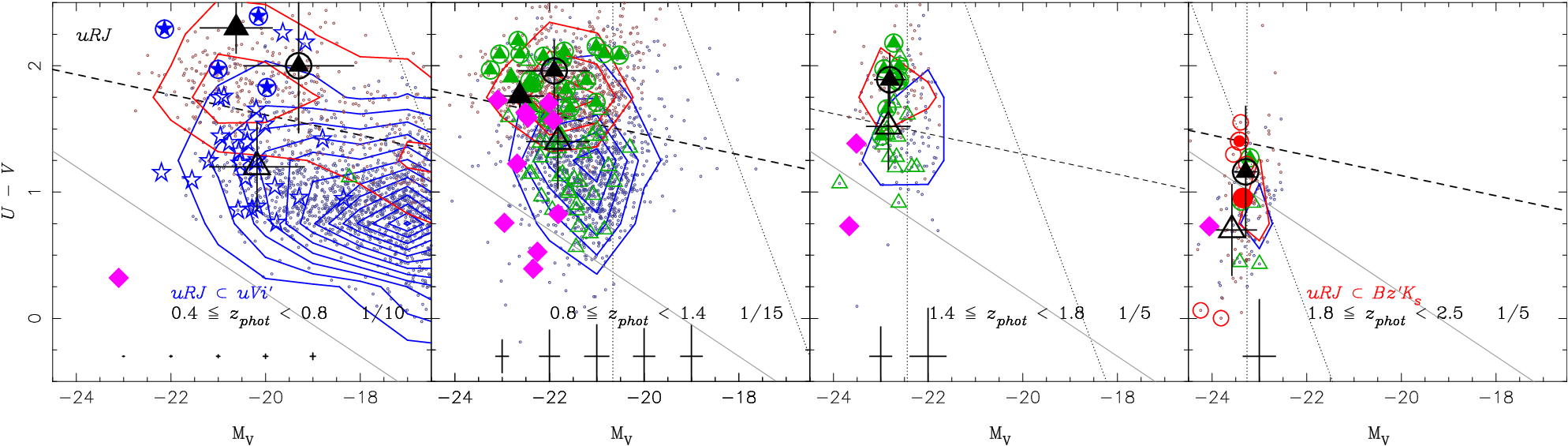}
\end{minipage}
\begin{minipage}{0.99\linewidth}
\includegraphics[angle=0,width=\linewidth]
{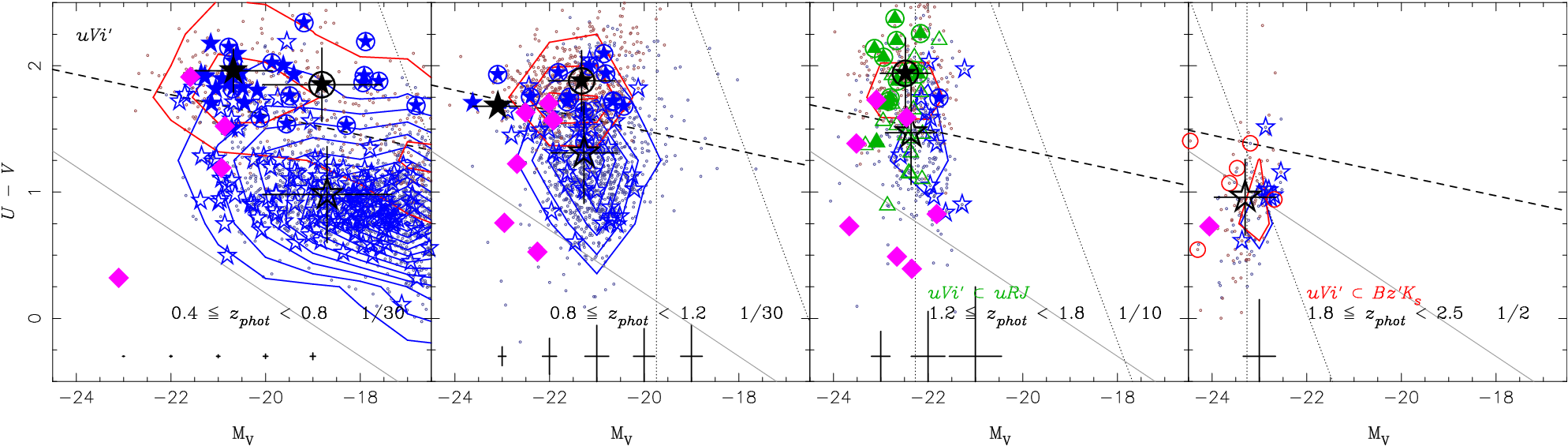}
\end{minipage}
\caption{The same as figure~\ref{fig:mv_uvcol_zdet_spec} for classical BBGs.  
Typical errors of the magnitude and color are represented with crosses at 
the bottom.  The vertical dotted and steep slope dotted lines represent 
the limiting absolute $V$ and $U$-band magnitudes at the mean redshift, 
respectively.  
In general, colored open, solid, and open-solid symbols 
represent objects classified as s-, p-, and qp-BBGs, respectively. 
Basically, red circles, green triangles, and blue stars  
represent objects selected with the two-color criteria 
as BzKs, uRJs, and uVis, respectively. 
Black-large open, solid, and open-solid symbols represent mean values 
for the s-, p-, and qp-BBGs, respectively.  
Crosses on the black symbols represent their standard 
deviations.  
Only a part of samples are plotted to reduce crowding in the diagram, whose 
fractions are presented as numbers in the right-bottom region of all figures.
Top: The BzKs at $0.4 \le z<1.2$, $1.2 \le z<1.4$, $1.4 \le z<2.5$, and 
$z \ge 2.5$ are plotted from left to right.  
For BzKs also classified as uRJs at $1.2 \le z<1.4$, 
their symbols are changed to green triangles from red circles.  
For BzKs also classified as uRJs at $0.4 \le z<1.2$, 
the symbols are changed to blue stars from red circles.  
Middle: The uRJs at $0.4 \le z<0.8$, $0.8 \le z<1.4$, $1.4 \le z<1.8$, and 
$1.8 \le z<2.5$ are plotted from left to right.  
For uRJs also classified as BzKs at $1.8 \le z<2.5$, 
their symbols are changed to red circles from green triangles.  
For uRJs also classified as uVis at $0.4 \le z<0.8$, 
their symbols are changed to blue stars from green triangles.  
Bottom: The uVis at $0.4 \le z<0.8$, $0.8 \le z<1.2$, 
$1.2 \le z<1.8$, and $1.8 \le z<2.5$ are plotted from left to right.  
For uVis also classified as uRJs at $1.2 \le z<1.8$, 
their symbols are changed to green triangles from blue stars.  
For uVis also classified as BzKs at $1.8 \le z<2.5$, 
their symbols are changed to red circles from blue stars.  
}
\label{fig:mv_uvcol_bbg}
\end{figure}  

\subsection{The $BzK$, $uRJ$, and $uVi$ color-color diagram}
\label{app_subsec:bbg}

Daddi et al. (2004) proposed that $z > 1.4$ star-forming BBGs (s-BBGs)
should occupy the upper-left region in a two-color $B-z$ and $z-K$
($BzK$) diagram, as the bluer and redder colors trace 
UV excess and the Balmer/4000 \AA\ break feature, representing 
the star forming activity and the old stellar
component, respectively. We could also show that the $BzK$ technique
is efficient for selecting galaxies at $1.4 < z_{phot} < 2.5$, as
illustrated in figure~\ref{fig:bzk_spec_phot}.

We will argue that this kind of two-color technique can be extended
to selecting s-BBGs at certain redshift intervals with other sets
of three photometric observed broad bands, in which their wavelengths
are nearly in an equal ratio as a geometric series. By introducing ${\cal
B}$, $\mu$, and ${\cal R}$ as their AB magnitudes of general bluer,
medium, and redder bands, resepectively, and defining a two-color index ${\cal
B}\mu{\cal R}$ generalized in $BzK$: 
\begin{eqnarray}
{\cal B}{\mu}{\cal R} 
\equiv ({\mu}-{\cal R})  -
 h_{\scriptsize{{\cal B}\mu{\cal R}}} ({\cal B}-{\mu}), 
\label{eq:BmR}
\end{eqnarray}
it follows that s-BBGs at certain redshift intervals can be selected
by a criterion: 
\begin{eqnarray}
{\cal B}{\mu}{\cal R} \geq f_{\scriptsize{{\cal B}\mu{\cal R}}} ,
\label{eq:s-bbg}
\end{eqnarray} 
and passively evolving BBGs (p-BBGs) can be selected with a criterion:   
\begin{eqnarray}
{\cal B}{\mu}{\cal R} < f_{\scriptsize{{\cal B}\mu{\cal R}}}\ 
\cap\ ({\mu}-{\cal R})_{\mbox{\scriptsize{AB}}} > 
g_{\scriptsize{{\cal B}\mu{\cal R}}} \; . 
\label{eq:p-bbg}
\end{eqnarray} 
Even though we can obtain only the upper limit 
in their ${\cal B}$ magnitudes, 
we can select quasi passively evolving BBGs (qp-BBGs) 
with a criterion:  
\begin{eqnarray}
({\mu}-{\cal R})_{\mbox{\scriptsize{AB}}} > 
g_{\scriptsize{{\cal B}\mu{\cal R}}} \; . 
\label{eq:qp-bbg}
\end{eqnarray} 

We confirm the validity of this generalized BBG technique as shown in
figures~\ref{fig:urj_spec_phot} and \ref{fig:uvi_spec_phot}. These
generalized criteria are efficient ways to select photometric s/p/qp-BBGs
at $0.8 < z < 1.8$ and $0.4 < z < 1.2$ in the field for the $uRJ$ and
$uVi$ filter sets.  These particular color criteria also allow
efficient selections of the BBGs not only for $1.4 < z < 2.5$, but
also $0.8 < z < 1.8$ and $0.4 < z < 1.2$.  

In order to study the validity and physical meaning of the two-color
criteria that are generalized from the phenomenologically established
$BzK$ technique of \citet{daddi_new_2004}, we used the BC03 models to 
reproduce the blue ${\cal B}-{\mu}$ and red colors ${\mu}-{\cal R}$ of 
model galaxies, which are described and shown in 
figures~\ref{fig:urjmodel} and \ref{fig:uvimodel} 
in appendix~\ref{app_subsec:track_bbg}.  
The result from the stellar population synthesis study 
is essentially similar to 
the cases of BzKs, as summarized below.

In the redshift ranges for each BBG selection in table~\ref{tab:bbg_select}, 
the galaxies with ongoing star formation appear in the upper-left region 
${\cal B}{\mu}{\cal R} \geq f_{\scriptsize{{\cal B}\mu{\cal R}}}$ of the 
generalized ${\cal B}{\mu}{\cal R}$ color diagrams.  The duration of 
the star-formation (age) has little influence on the blue color 
${\cal B}-{\mu}$, while the redder color ${\mu}-{\cal R}$ increases 
with age.  Only very young starbursts with ages less than 10 Myr, and 
without underlying older stellar populations, are located around the 
lower-left region, 
falling just below the line of ${\cal B}{\mu}{\cal R}=f_{{\cal
B}{\mu}{\cal R}}$.   On the other hand, the colors of normal galaxies, 
obtained from the CWW SED templates,   
fall outside the color area of the BBGs in the redshift ranges for
each of the BBG selections.  This means that the generalized
two-color criteria can also be used to exclude the lower-$z$ objects.

The reddening direction lies approximately parallel to the line
defined by ${\cal B}{\mu}{\cal R} = f_{{\cal B}{\mu}{\cal R}}$, since
the dust extinction can be approximately represented by a power-law
function of wavelength, implying that the reddening is similar in the
${\cal B}-{\mu}$ and ${\mu}-{\cal R}$ colors for star-forming
galaxies in the various redshift ranges.  This means that the
generalized criteria $ {\cal B}{\mu}{\cal R} \geq f_{{\cal
B}{\mu}{\cal R}}$ for selecting BBGs are also robust from their dust
reddening.  

Only a few galaxies are found as faint blue objects at ${\cal B}-{\mu}
\simeq 0$, suggesting that purely un-reddened, star-forming galaxies
are rare, not only for s-BzKs at $z \simeq 2$, but also s-uRJs at $
z\simeq 1$.  On the other hand, we could find some fraction of these
nearly extinction-free populations in s-uVis at $z\simeq 0.6$. 

Figure~\ref{fig:photzhist_bbg_exbbg} shows the distributions of 
photometric redshifts $z_{phot}$ for all the $z'$-detected galaxies and
the BBGs.  Total numbers of detected uRJs and uVis as the lower-z BBGs 
are greater than that of detected BzKs while the fractions of low-z interlopers 
in the uRJs and uVis are in the same order as those in the BzKs as shown in 
figure~\ref{fig:photzhist_bbg_exbbg}.  Thus, all the BBG selections can be 
effective to exclude low-z interlopers at $z<0.4, z<0.6$, and $z<1.2$ 
with the $uVi$, $uRJ$, and $BzK$ criteria.  
These selected-out redshift ranges are consistent with the 
expected values of $z<0.4, z<0.8$, and $z<1.2$ from the stellar
population models (see in appendix~\ref{app_sec:model}).  
While the uVis are selected out at
$z>1.2$ as expected with the model tracks 
(see in appendix~\ref{app_sec:model}), the uRJs
and BzKs gradually decrease in population at $z>1.2$, and $>1.6$ lower
than expected.  Low efficiency for BBG selections at the high redshift
side may be caused from the shallowness of our ground-based NIR
photometry.  The limiting magnitudes of $J<21$ and $K_s<22$ with the
KPNO/FLMG are shallow compared with studies for the BzKs in other
survey fields \citep{kong_wide_2006, lane_colour_2007,
quadri_multiwavelength_2007, blanc_multiwavelength_2008}.  Thus, we
take into account this limitation of the ground-based NIR photometry in the
dataset when we study galaxies at $z\simeq 0.8-2$ as uRJs and BzKs.  
We try to overcome the limitation due to the shallow  
ground-based NIR photometry for the uRJs and BzKs.  
We further extend the two-color criteria used for detecting the Balmer 
in combination with AKARI NIR photometry as discussed in 
appendix~\ref{app_subsec:exbbg}.  

\subsection{Rest frame color  of  the BBGs} 

It is worth comparing the classification with the two-color criteria 
detecting the Balmer break for the BBGs and their rest frame colors.   
As shown in the top and middle of figure~\ref{fig:mv_uvcol_bbg}, 
most of the BzKs at $1.2 \le z_{phot}<1.4$ are classified  
as uRJs while only a few of the uRJs at $1.4 < z_{phot}<2.5$ are classified as 
BzKs.  This is related to the shallowness in our ground based NIR photometry, 
which is not deep enough to select the population around $z\simeq 2$ as BzKs 
as remarked in the previous section.  
For our sample up to $z\simeq 2$, the uRJ criteria 
are more useful than the BzK criteria.  
We can see the bimodality even for the BBGs again as 
the boundary between the blue cloud and the red sequence 
is consistent with that by \citet{bell_nearly_2004}, 
which are plotted as nearly horizontal dotted lines 
in figure~\ref{fig:mv_uvcol_bbg}.  Even though the bimodality of BBGs 
at $z>1.2$ is not so clear compared with that at $z<1.2$, 
the mean colors of the s-BBGs and the p- and qp-BBGs, represented as 
black open symbols and solid and open-solid symbols 
in figure~\ref{fig:mv_uvcol_bbg},  
are in the blue cloud region and in the red sequence, respectively. 
Thus, the uRJ and uVi criteria can reasonably classify the   
the star forming and the passively evolving populations as the s-BBGs and  
the p- and qp-BBGs at $z < 1.8$ in the field at least.  
It confirms that the generalized BBG criteria are also useful 
in order to distinguish a star-forming population from a passively evolving 
one.  

\subsection{Salvaging BBGs from N3Rs at $z >0.8$} 
\label{app_subsec:exbbg}  

\begin{figure}[ht]
\begin{minipage}{0.49\linewidth}
\includegraphics[width=\linewidth]
{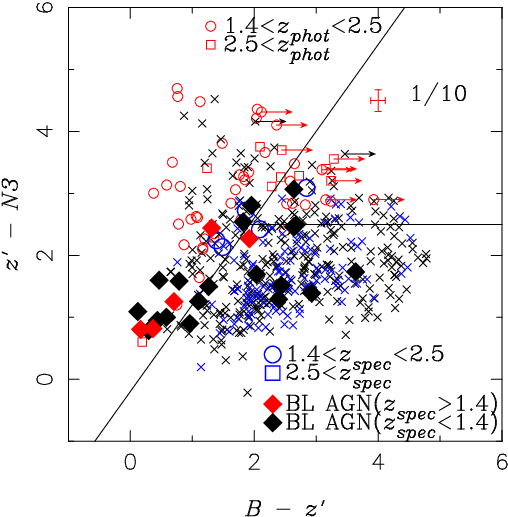}
\end{minipage}
\begin{minipage}{0.49\linewidth}
\includegraphics[width=\linewidth]
{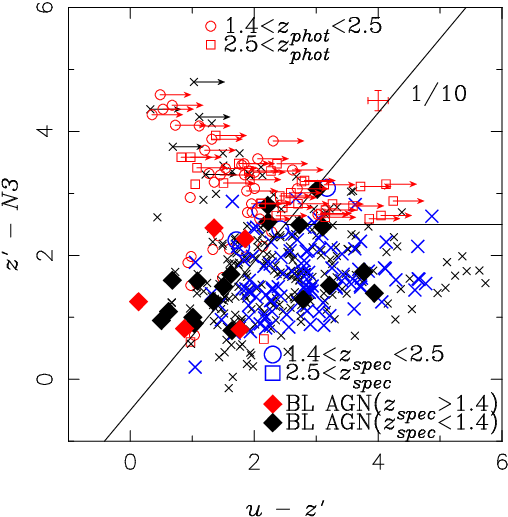}
\end{minipage}
\begin{minipage}{0.49\linewidth}
\includegraphics[width=\linewidth]
{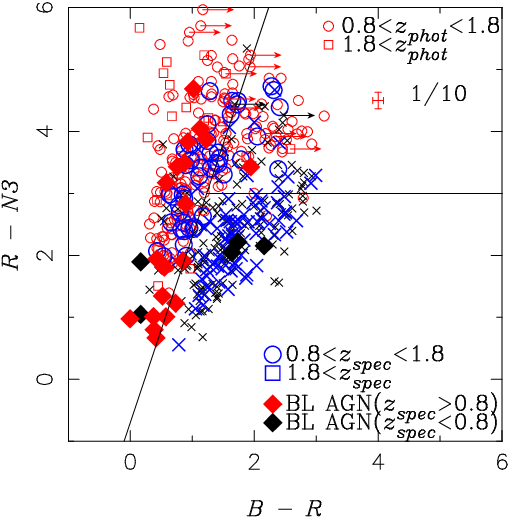}
\end{minipage}
\begin{minipage}{0.49\linewidth}
\includegraphics[width=\linewidth]
{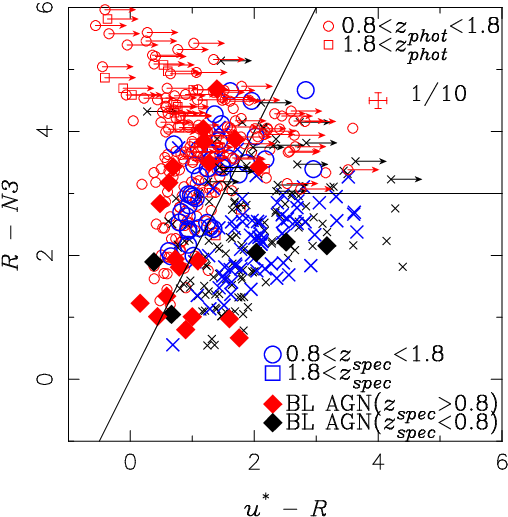}
\end{minipage}
\caption{Two-color diagrams for the extended BBGs selected from the N3Rs, 
in which the $N3$ band is taken as the reddest band ${\cal R}$ 
in the BBG criteria.  
Red crosses at $({\cal B}-\mu,\mu-{\cal R})=(4,4.5)$ represent  
typical errors of the color.  
Only a part of the N3Rs are plotted to reduce crowding in the diagram, whose 
fractions are presented as numbers at $({\cal B}-\mu,\mu-{\cal R})=(5,4.5)$ 
in all figures.  Diamonds represent BL AGNs.  
Large and small symbols represent the spectroscopic and the photometoric 
samples, respectively.  
Top-Left: The same as figure~\ref{fig:bzk_spec_phot} for the BzN3s.  
Circles and squares represent objects at $z=1.4-2.5$ and $z>2.5$, 
respectively.  
Top-Right: The same as figure~\ref{fig:bzk_spec_phot} for the uzN3s.  
Circles and squares represent objects at $z=1.4-2.5$ and $z>2.5$, 
respectively.  
Bottom-Left: The same as figure~\ref{fig:urj_spec_phot} for the BRN3s.  
Circles and squares represent objects at $z=0.8-1.4$ and $z \ge 1.4$, 
respectively. 
Bottom-Right: The same as figure~\ref{fig:urj_spec_phot} for the uRN3s.  
Circles and squares represent objects at $z=0.8-1.4$ and $z \ge 1.4$, 
respectively. 
}
\label{fig:bbg2_phot}
\end{figure}  

\begin{figure}[ht]
\begin{minipage}{0.98\linewidth}
\includegraphics[angle=0,width=\linewidth]
{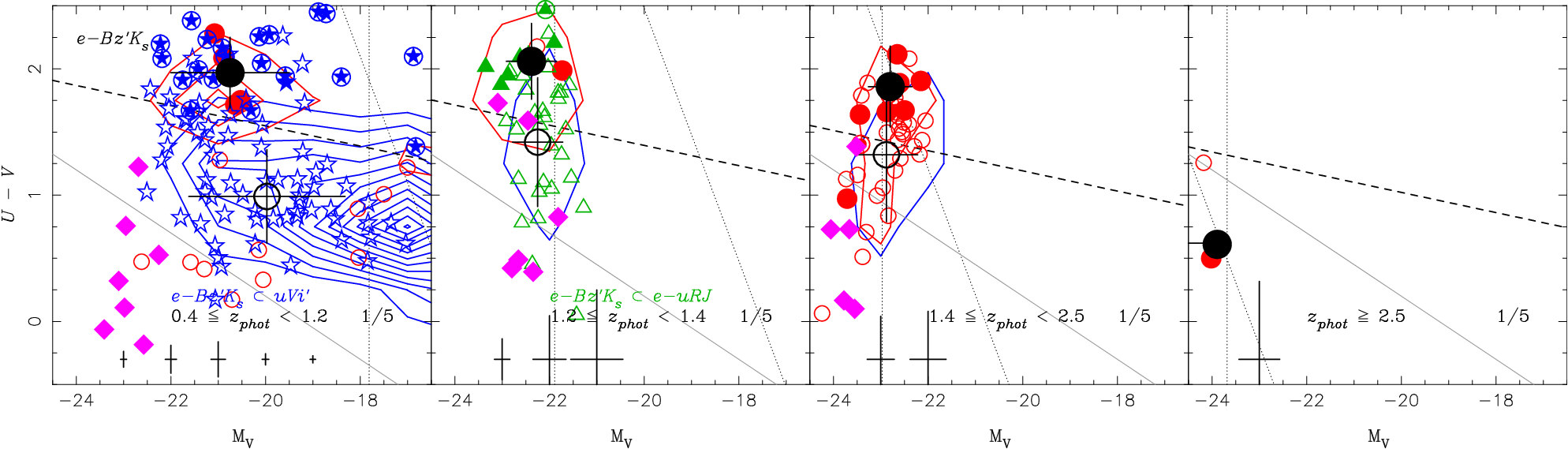}
\end{minipage}
\begin{minipage}{0.98\linewidth}
\includegraphics[angle=0,width=\linewidth]
{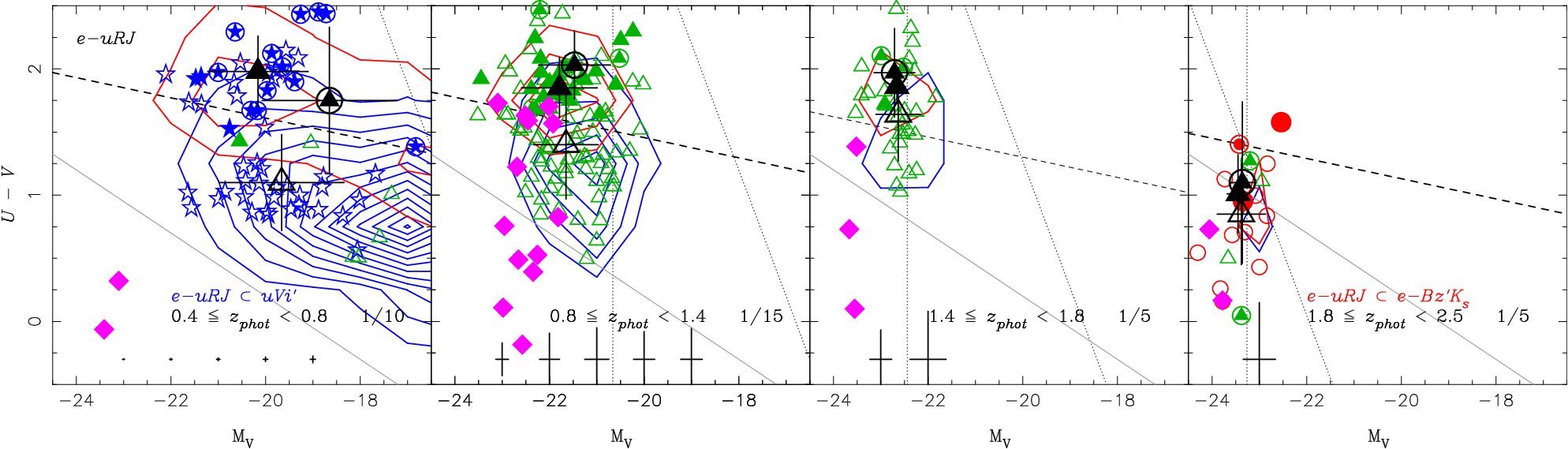}
\end{minipage}
\begin{minipage}{0.98\linewidth}
\includegraphics[angle=0,width=\linewidth]
{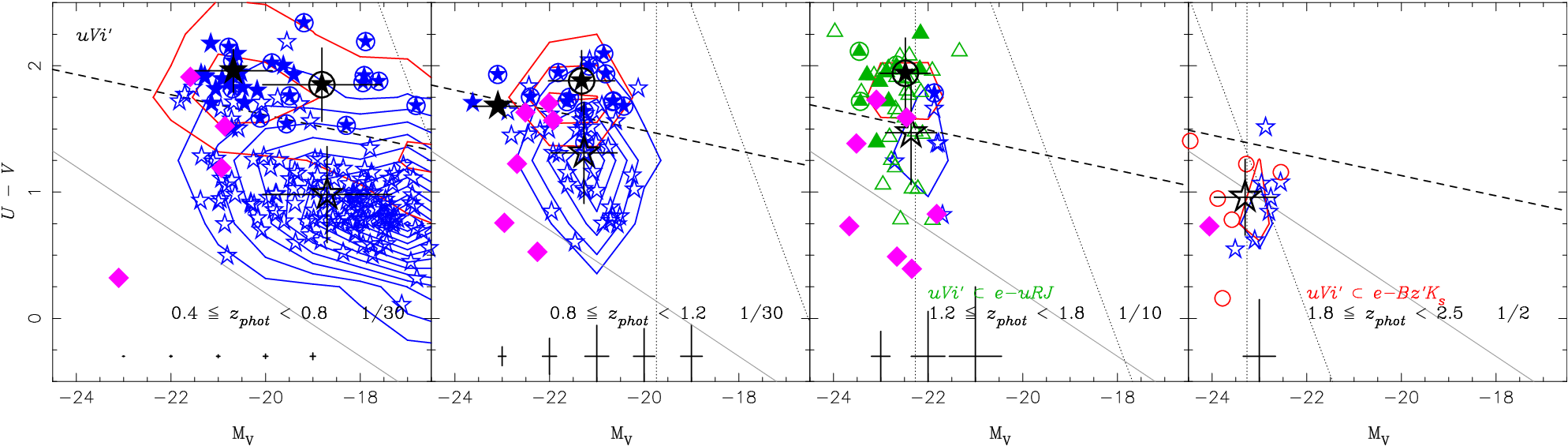}
\end{minipage}
\caption{The same as figure~\ref{fig:mv_uvcol_bbg} for all the BBGs, 
including the extended BBGs.  
Top: For all the BzKs, including uzN3s and BzN3s.  
For BzKs+uzN3s+BzN3s also classified as uRJs+uRN3s+BRN3s at 
$1.2 \le z<1.4$, 
their symbols are changed to green triangles from red circles.  
For BzKs+uzN3s+BzN3s also classified as uVis at 
$0.4 \le z<0.8$, their symbols are changed to blue stars from red circle.  
Middle: The same as on the left for all the uRJs, 
including uRN3s and BRN3s, as the Top.   
For uRJs+uRN3s+BRN3s also classified as BzKs+uzN3s+BzN3s at 
$1.4 \le z<2.5$, their symbols are changed to red circles 
from green triangles.   
For uRJs+uRN3s+BRN3s also classified as uVis at 
$0.4 \le z<0.8$, their symbols are changed to blue stars 
from green triangles.  
Bottom: The same as the Top for the uVis. 
For uVis also classified as uRJs+uRN3s+BRN3s at $1.2 \le z<1.8$, 
their symbols are changed to green triangles from blue stars.   
For uVis also classified as BzKs+uzN3s+BzN3s at $1.8 \le z<2.5$, 
their symbols are changed to red circles from blue stars.   
}
\label{fig:mv_uvcol_exbbg}
\end{figure}  

As shown in appendix~\ref{app_sec:bbg}, 
the BBG techniques employing not only
the original $BzK$ color but also the $uRJ$ color can select
galaxies around the redshift desert.  However, we
could not completely select these BBGs from the ground-based
photometric dataset in the field since the ground-based $JK_s$
photometry with the KPNO/FLMG is not deep enough to pick up the IRBGs
detected with the AKARI as seen in table~\ref{tab:photmt}.  
Since BBGs and IRBGs are almost the same populations 
at $z>0.4$ as discussed above, 
their statistics can confirm the limitation of the $JK$ depth.  
Roughly one third of N4Rs and N34 bumpers could not be
identified as BzKs or uRJs as seen in the comparison between the left plot in 
figure~\ref{fig:photzhist_bbg_exbbg} and the top-left plot in 
figure~\ref{fig:photzhist_bbgn3r}.  
Even though the limiting magnitude in the N3 band is 
comparable to those in J and K photometry as seen in table~\ref{tab:photmt}, 
the N3 band is effective to detect emissions around the rest-frame 
1.6-$\mu$m IR bump for the objects at $z>1$ more than the J and K bands. 
In order to salvage these missing BBGs as
non-BzK(non-uRJ) in N3Rs around the redshift desert, we 
used $N3$ photometry instead of the ground-based NIR ones for BBG
selections with $BzN3/uzN3$ ($uRN3/BRN3$) colors as shown in 
figure~\ref{fig:bbg2_phot}.  The numbers of 
salvaged BzN3s and uzN3s (uRN3s and BRN3s) were greater than those of
BzKs(uRJs) even in this dataset as shown in
figure~\ref{fig:photzhist_bbg_exbbg}.  Hereafter, we often
treat BzN3s and uzN3s (uRN3 and BRN3) as an extended BBG class of
BzKs(uRJs) at $z \simeq 2$($z \simeq 1$) and call 
these BBGs selected by including the AKARI NIR photometry 
as extended BBGs while the BBGs selected only 
with the ground-based photometry as classical BBGs.  

\section{Summary of classifications} 
\label{app_sec:sum_class}

Hereafter, we will summarize the photometric 
classification schemes introduced in the paper 
and show their outline as following tree diagrams, 
in which asterisk $\ast$, circle $\circ$, and 
star $\star$ at the front of a category
represent a classification without using 
redshift information, with using redshift information, 
and with substituting photometric color conditions 
for redshift selections.  

\subsection{Optical classifications} 
\label{subsec_app:opt_class}

\begin{classify}{$z'$-detected galaxy}
\class{
  \begin{classify}{$\circ$ EBO
}
  \end{classify} 
      }
\class{
  \begin{classify}{$\circ$ Rest-frame $V$-detected galaxy }  
     \class{red sequence}  
     \class{blue cloud}  
  \end{classify} 
      } 
\class{
  \begin{classify}{BBG}
         \class{  
             \begin{classify}{uVi} 
                \class{$\ast$s-uVi}  
                \class{$\ast$p-uVi}  
                \class{$\ast$qp-uVi}  
              \end{classify}
               }
           \class{
              \begin{classify}{uRJ+(BRN3,uRN3)} 
                \class{$\ast$s-uRJ} 
                \class{$\ast$p-uRJ}  
                \class{$\ast$qp-uRJ}  
              \end{classify}
               }
           \class{
              \begin{classify}{BzK+(BzN3,uzN3)} 
                \class{$\ast$s-BzK} 
                \class{$\ast$p-BzK}  
                \class{$\ast$qp-BzK}  
              \end{classify}
               }
  \end{classify} 
       }
\end{classify}  

\begin{itemize}
\item $z'$-detected galaxies: 
As shown in subsection~\ref{subsec:ground}, we selected $\sim$56000 of them, 
performed their photometry as summarized in table~\ref{tab:photmt}, and 
estimated $z_{phot}, M_{\ast}, A_V$, and $SFR_{opt;corr}$ as described 
in section~\ref{sec:photz} and appendix~\ref{app_sec:photz_var}.   
\item EBOs: The Extremely Blue Objects were selected with the criteria of 
equation~(\ref{eq:ebo}), which are possibly AGNs 
as discussed in subsection~\ref{subsec:mv_uvcol} and 
appendix~\ref{app_subsec:track_ebo}.      
\item Rest-frame $V$-detected galaxy: We studied their stellar populations, 
which are subclassified as red sequence and blue cloud in the CMD 
with criteria~(\ref{eq:mv_uvcol_red_blue}) as 
discussed in subsection~\ref{subsec:mv_uvcol}.  
\item BBGs: The Balmer Break Galaxies were introduced for  
redshift subclassifications of uVis, (e-)uRJs, and (e-)BzKs and 
distinction between star-forming (s-BBGs), passive (p-BBGs), and 
quasi-passive (qp-BBGs) 
populations with the criteria (\ref{eq:s-bbg}), (\ref{eq:p-bbg}), and 
(\ref{eq:qp-bbg}) in appendix~\ref{app_sec:bbg}.   
\end{itemize}

\subsection{NIR classifications} 
\label{subsec_app:nir_class}

\begin{classify}{$z'$-detected galaxy}
\class{             
  \begin{classify}{$\ast$N3R}  
       \class{ 
           \begin{classify}{IRBG} 
                 \class{$\ast$N23 bumper} 
                 \class{$\ast$N34 bumper} 
                 \class{$\ast$N4R} 
           \end{classify}
             } 
        \class{         
           \begin{classify}{N3N4R} 
           \end{classify}
             } 
  \end{classify} 
      }
\end{classify} 

\begin{itemize}
\item N3Rs: The N3 Red galaxies are selected from the $z'$-detected galaxies 
by the criterion~(\ref{eq:n234}) in subsection~\ref{sec:irbg}, 
which roughly correspond to populations at $z>0.4$.  
\item IRBGs: The N3Rs were subclassified to the N23, N34 bumpers, 
and N4 Red galaxies (N4Rs) by the criterion~(\ref{eq:n23_n34_n4})  
as InfraRed Bump Galaxies with 1.6~$\mu$m IR bump features from their 
stellar emissions.  
\item N23, N34 bumpers, and N4Rs: They approximately 
correspond to the uVis, e-uRJs, and e-BzKs, respectively. 
\item N3N4Rs: The N3N4 Red galaxies were classified by 
equation~(\ref{eq:n3n4r}), which are possibly bright AGN candidates 
as discussed in subsection~\ref{subsec:nir_mbn3r}.   
\end{itemize}

\subsection{MIR classifications} 
\label{subsec_app:mir_class}

\begin{classify}{$z'$-detected galaxy}
\class{             
  \begin{classify}{$\ast$N3 Red galaxy(N3R)}  
       \class{ 
           \begin{classify}{MbN3R} 
                  \class{ 
                      \begin{classify}{$\star$s-MbN3R}
                         \class{sb-MbN3R}  
                         \class{s/a-MbN3R} 
                      \end{classify} 
                        }  
                  \class{
                      \begin{classify}{$\star$agn-MbN3R}
                      \end{classify}
                        } 
           \end{classify} 
               }
        \class{  
           \begin{classify}{\mbox{$\star$ MmN3R$^{\#}$:\ref{subsec:mcn3r}}} 
              \class{
                 \begin{classify}{$\star$pP-MmN3R} 
                    \class{sb-pP-MmN3R}  
                    \class{non sb-pP-MmN3R}
                 \end{classify}
                     }   
              \class{
                 \begin{classify}{$\star$fP-MmN3R} 
                    \class{sb-fP-MmN3R}  
                    \class{non sb-fP-MmN3R}
                 \end{classify}
                     }   
              \class{ 
                 \begin{classify}{$\star$MmN4R} 
                 \end{classify}   
                    } 
           \end{classify}   
               }
        \class{  
           \begin{classify}{$\star$MfN3R}     
           \end{classify}   
               }
  \end{classify} 
      }
\end{classify}

\begin{itemize}
\item MbN3Rs: The MIR bright N3Rs were selected from the N3Rs with the 
detection in more than two MIR bands, as described in subsection
~\ref{subsec:mir_sed_sb_agn}.  
\item s- and agn-MbN3R:  The MbN3Rs were subclassified 
to starbursts (s-MbN3Rs) and AGNs (agn-MbN3Rs) 
by the MIR color criteria~(\ref{eq:lirg1}), (\ref{eq:lirg2}), 
and (\ref{eq:lirg3}) at $z=0.4-0.8, 0.8-1.2$, and $>1.2$, respectively, 
as shown in figure~\ref{fig:col_col_mir_mbn3r}.   
\item sb- and s/a-MbN3Rs: The s-MbN3Rs were subclassified 
to StarBurst-dominant (sb-MbN3Rs) and Starburst/AGN mixture(s/a-MbN3Rs) 
by their IR SED fittings.  
\item MmN3Rs: The MIR marginally-detected N3Rs were detected 
only in one or two MIR bands with SN3, which may be mimics of MbN3Rs
with fainter MIR emission. They were subclassified 
the pP-, fP-, sb- and non sb-MmN3Rs as discussed in 
subsection~\ref{subsec:mcn3r}.  
\item pP- and fP-MmN3Rs: The MmN3Rs were subclassified 
to the Possibly PAH emitting (pP-MmN3Rs) and the faint 
PAH (fP-MmN3Rs) populations as the PAH 7.7 um emission 
from the former is possibly dominant in $S11, L15$, and $L18$ bands 
at $z=0.4-0.65, 0.65-1.5$, and $>1.5$ while that from the latter is not.  
\item sb- and non sb-MmN3Rs: The StarBurst-dominant populations (sb-MmN3Rs) 
have MIR SEDs which are reproduced by the S\&K model with a low AGN 
mixture rate $<50\%$ while non sb-MmN3Rs do not.  
\item MfN3Rs: The MIR faint N3Rs were not detected in any MIR bands 
with SN3 as discussed in subsection~\ref{subsec:mfn3r}.  
\end{itemize}

\subsection{Combining classifications}

We have sometimes introduced subgroups by combining these optical, 
NIR, and MIR classifications. For example, the N3Rs, classified 
by the criteria of the IR bump in the NIR 
photometry, can be also subclassified into s-,p-, and qp-BBGs 
with the criteria of the Balmer break in the optical 
photometry, which were applied for the N3 Red BBGs as the 
exclusively classified BBGs from the N3Rs as seen 
in subsection~\ref{subsec:irbg_bbg}.  
Thus, the combination of classifications at different 
wavelengths can help us in quickly studying galaxies.  

\section{Photometric redshifts of various populations}
\label{app_sec:photz_var}

\begin{figure}[ht]
\begin{minipage}{0.99\linewidth}
\includegraphics[width=\linewidth]
{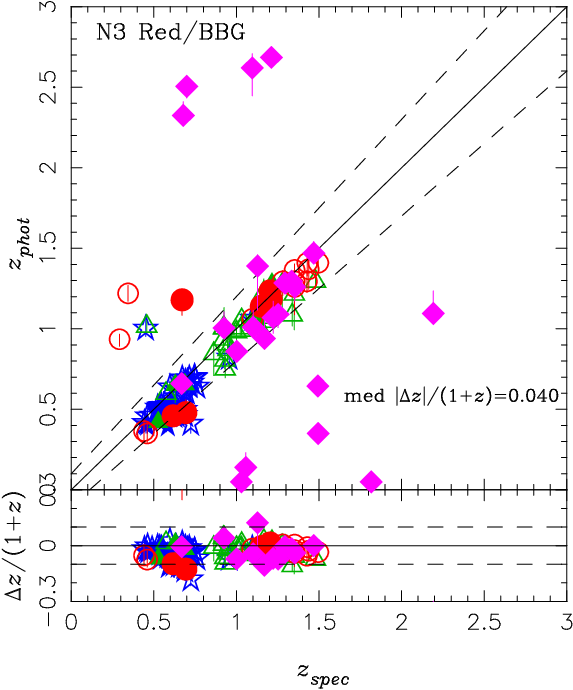}
\end{minipage}
\begin{minipage}{0.49\linewidth}
\includegraphics[width=\linewidth]
{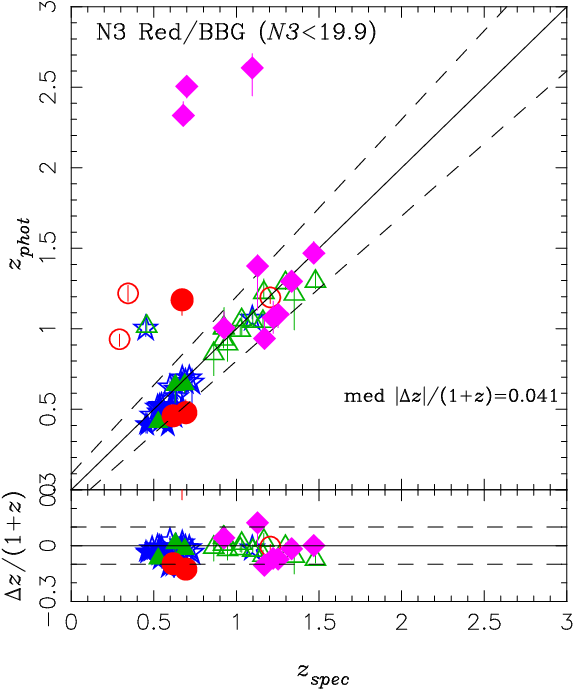}
\end{minipage}
\begin{minipage}{0.49\linewidth}
\includegraphics[width=\linewidth]
{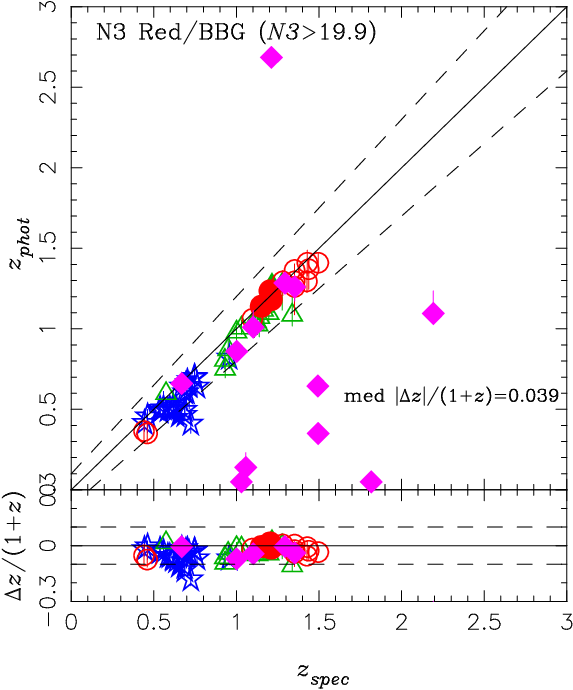}
\end{minipage}
\caption{The same as figure~\ref{fig:specz_photz} for the N3 Red BBGs.   
The colors and symbols are the same as figure~\ref{fig:col_col_mir_mbn3r}.  
The magenta square means BL AGN classified with the spectroscopy.   
Top: For all the spectroscopic N3Rs. 
Bottom-Left: For reltaively bright N3Rs with $N3<19.9$.  
Bottom-Right: For relatively not-bright N3Rs with $N3 \ge 
19.9$. }
\label{fig:specz_photz_irbg}
\end{figure}  

\begin{figure}[ht]
\begin{minipage}{0.32\linewidth}
\includegraphics[width=\linewidth]
{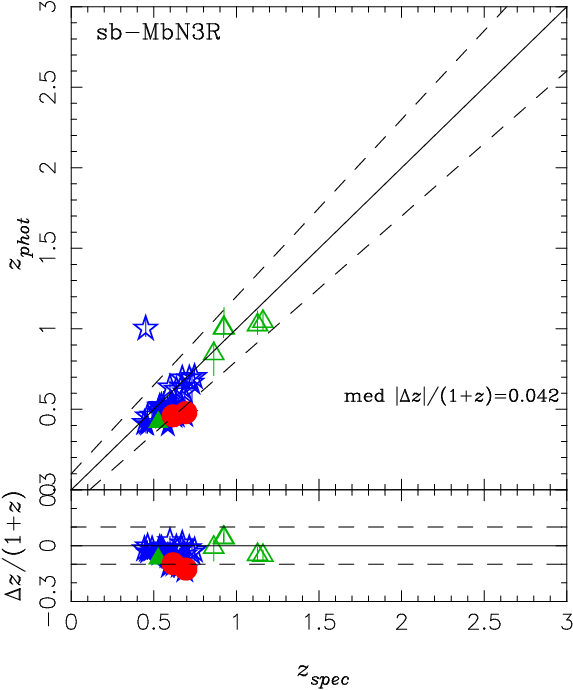}
\end{minipage}
\begin{minipage}{0.32\linewidth}
\includegraphics[width=\linewidth]
{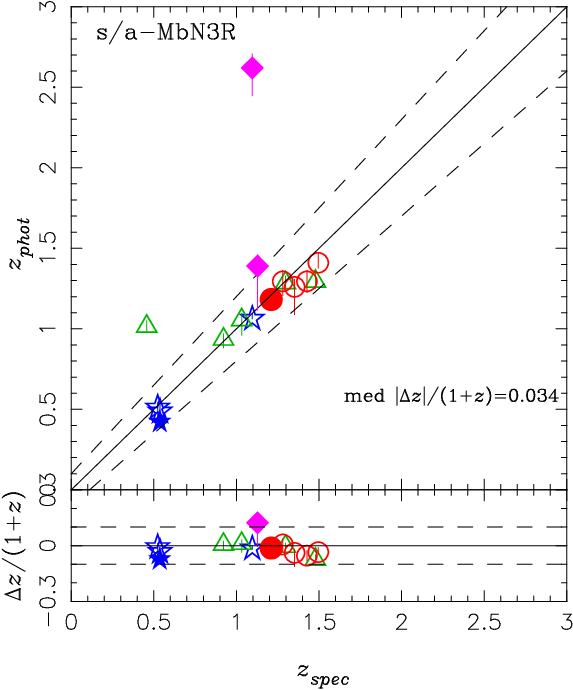}
\end{minipage}
\begin{minipage}{0.32\linewidth}
\includegraphics[width=\linewidth]
{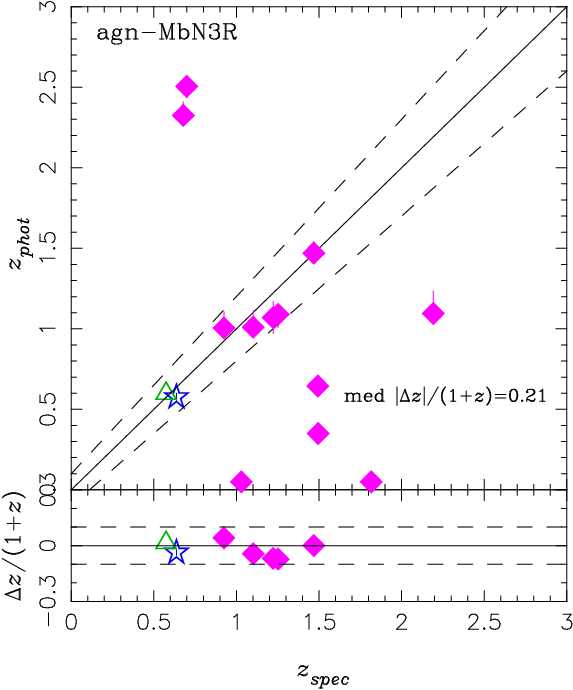}
\end{minipage}
\caption{
The same as figure~\ref{fig:specz_photz_irbg} for the MbN3Rs; 
the sb-, s/a-, and agn-MbN3Rs from left to 
right, which are subclassified on the basis of their MIR SED features as shown 
in subsection~\ref{subsec:mir_sed_sb_agn}. }
\label{fig:specz_photz_mbn3r}
\end{figure}  

We obtained the spectra of 83 N3Rs, in which MIR bright sources 
are subclassified into 35 sb-MbN3Rs, 9 s/a-MbN3Rs, and 14 agn-MbN3Rs.  
For these spectroscopic N3Rs and MbN3Rs samples, 
figures~\ref{fig:specz_photz_irbg} and 
~\ref{fig:specz_photz_mbn3r} show the redshift discrepancy 
$\Delta z= z_{phot} -z_{spec}$.  The median accuracies of 
$\langle \Delta z \rangle/(1+z_{spec})$ 
are 0.040, 0.041, 0.039, 0.042, 0.034, and 0.21 for 
all the N3Rs, relatively bright N3Rs ($N3<19.9$), relatively faint N3Rs 
 ($N3>19.9$), sb-MbN3Rs, s/a-MbN3Rs, and agn-MbN3Rs, respectively.   

We can see no essential difference in the median accuracy between the 
relatively bright and faint N3Rs, which means that 
the brightness does not essentially affect the photometric 
redshift estimations.  We can also see that sb-MbN3Rs, preselected for $z>0.4$ 
with the IR bump criterion $N2-N3>-0.3$, show good agreement 
between $z_{spec}$ and $z_{phot}$, which means that the photometric 
redshift estimation is still robust even under heavy extinctions 
in these MbN3Rs as long as one excludes low redshift dusty starbursts with 
the IR bump detection.     

The agn-MbN3Rs show the largest discrepancy in all the subclasses in the 
N3Rs.  It confirms that AGN is one major cause for inducing deviations in 
the photometric redshifts, which is reasonable since we used only stellar 
population synthesis modeled SEDs as BC03 without including AGN SEDs.    
Even though 14 spectroscopically observed agn-MbN3Rs, as candidates harboring 
AGNs, include 8 outliers, all the outliers are spectroscopically 
identified as BL AGNs and the remaining 6 non-outliers out of BL AGNs 
show that their median 
accuracy is comparable to that of the sb-MbN3Rs as 
$\langle \Delta z \rangle/(1+z_{spec}) =0.042$.   
It suggests that their photometric redshifts can be accurately 
estimated from the SED fitting with their major stellar emission 
of the host galaxies even harboring AGNs.  

Thus, the photometric redshifts for N3Rs and MbN3Rs 
can be used to reconstruct their redshifts.  

\section{Redshift distributions of galaxies}
\label{app_sec:zdist}

\begin{figure}[ht]
\begin{minipage}{0.49\linewidth}
\includegraphics[angle=0,width=\linewidth]
{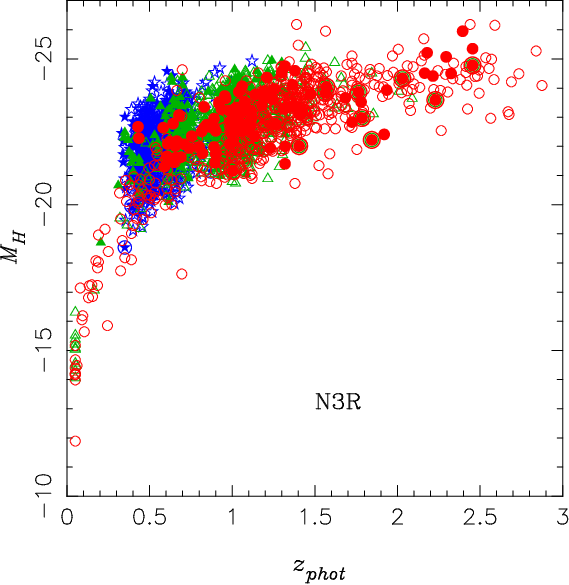}
\end{minipage}
\begin{minipage}{0.49\linewidth}
\includegraphics[angle=0,width=\linewidth]
{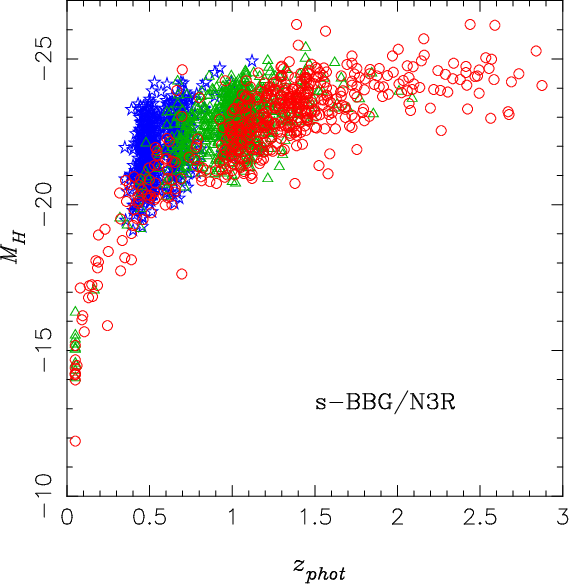}
\end{minipage}
\begin{minipage}{0.49\linewidth}
\includegraphics[angle=0,width=\linewidth]
{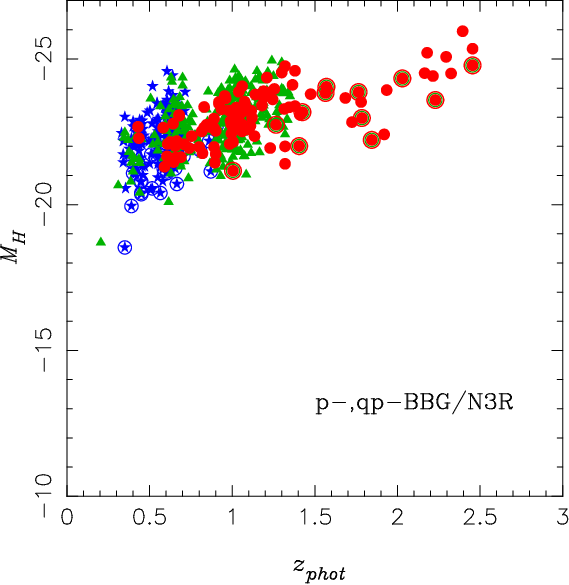}
\end{minipage}
\begin{minipage}{0.49\linewidth}
\includegraphics[angle=0,width=\linewidth]
{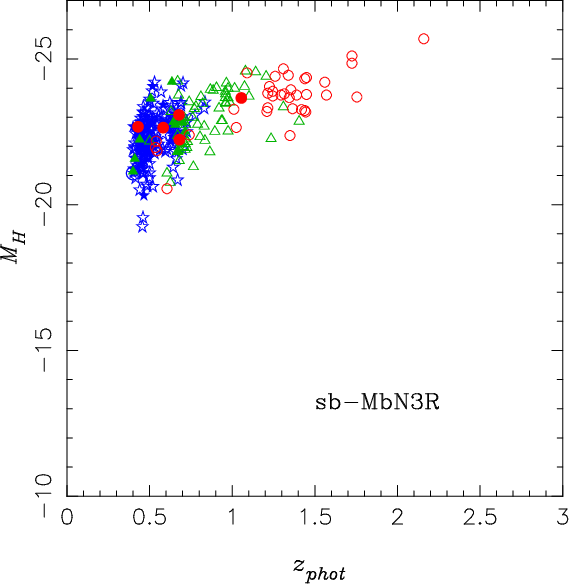}
\end{minipage}
\begin{minipage}{0.49\linewidth}
\includegraphics[angle=0,width=\linewidth]
{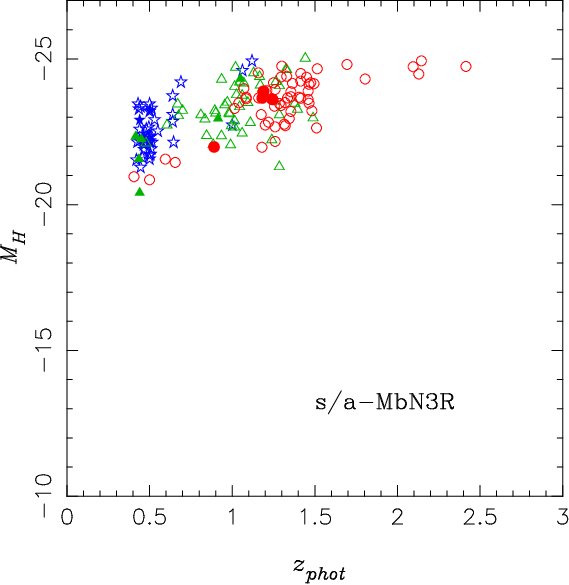}
\end{minipage}
\begin{minipage}{0.49\linewidth}
\includegraphics[angle=0,width=\linewidth]
{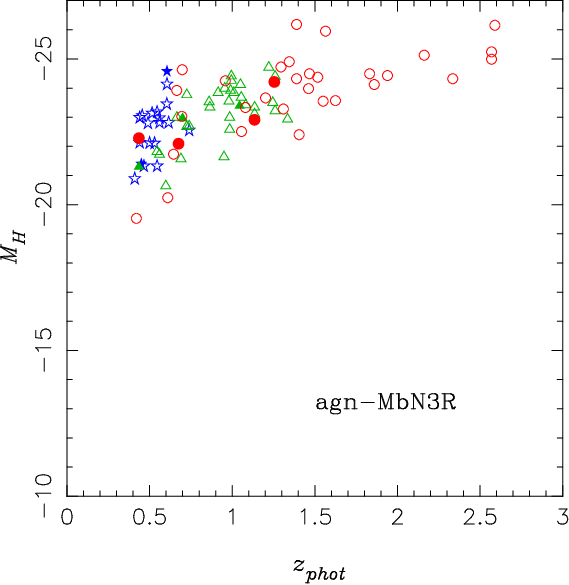}
\end{minipage}
\caption{
Photometric redshift $z_{phot}$
vs. rest-frame $H$ absolute magnitudes for N3Rs.  
The colors and symbols are the same as figure~\ref{fig:col_col_mir_mbn3r}.  
Upper-Left: For all the N3Rs. Upper-Right: 
The same as the Upper-Left, for the IRBGs classified
as s-BBGs. Middle-Left: The same as the Upper-Left, for the IRBGs
classified as p-BBGs. Middle-Right: The same as the Top-Left, for the
AKARI MIR detected IRBGs classified as sb-MbN3Rs. Bottom-Left: The same
as the Top-Left, for the MIR detected IRBGs classified as
s$/$a-MbN3Rs. Lower-Right: The same as the Upper-Left, for the MIR detected
IRBGs classified as agn-MbN3Rs.
}
\label{fig:zphot_hmag_n3r_mbn3r}
\end{figure} 

\begin{figure}[ht]
\begin{minipage}{0.49\linewidth}
\includegraphics[angle=0,width=\linewidth]
{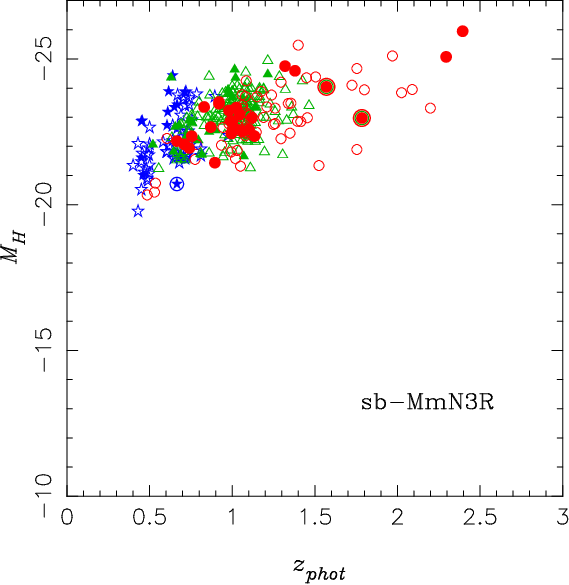}
\end{minipage}
\begin{minipage}{0.49\linewidth}
\includegraphics[angle=0,width=\linewidth]
{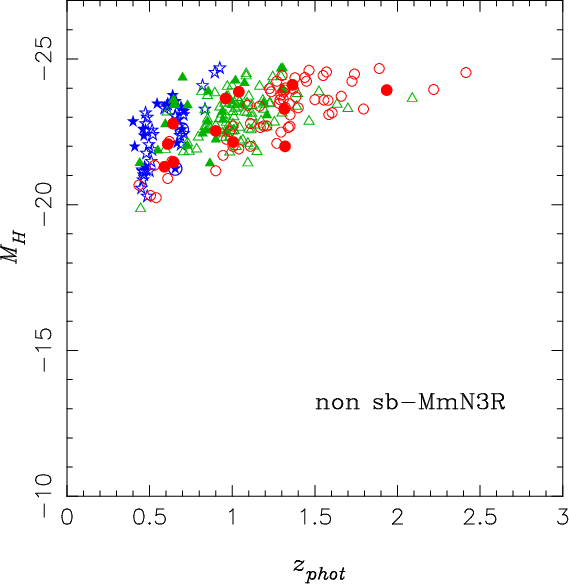}
\end{minipage}
\caption{
Photometric redshift $z_{phot}$
vs. rest-frame $H$ absolute magnitudes for the MmN3Rs.  
The colors and symbols are the same as figure~\ref{fig:col_col_mir_mbn3r}.  
Left: For the sb-MmN3Rs. Right: For the non ab-MmN3Rs.   
}
\label{fig:zphot_hmag_mcn3r}
\end{figure} 

\begin{figure}[ht]
\begin{minipage}{0.49\linewidth}
\includegraphics[angle=0,width=\linewidth]
{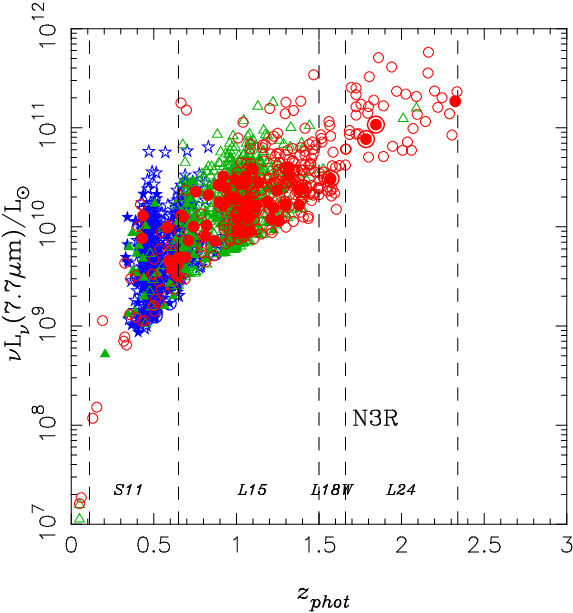}
\end{minipage}
\begin{minipage}{0.49\linewidth}
\includegraphics[angle=0,width=\linewidth]
{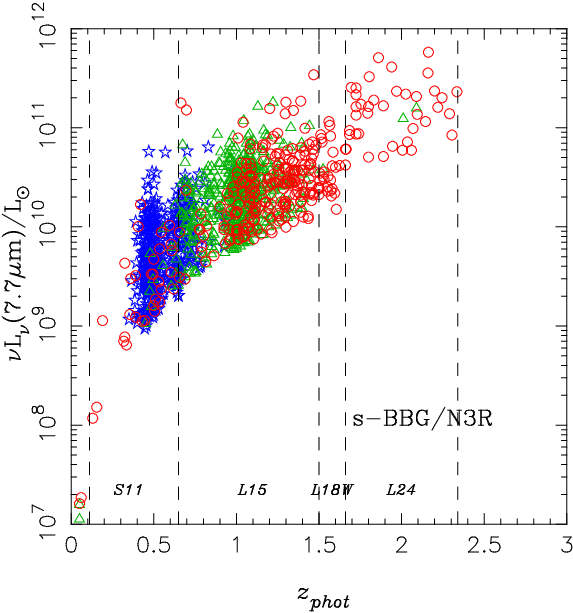}
\end{minipage}
\begin{minipage}{0.49\linewidth}
\includegraphics[angle=0,width=\linewidth]
{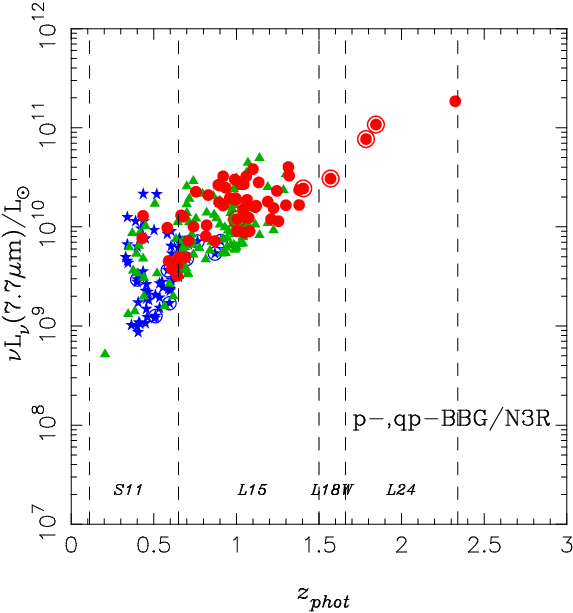}
\end{minipage}
\begin{minipage}{0.49\linewidth}
\includegraphics[angle=0,width=\linewidth]
{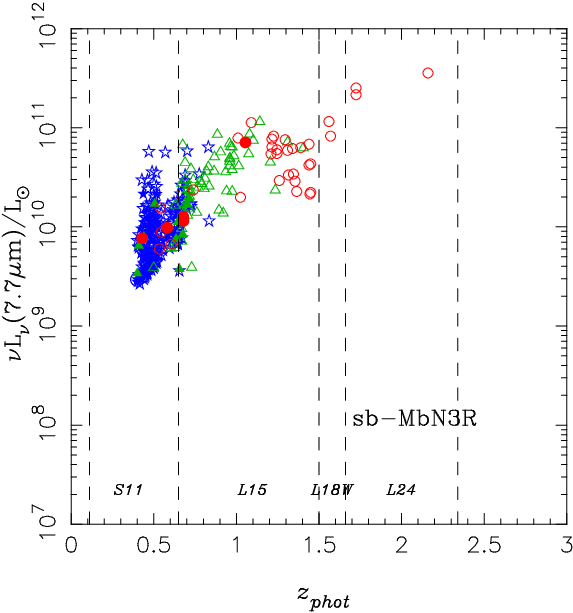}
\end{minipage}
\begin{minipage}{0.49\linewidth}
\includegraphics[angle=0,width=\linewidth]
{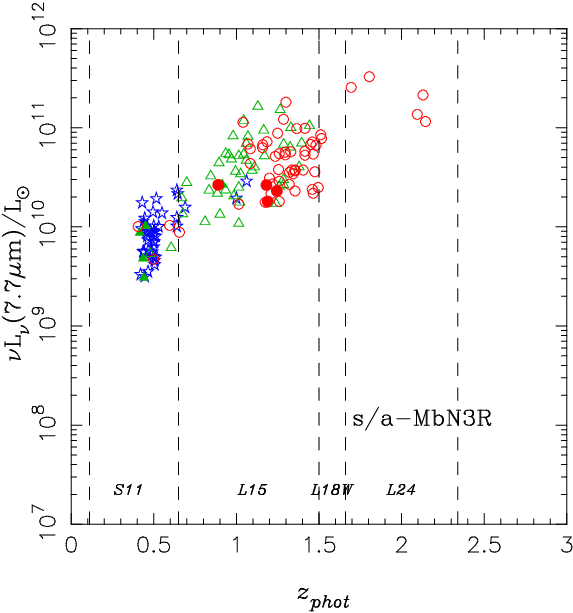}
\end{minipage}
\begin{minipage}{0.49\linewidth}
\includegraphics[angle=0,width=\linewidth]
{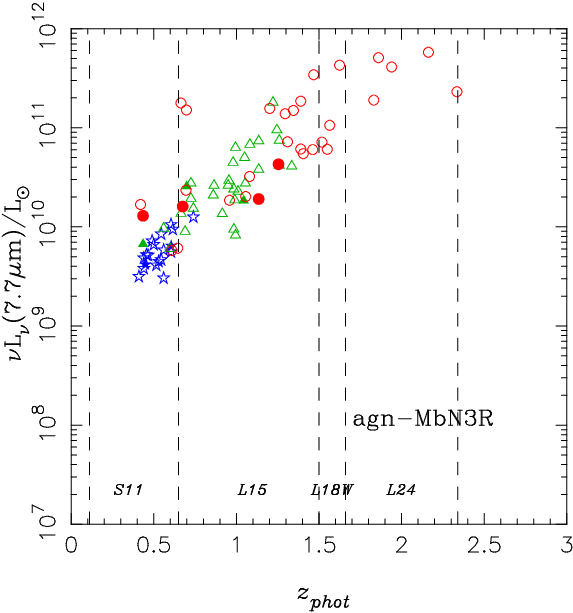}
\end{minipage}
\begin{minipage}{0.49\linewidth}
\includegraphics[angle=0,width=\linewidth]
{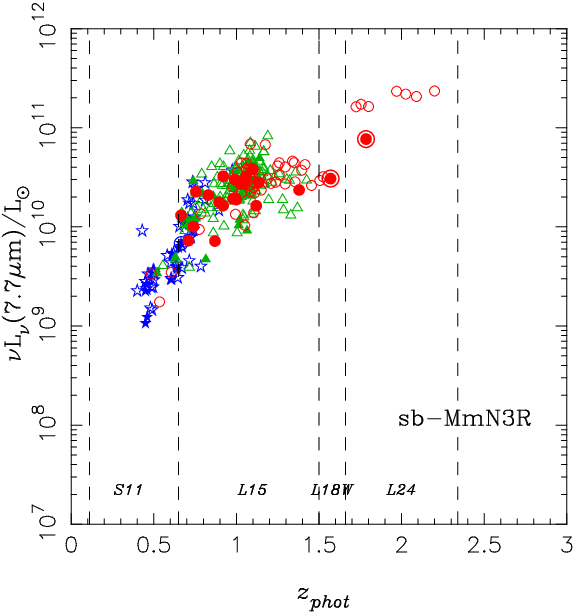}
\end{minipage}
\begin{minipage}{0.49\linewidth}
\includegraphics[angle=0,width=\linewidth]
{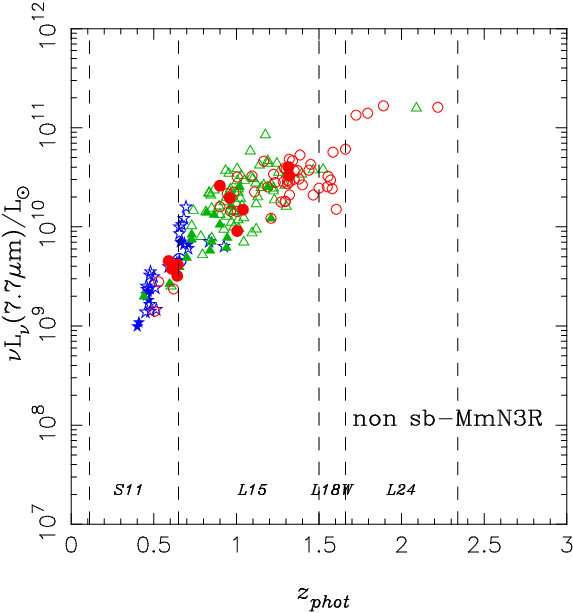}
\end{minipage}
\caption{
Photometric redshifts $z_{phot}$ vs. 
monochromatic luminosities $\nu L_{\nu} (7.7~\mu\mbox{m})$ 
at the rest frame 7.7 $\mu$m.  The colors and symbols are the same 
as figure~\ref{fig:col_col_mir_mbn3r}.   
}
\label{fig:photz_lum7.7}
\end{figure} 

Figures~\ref{fig:zphot_hmag_n3r_mbn3r}, ~\ref{fig:zphot_hmag_mcn3r}, 
and ~\ref{fig:photz_lum7.7} show 
the photometric redshift $z_{phot}$ vs. rest-frame $H$ absolute magnitudes 
and monochromatic luminosities $\nu L_{\nu} (7.7~\mu\mbox{m})$.  
Following the conversion from rest-frame $H$ absolute 
magnitude to $M_{\ast}$ 
with equation~(\ref{eq:mag_smass}) in appendix~\ref{app_sec:smass}, 
we can see that galaxies with $>10^{10}$M$_{\odot}$ are detected as the N3Rs 
at $z>1$ in figire~\ref{fig:zphot_hmag_n3r_mbn3r}.  
The standard nomenclature LIRG refers to a
luminosity class of galaxies with $L\simeq 10^{11-12} L_{\odot} $ 
and is distinct from starbursts $L\simeq 10^{10-11} L_{\odot}$ and ULIRG 
$ >10^{12}L_{\odot}$.  
Following the conversion from $\nu L_{\nu} (7.7~\mu\mbox{m})$ to 
TIR luminosity with equation~(\ref{eq:lum7_lum10_lumir}) 
in subsection~\ref{sec:ir_lum}, we can see 
that most of the AKARI MIR-detected galaxies, classified as the 
s- and s/a-MbN3Rs and sb-MmN3Rs, have typically a TIR 
luminosity of $10^{11-12} L_{\odot}$, which corresponds 
to those of local LIRGs.  

We can see also that the trend of agn-MbN3Rs is different 
from those of sb- and s/a-MbN3Rs.    

\section{Models for SEDs and colors}
\label{app_sec:model}

\subsection{Model tracks for extremely blue objects} 
\label{app_subsec:track_ebo} 

\begin{figure}[ht]
\centering 
\begin{minipage}{0.49\linewidth}
\includegraphics[angle=0,width=\linewidth]
{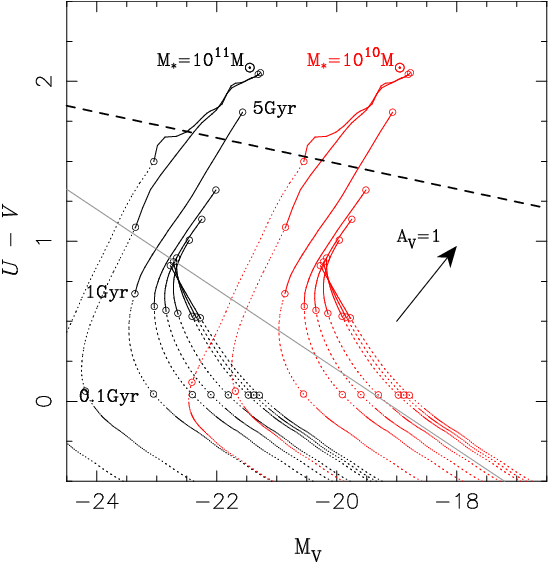}
\end{minipage}
\begin{minipage}{0.49\linewidth}
\includegraphics[angle=0,width=\linewidth]
{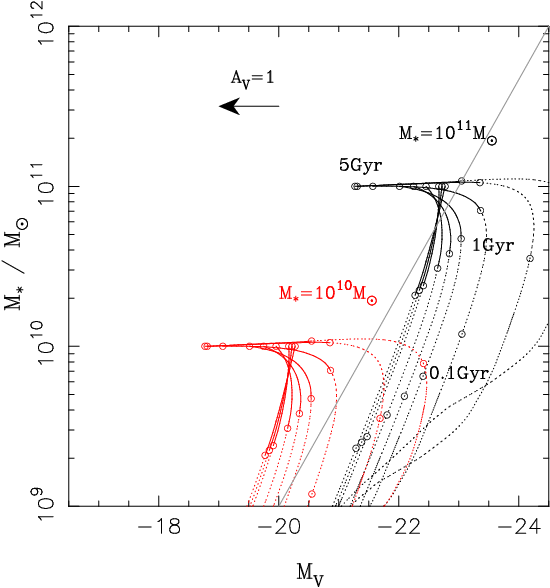}
\end{minipage}
\caption{Left: The evolutionary tracks on the color-magnitude diagram of 
$M_V$ vs. $(U-V)$ from the BC03 models.  
Red  and black lines correspond 
to systems with M$_{\ast}=10^{10}$M$_{\odot}$ and  
 M$_{\ast}=10^{11}$M$_{\odot}$ at 5 Gyr, repsectively,  
which are evolving 
with exponential decaying SF with time scales of 
$\tau=$0.1, 0.3, 1,2, 3,5, 15, 30 Gyr and Constant Star Formation (CSF) 
from the left to the right on the diagram.  All of them are 
in the cases without extinctions.  The open circles represent 
ages of the systems of 0.1, 1, and 5 Gyr for all the models.  
An Arrow at the $(M_V,U-V)=(-18,1)$ represents   
color-magnitude correction with a mean extinction $A_V \simeq 1$.  
Right: The same as on the left for the mass-magnitude diagram of 
$M_V$ vs. $M_{\ast}$ from the BC03 models.  
}
\label{fig:mv_uvcol_sm_model}
\end{figure}  

As remarked in subsection~\ref{subsec:photz}, all the spectroscopic BL AGNs 
in the $z_{phot}$ outliers are also the EBOs with 
$(U-V) < -0.25(M_V+22.0) +0.7$, which suggests that most of the EBOs are 
possibly AGN candidates as long as the redshift estimation is unaffected 
even by SEDs with a mixture of stellar and AGN emissions.  
In order to confirm this assumption, 
we reconstruct the 
evolutionary tracks with the BC03 models on the CMD and the Magnitude vs. 
Stellar Mass diagram as shown in figure~\ref{fig:mv_uvcol_sm_model}.  
Most of the 
observed star-forming galaxies suffer from extinctions $A_V \simeq 1$ 
as discussed in subsection~\ref{subsec:extinctions}.  Thus, most normal 
star-forming galaxies with $A_V \simeq 1$, except young phase ones with age 
$<0.1$ Gyr, cannot appear in the EBO region of $(U-V) < -0.25(M_V+22.0) +0.7$ 
on the CMD.  Even without any extinctions, only objects in an 
early star forming epoch $<1$ Gyr can have the color of the EBOs.  
It is not realistic 
to ingnore extinctions.  Indeed, we found evidence for existing 
dust even in the EBOs with MIR emission as reported in 
subsection~\ref{subsec:photz}.  

\subsection{Model tracks for Balmer break galaxies} 
\label{app_subsec:track_bbg} 

\begin{figure}[ht]
\centering 
\begin{minipage}{0.49\linewidth}
\includegraphics[angle=0,width=\linewidth]
{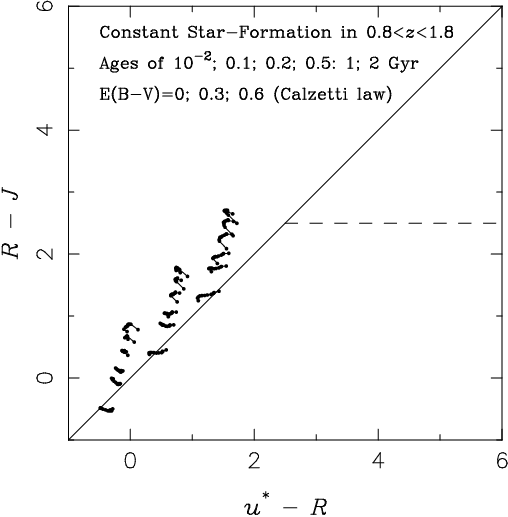}
\end{minipage}
\begin{minipage}{0.49\linewidth}
\includegraphics[angle=0,width=\linewidth]
{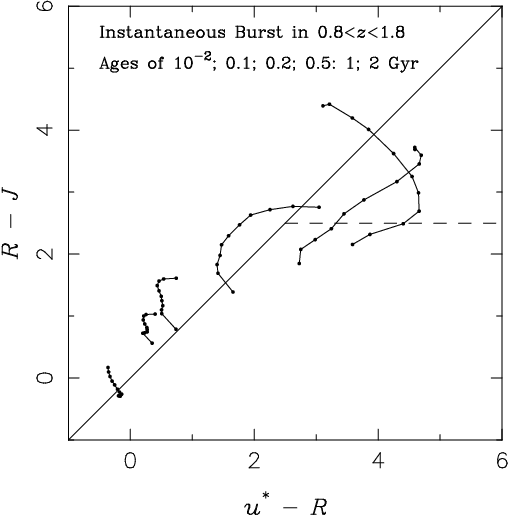}
\end{minipage}
\begin{minipage}{0.49\linewidth}
\includegraphics[angle=0,width=\linewidth]
{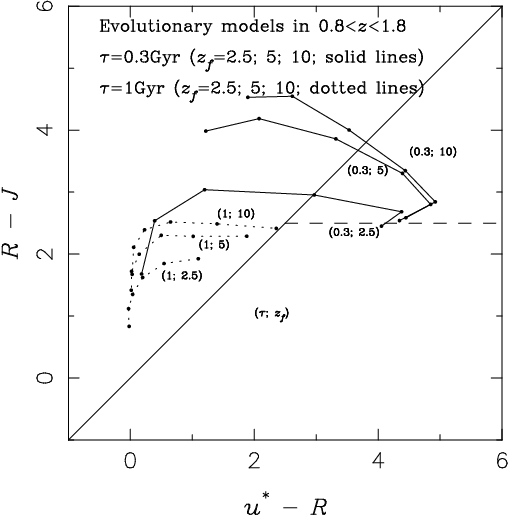}
\end{minipage}
\begin{minipage}{0.49\linewidth}
\includegraphics[angle=0,width=\linewidth]
{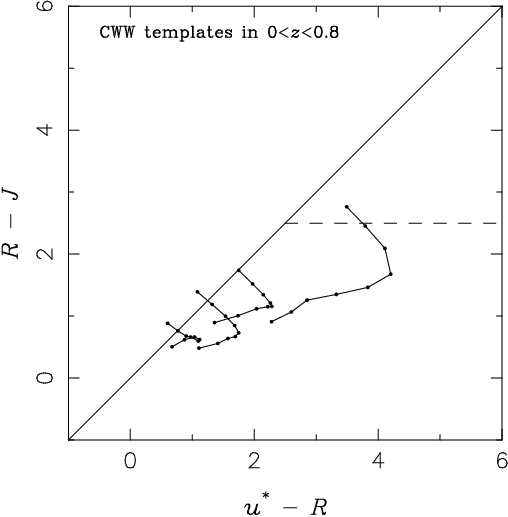}
\end{minipage}
\caption{The evolutionary tracks in the $uRJ$ diagram from theoretical
models the same as the Fig. 8 for the $BzK$ diagram in
\citet{daddi_new_2004}.  The top-left panel shows continuous star
formation model tracks for ages from 1 Myr to 2 Gyr and for E(B-V)=0,
0.3, 0.6.  Top right panel has simple stellar population models for
ages from 0.1 to 2 Gyr and no reddening. The bottom-left panel shows
evolutionary models with various formation redshifts and SFR
timescales and no reddening.  At decreasing redshifts the tracks
generally turn from bottom-left to top-right. The bottom-right panel
show colors for the local templates of various galaxy types from
\citet{coleman_colors_1980}.  All models are plotted for the range
$0.8<z<1.8$, except in the bottom-right panel where plots are for
$0<z<0.8$. The limits and color ranges of all four panels reproduce
those of figure~\ref{fig:urj_spec_phot}, for direct reference.}
\label{fig:urjmodel}
\end{figure}

\begin{figure}[ht]
\centering 
\begin{minipage}{0.49\linewidth}
\includegraphics[angle=0,width=\linewidth]
{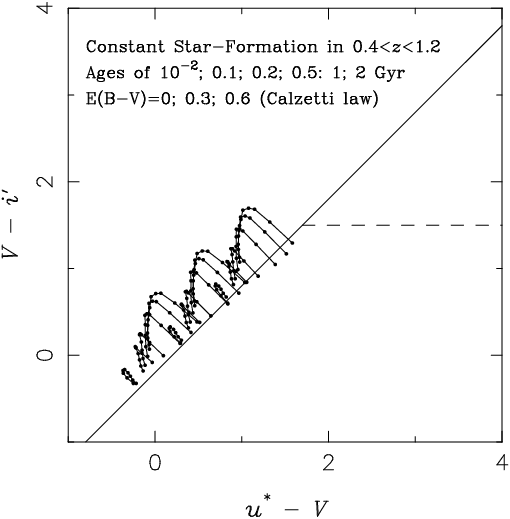}
\end{minipage}
\begin{minipage}{0.49\linewidth}
\includegraphics[angle=0,width=\linewidth]
{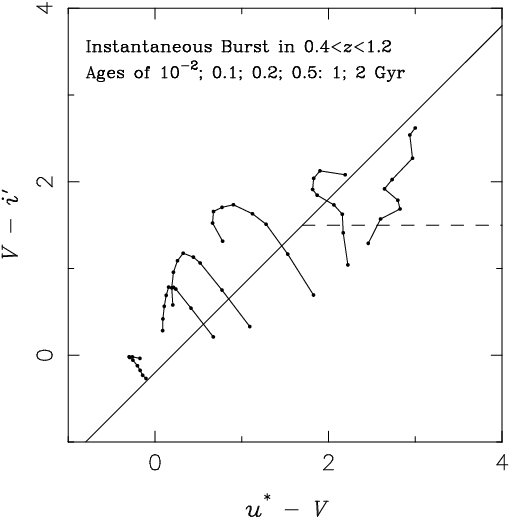}
\end{minipage}
\begin{minipage}{0.49\linewidth}
\includegraphics[angle=0,width=\linewidth]
{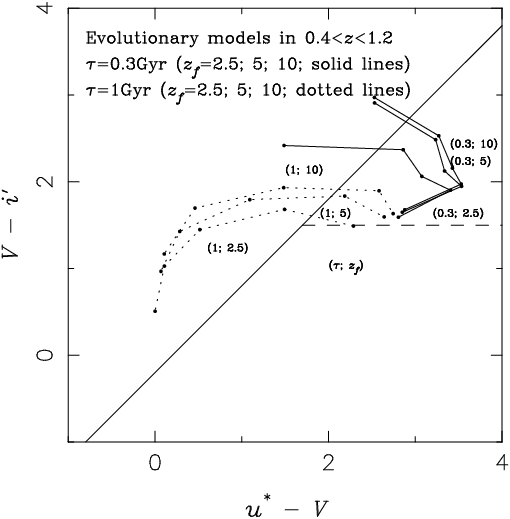}
\end{minipage}
\begin{minipage}{0.49\linewidth}
\includegraphics[angle=0,width=\linewidth]
{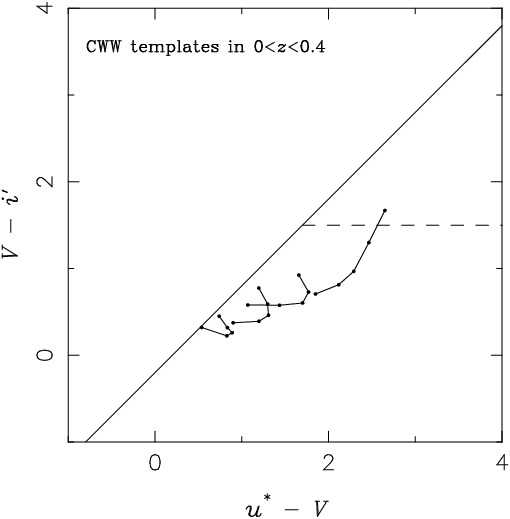}
\end{minipage}
\caption{The $uVi$ diagram the same as figure~\ref{fig:urjmodel}
except the redshift range.  All models are plotted for the range
$0.4<z<1.2$, except in the bottom-right panel where plots are for
$0<z<0.5$.  The limits and color ranges of all four panels reproduce
those of figure~\ref{fig:uvi_spec_phot}, for direct reference.}
\label{fig:uvimodel}
\end{figure}  

\begin{figure}[ht]
\centering 
\begin{minipage}{0.49\linewidth}
\includegraphics[angle=0,width=\linewidth]
{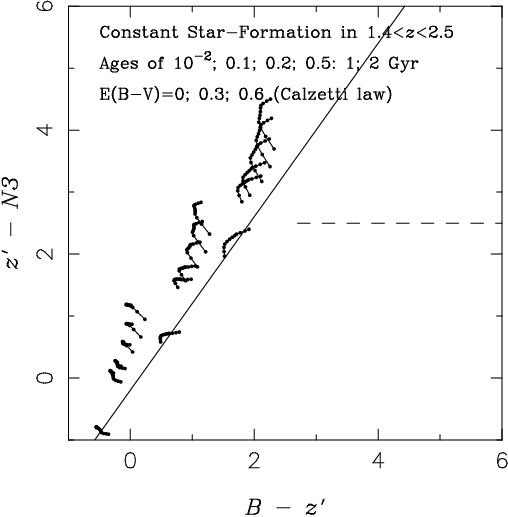}
\end{minipage}
\begin{minipage}{0.49\linewidth}
\includegraphics[angle=0,width=\linewidth]
{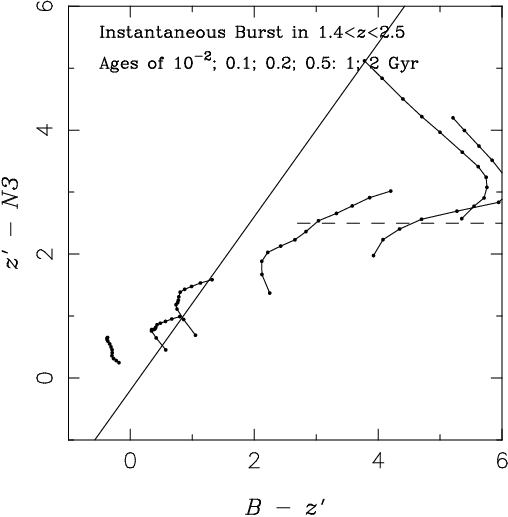}
\end{minipage}
\begin{minipage}{0.49\linewidth}
\includegraphics[angle=0,width=\linewidth]
{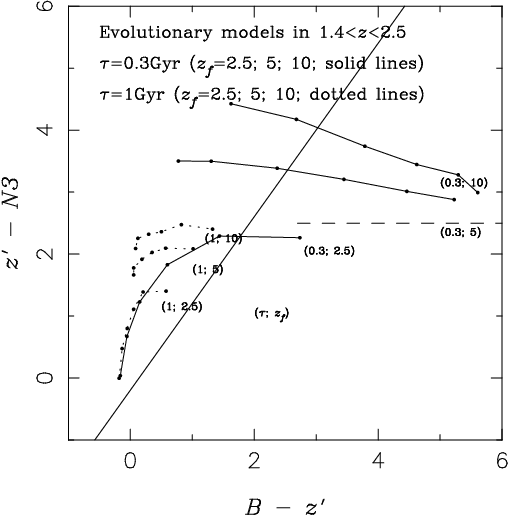}
\end{minipage}
\begin{minipage}{0.49\linewidth}
\includegraphics[angle=0,width=\linewidth]
{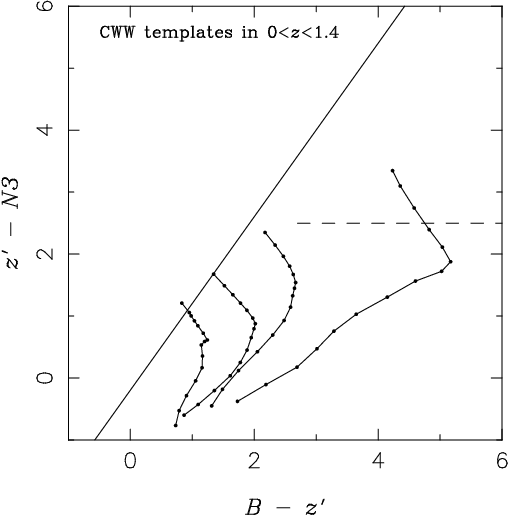}
\end{minipage}
\caption{The $BzN3$ diagram the same as figure~\ref{fig:urjmodel}
except the redshift range.  All models are plotted for the range
$1.4<z<2.5$, except in the bottom-right panel where plots are for
$0<z<1.4$.  The limits and color ranges of all four panels reproduce
those of figure~\ref{fig:bbg2_phot}, for direct reference.}
\label{fig:bzn3model}
\end{figure}  

\begin{figure}[ht]
\centering 
\begin{minipage}{0.49\linewidth}
\includegraphics[angle=0,width=\linewidth]
{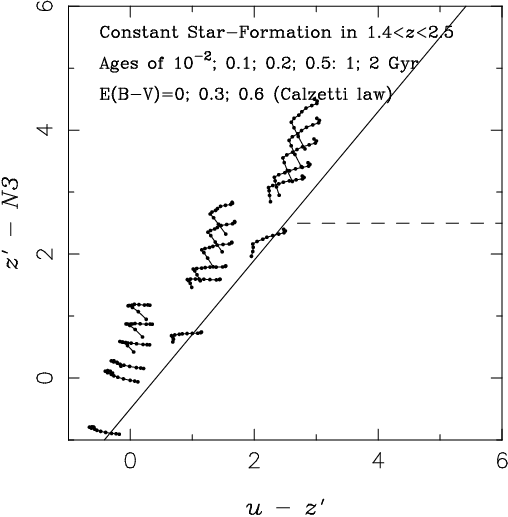}
\end{minipage}
\begin{minipage}{0.49\linewidth}
\includegraphics[angle=0,width=\linewidth]
{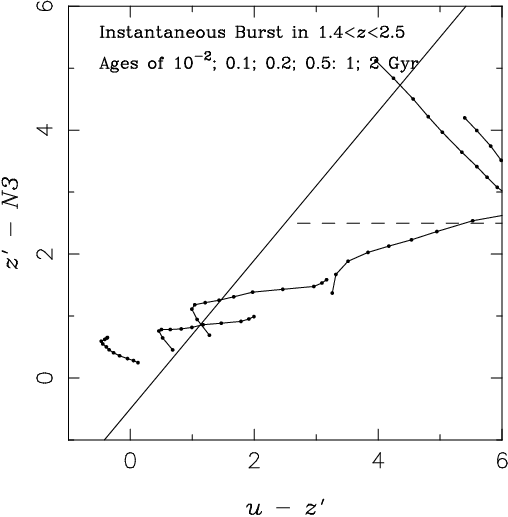}
\end{minipage}
\begin{minipage}{0.49\linewidth}
\includegraphics[angle=0,width=\linewidth]
{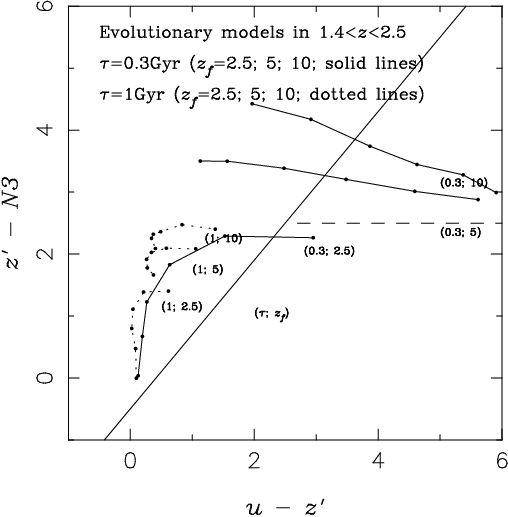}
\end{minipage}
\begin{minipage}{0.49\linewidth}
\includegraphics[angle=0,width=\linewidth]
{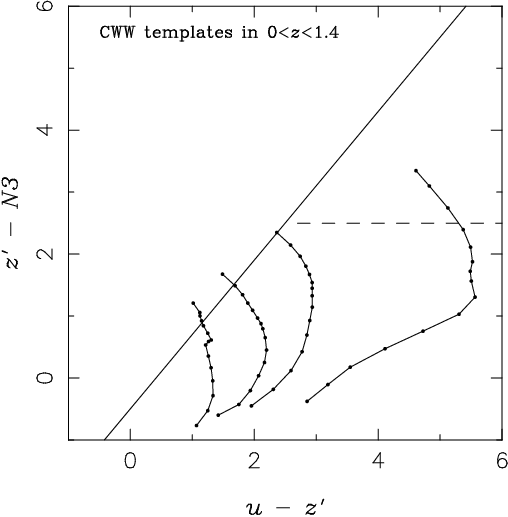}
\end{minipage}
\caption{The $uzN3$ diagram the same as figure~\ref{fig:bzn3model}
except the redshift range.  All models are plotted for the range
$1.4<z<2.5$, except in the bottom-right panel where plots are for
$0<z<1.4$.  The limits and color ranges of all four panels reproduce
those of figure~\ref{fig:bbg2_phot}, for direct reference.}
\label{fig:uzn3model}
\end{figure}  

\begin{figure}[ht]
\centering 
\begin{minipage}{0.49\linewidth}
\includegraphics[angle=0,width=\linewidth]
{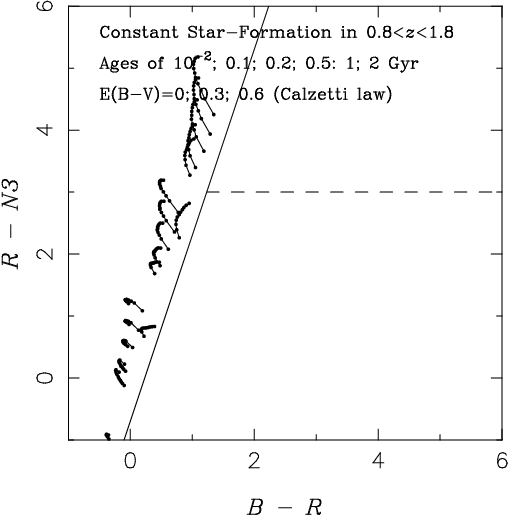}
\end{minipage}
\begin{minipage}{0.49\linewidth}
\includegraphics[angle=0,width=\linewidth]
{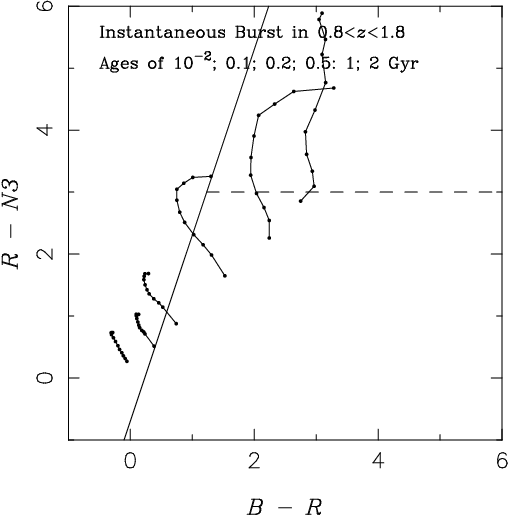}
\end{minipage}
\begin{minipage}{0.49\linewidth}
\includegraphics[angle=0,width=\linewidth]
{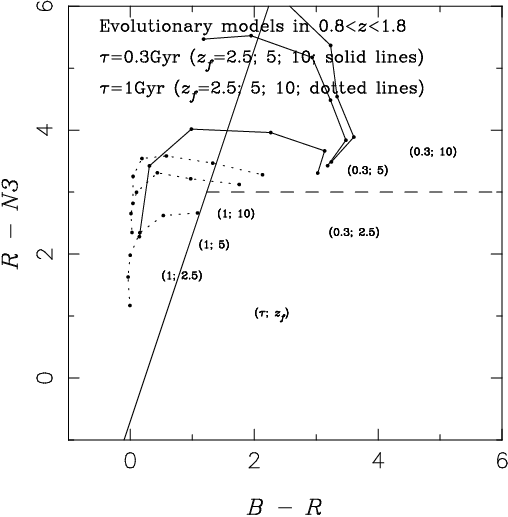}
\end{minipage}
\begin{minipage}{0.49\linewidth}
\includegraphics[angle=0,width=\linewidth]
{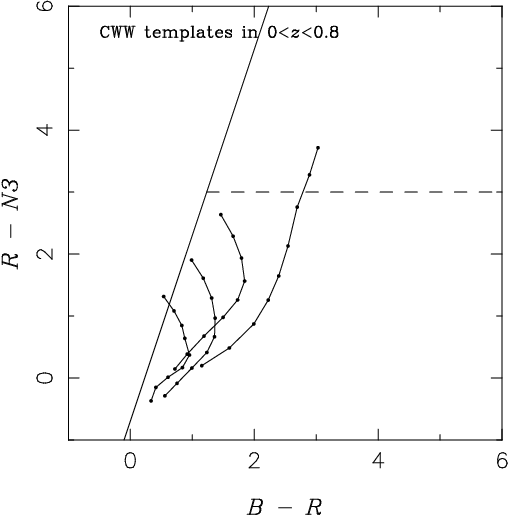}
\end{minipage}
\caption{The $BRN3$ diagram the same as figure~\ref{fig:urjmodel}
except the redshift range.  All models are plotted for the range
$0.8<z<1.8$, except in the bottom-right panel where plots are for
$0<z<0.8$.  The limits and color ranges of all four panels reproduce
those of figure~\ref{fig:bbg2_phot}, for direct reference.}
\label{fig:brn3model}
\end{figure}  

\begin{figure}[ht]
\centering 
\begin{minipage}{0.49\linewidth}
\includegraphics[angle=0,width=\linewidth]
{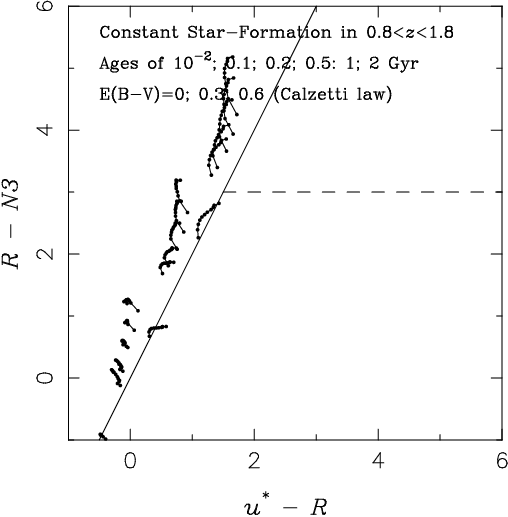}
\end{minipage}
\begin{minipage}{0.49\linewidth}
\includegraphics[angle=0,width=\linewidth]
{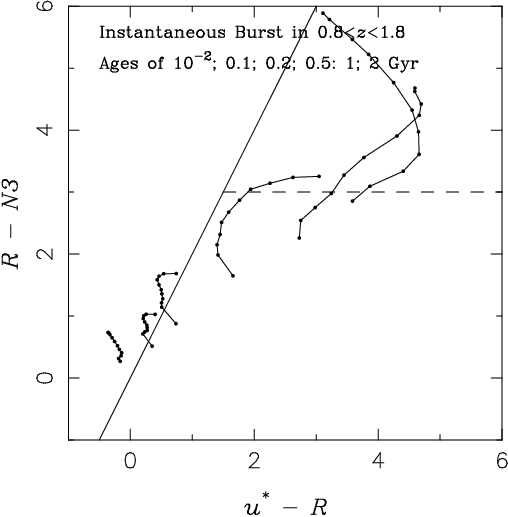}
\end{minipage}
\begin{minipage}{0.49\linewidth}
\includegraphics[angle=0,width=\linewidth]
{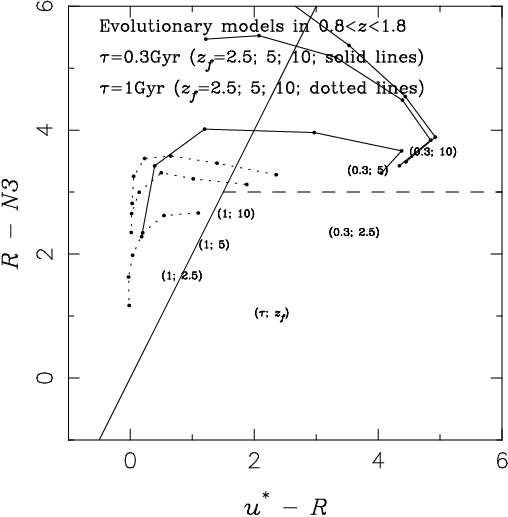}
\end{minipage}
\begin{minipage}{0.49\linewidth}
\includegraphics[angle=0,width=\linewidth]
{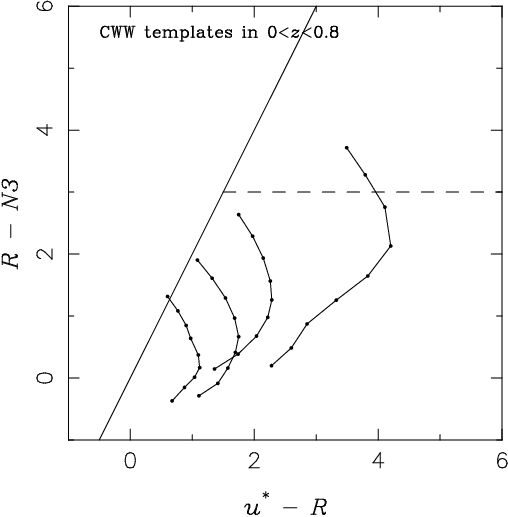}
\end{minipage}
\caption{The $uRN3$ diagram the same as figure~\ref{fig:bzn3model}
except the redshift range.  All models are plotted for the range
$0.8<z<1.8$, except in the bottom-right panel where plots are for
$0<z<0.8$.  The limits and color ranges of all four panels reproduce
those of figure~\ref{fig:bbg2_phot}, for direct reference.}
\label{fig:urn3model}
\end{figure}  

In order to determine the criteria for selecting the 
Balmer break galaxies as shown in appendix~\ref{app_subsec:bbg}, 
we studied the model tracks on two-color diagrams of $uRJ$, $uVi$, $BzN3$, 
$uzN3$, $BRN3$, and $uRN3$ as shown 
in figures~\ref{fig:urjmodel}, ~\ref{fig:uvimodel},
~\ref{fig:bzn3model}, ~\ref{fig:uzn3model}, ~\ref{fig:brn3model}, and 
~\ref{fig:urn3model}, respectively.   
 
Top-left panels of all the figures 
represent the tracks of constant star-formation (CSF) 
for ages of $10^{-2}$, $0.1, 0.2, 0.5, 1$,  and 2 Gyr, 
various reddening with Calzetti's extinction law, and solar metalicity.  
We can see that galaxies in such a redshift range with
ongoing star formation are indeed expected to lie in the left-upper
side region ${\cal B}{\mu}{\cal R} \geq f_{\scriptsize{{\cal
B}\mu{\cal R}}}$ of the two-color diagrams.  The duration of the
star-formation (age) has little influence on the bluer color ${\cal
B}-{\mu}$, while the redder color ${\mu}-{\cal R}$ increases with
age due to the development of strong Balmer/4000 \AA\ 
breaks falling in the ${\mu}$-band around these selecting redshift 
$z \simeq \langle z\rangle_{{\cal B}{\mu}{\cal R}}$.  
Very young bursts with ages less
than 10 Myr without underlying older stellar populations would be
located around the left-bottom region just below the line of ${\cal
B}{\mu}{\cal R}=f_{{\cal B}{\mu}{\cal R}}$.

We will note that 
the reddening direction is approximately parallel to the critical line
of criteria ${\cal B}{\mu}{\cal R} = f_{{\cal B}{\mu}{\cal R}}$ since
dust extinction is approximately represented as a power-law function
of the wavelength and it is implying that the reddening is similar 
in the ${\cal B}-{\mu}$ and ${\mu}-{\cal R}$ colors for
star-forming galaxies in the selected redshift range. It means that the
generalized criteria ${\cal B}{\mu}{\cal R} \geq f_{{\cal B}{\mu}{\cal
R}}$ for selecting BBGs are also robust from their dust reddening.
  
Top-right panels of all the figures represent the tracks of
simple stellar population (SSP) models 
after an instantaneous burst, in the two-color diagrams, for
ages of $10^{-2},0.1, 0.2, 0.5, 1$, and 2 Gyr, with no reddening, 
and solar metalicity. Their
features are also similar to those in the BzKs as the tracks are similar
to those of star-formings at young ages and begin 
to move to the region of p-BBGs at older ages of $\geq 1$Gyr. 
Even though some tracks for intermediate ages of $\sim0.5-1$ Gyr 
are outside of the s/p-BBG regions, the extinction tends 
to make them redder as those found in the s/p-BBG region.  Even though
we should be cautious when applying SSP models with no reddening to the
analysis for real galaxies, the results support that the generalized
schemes for selecting s/p-BBGs might still be robust.

The left-bottom panels of all the figures show that the ${\cal
B}{\mu}{\cal R}$ colors for galaxies with various formation redshifts
are exponentially declining SFRs ($\tau=0.3$ and 1 Gyr), with no
reddening and solar metalicity. The color evolution of galaxies
formed at high redshifts is such that most objects move directly from the
star-forming galaxy region to the passive galaxy region without 
crossing the bluer regions populated by $z<1.4$ objects.

The right-bottom panel of all the figures shows that the colors of 
normal galaxies are derived
from the CWW templates of E-Sbc-Scd-Irregular galaxies
\citep{coleman_colors_1980}, which fall outside the color area of
the BBGs for $z>0.8$($z>0.4$), respectively.  It means that the
generalized two-color criteria can work to exclude the lower-z
interlopers.

\subsection{Model Tracks for dusty starbursts and AGNs}
\label{app_subsec:track_mbn3r}

\begin{figure}[ht]
\centering 
\begin{minipage}{0.49\linewidth}
\includegraphics[angle=0,width=\linewidth]
{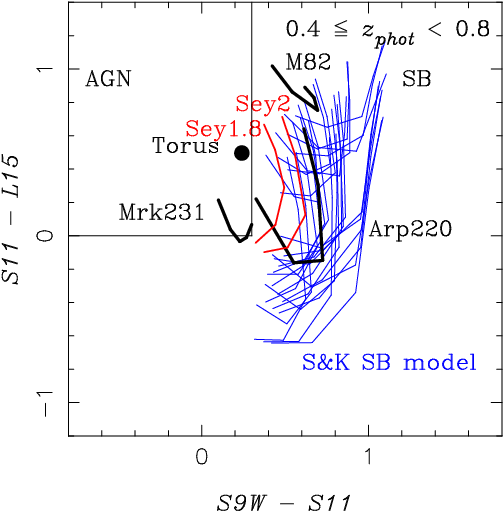}
\end{minipage}
\begin{minipage}{0.49\linewidth}
\includegraphics[angle=0,width=\linewidth]
{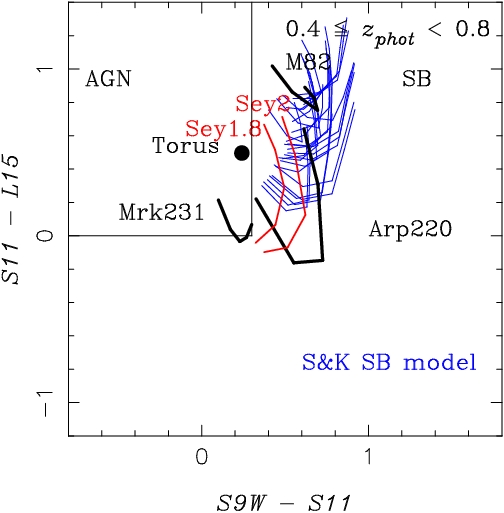}
\end{minipage}
\begin{minipage}{0.49\linewidth}
\includegraphics[angle=0,width=\linewidth]
{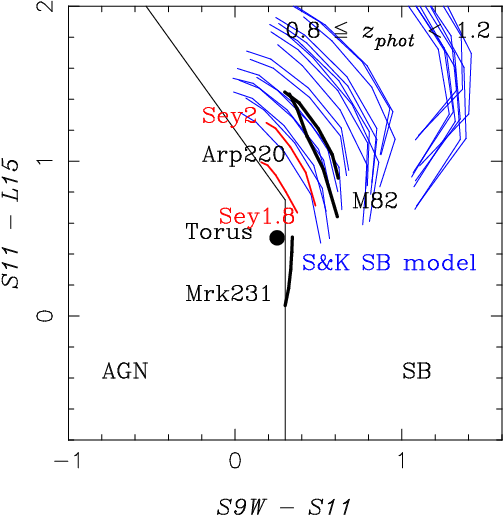}
\end{minipage}
\begin{minipage}{0.49\linewidth}
\includegraphics[angle=0,width=\linewidth]
{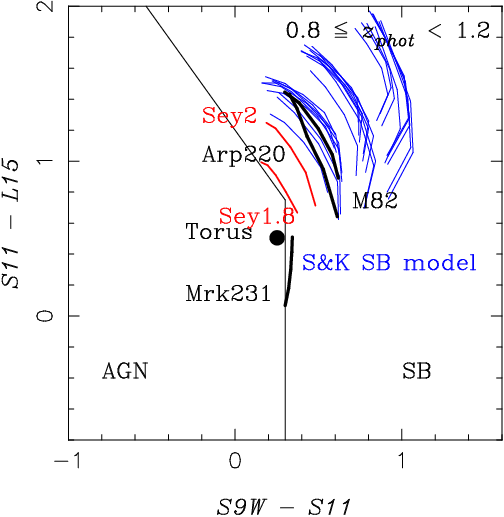}
\end{minipage}
\begin{minipage}{0.49\linewidth}
\includegraphics[angle=0,width=\linewidth]
{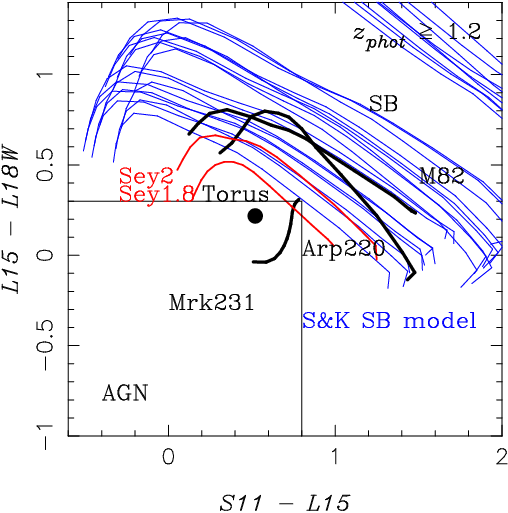}
\end{minipage}
\begin{minipage}{0.49\linewidth}
\includegraphics[angle=0,width=\linewidth]
{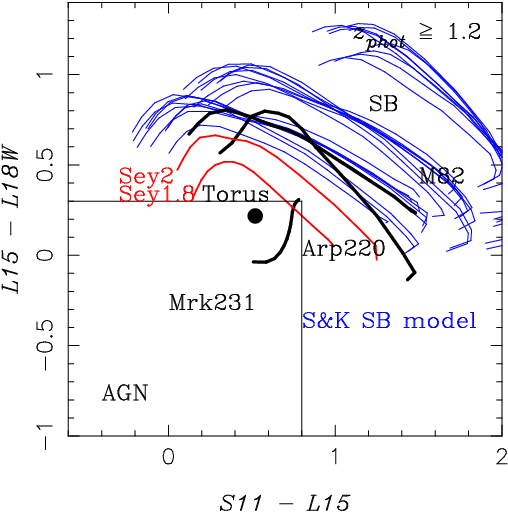}
\end{minipage}
\caption{The MIR color-color diagrams for the model SEDs of LIRGs.  
Thin blue and solid black lines 
represent tracks for dusty starburst 
models, and observed (U)LIRGs: M82, Arp 220, and Mrk231.  
Blacked filled circle represents the dusty torus model of QSO2 
in SWIRE SED templates.  Objects in the left regions
bounded with a thin solid line are excluded from starburst sample.  
Left: For the S\&K model SEDs with a hot spot parameter of  
$n^{hs}=10^2 (10^4)\mbox{cm}^{-3}$.  
Right: The same as on the Left for $n^{hs}=10^4\mbox{cm}^{-3}$.   
Top: The $S9W-S11$ vs. $S11-L15$ diagram with the model 
tracks in the interval $0.4 <z <1.0$.  
Middle: The same as the Top-Left, with the tracks in $0.8 < z<1.4$.  
Bottom: The $S11-L15$ vs. $L15-L18W$ diagram with the tracks in $1.0<z<2.8$.}
\label{fig:col_col_mir_model}
\end{figure}  

In figure~\ref{fig:col_col_mir_model}, the thin blue
curves, thick solid black curves, and black filled 
circles represent the
redshifted tracks of dusty starburst SEDs of models by
\citet{siebenmorgen_dust_2007}, SEDs of Arp220, M82 and Mrk231, 
and a model torus SED used to fit the SED of a heavily obscured type 2 QSOs 
SWIRE J104409.95+585224.8 (Polletta et al. 2006), respectively.  As shown 
in figure~\ref{fig:col_col_mir_model},   
all the dusty starbursts of S\&K models, Arp220, and M82 
appear in the right area 
while the AGNs,  including Mrk231 as an obscured AGN, 
appear in the left area.  Thus, 
dusty starburst dominant LIRGs can be distinguished from AGN dominant LIRGs 
on the MIR color diagrams with a boundary represented as
solid lines, which are introduced with equation~(\ref{eq:lirg1}) 
at $z=0.4-1.0$, equation~(\ref{eq:lirg2}) at $z=0.8-1.4$,  
and equation~(\ref{eq:lirg3}) at $1.0<z<2.8$ in section~\ref{sec:pah}.  

\subsection{IR SED models}
\label{app_subsec:ir_sed_models}

\begin{figure}[ht]
\begin{minipage}{0.49\linewidth}
\includegraphics[angle=0,width=\linewidth]
{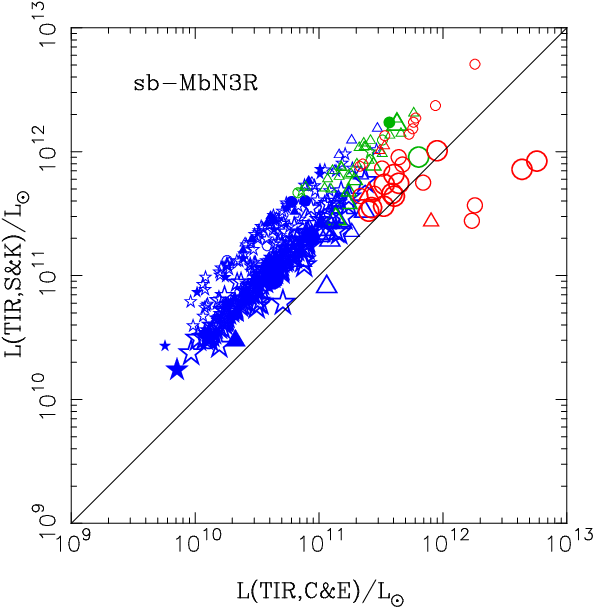} 
\end{minipage}
\begin{minipage}{0.49\linewidth}
\includegraphics[angle=0,width=\linewidth]
{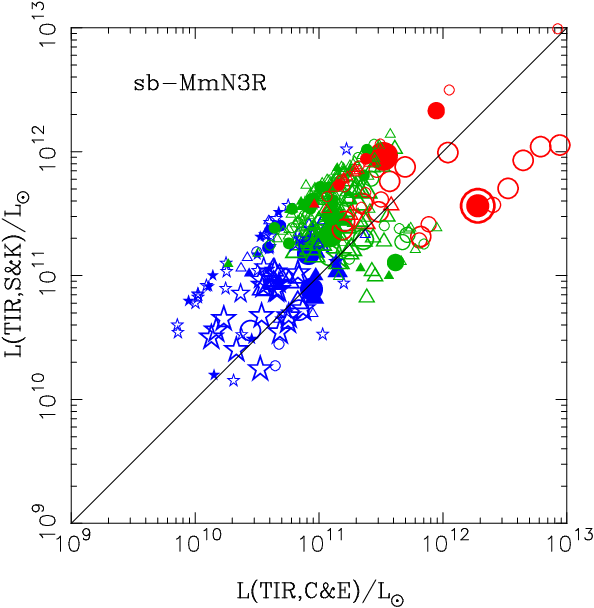} 
\end{minipage}
\caption{
Comparison between TIR luminosities $L_{IR}$
estimated from the C\&E model and the S\&K model for the sb-MbN3Rs and 
the sb-MmN3Rs. 
Symbols are the same as in figure~\ref{fig:col_col_mir_mbn3r} except their 
colors.  Blue, green, and red symbols represent the redshifts of $0.4 \le
z <0.8$, $0.8 \le z < 1.2$, and $z \ge 1.2$, respectively. 
Large, medium, and small symbols indicate the results which fit well 
with the S\&K model SEDs of $A_{V;SK}=2.2, 6.7$, and
$17.9$, respectively.
Left: For the sb-MbN3Rs.  
Right: For the sb-MmN3Rs.   
}
\label{fig:tir_sk_ce}
\end{figure}  

With integrating a fitted model SED in rest-frame 8-1000 $\mu$m wavelength 
and correcting for the photometric redshift $z_{phot}$,  
we can obtain the total IR (TIR) luminosity $L_{IR}$ of the emission 
from the dusty star forming regions.  There are various IR SED models.  
IR SED libraries can be prepared with three different 
kinds of schemes: 1) using SED
templates of known galaxies \citep{laurent_mid-infrared_2000,
lutz_iso_2003, spoon_mid-infrared_2004}, 2) reproducing the SED with
optical thin emission from the dusts assumed to be heated in a given
radiation field without radiative transfer calculations
\citep{chary_interpretingcosmic_2001, dale_infrared_2002,
lagache_modelling_2003}, and 3) calculating radiative transfer to evaluate
the emission reprocessed with dusty medium in a spheroidal
shape~\citep{silva_modelingeffects_1998, efstathiou_massive_2000,
takagi_evolutionary_2003, dopita_modelingpan-spectral_2006,
piovan_modelling_2006, piovan_modelling_2006-1,
siebenmorgen_dust_2007}, or a general three dimensional
shape~\citep{kylafis_dust_1987, popescu_modellingspectral_2000},  
using ray tracing or Monte Carlo
techniques \citep{bianchi_monte_2000}.  
Even in scheme 3), the interaction between
dust and radiation is treated consistently in
\citet{takagi_evolutionary_2003} and \citet{siebenmorgen_dust_2007}.
Even though empirical conversions from a monochromatic IR luminosity
have been sometimes used, their empirical laws have been frequently
derived from the SED fitting \citep{bavouzet_estimatingtotal_2008}.
When we choose to use them, we should take into account their
conceptional differences and applicable situations along with their
limitations.  

In order to extract information related to the 
physical properties in a dusty star-forming nucleus such as optical depth and 
dust density from the MIR SEDs with PAH emissions and Si absorption, 
in this paper, we have taken the S\&K model.  This can be used to prepare 
prepare MIR SEDs not only with various PAH emssion strengths but also 
Si absorption depth mainly parametrized with radiation field strength and 
optical depth, respectively.  
Individual SED features in the S\&K model depend on the following parameters; 
total luminosity $L^{tot}$ L$_\odot$, size of starbursting nucleus $R$ kpc
visual extinction $A_{V;SK}$, ratio of OB star luminosity with hot spots 
to the total luminosity $L_{OB}/L^{tot}$, and 
hydrogen number density in the hot spots $n^{hs}$.  
Even though the S\&K model depends on the luminosity $L^{tot}$, we have scaled 
luminosities and used only their features in the MIR SED fittings.  
The parameters $L^{tot}$, $R$, and $A_{V;SK}$ have some degeneracy for 
variations of the MIR SED features.  
As shown in the comparison between 
figures~\ref{fig:col_col_mir_model}, ~\ref{fig:col_col_mir_mbn3r}, 
and  
~\ref{fig:col_col_mir_ppmcn3r}, 
the models with a hot spot parameter $n^{hs}=10^2 \mbox{cm}^{-3}$
can cover typical MIR colors of MbN3Rs and MmN3Rs 
while the model with $n^{hs}=10^4 \mbox{cm}^{-3}$ cannot.   
Thus, we have fixed the hot spot parameter with 
 $n^{hs}=10^2 \mbox{cm}^{-3}$.  
And, we have used its 27 models fixed with $R=3$kpc and parametrized 
with $L^{tot}=10^{10.1}, 10^{11.1}$, and $10^{12.1}$L$_\odot$, 
$A_{V;SK}=2.2, 6.7$, and $17.9$, and $L_{OB}/L^{tot}=0.40, 0.60$, 
and $0.90$, 
which are treated as only qualitative parameters to reproduce 
varieties in the MIR SEDs, not as quantitative parameters describing 
an exact physical state in the starburst nucleus.      

In fact, the conversion from $\nu L_{\nu 7.7}$ 
to $L_{IR}$ depends on the IR SED models used in the fitting.  
For an example, as shown
in figure~\ref{fig:tir_sk_ce}, for most of the sb-MbN3Rs and MmN3Rs, 
we can see that the estimated $L_{IR;SK}$ from the SED fitting 
with the S\&K model is approximately larger by a factor of three  
than those with the Chary \& Elbaz
model (hereafter C\&E model).  The IR emission at $>10~\mu$m in most
of the S\&K model SEDs dominate more than those in the C\&E model,
which causes the TIR luminosity $L_{IR;SK}$ estimated from the S\&K model
to be larger than the $L_{IR;CE}$ from the C\&E model.
The S\&K model tends to derive larger TIR luminosity ratio of 
$L_{IR;SK}/\nu L_{\nu 7.7}$ when SEDs are fitted well with larger $A_{V;SK}$ 
even for samples with the same $\nu L_{\nu 7.7}$.  On the other hand, 
the C\&E model derives almost the same TIR luminosity ratio 
$L_{IR;CE}/\nu L_{\nu 7.7}$ for them.  Thus, 
the difference among the three model parameters 
for $A_{V;SK}$ appears as the split multi-sequences 
in figure~\ref{fig:tir_sk_ce}. 
Figure~\ref{fig:tir_sk_ce} shows multi-sequences with different
ratios $L_{IR;SK}/L_{IR;CE}$ and $A_{V;SK}$ 
more clearly than figure~\ref{fig:lum77_tir}.  

In order to reproduce the dependence of the optical depth to derive 
TIR luminosity in the paper, we have used the templates in the public library by
\citet{siebenmorgen_dust_2007} while the analysis with
\citet{takagi_evolutionary_2003}, for MIR detected objects in the
field, are also reported in other papers
\citep{takagi_multi-wavelength_2007, takagi_polycyclic_2010}.

\section{Error estimations}
\label{sec:error}  

With photometric redshifts, we have derived physical values for the 
detected galaxies as absolute magnitudes $M$, 
monochromatic luminosities $\nu L_{\nu}$, 
stellar mass $M_{\ast}$, and SFR.  
The physical values include errors from 
not only their photometric fluctuation $\Delta m$ in the observed band  
but also uncertainty of their photometric redshift.  
In their error estimations, we have assumed that 
the former and the latter are independent of each other.  

Absolute magnitudes $M$ is defined with luminosity $L_{\nu}$, observed  
magnitude $m_{\nu}$ and flux $F_{\nu}$  at the observed frequency $\nu$ as 
\begin{eqnarray} 
M & = & -2.5 \log_{10} \frac{L_{\nu}}{4\pi d_L(10 \mbox{pc})}  -48.6 \; , \\
L_{\nu} & = & 4\pi d_L(z)^2  \frac{F_{\nu}}{(1+z)} \; , \\
F_{\nu} & = & 10^{-0.4(m_{\nu}+48.6)} \; , 
\end{eqnarray} 
where $d_L$ is the luminosity distance.  
Thus, since  
$
M = -2.5 \log_{10} F_{\nu} -2.5  \log_{10} \frac{d_L(z)^2}{(1+z)} + \mbox{const} 
$ with $f(z)=d_L(z)^2/(1+z)$, 
we estimate the error of absolute magnitudes $M$ as 
\begin{equation} 
\Delta M = \sqrt{ \Delta m_{\nu}^2 + 
\left( 1.09 \frac{\Delta f(z)}{f(z)} \right)^2} \; , 
\end{equation} 
where we used $ d (\log_{10} x) = (\log_e 10)^{-1} dx/x = 0.434 dx/x$ and 
$\Delta m_{\nu}= 1.09 \Delta F_{\nu}/F_{\nu}$.  

For the monochromatic luminosities $\nu L_{\nu}$ and the 
stellar mass $M_{\ast}$, with the similar schemes, 
their errors can be derived as 
\begin{eqnarray} 
\Delta (\log_{10} \nu L_{\nu}) & = & 0.4 
\sqrt{ \Delta m_{\nu}^2 + 
\left( 1.09 \frac{\Delta f(z)}{f(z)} \right)^2} \; , \\
\Delta (\log_{10} M_{\ast} ) & = & 0.4 
\sqrt{ \Delta m_{K}^2 + 
\left( 1.09 \frac{\Delta f(z)}{f(z)} \right)^2} \; ,  
\end{eqnarray} 
where we used the error $\Delta m_{K}$ in $K$ band for the stellar mass error 
estimation.   

In the error estimations for $SFR_{UV}$ and $SFR_{IR}$, 
we have used errors $\Delta m_{\nu}$ 
in the $u$ band and the IRC band related to 
the rest-frame 7.7 $\mu$m, respectively.   

\section{Stellar mass from Optical/NIR SEDs}
\label{app_sec:smass}  

\begin{figure}[ht]
\begin{minipage}{0.99\linewidth}
\includegraphics[angle=0,width=\linewidth]
{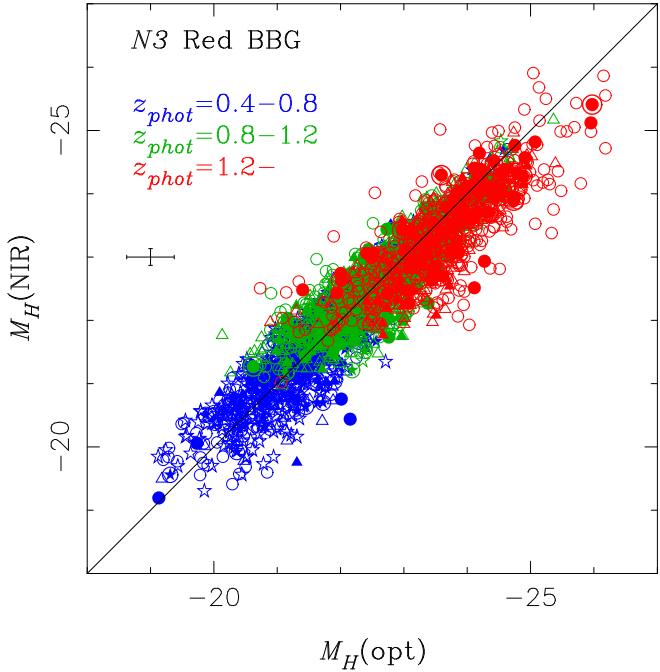}
\end{minipage}
\caption{Comparison of the rest-frame $H$-band magnitudes for the N3 Red BBGs 
between $M_{H;NIR}$ and $M_{H;opt}$, which are derived from the AKARI NIR 
photometry and the ground-based optical-NIR SED fitting,
respectively.  Symbols are the same as in figure~\ref{fig:col_col_mir_mbn3r} 
except their colors.  
Blue, green, and red symbols represent the redshifts of $0.4 \le
z <0.8$, $0.8 \le z < 1.2$, and $z \ge 1.2$, respectively.   
}
\label{fig:n234_rest_h}
\end{figure}  

\begin{figure}[ht]
\begin{minipage}{0.49\linewidth}
\includegraphics[angle=0,width=\linewidth]
{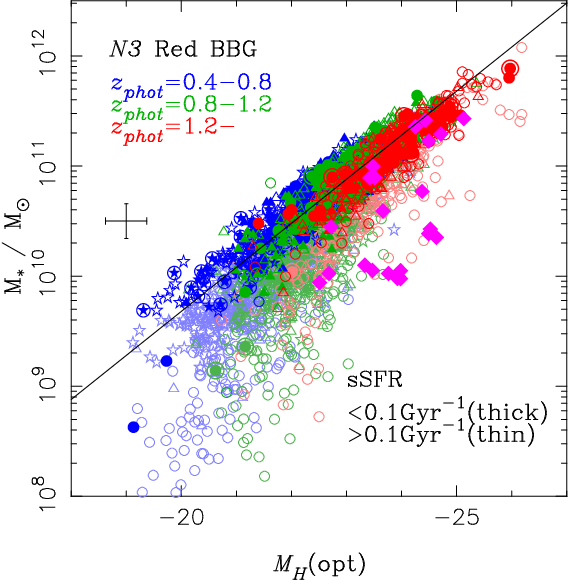}
\end{minipage}
\begin{minipage}{0.49\linewidth}
\includegraphics[angle=0,width=\linewidth]
{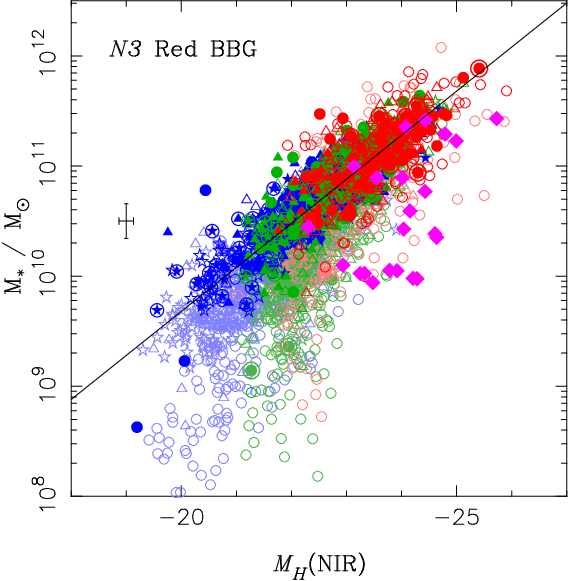}
\end{minipage}
\caption{The absolute rest-frame $H$-band magnitude $M_H$ vs. stellar
mass $M_{\ast}$ derived from the optical-NIR SED fitting, for all the
N3 Red BBGs.  
Symbols are the same as in figure~\ref{fig:n234_rest_h}.  
The dark and faint colored symbols
represents the subpopulation with sSFR $<0.1\mbox{Gyr}^{-1}$ and 
sSFR $>0.1\mbox{Gyr}^{-1}$.  
Left: The $M_{H;opt}$ reproduced from the BC03 model fitting vs. $M_{\ast}$.  
Right: The $M_{H;NIR}$ interpolated from the AKARI NIR photometry 
vs. $M_{\ast}$. 
 }
\label{fig:habs_smass_irbg}
\end{figure}  

\begin{figure}[ht]
\begin{minipage}{0.49\linewidth}
\includegraphics[angle=0,width=\linewidth]
{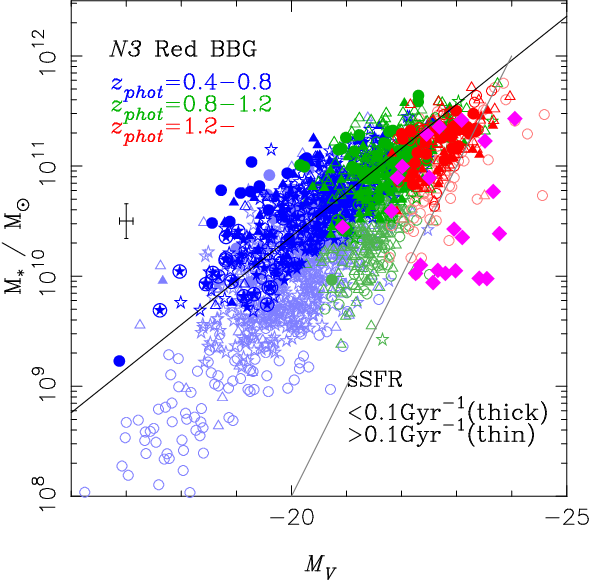}
\end{minipage}
\begin{minipage}{0.49\linewidth}
\includegraphics[angle=0,width=\linewidth]
{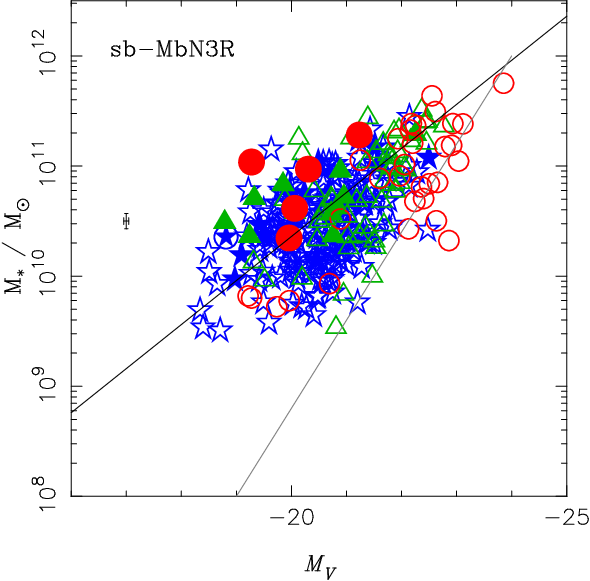}
\end{minipage}
\begin{minipage}{0.49\linewidth}
\includegraphics[angle=0,width=\linewidth]
{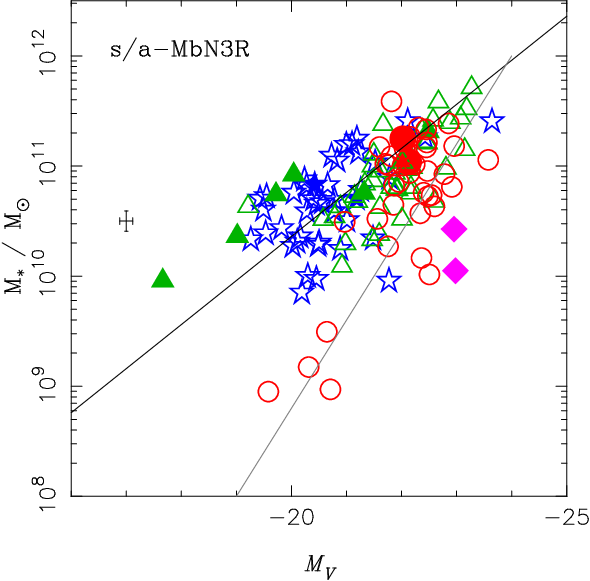}
\end{minipage}
\begin{minipage}{0.49\linewidth}
\includegraphics[angle=0,width=\linewidth]
{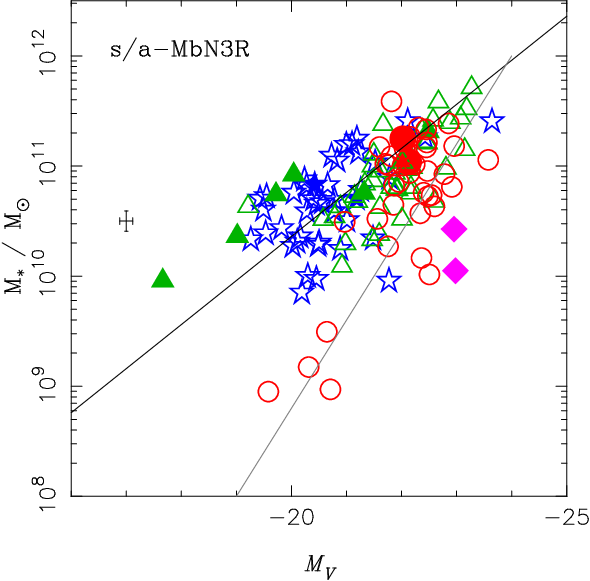}
\end{minipage}
\caption{The absolute rest-frame $V$-band magnitude $M_V$ vs. stellar
mass $M_{\ast}$ derived from the optical-NIR SED fitting.  
Symbols are the same as in figure~\ref{fig:col_col_mir_mbn3r} except their 
colors.  Blue, green, and red mean the redshifts of $0.4 \le
z <0.8$, $0.8 \le z < 1.2$, and $z \geqq 1.2$, respectively.    
Magenta diamond represents the spectroscopic BL AGN.  
Steep line at the right side represents 
the boundary of the EBOs.  Top-Left: All the N3 Red BBGs with 
$sSFR <0.1\mbox{Gyr}^{-1}$ and $>0.1\mbox{Gyr}^{-1})$ 
are represented with dark and thin symbols, respectively. Top-Right: 
For the sb-MbN3Rs.  Bottom-middle: For the s/a-MbN3Rs.  
Bottom-Right: For the agn-MbN3Rs.  
}
\label{fig:mv_smass_bbg_irbg_mbn3r}
\end{figure}  

The stellar mass $M_{\ast}$ is estimated in a standard way 
with the optical-NIR SED fitting for all the $z'$-detected galaxies and an
alternative way from the AKARI/NIR photometry for the IRBGs as a
subclass in the $z'$-detected galaxies.  

The rest-frame $H$ band wavelength corresponds to the rest-frame 
1.6-$\mu$m IR bump where the emission is dominated 
by the stellar components in
galaxies.  It means that the absolute magnitude in rest-frame $H$-band
$M_H$ represents the underlying stellar mass content $M_{\ast}$.  
This provided us with two ways to estimate the $M_H$ with our dataset; 
1) the direct estimation from the AKARI/IRC photometry in the rest-frame 
$H$-band at 1.6 $\mu$m,   
and 2) the SED fitting with the ground-based photometric
data.

First, the $N2, N3$, and $N4$ photometry can directly detect the
1.6-$\mu$m bump from the AKARI/IRC detected IRBGs even at $z>0.4$. 
Using the interpolation scheme with equation~(\ref{eq:mag_intr}) 
for the $N2, N3$, and $N4$ photometry with
$\lambda_0 = 1.6~\mu$m and $z = z_{phot}$, we can estimate the absolute
magnitude $M_{H;NIR}(z_{phot})$ corresponding to the rest-frame
$H$-band directly from the $N2,N3, \mbox{and} N4$ magnitudes of the
AKARI/IRC detected IRBGs.  

Second, the model fitting of \citet{bruzual_stellar_2003}, with the
whole ground-based $u^{\ast}BV$$Ri'z'$$JK_s$ SED, derives 
the absolute magnitude $M_{H;opt}$ with our small modified 
version of the {\it hyperz} as instantaneously deriving 
the photometric redshift $z_{phot}$ in subsection~\ref{subsec:photz}.  
It can apply for all the $z'$-detected galaxies.  Thus, the derived 
$M_{H;opt}(z_{phot})$ from the optical-NIR SED fitting with the 
ground-based photometric data is an alternative to determine the stellar 
mass $M_*$.

The absolute magnitude $M_{H;opt}(z_{phot})$ from the optical-NIR SED fitting 
is well correlated with that $M_{H;NIR}(z_{phot})$ from the $N2,N3$, and  
$N4$ photometry as shown in figure~\ref{fig:n234_rest_h}.  Note this
rest-frame $M_{H;NIR}(z_{phot})$ from the AKARI/NIR data can be
approximately independent to the $M_{H;opt}(z_{phot})$ from the
ground-based optical/NIR data, except for the photometric redshift
$z_{phot}$ derived from the optical dataset.
Figure~\ref{fig:n234_rest_h} presents not only the consistency
between the ground-based NIR and the AKARI/NIR observations, but also
accuracy for the photometric redshift from the SED fitting only with
the ground-based data.  

Weak age dependence between $M_{\ast}$ and $M_{H}$ can be seen in    
figure~\ref{fig:habs_smass_irbg},   
in which the deep dark and faint symbols represent old and young 
populations in the red sequence and blue cloud), respectively, 
in the $(U-V)-\Delta_{UV}$ as discussed in 
subsection~\ref{subsec:smass_uvcol}.  
  
Even though the relation between an absolute magnitude $M$ in a band 
and the stellar mass $M_*$ depends on the Initial
Mass Function (IMF) and the star formation histories, they can be
approximately described as
\begin{equation}
{\rm log}\left( \frac{M_*}{10^{11} M_\odot} \right) 
= -0.4 (M-M^{11}) \; ,  
\label{eq:mag_smass}
\end{equation}
where $M^{11}$ is the rest-frame absolute magnitude corresponding to 
the stellar mass of $10^{11}$M$_{\odot}$. 
As shown in figure~\ref{fig:habs_smass_irbg},  
for the $M_H$ in the $H$ band, we can take $M^{11}=M_H^{11} =-23.6$.   
Even though we mainly used the stellar mass $M_{\ast}$ 
derived directly from the optical-NIR SED fittings in the following,  
the conversion from $M_H$ to $M_{\ast}$ is useful as a quick estimation of 
stellar mass $M_{\ast}$.  

We also checked the relation between absolute rest-frame $V$-band magnitude 
$M_V$ and stellar mass $M_{\ast}$ ($M_V$-$M_{\ast}$ diagram) derived from 
the optical-NIR SED fitting as shown in 
figure~\ref{fig:mv_smass_bbg_irbg_mbn3r}.  The EBOs and the BL AGNs appear in  
the right region in the $M_V$-$M_{\ast}$ diagram, where normal star forming 
galaxies can appear only at the early phase $<1$ Gyr.  Their results in 
$M_V$-$M_{\ast}$ is consistent with their one on the CMD in subsection 
~\ref{subsec:mv_uvcol} as the EBOs are candidates of AGNs 
(see also appendix~\ref{app_sec:model}).  
After excluding these EBOs, we obtained a relation 
between $M_V$ and $M_{\ast}$, approximated as equation~(\ref{eq:mag_smass}) 
with taking $M_V^{11} =-22$, 
as shown in figure~\ref{fig:mv_smass_bbg_irbg_mbn3r}.  
This conversion from $M_V$ to $M_{\ast}$ is useful in 
making mass-color diagrams 
in section~\ref{sec:evol_pop}.  

%
%

\end{document}